\PassOptionsToPackage{unicode,bookmarksnumbered}{hyperref}

\documentclass[11pt]{article} 
\usepackage[page, toc, title, titletoc]{appendix}

\usepackage[margin={0.8in}]{geometry}
\usepackage{rotating, lscape}
\usepackage{graphicx, color} 
\usepackage{chngcntr}
\usepackage{threeparttable, booktabs, longtable, threeparttablex, tabularx}
\usepackage{pdflscape} 
\usepackage[]{tocbibind} 
\usepackage{natbib}
\usepackage{authblk}
\usepackage{mathtools}
\usepackage{dsfont}
\usepackage{lmodern} 
\usepackage[sc]{mathpazo}
\usepackage{courier}
\usepackage[utf8]{inputenc}
\usepackage[T1]{fontenc}
\usepackage{setspace}
\usepackage{amssymb,amsmath,mathrsfs,amsthm, bbm}
\usepackage{microtype}
\usepackage{float}
\usepackage{comment}
\usepackage{multirow}
\usepackage{tikz}
\usepackage{soul}
\usetikzlibrary{patterns}
\usetikzlibrary{decorations,arrows}
\usetikzlibrary{decorations.pathmorphing}
\usepgflibrary{decorations.pathreplacing}
\usetikzlibrary{arrows,positioning}

\usepackage{xfrac}

\usepackage{subcaption}
\usepackage{alphalph} 

\usepackage[colorlinks = true,
            linkcolor = blue,
            urlcolor  = blue,
            citecolor = black,
            anchorcolor = black]{hyperref}

\usepackage[nameinlink, noabbrev]{cleveref} 

\usepackage{algorithm}
\usepackage{algpseudocodex}

\newcommand{\mortality}{\ensuremath{{Mortality}_{s, 1918}}}

\usepackage{xspace}
\newcommand{\HighNPI}{\ensuremath{High \, NPI}\xspace}
\newcommand{\NPISpeed}{\ensuremath{NPI \, Speed}\xspace}
\newcommand{\NPIIntensity}{\ensuremath{NPI \, Intensity}\xspace}

\newcommand{\RSq}{\ensuremath{R^2}\xspace}
\newcommand{\Oster}{Oster bound (2019)\xspace}

\newcommand{\CheckYes}{{\tiny\ensuremath{\blacksquare}}}

\newcommand{\CitiesCaption}{\caption{\textbf{Mortality and NPIs by city.} 
This figure shows two measures of excess daily death rates during the pandemic, as well as the range of dates where three different NPI measures were active.}}

\providecommand{\tightlist}{\setlength{\itemsep}{0pt}\setlength{\parskip}{0pt}}

\newcommand\minput[1]{%
  \input{#1}%
  \ifhmode\ifnum\lastnodetype=11 \unskip\fi\fi}

\graphicspath{{/JEH/revised_draft/}}

\title{ \vspace{1.5cm} Pandemics Depress the Economy, Public Health Interventions Do Not: Evidence from the 1918 Flu
 \vspace{.5cm}
}

\author{Sergio Correia, Stephan Luck, and Emil Verner\textsuperscript{*} \vspace{.5cm}} 
\date{May 2022}

\begin{document}
\maketitle

\begin{abstract}
We study the impact of non-pharmaceutical interventions (NPIs) on mortality and economic activity across U.S. cities during the 1918 Flu Pandemic. The combination of fast and stringent NPIs reduced peak mortality by 50\% and cumulative excess mortality by 24\% to 34\%. However, while the pandemic itself was associated with short-run economic disruptions, we find that  these disruptions were similar across cities with strict and lenient NPIs. NPIs also did not worsen medium-run economic outcomes. Our findings indicate that NPIs can reduce disease transmission without further depressing economic activity, a finding also reflected in discussions in contemporary newspapers.
 
\end{abstract}

\let\oldthefootnote\thefootnote
\renewcommand{\thefootnote}{\fnsymbol{footnote}}
\footnotetext[1]{Correia: Federal Reserve Board, \href{mailto:sergio.a.correia@frb.gov}{sergio.a.correia@frb.gov}; Luck: Federal Reserve Bank of New York, \href{mailto:stephan.luck@ny.frb.org}{stephan.luck@ny.frb.org}; Verner: MIT Sloan School of Management, \href{mailto:everner@mit.edu}{everner@mit.edu}. 

The authors thank seminar participants at the Federal Reserve Board, SAECEN, Virtual Finance and Economics Conference, VMACS, Banco Central de Chile, Banco Central de Reserva del Per\'{u}, Stanford GSB, University of Cologne, European Macro History, MIT, HELP Webinar, ASSA 2021 Annual Meeting, and the World Bank, as well as Natalie Cox, Andrew Bossie, Casper Worm Hansen, Eric Hilt, Simon Jaeger, Aart Kraay (discussant), Sam Langfield, Atif Mian, Kris Mitchener, Karsten M\"{u}ller, Alex Navarro, Michala Riis-Vestergaard, Ole Risager, Hugh Rockoff, Paul Schempp, Francois Velde, and Dorte Verner for valuable comments.  Outstanding research assistance was provided by Fanwen Zhu, and we thank Hayley Mink for her assistance in researching historical newspaper articles. We thank Alex Navarro and Howard Markel for sharing their city-level dated list of NPIs, as well as Casper Worm Hansen for sharing city-level public spending data.

The opinions expressed in this paper do not necessarily reflect those of the Federal Reserve Bank of New York or the Federal Reserve Board.}

\let\thefootnote\oldthefootnote

\doublespacing


\clearpage
\pagenumbering{arabic}

\clearpage
\section{Introduction}


Can non-pharmaceutical interventions (NPIs), such as social distancing, reduce mortality during a pandemic? If so, do NPIs benefit public health at the expense of the economy, or can NPIs intended to contain the spread of a pandemic also reduce its economic severity?

The 1918 Flu Pandemic was the   most severe influenza pandemic in U.S. history and killed 675,000 people in the U.S., or about 0.66 percent of the population. Most deaths occurred during the second wave in the fall of 1918. In response, major U.S. cities implemented a range of NPIs. These included school, theater, and church closures, public gathering bans, quarantine of suspected cases, and restricted business hours. In this paper, we study the impact of these NPIs on both mortality and economic activity.

Our analysis proceeds in three steps. First, we study contemporary newspaper accounts of the pandemic to understand what drove the variation in speed and stringency of NPI implementations across U.S. cities. We find that an important driver of the NPIs were  differences in the information available to local policymakers, which in turn is related to the geography of cities. The flu moved from east to west and thus cities further west had more time to prepare for the arrival of the virus. Our analysis further reveals that NPIs were heavily debated by public health officials, business owners, and other local actors. Beyond the geographic factors, the implementation of NPIs hence often came down to local policy preferences and political economy factors. 

Second, we empirically evaluate the effects of NPIs on mortality. We use data on the timing and intensity of NPIs for 46 cities based on \citet{Markel2007}, augmented with information from \citet{Berkes2022} and new hand-collected data. We find that  NPIs were most successful in reducing mortality if they were both  implemented sufficiently quickly after the arrival of the virus and upheld for long enough. Cities that were both fast and aggressive in implementing NPIs achieved reductions in peak influenza and pneumonia mortality of about 50\%, thereby \emph{flattening the mortality curve}.  Moreover, these cities experienced a reduction in cumulative excess influenza and pneumonia mortality of up to 34\%. We also show that NPIs not only reduced mortality due to influenza and pneumonia but also all-cause mortality. Our findings suggest that NPIs were successful in slowing the rate of disease transmission and, to a lesser extent, lowering cumulative mortality, potentially by mitigating epidemic overshoot \citep{Bootsmaa2007}.


Third, we ask: is there a trade-off between NPIs that reduced mortality and economic activity? In theory, the economic effects of NPIs could be either positive or negative. All else equal, NPIs constrain social interactions and thus economic activity that relies on such interactions. However, economic activity in a pandemic is also reduced in the absence of such measures, as households reduce consumption and labor supply to lower the risk of becoming infected, and firms cut investment in response to increased uncertainty. While the direct effect of NPIs is to lower economic activity, they also mitigate the impact of the original shock: the pandemic itself. By containing the pandemic, NPIs can thus also mitigate the pandemic-related economic disruptions such as the contraction in labor supply from voluntary distancing and illness. 

To study the short-term impact of NPIs on local economic activity, we construct a city-level index of business disruptions at a monthly frequency from information in \textit{Bradstreet's}, a contemporary trade journal. Our index implies that the pandemic itself was associated with an increase in business disruptions in the fall of 1918. However, comparing cities with strict and lenient NPIs, we find that the increase in business disruptions was quantitatively similar across the two sets of cities. 



Further, we examine the economic impact of NPIs in the medium run using data on city-level employment and output from the Census of Manufactures. We find no evidence that cities that intervened earlier and more aggressively perform worse in the years after the pandemic.  At a minimum, our estimates reject that cities with stricter NPIs experienced a large decline in employment and output in the years following the pandemic, relative to cities with lenient NPIs. Altogether, our findings suggest that, while the pandemic was associated with  economic disruptions, NPIs reduced disease transmission without exacerbating the pandemic-induced downturn.

Our empirical findings are subject to the concern that policy responses are endogenous and could be correlated with shocks to mortality or economic activity. To allay this concern, we show that the results are robust to controlling for a range of potential confounders including  longitude, the timing of the flu's arrival, city demographics, density, manufacturing employment-to-population ratio, exposure to WWI mortality and production, poverty, and air pollution. 


We further support our evidence on the negative short-term impact of the pandemic on the economy by accounts in contemporary newspapers. Newspaper articles suggest that there were significant declines in output and sales across a wide range of industries due to labor shortages. Moreover, there was significant voluntary distancing due to fear of the virus. Attendance at theaters, cafes, and places of public amusement declined, while absenteeism from schools and workplaces increased markedly. These accounts help understand the limited adverse economic effects of NPIs. Demand and labor supply were depressed before NPIs were implemented and in places with less stringent NPIs, indicating that the worst economic disruptions were caused by the pandemic itself. 

Notably, our review of newspaper articles reveals that some contemporaries appreciated the absence of a trade-off between moderate NPIs and the economy. When studying debates about the costs and benefits of NPIs, we find that some businesses that were directly affected by closures opposed these measures. However, we also discuss examples of business owners who supported stricter NPIs because they believed these policies would mitigate the most disruptive effects of the pandemic on the economy. 


We emphasize caution when generalizing these results to the  COVID-19 pandemic. The 1918 Flu Pandemic was significantly deadlier than COVID-19, especially for working-age individuals, and effective vaccines were not available. Thus, the economic merits of NPIs may have been greater in 1918. NPIs implemented in 1918 were also less extensive than those used during the COVID-19 outbreak. Moreover, the structure of the U.S. economy and society has evolved substantially over a century. Nevertheless, our results suggest that it is not a foregone conclusion that there is a trade-off between reducing disease transmission and stabilizing economic activity in a pandemic.

The paper proceeds as follows. We first briefly discuss the related literature. In \cref{sec:background} we describe the historical background of the 1918 flu pandemic and NPIs and discuss the sources of variation in NPIs across U.S. cities. \Cref{sec:data} describes our data. \Cref{sec:mortality} and \cref{sec:economy} present our results on the effect of NPIs on mortality and economic activity, respectively. We also discuss newspaper accounts of how the pandemic and NPIs affected the economy and broader society. Finally, \cref{sec:conclusion} concludes.

\subparagraph{Related Literature}
This study is most closely related to the literature studying the impact of NPIs in the 1918 flu pandemic, to which we make several contributions. First, we expand the coverage of NPI measures to a larger sample of cities with information from \cite{Berkes2022} and with newly collected data. Second, we provide new insights into the effect of NPIs on mortality. Our findings on mortality are consistent with evidence from the epidemiology literature \citep{Markel2007,Hatchett2007}, but point to larger and more robust reductions in cumulative mortality than found in other work \citep{Clay1918,BarroNPI2020}. An important new insight from our analysis is that the most robust reductions in mortality occurred in cities that were both timely and aggressive in implementing NPIs.
Our results also extend to all-cause excess mortality, which we also analyze as it is a potentially less biased alternative to mortality explicitly attributed to influenza and pneumonia. Third, while the existing literature on NPIs during the 1918 flu pandemic has primarily focused on mortality outcomes, our paper also explores the economic effects of NPIs. An exception is \citet{Velde2020}, who finds that NPIs reduced mortality at limited economic cost.\footnote{\citet{Velde2020} focuses on the impact of business closings, while we analyze all NPIs considered in \cite{Markel2007}. \citet{Velde2020} also examines the effect on trade conditions from \textit{Bradstreet's}, while we also analyze the impact on manufacturing activity in the medium run.} Finally, we provide extensive narrative evidence on what factors drove the implementation of NPIs at the city level and how their impact on mortality, the economy, and the broader society was perceived by contemporaries.

More broadly, our paper is related to research on the economic impact of the 1918 Flu Pandemic and other pandemics. \citet{Velde2020} presents a comprehensive account of the economic impact of the 1918 Flu Pandemic in the U.S.\ and documents that it was associated with a short and moderate recession in the aggregate. \citet{Garrett2009} finds that geographic areas with higher influenza mortality saw a relative increase in wages between the 1914 and 1919 manufacturing census years, consistent with labor shortages.  \citet{Dahl2022} find that the 1918 pandemic resulted in a V-shaped recession in Denmark. Using regional data from Sweden, \citet{Karlsson2014} find that the 1918 pandemic led to a persistent increase in poverty rates and a reduction in the return on capital. \citet{Barro2020} uses country-level data and find that the 1918 Flu Pandemic lowered real GDP by 6-8 percent in the typical country. Using more disaggregated variation, \citet{Guimbeau2019} find negative effects of the 1918 flu on long-term health and productivity in Brazil, and \citet{Almond2006} finds that cohorts in utero during the pandemic displayed worse education and labor market outcomes in adulthood.


    
\section{Non-pharmaceutical interventions during the 1918 flu pandemic}

\label{sec:background}

\subsection{Historical background on the 1918 flu pandemic}

\begin{figure}[htbp]
    \centering
    \includegraphics[width=0.8\textwidth]{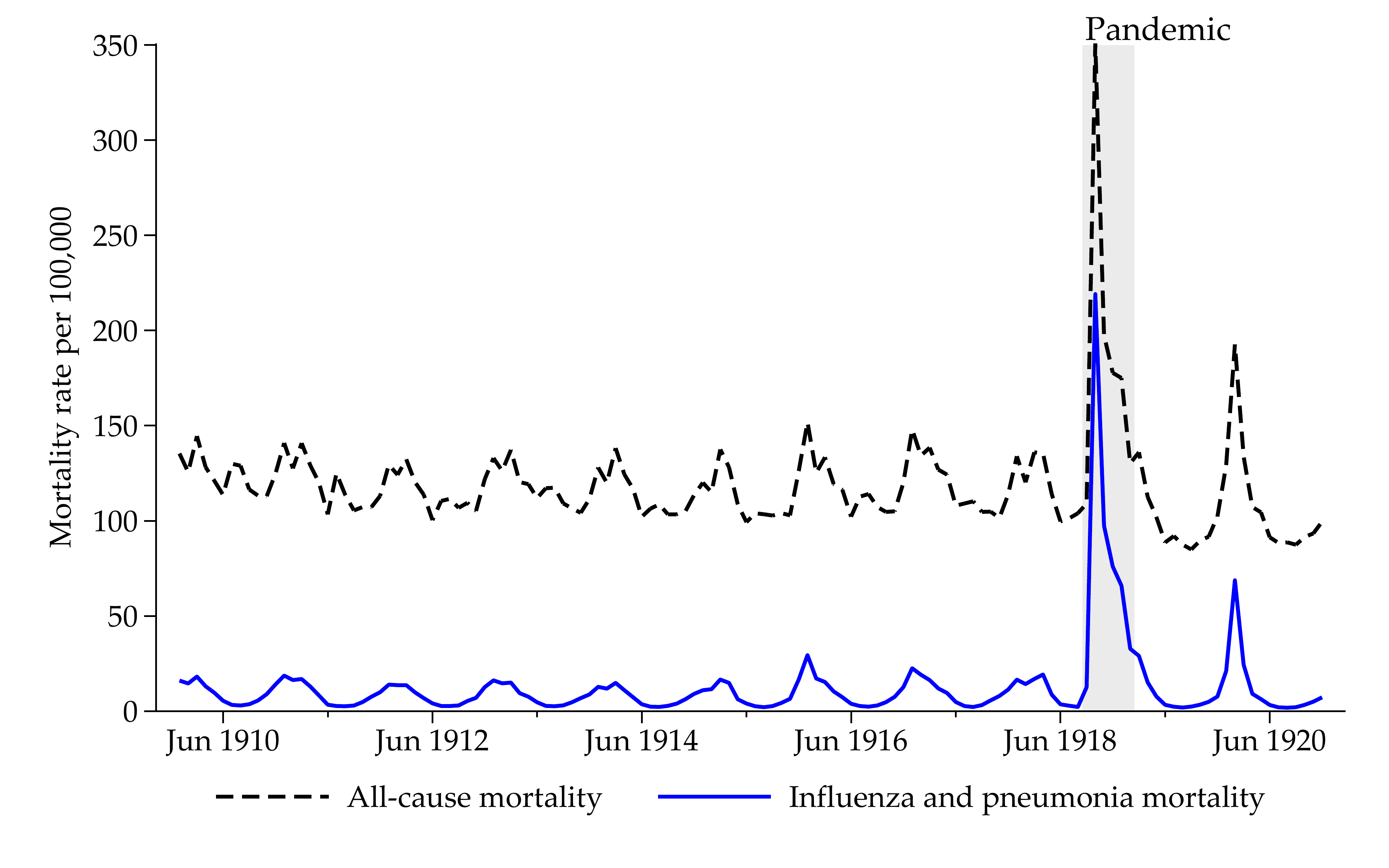}
    \caption{{\bf Monthly mortality rates in the United States, 1911-1920.} \footnotesize  Monthly death counts are obtained from the Census Bureau's \textit{Mortality Statistics} (1910-1920). Annual population estimates are taken from the 1939 edition of the \textit{``Vital Statistics of the United States"} and interpolated to a monthly frequency. Information is based only on the so-called ``registration areas,'' as the Census Bureau began collecting mortality statistics for the entirety of the United States only in 1933. See \citet{Doshi2008} for an overview on the evolution of registration areas, which by 1920 encompassed 82.3\% of the U.S. population.}\label{fig:agg_mortality}
\end{figure}

The 1918 Influenza Pandemic spread worldwide and lasted from January 1918 through 1920. Its precise origins are unknown, though it is believed that it originated in either France, China, or the United States. The number of deaths caused by the pandemic is estimated to be at least 50 million globally, with about 550,000 to 675,000 occurring in the U.S between September 1918 and June 1919 \citep{Johnson2002}.  The pandemic thus killed about 0.66 percent of the U.S. population. \Cref{fig:agg_mortality} shows the sharp spike in mortality in the U.S at the time.

In the United States, the pandemic came in three waves, starting with the first wave in spring 1918, a second wave in fall 1918, and a third wave in the winter of 1918-19 and spring of 1919.  The spring 1918 wave was mild. The pandemic peaked during the second wave, with the highest death toll occurring in October of 1918. This highly fatal second wave was responsible for most of the deaths attributed to the pandemic in the U.S. The increase in the severity of the pandemic in the second wave is believed to have been caused by a mutation that made the virus significantly deadlier \citep{Barry2004}. 

\begin{figure}[htbp]
    \centering
    \includegraphics[width=0.8\textwidth]{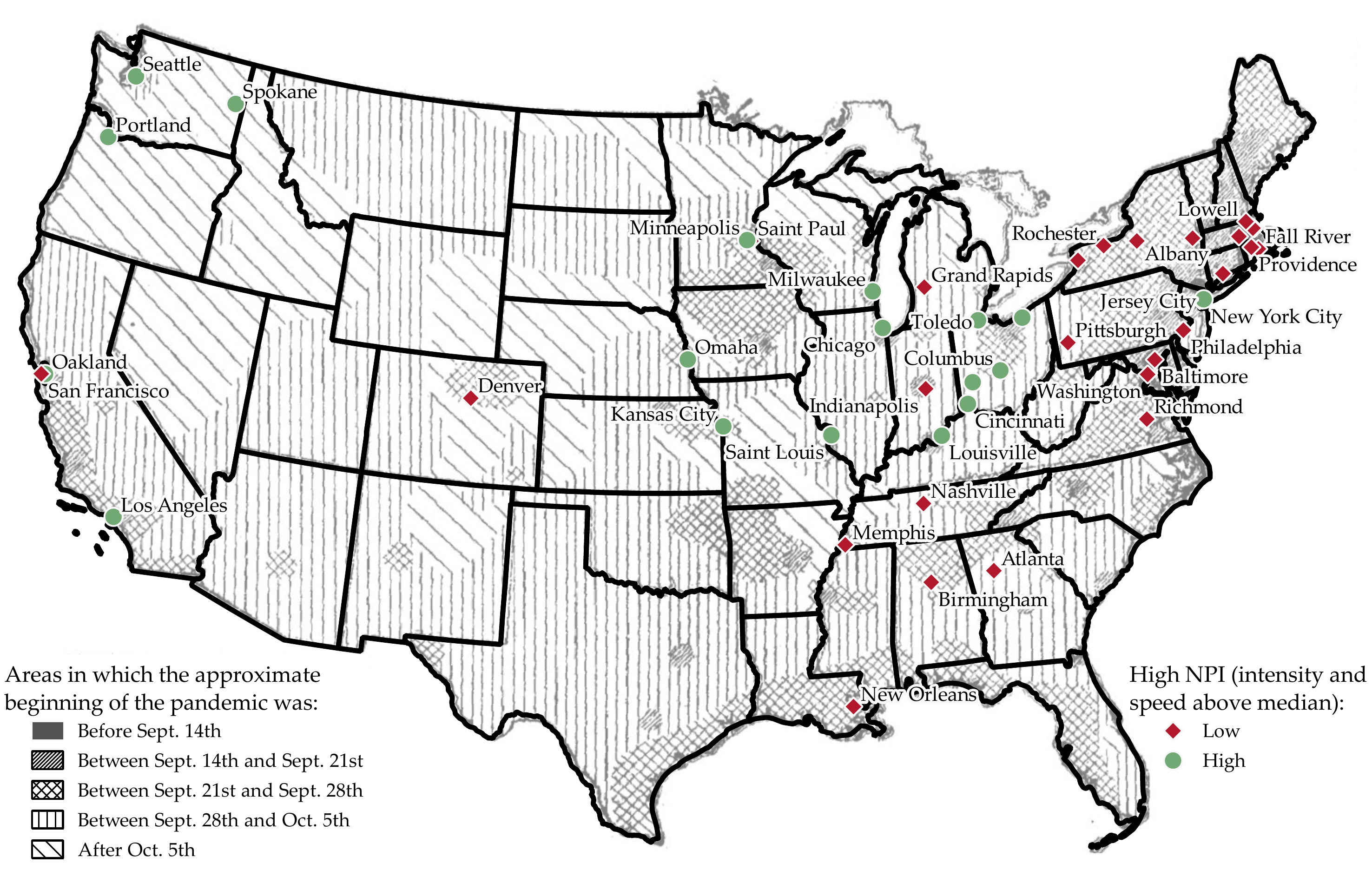}
    \caption{\textbf{Approximate start date of the pandemic by region, and city classification by NPI measures.} \footnotesize This figure shows the map by \citet{Sydenstricker1918} indicating the approximate start date of the pandemic by area. We georeference it and overlay on top of it the 46 cities in our sample that have fully-available NPI information. Cities with above-median NPI implementations in terms of both speed and intensity are labelled as \emph{High NPI}, and the rest as \emph{Low NPI}. See \cref{sec:data} for the formal definition of each measure. Data on NPIs is from \citet{Markel2007}, \citet{Berkes2022}, and hand-collected by the authors.}\label{fig:npi_map}
\end{figure}

The second wave began in Boston with the first cases reported in late August 1918, and it ``exploded'' in Camp Devens in early September 1918 \citep{Barry2004}. It then spread down the eastern seaboard to New York, Philadelphia, all the way to New Orleans, and around to Washington state, as shown in \cref{fig:npi_map}, based on a map by \citet{Sydenstricker1918}. Meanwhile, the disease followed the rivers and railroads to the interior of the country.  Mass troop movements during the closing stages of WWI possibly contributed to the spread of the virus in the U.S. and around the world \citep{Crosby2003}. Crowded army camps spread the disease among troops, and troop movements disseminated the disease geographically.

In most cases, patients experienced a moderate-to-severe influenza, but 10 to 20\% of cases in the U.S. were very severe with a virulent virus or pneumonia \citep{Crosby2003}. Symptoms included fever, chills, joint pain, pain in the diaphragm, vomiting, headache, bleeding from oral cavities, damage to internal organs, and impacts on a patient's mental state \citep{Starr,Barry2004}. A distinct feature of the pandemic was a high death rate for adults in the 18-to-44 age range, as well as for healthy adults, resulting in a W-shaped mortality age profile. Various treatments and vaccines were tried, but effective treatment was limited.\footnote{Some hospitals had oxygen ``but no effective way of administering it'' \citep{Starr}.} Instead, nursing care, such as keeping a patient hydrated, nourished, warm, and providing access to fresh air, was the only scope for increasing the chances of recovery \citep{Crosby2003}. 



\subsection{Overview of non-pharmaceutical interventions}

\label{sec_npi}

Most major U.S. cities applied a range of non-pharmaceutical interventions (NPIs) during the fall and winter of 1918.  While the exact cause of the disease was unknown at the time, public health officials understood that it was more likely to spread in crowded areas, by coughing and sneezing, and from the use of common drinking cups \citep{Barry2004}. For instance, the \textit{New York Times} reported on October 7, 1918: ``Under adverse conditions the health authorities of American communities are now grappling with an epidemic that they do not understand very well. But they understand it well enough to know that it spreads rapidly where people are crowded together in railway trains, in theatres and places of amusement, in stores and factories and schools.''\footnote{``The Spanish Influenza.'' {\normalfont\emph{New York Times}, October 7, 1918.} See also \cref{sec:newspapers_NPIgeneral}. Longer excerpts and additional narrative evidence from newspapers are provided in \cref{appendix:narrative}.} 

The NPIs implemented in U.S. cities included social distancing measures such as the closure of schools, theaters, and churches, and the banning of mass gatherings, such as parades, public funerals, and meetings of political parties and unions. NPIs also included other measures such as mandated mask-wearing, case isolation, making influenza a notifiable disease, and public disinfection/hygiene measures.  Measures in 1918 were not as extensive as measures used to combat COVID-19 in terms of closing non-essential businesses. For instance, rather than closing businesses altogether, in many cities staggered business hours were introduced to avoid crowding in public transportation. Some cities did, however, close libraries, colleges, clubs, saloons, dance halls, pool halls, bowling alleys, movie theaters, skating rinks, and other places of public amusement.

\subsection{Variation across U.S. cities in non-pharmaceutical interventions}

Our empirical analysis exploits variation in the speed and intensity of the implementation of NPIs across major U.S. cities in fall 1918 to estimate the impact of NPIs on mortality and economic activity. This raises an important question: What are the sources of variation in NPIs across cities?

In this section, we discuss evidence on the variation of NPIs from contemporary newspaper articles. We find that there are two broad sources of variation in local policymakers' decisions to implement NPIs. The first source is differences in the information available to local policymakers at the time of arrival at the flu, which is primarily driven by differences in city location -- essentially, distance to the East Coast. The second source is differences in the policy preferences, beliefs, knowledge, and incentives of local policymakers, as well as local political economy factors. The latter results in variation conditional on city location. We next discuss each in more detail.

\subsubsection{City location}

Distance to the East Coast explains a substantial part of the variation in NPIs across cities (see \cref{fig:npi_map}). As the fall wave of the pandemic swept the country from east to west, cities in the Midwest and West were affected later. These cities generally implemented NPIs faster, as they had more time to prepare and were able to learn from cities in the east that were affected earlier. 

There are numerous examples of this information channel driven by city location.\footnote{See appendix \ref{sec:newspapers_NPIvariation} for additional examples on the sources of variation of NPIs and longer excerpts of the quotes provided here.} The Health Commissioner of Omaha argued that public gathering bans were introduced preemptively based on the experience of eastern cities: ``The condition in Omaha is by no means as serious as in eastern cities... We are taking this drastic step to keep it from becoming so. I would rather be blamed for being over cautious than be responsible for a single death. Prohibition of public gatherings is the only way known to medical science for checking the spread of the disease.''\footnote{``No Epidemic, But Omaha is Near Closed.'' {\normalfont \emph{The Omaha World-Herald}, October 5, 1918, p. 1, 2.}} The mayor of Cincinnati emphasized that NPIs were introduced based on the experience of other cities that were hit earlier: ``Cincinnati is endeavoring to prevent an epidemic of Spanish influenza. There is no epidemic here. We are doing what other cities should have done---we are preventing.''\footnote{``No Quarantine!'' {\normalfont\emph{The Cincinnati Enquirer}, October 7, 1918, p. 14.}.} \textit{The Seattle Star} admonished citizens to not ``grumble because you can't see a movie'' as ''[t]he health of the city is more important than all else. An ounce of prevention now is worth a thousand cures. In Boston, influenza has taken a toll of thousands. We do not want to court that situation here.''\footnote{``Halls and Churches to be Flu Hospitals.'' {\normalfont\emph{The Seattle Star}, October 07, 1918, p. 1.}}

Some cities in the mid-west and west also implemented longer-lasting NPIs based on the experience of eastern cities. As new cases declined in Milwaukee, health officials warned the public not to be overconfident that the epidemic was abating, as ``some eastern cities made this mistake and later found the decrease due to physicians being too busy to turn in their reports.''\footnote{``Weather Cause of Deaths’'{\normalfont \emph{The Milwaukee Journal} October 26, 1918, p. 2.}} The Health Commissioner believed Milwaukee had a lower death rate than many other cities in part because people were ready to comply with NPIs ``for which they had been prepared by reports of what happened in other cities'' and because NPIs were implemented quickly and widely publicized.

The close relation of NPIs and city location, however, raises concerns for the empirical analysis that exploits variation of NPIs across cities. For instance, western status may also be correlated with other city-specific characteristics that interact with the arrival of the Flu and drive both policy and outcomes such as city density, demographics and city poverty, and air pollution. Further, the virus may have become weaker over time \citep{Crosby2003,Almond2006}.\footnote{While declining virulence is a potential concern, \citet{Crosby2003} (p. 64) argues that ``the decline [in virulence] was too slow for a week or two or three to make much difference... To take advantage of the decline in virulence, a community had to lock the door against the disease for many weeks, even months, as did Australia by means of a strict maritime quarantine.''} A decline in virulence may generate a spurious correlation between faster and more stringent NPIs and lower mortality. The relatively later arrival of the flu  in the west may have also given the local population more time to prepare,  making it a  priori difficult to distinguish between the effects of NPIs and the effects of self-distancing by the local population.

To address these important concerns, in our empirical analysis below, we ensure that all our main  results are robust to controlling for various city characteristics such as density, longitude, and the timing of the arrival of the flu based on \citet{Sydenstricker1918}. We also present results excluding the most western cities. After controlling for these characteristics that are associated with the timing of the flu's arrival, we argue that most remaining variation stems from factors associated with the local policymakers that happened to be in charge when the second wave hit and are thus not directly related to local health or economic fundamentals. We discuss this second source of variation in NPIs next.

\subsubsection{Local policy preferences and political economy factors}

A second source of variation in NPIs across cities is differences in the preferences, beliefs, and knowledge of local health officials, as well as local political economy factors. Views on the need for and effectiveness of NPIs differed widely across public health officials and other local actors. Moreover, there were widespread debates about NPIs involving local public officials, doctors, business owners, local civil society, and other local residents. These debates informed local policy decisions. We next discuss in detail how these factors shaped local NPIs. 

\paragraph{The role of local public health officials.} Local responses to the pandemic were not driven by a federal response, as no coordinated pandemic plans existed.\footnote{On October 7, 1918, Surgeon General Rupert Blue sent out a recommendation to enact social distancing measures to prevent the spread of influenza \citet{Crosby2003}. This was relatively late in the outbreaks of many cities, and many cities had already responded, though some cities only acted after this recommendation was issued.} Instead, local health officials had discretion to implement NPIs, and individual public health officials were central to the implementation of NPIs. 

In Oakland, California, for example, the mayor initially pursued a light-touch approach. However, the appointment of a new health commissioner who believed that more stringent NPIs were necessary, led the mayor to reverse course and close places of public amusement, schools, and churches.\footnote{``Theaters, Churches Are Closed by Mayor,'' \emph{Oakland Tribune}, 18 October 1918.} Officials in Des Moines introduced NPIs to slow the spread of disease and avoid overburdening local hospitals, a particularly sophisticated understanding of how to handle the epidemic \citep{InfluenzaArchive}.

St. Louis is often cited as an example of quick and effective implementation of social distancing measures to manage the epidemic. This included canceling the Liberty Loan parade, in contrast to other cities such as Charleston and Philadelphia, where the exigency to meet quotas for liberty loans to finance WWI led local governments to ignore the recommendation of health officials   \citep{Hatchett2007,Barry2004}.\footnote{In Charleston, South Carolina, the local health officer did not cancel the fourth Liberty Loan parade, despite acknowledging that it was dangerous for the spread of influenza, because those citizens who could not go to war ``must make martyrs of ourselves, if it can be called that, by facing possible illness in order to help our country in time of need.'' The health officer explained: ``That is how we can explain the meetings of the fourth liberty loan being held at the time. We may contract influenza. But if we do, we shall have done so in a splendid cause.'' See ``Health Officer Explains Order—Best of Reasons for Exempting the Liberty Loan,'' {\emph{Charleston News and Courier}, October 9, 1918, p. 8}.} St. Louis even implemented a sweeping four-day closure of all non-essential stores and factories on November 9, 1918. The leadership of the ``strong-willed and capable'' Health Commissioner Max C. Starkloff was crucial to the policy response in that city \citep{InfluenzaArchive}. Medical doctors in St. Louis believed that the timely closing of public places had prevented the high mortality experienced by eastern cities such as Boston.\footnote{``Mayor Voted Down in Effort to Take Off Influenza Ban. Medical Men Show St. Louis' Precautions Kept Disease from Spreading.'' {\normalfont \emph{St. Louis Globe Democrat}, October 25, 1918, p. 9.}.} 

In many other cities, however, public health officials did not take the virus seriously at first, played down the disease as nothing other than ``old fashioned grippe,'' incorrectly claimed they had the disease under control, or were unwilling to take charge of the response. In most cities, influenza was not a reportable disease until after the local epidemic was already underway, so policymakers did not have a firm understanding of how bad a local outbreak was.\footnote{See the discussions in \citet{InfluenzaArchive} and \citet{SF1918} for examples from Providence, Rochester, and San Francisco.} The Baltimore Health Commissioner Dr. John D. Blake played down the disease as ``same old influenza.''\footnote{``Meade is quarantined,'' \emph{Baltimore Sun}, 26 September 1918.}

The Twin Cities of Minneapolis and St. Paul provide an interesting case study of differing responses to the epidemic. While the flu arrived in the Twin Cities around the same time, health commissioners in Minneapolis and St. Paul disagreed on whether closing public spaces was the best course of action in the epidemic. Local officials in Minneapolis moved quickly to ban public gatherings and close schools in early October. Right across the Mississippi River, St. Paul remained largely open into November, as its leaders were confident they had the epidemic under control and believed NPIs would not be effective.\footnote{See   \href{https://www.minnpost.com/health/2020/03/a-look-back-at-the-1918-flu-pandemic-and-its-impact-on-minnesota/}{``A look back at the 1918 flu pandemic and its impact on Minnesota,'' \textit{MinnPost}, March 4, 2020} and \citet{Ott}.}

\paragraph{Debates over the cost-effectiveness of NPIs.} Public officials and other actors debated whether NPIs were effective in reducing the spread of the disease and mortality, as well as the extent of a tradeoff between public health and economic activity.\footnote{Examples from newspapers of these debates are provided in Appendix \ref{sec:newspapers_debate}.} Policymakers and public health officials in some cities argued that NPIs were effective in protecting public health. Some also argued NPIs would benefit the economy. 

For example, following a meeting of public local health officials in which Denver closed all places of public assembly, \emph{The Denver Post} reported: ``From an economic viewpoint, the doctors were agreed that one or two or three weeks closing of public assemblies now would save many dollars in the long run, for they confidently predicted that 40 percent of the population would be stricken if strict measures were not taken to prevent the spread of the contagion and that another 40 percent of the population would be caring for the great mass afflicted.''\footnote{``Denver Closes Churches And Theaters—All Civic and Business Interests Unite to Safeguard Lives of People and Halt Plague.'' {\normalfont\emph{The Denver Post}, October 6, 1918}.}

After introducing an order restricting opening hours for downtown stores,  the mayor of Portland explained: ``our belief that more stringent regulations during the next few days will have a direct tendency to shorten the period during which regulations of any sort will be needed... In other words, it will save lives, prevent suffering and lessen economic hardships if all of us for a short time do our utmost to stamp out this epidemic than to use only halfway measures over a long period of time.''\footnote{``Drastic Rules to Combat Influenza.'' {\normalfont \emph{The Oregonian}, November 3, 1918, p 22.}} 

When announcing that restrictions in New Orleans would be removed within two weeks, local U.S. Public Health Service representative Dr. Gustave Corput warned residents to continue avoiding crowds. He emphasized that ``patience and compliance on the part of the public for the next few weeks means the wiping out of the epidemic in the state. Failure to do these things undoubtedly means the loss of many lives and an inestimable damage to business conditions.''\footnote{``Nov. 16 Day Fixed For Raising Ban Against Crowds'' {\normalfont \emph{New Orleans Times-Picayune}, November 7, 1918.}}

    
On the other hand, other public health officials and interest groups believed that closures were not effective and would cause a panic that was worse than influenza. The Minnesota State health officer Dr. Henry M. Bracken argued that closures in Minneapolis were ``unnecessary and inadvisable'' and would have no effect on the epidemic.\footnote{``Business Hours May Be Changed to Curb Epidemic,'' \emph{Minneapolis Morning Tribune,} October 15, 1918.} Bracken preferred to rely on isolation and quarantine to limit the spread of influenza, but Minneapolis city officials took issue with Bracken's position.

In Detroit, representatives from the Red Cross, local businesses, and doctors concluded that ``nothing would be accomplished--save increasing hysteria--by closing the schools, amusements and places of public gatherings.''\footnote{``Be Calm, Cool; Check Disease—Public Given Advice on Influenza at Meeting of Doctors and Laymen. City Not To Be Closed.'' {\normalfont \emph{Detroit News} October 18, 1918, p. 1, 2.}, see also appendix \ref{sec:newspapers_debate}.} A few cities such as Newark, Pittsburgh, and Philadelphia were only closed after closure mandates from the state board of health, against the wishes of city health officials.\footnote{See Appendix \ref{sec:newspapers_NPIvariation}.} 

The \textit{Philadelphia Inquirer} ran an editorial opposing closures for causing panic. ``Since crowds gather in congested eating houses and press into elevators and hang to the straps of illy-ventilated streetcars, it is a little difficult to understand what is to be gained by shutting up well-ventilated churches and theaters. The authorities seem to be going daft. What are they trying to do, scare everybody to death? \dots The fear of influenza is creating a panic, an unreasonable panic that will be promoted, we suspect, by the  drastic commands of the authorities.'' In contrast, the \textit{Philadelphia Evening Bulletin} believed closures were necessary to limit the spread of the disease.\footnote{See Appendix \ref{sec:newspapers_debate} for excerpts of both editorials, as well as \cite[p. 85]{Crosby2003}.} 


The question of whether to close schools was also contentiously debated in many cities among health officials, school officials, and parents. Health officials and school superintendents in some cities argued for keeping schools open, since children could be monitored by teachers and nurses instead of playing in the streets (e.g., Minneapolis and Seattle). The \textit{New York Times} questioned whether it was wise to keep schools open in New York City, writing that ``Dr. Copeland's reasons for keeping them open are not altogether convincing.''\footnote{``The Spanish Influenza.'' {\normalfont\emph{New York Times}, October 7, 1918, p. 12.}} When schools were not closed, attendance was often low due to illness and cautious parents keeping their children home, in some cases on the advice of local doctors. 


There were also debates on the efficacy of masks in preventing the spread of influenza, and on whether mask-wearing should be compulsory or not.
These debates were most prevalent in western cities.  San Francisco and Oakland implemented mask ordinances. In San Francisco, mask use was widespread even before the mask ordinance, as mask-wearing was encouraged by public health officials \citep{Crosby2003}. Compliance with mask-wearing was high at about 80\%.\footnote{See ``Everyone Is Compelled to Wear Masks by City Resolution; Great Variety in Styles of Face Adornment in Evidence.'' {\normalfont \emph{San Francisco Chronicle}, October 25, 1918 and Appendix \ref{sec:newspapers_noncompliance}}.} 

However, various groups opposed the mask ordinance. A cigar dealer in Oakland noted that the ordinance would negatively affect sales at cigar stores because masks made it difficult to smoke.\footnote{See ``2779 Cases of Influenza Now on Hand.'' {\normalfont \emph{Oakland Tribune}, October 26, 1918, p. 7.} and Appendix \ref{sec:newspapers_masks}.} A second wave of cases in the January of 1919 led to renewed calls for a mask ordinance in Oakland and San Francisco. In Oakland, the ordinance was opposed by Christian Scientists, labor unions, and the mayor of Oakland, citing research from the California State Board of Health that masking did not reduce the spread of influenza.\footnote{California State Board of Health Executive Wilfred Kellogg argued masks were ineffective in practice because they were often made from low quality material and because people wore them mostly outside when needed least and then took them off inside when they were more valuable \cite[][p. 11-13]{Kellogg}. \citet{Kellogg} also argued that comparing city epidemic death curves indicated ``closing, at least in large, cities, avails little or nothing,'' though he did continue to recommend that local health authorities close schools and places of public amusement \citep[][p. 24]{Kellogg}. Other local doctors disagreed, arguing that masks had been successful in reducing the spread of influenza (Appendix \ref{sec:newspapers_masks}).} Following the reintroducing of the masking ordinance, opponents of masking formed an Anti-Mask League, which pushed for the ordinance to be rescinded \citep{Crosby2003}.

\paragraph{Lobbying by business interests and other groups.} Businesses lobbied both for and against NPIs. In many cases, businesses directly affected by closures opposed them, while businesses who were adversely affected by spread of influenza urged more closures. In some cases, businesses directly affected by closures, such as theaters, demanded that closures should also apply to other places where people gathered.

In Buffalo, New York, theater owners supported the closure, as they believed that ticket sales would be depressed until the epidemic was brought under control.\footnote{``City Closes Today To Fight Influenza, Which Is Spreading,'' \emph{Buffalo Courier,} October 11, 1918.} In Denver, a manager of a theater supported the closures, saying, ``I shall sacrifice gladly all that I have and hope to have, if by so doing I can be the means of saving one life.''\footnote{``Denver Closes Churches And Theaters—\dots—All Civic and Business Interests Unite to Safeguard Lives of People and Halt Plague.'' \emph{The Denver Post}, October 6, 1918, p. 1, 2. See appendix \ref{sec:newspapers_notradeoff} for an extended quote.}  Following a second closure order in late November 1918 that did not affect stores, theater managers said they supported efforts to stamp out the disease, but they complained that the order did not affect other crowded places. In Los Angeles, the Theater Owners' Association felt they were being treated unfairly by the closure of theaters. They unsuccessfully lobbied for stricter closure orders for other businesses to stamp out disease more quickly, which businesses in other sectors opposed. In Worcester, theater owners protested the closure of theaters. At the same time, they requested that the board of health also close other places of amusement such as saloons and bowling alleys, which they argued posed a greater health risk.\footnote{``Theater Men Protest On New Order. Claim Other Amusement Places Should Also Be Closed.''  \emph{Worcester Evening Post}, October 4, 1918, p. 1, 10. See also appendix \ref{sec:newspapers_business}.}

A particularly interesting illustration of debates over NPIs comes from San Francisco. On October 17, the mayor met with members of the  San Francisco board of health and other health officials, the Red Cross, the U.S. Shipping Board, and local business owners to discuss the epidemic and policy response. At the meeting, representatives of the U.S. Shipping Board emphasized that the ``influenza epidemic in the East has seriously hampered the ship building program'' and that ``the East is now looking to California and the western states to carry on the work of ship building.'' They therefore urged health officials to close public places of amusement to limit the severity of the epidemic \citep{SF1918}.\footnote{Similarly, three shipbuilding firms in Staten Island, NY sent a letter to New York City Health Commissioner Dr. Royal S. Copeland asking that public places in Staten Island be closed to mitigate the spread of the disease, as the influenza epidemic had decreased efficiency at the shipbuilding plants by 40\% (see Appendix \ref{sec:newspapers_voluntary}). } Theater and movie theater owners in attendance said revenues had declined sharply, at one theater by 40\%, due to fear of the virus. The theater owners therefore supported a closure to bring the epidemic under control. The owner of one theater said: ``I have interviewed the managers and owners of many theaters and show houses and they have all stated that as the people appear to be voluntarily staying away anyhow, that the amusement managers would not suffer much more if all theaters were ordered closed'' \citep{SF1918}. However, not all present supported closures. A doctor from the Navy stated: ``I do not think the closing of these activities will be successful in controlling the epidemic. \dots I think forcible closing would do more harm than good.'' This opinion was in the minority, and the board of health voted to close all places of public amusement, ban public gatherings, and close schools \citep{SF1918}. 
        

Lobbying was, in some cases, directly effective in changing policy. For example, in St. Louis, opposition to restricted business hours led the health commissioner to rescind the order because ``he had been convinced that the order was working a hardship on small businesses, such as cigar dealers, and that it was accomplishing little in preventing gathering of persons, because there were no congregations of size in small stores.''\footnote{``Order Fixing 9:30 To 4:30 As Business Hours For Downtown Stores Rescinded.''  \emph{St. Louis Post-Dispatch}, October 23, 1918, p. 3.} In Pittsburgh, saloon owners and other businesses affected by the state-level closure order lobbied to reopen the city. In response, the mayor announced that he would not enforce Pennsylvania's statewide closure order after it was extended an additional week in early November 1918.\footnote{See ``Influenza Ban Abrogated by City Officials. Mayor Babcock Advises Disregard of State Health Authorities’ Ruling. Churches May Open. Schools to Resume and Other Activities Expected to Become Normal Again.''  {\normalfont \emph{The Pittsburgh Gazette Times}, November 2, 1918, p. 1, 5} and Appendix \ref{sec:newspapers_debate}.}

Finally, other interest groups such as churches also lobbied on NPIs. Churches were closed in some cities. In many of these cities, clergy opposed the closure of churches, and more so in cities where saloons were allowed to remain open, such as Columbus and Milwaukee.\footnote{\emph{The Milwaukee Journal} reported: ```Why are saloons permitted to remain open while churches must be closed.' These are the questions that will be asked of Health Commissioner Ruhland by about 100 members of the Milwaukee federation of churches,'' (``Nurses Bureau Is Opened,'' \emph{The Milwaukee Journal,} October 20, 1918).} Catholic Churches in Lowell, Massachusetts went further and defied the closures. Fear of infection made attendance at services lower than usual, though many did still attend the masses. Churches in Seattle held services out of town in localities not affected by the closures.\footnote{See ``Gloomy Sunday Is Result Of the Influenza Ban On All Places of Amusement.'' \emph{Seattle Post-Intelligencer}, October 7, 1918 and Appendix \ref{sec:newspapers_noncompliance}.} At the same time, in New Haven churches cancelled mass voluntarily in a contribution to reduce the spread of influenza.\footnote{See ``Little Change In Local Epidemic,'' {\normalfont \emph{New Haven Journal - Courier}, October 21, 1918} and Appendix \ref{sec:newspapers_voluntary}.}



\section{Data}
\label{sec:data}

The core data underlying this paper consists of city-level information on non-pharmaceutical interventions, influenza mortality, and local economic activity, for the period around the 1918 pandemic. This section outlines each of these three components. Further details are provided in \cref{sec:data-appendix}.

\subsection{City-level measures of non-pharmaceutical interventions}

Our starting point for constructing city-level measures of the intensity and speed of NPIs is the work by \citet{Markel2007}, who gather detailed information on NPIs for 43 major U.S. cities from municipal health department bulletins, local newspapers, and reports on the pandemic.

\citet{Markel2007} measure city-level NPIs in two ways. First, they measure the intensity of NPIs by the cumulative number of days where three types of NPIs were active (school closure, public gathering bans, and a collection of other measures that included quarantine/isolation of suspected cases) from September 8, 1918 through February 22, 1919. This measure, which we denote as \NPIIntensity, can thus take values between 0 and 504---three times the number of days in the sample.\footnote{\citet{Markel2007} refer to this intensity measure as the ``Total Number of Days of NPIs."}

Second, they measure how quickly NPIs were implemented by the number of days elapsed between the ``mortality acceleration date''---the date when the excess death rate surpassed twice the baseline death rate---and the date where city officials first enforced a local NPI.
We multiply this day count by minus one, so that higher values indicate a faster response, and denote this measure as \NPISpeed.

To extend the NPI measures to additional cities, we first add information for Atlanta by using the raw NPI data provided by Markel and coauthors.
Second, we add NPI information for a further six cities from \citet{Berkes2022}.
Third, we hand-collect NPI information on four additional cities, extending the total sample from 43 to 54 cities.\footnote{We use the start date and length of NPIs as reported in the `The Commercial Appeal' for Memphis, TN, the `Jersey Journal' for Jersey City, NJ, `The Morning Call' for Paterson, NJ, and The Tribune for Scranton, PA.}
Mortality data is available only for 46 out of the 54 cities. Other outcomes variables such as manufacturing employment is available for a wider set of cities, although only the \NPIIntensity is available for them, as computing the \NPISpeed measure requires knowing the ``mortality acceleration date.''
See \cref{tab:samplecities} for a list of the cities in each sample, \cref{tab:npi-sources} for the values of the NPI variables for each city, \cref{appendix:data-npi} for more details on how the NPI measures were computed, and \cref{appendix:city-mortality-npi} for city-level figures indicating the daily excess mortality rates, and the period when each type of NPI measure was active.

Reflecting that NPIs are likely to be most effective if they are both implemented relatively early and aggressively, our preferred measure of NPIs  is an indicator variable equal to one for cities with both \NPISpeed and \NPIIntensity above their medians. We refer to this indicator variable as \HighNPI. \HighNPI equals one for 18 cities and zero for 28 cities in our main sample of 46 NPI cities with available mortality data.

 All the cities in our sample eventually adopted at least one of the three types of NPIs. School closures and cancellation of public gatherings were the most common. However, there was considerable variation across cities in the speed and aggressiveness of these measures. The median duration was four weeks, with the longest lasting ten weeks. \HighNPI cities on average implemented the first NPI about 1.5 days after the mortality rate reached twice its baseline level, whereas $Low \, NPI$ cities reacted on average only after twelve days (see \cref{tab:npi_diff} in the Appendix). Similarly, \HighNPI cities had an average NPI intensity of 133, compared to 56 for the $Low \, NPI$ cities. Further, \cref{fig:npi_map} provides a visual illustration of the geographic variation of NPIs for the 46 cities that constitute our main sample.

\subsection{Mortality data} 

Besides using mortality data to compute the ``mortality acceleration date'' and thus the \NPISpeed variable, we also study two mortality outcomes, peak excess mortality and cumulative excess mortality.\footnote{Excess mortality is defined as the difference between the deaths that took place and the number expected in the absence of the pandemic, i.e. the ``baseline mortality'' based on median mortality during the same time of year from 1910 to 1916. Throughout the paper, we report it as a rate, per 100,000 inhabitants.} In addition to studying excess mortality due to influenza and pneumonia (denoted as I\&P occasionally throughout the paper), we also analyze all-cause excess mortality. This is because of the possibility of systematic biases in the influenza and pneumonia mortality statistics, which, at least for the case of COVID-19, have been shown to suffer from poor accuracy and systematic measurement errors \citep{WhyAllCause}.

To compute daily excess mortality rates, we first collect monthly mortality rates between 1910 and 1916 from the annual \emph{Mortality Statistics} \citep{MortalityStatistics}, and use these rates to compute monthly median mortality rates, denoted as the ``baseline mortality rates.'' Second, we obtain 1918 population estimates from \citet{PopCensus1920}. Third, we collect weekly death counts during the pandemic from the \emph{Weekly Health Index} published by the U.S. Census Bureau, retrieved from \citet{InfluenzaArchive} and \citet{PublicHealthReports}. These weekly death counts occasionally have missing values, which we linearly interpolate. Lastly, we smooth out the weekly death counts and the monthly baseline mortality rates to a daily frequency, and compute the daily excess death rates as the difference between the daily mortality rates minus the daily baseline mortality rate. This approach follows \citet{Collins1930} and \citet{Markel2007}. Details of each step involved, including the algorithms employed, specific examples, comparisons between our estimates and those of \citet{Markel2007}, as well as comparisons between influenza and pneumonia and all-cause mortality rates, are provided in \cref{appendix:data-mortality}. 


\Cref{fig:cities} shows the weekly mortality rates for both influenza and pneumonia and all-cause mortality, as well as the mortality acceleration data and the start and end dates of the three different types of NPIs for eight different cities. Several of the patterns outlined in \cref{sec:background} can be seen. For instance, cities further east such as Boston and Philadelphia experienced  increases in mortality earlier and were slower to implement NPIs than cities further west such as Rochester or St.\ Louis. Moreover, there is variation in the speed and intensity of NPIs and the mortality outcomes in cities that are located close to each other such as San Francisco and Oakland and St.\ Paul and Minneapolis. The same type of figure for each city in our sample can be found in \cref{sec:data-appendix}.

\begin{figure}
    \centering
    \subfloat[Philadelphia]{
    \includegraphics[width=0.48\textwidth]{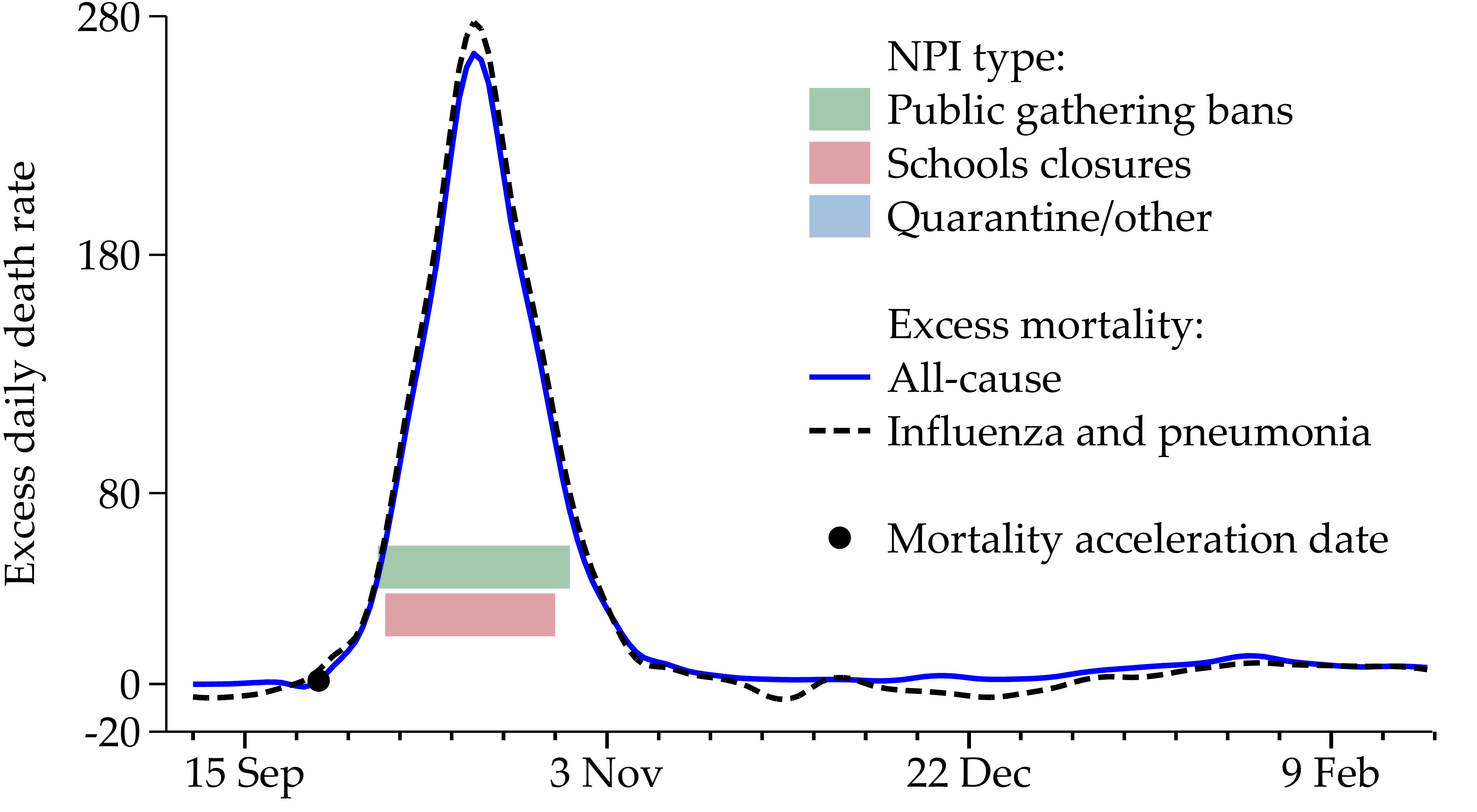}}
    \hfill
    \subfloat[Saint Louis]{
    \includegraphics[width=0.48\textwidth]{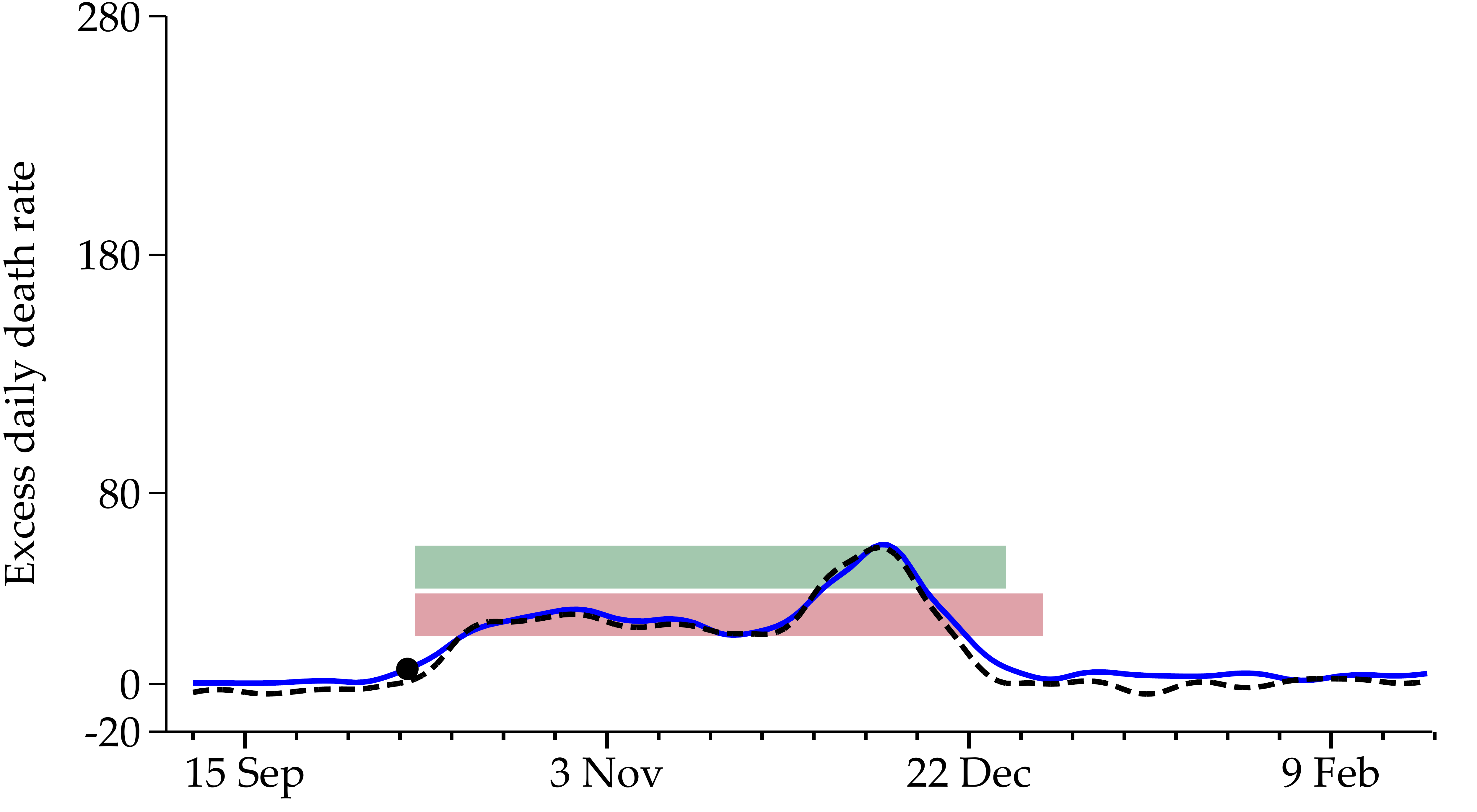}}
    
      \subfloat[Boston]{
    \includegraphics[width=0.48\textwidth]{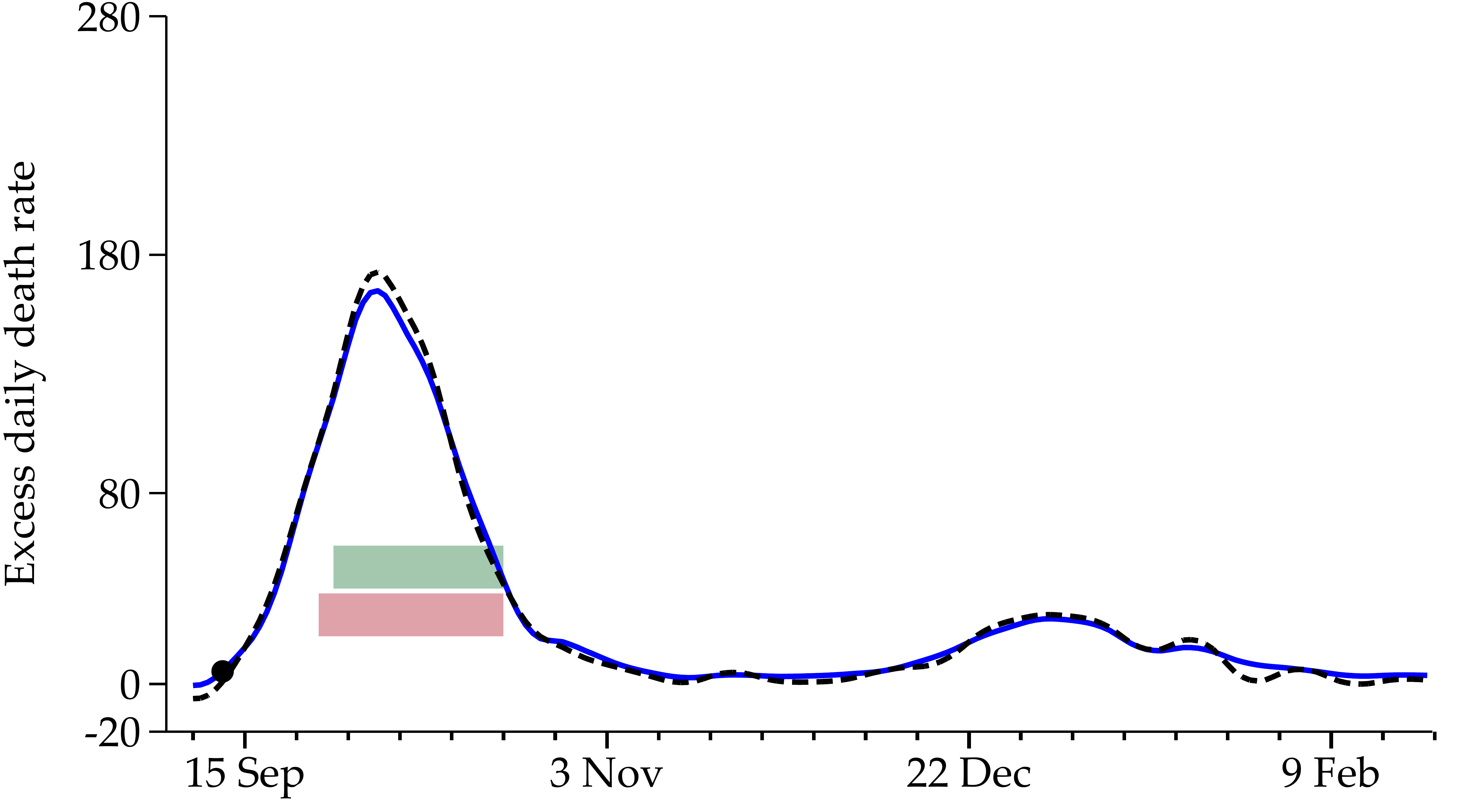}}
    \hfill
   \subfloat[Rochester]{
    \includegraphics[width=0.48\textwidth]{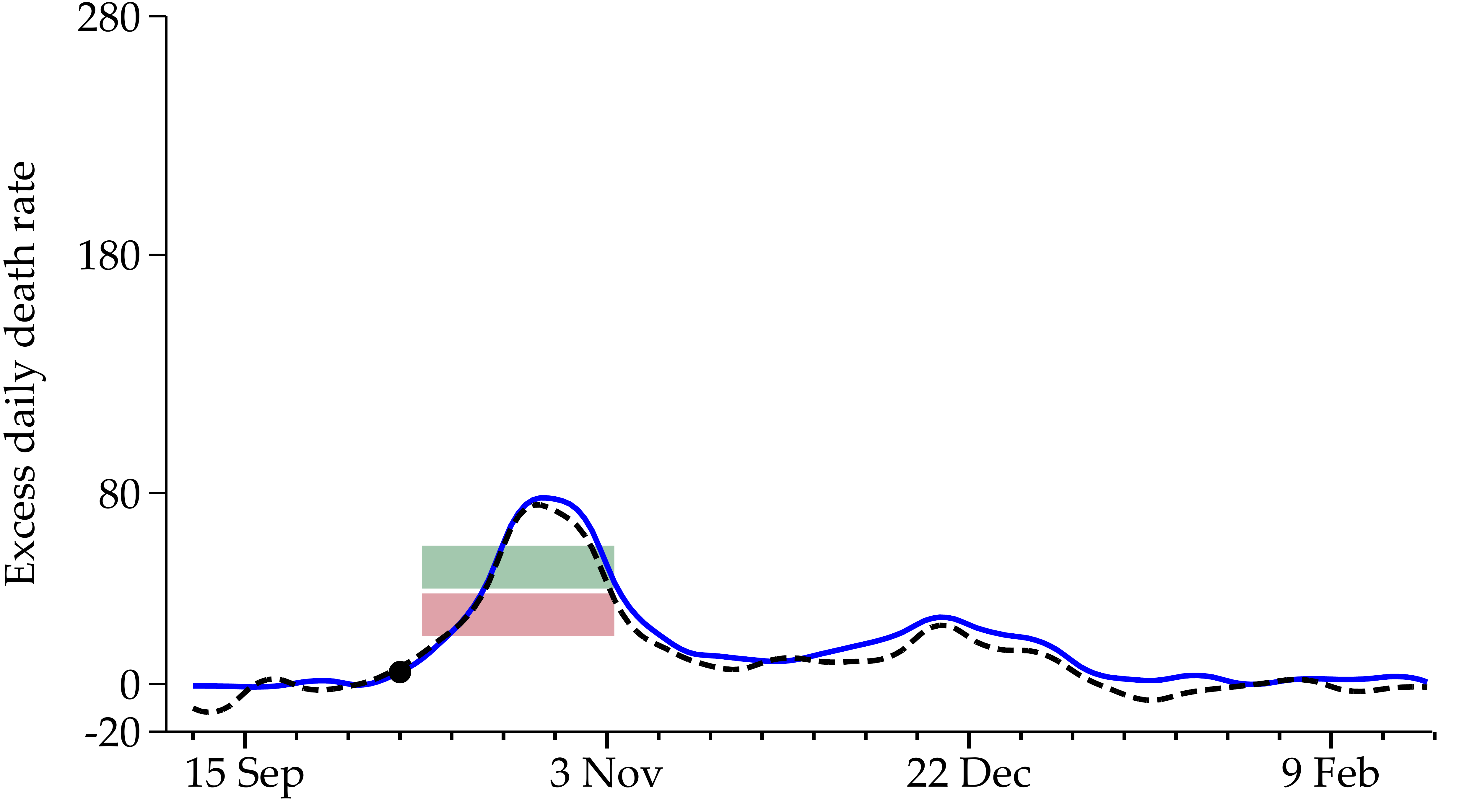}}
    
      \subfloat[San Francisco]{
    \includegraphics[width=0.48\textwidth]{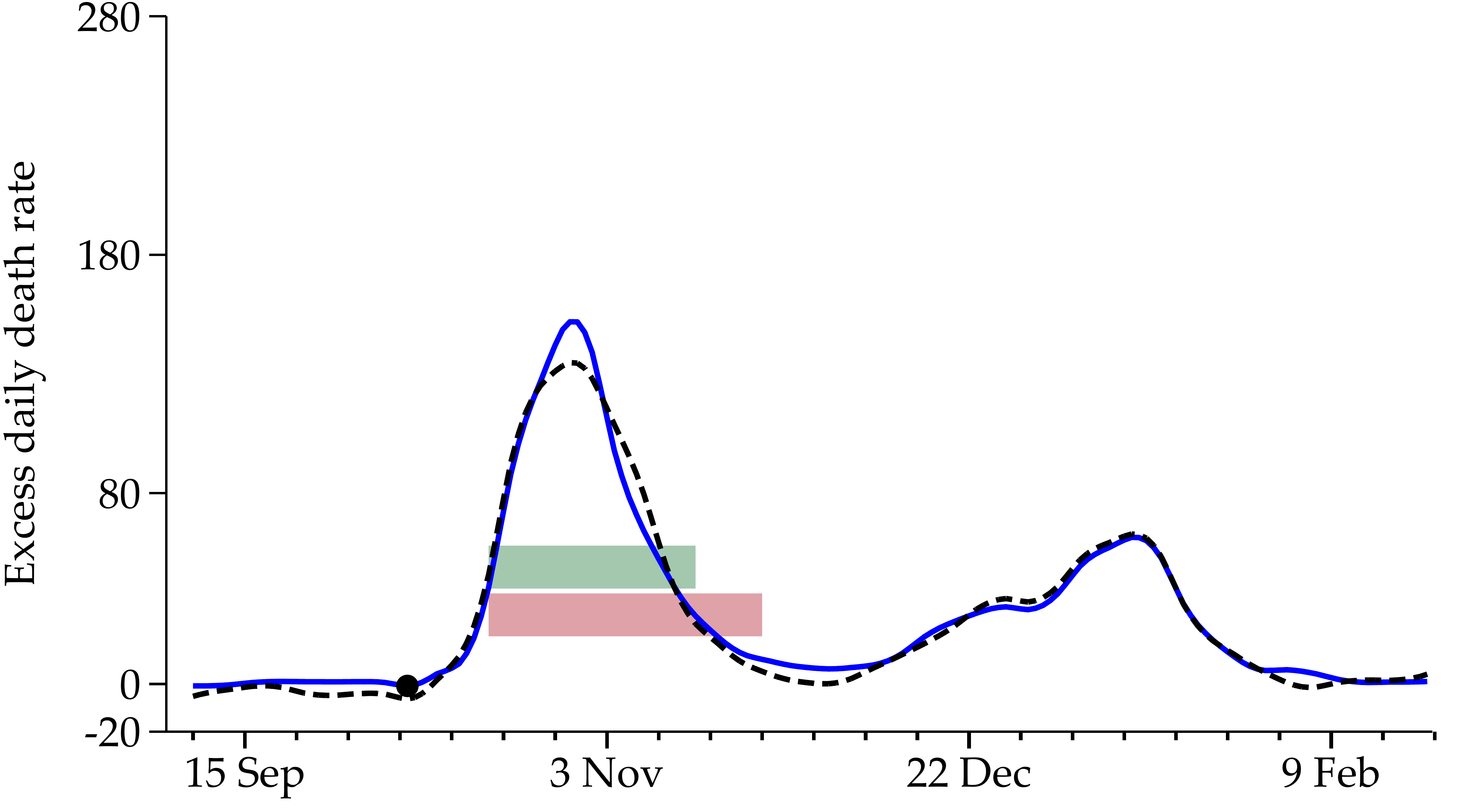}}
      \hfill
     \subfloat[Oakland]{
    \includegraphics[width=0.48\textwidth]{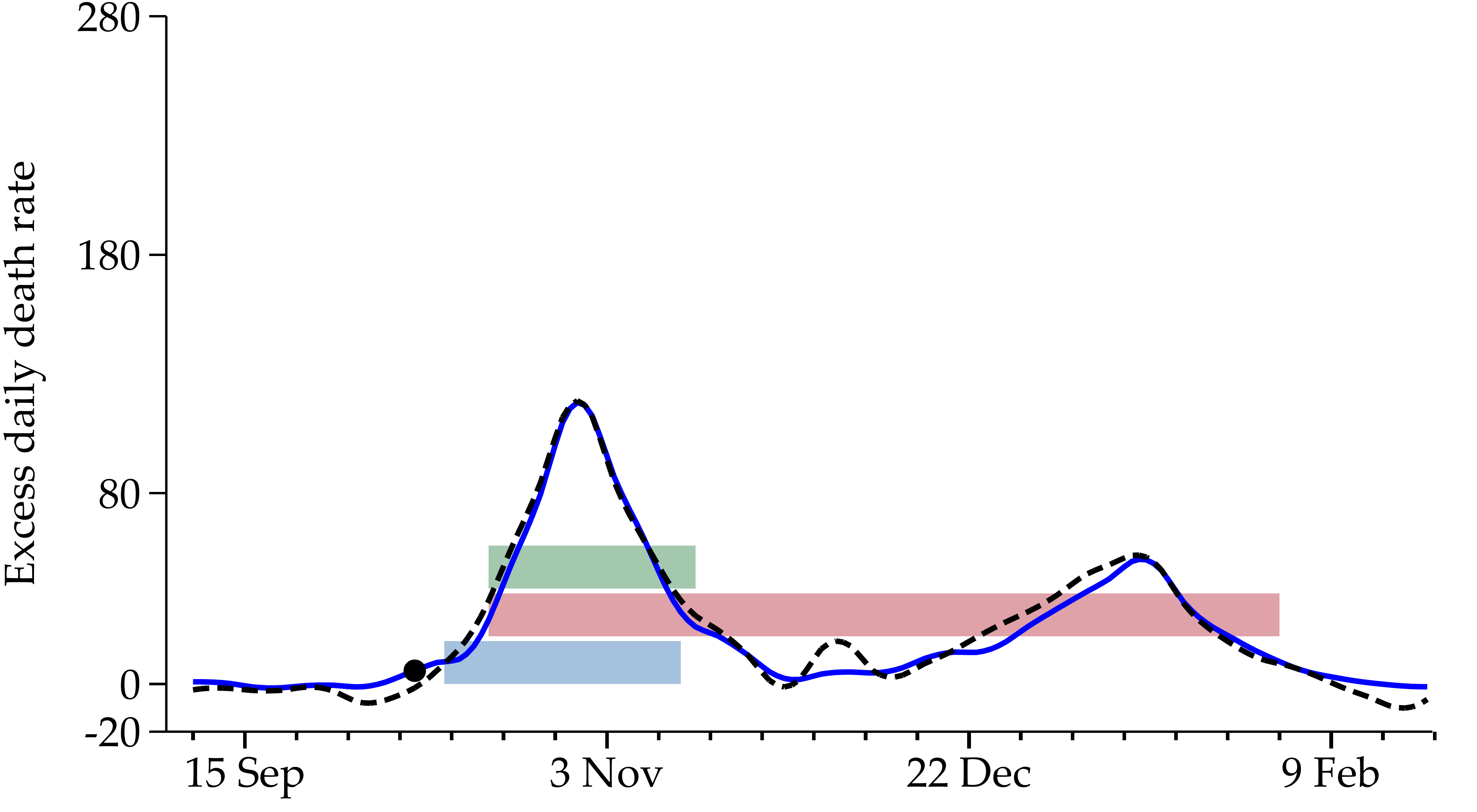}}
    
      \subfloat[Saint Paul]{
    \includegraphics[width=0.48\textwidth]{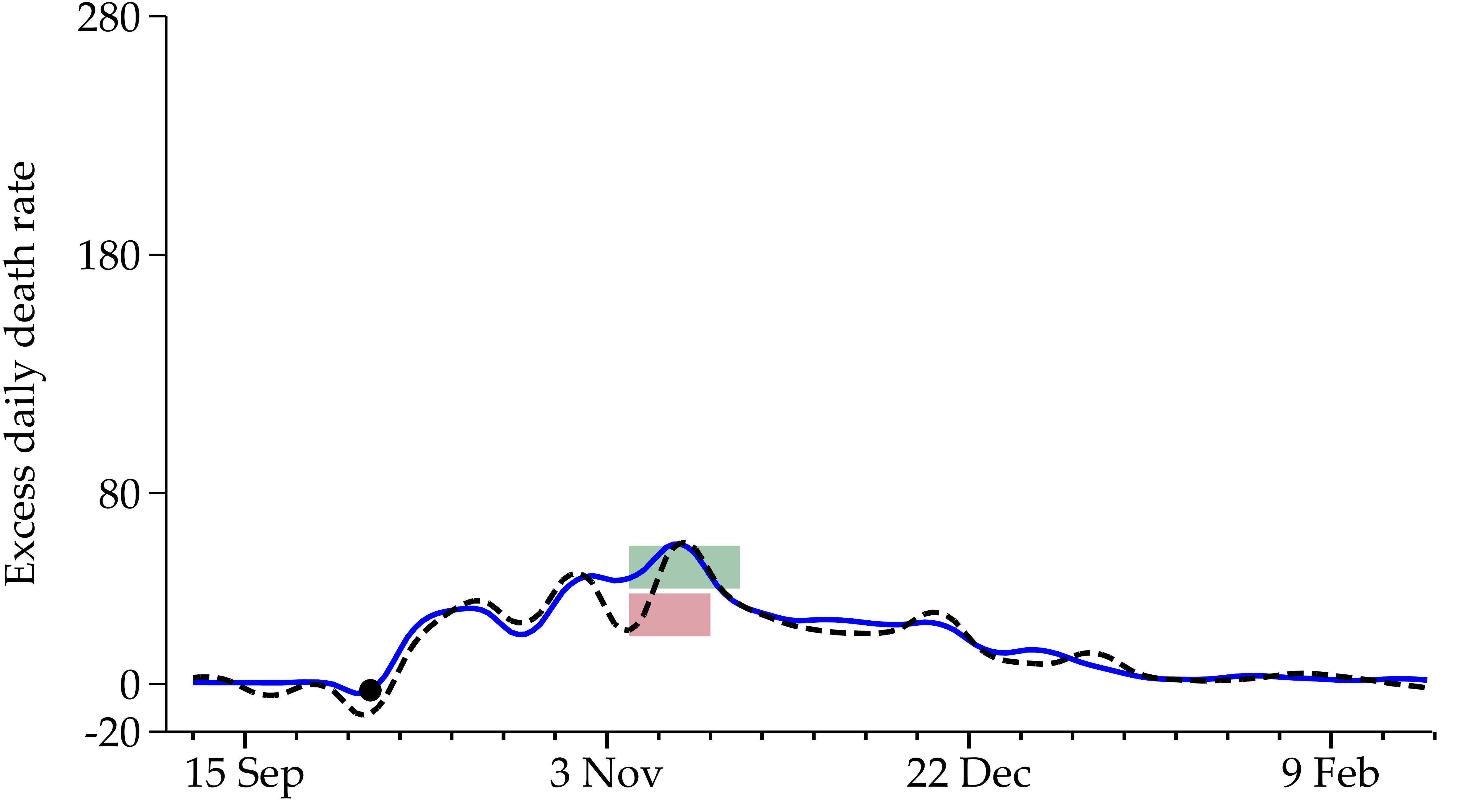}}
    \hfill
   \subfloat[Minneapolis]{
    \includegraphics[width=0.48\textwidth]{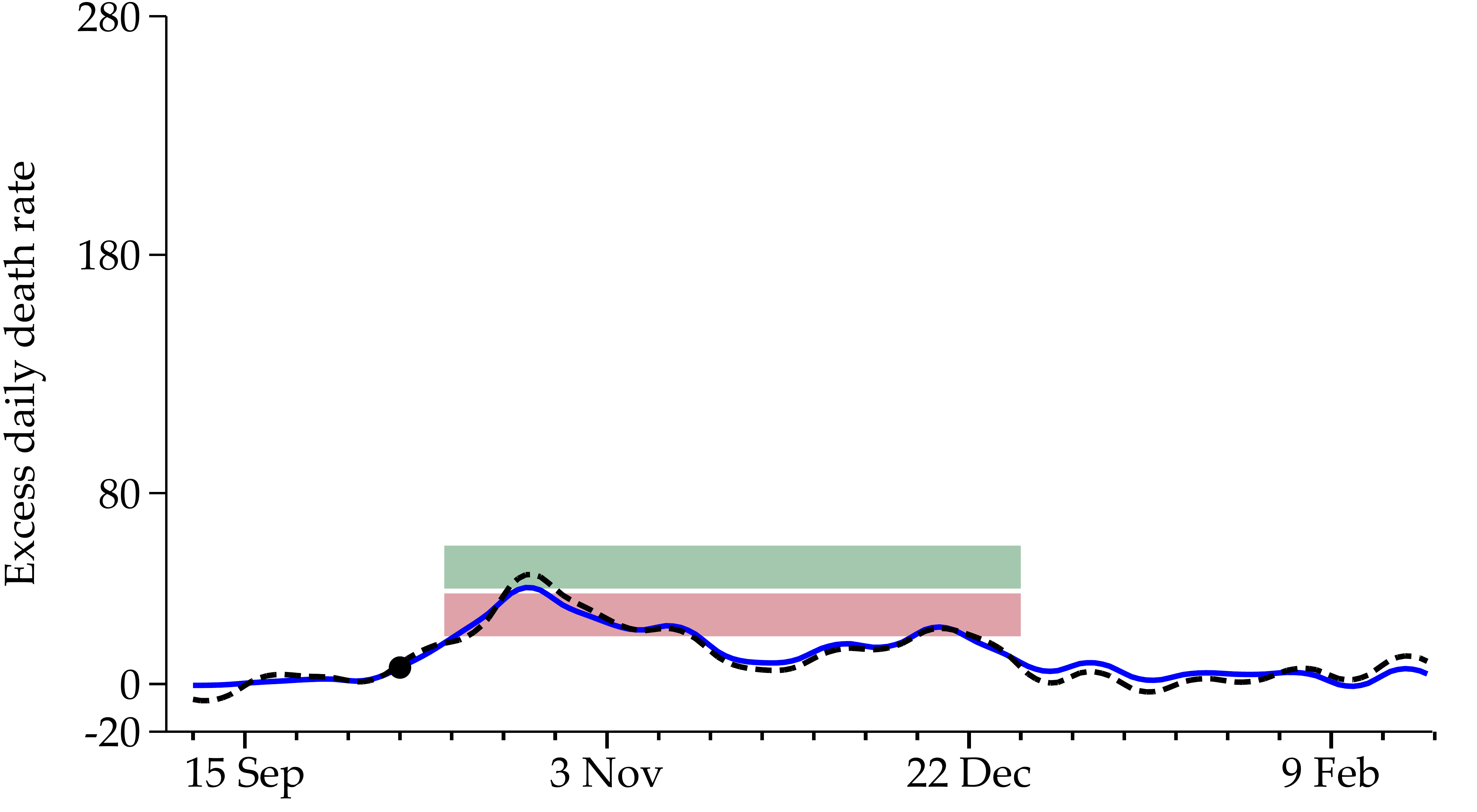}}
    \caption{\textbf{Non-pharmaceutical interventions and mortality.}
    This figure compares daily smoothed mortality rates against the dates where three types of NPIs were active (public gathering bans, school closures, and quarantine/isolation/etc.), for eight selected cities. See \cref{appendix:city-mortality-npi} for a full list of cities.}
    \label{fig:cities}
\end{figure}

\subsection{Economic outcomes} 

To study the short-run economic impact of the 1918 Flu Pandemic and associated NPIs, we construct a monthly city-level measure of business disruptions. We digitize information on business conditions from \textit{Bradstreet's} weekly ``Trade at a Glance'' tables.\footnote{\citet{Velde2020} also uses the Trade at a Glance tables to study the impact of mortality acceleration and business closures on local trade conditions.} These tables provide city-level one-word summaries of the conditions of wholesale trade, retail trade, and manufacturing. We categorize these words into an indicator variable of whether trade was ``Not disrupted'' or ``Disrupted.'' For robustness, we also construct a three-valued measure that ranks business conditions into ``Bad,'' ``Fair,'' and ``Good.'' We then aggregate this measure into monthly frequency, as information for some cities is not reported every week. This results in a monthly series of business disruptions for 27 cities with NPI measures from January 1917 to December 1922. Further details are provided in \cref{appendix:data-bradstreets}.

To study the medium-run impact of NPIs, we digitize information on city-level manufacturing activity from the Census of Manufactures. Manufacturing accounted for 32\% of nonfarm employment in the U.S. in 1910. The Census of Manufactures is available every five years until 1919 and every two years thereafter, so we use manufacturing data on employment and output for the years 1904, 1909, 1914, 1919, 1921, 1923, 1925, and 1927. See \cref{appendix:data-manufacturing} for a precise listing of the sources used, as well as a discussion of potential sources of measurement error in this data, such as methodological changes in the Census of Manufactures, as well as city annexations and changes in city boundaries.
We also use city-level annual bank assets as a proxy for local economic activity, digitized from the \textit{Annual Reports of the Comptroller of the Currency}. 

\subsection{Control variables} 

Finally, we collect various additional variables which we use to control for observable differences across cities. We use information on city population and city density from various decennial censuses, as well as city-level public health spending per capita from \citet{SwansonCurran1976}. Further, to more explicitly control for geography and the timing of the arrival of the flu, we use city longitude as well as information on the week of the arrival of the second wave of the flu as reported in  \citet{Sydenstricker1918}. We also draw on state-level WWI casualties from \citet{WWIcasualties}, state-level exposure to WWI production from \citet{Garrett2008}, and the distance of each city to military training camps. Further, to control for poverty and air pollution, we use city-level illiteracy shares, infant mortality, and reliance on coal-based energy production, following \citet{clay2018}.

\section{Non-pharmaceutical interventions and mortality}
\label{sec:mortality}

\begin{figure}[ht]\centering
\includegraphics[width=0.7\textwidth]{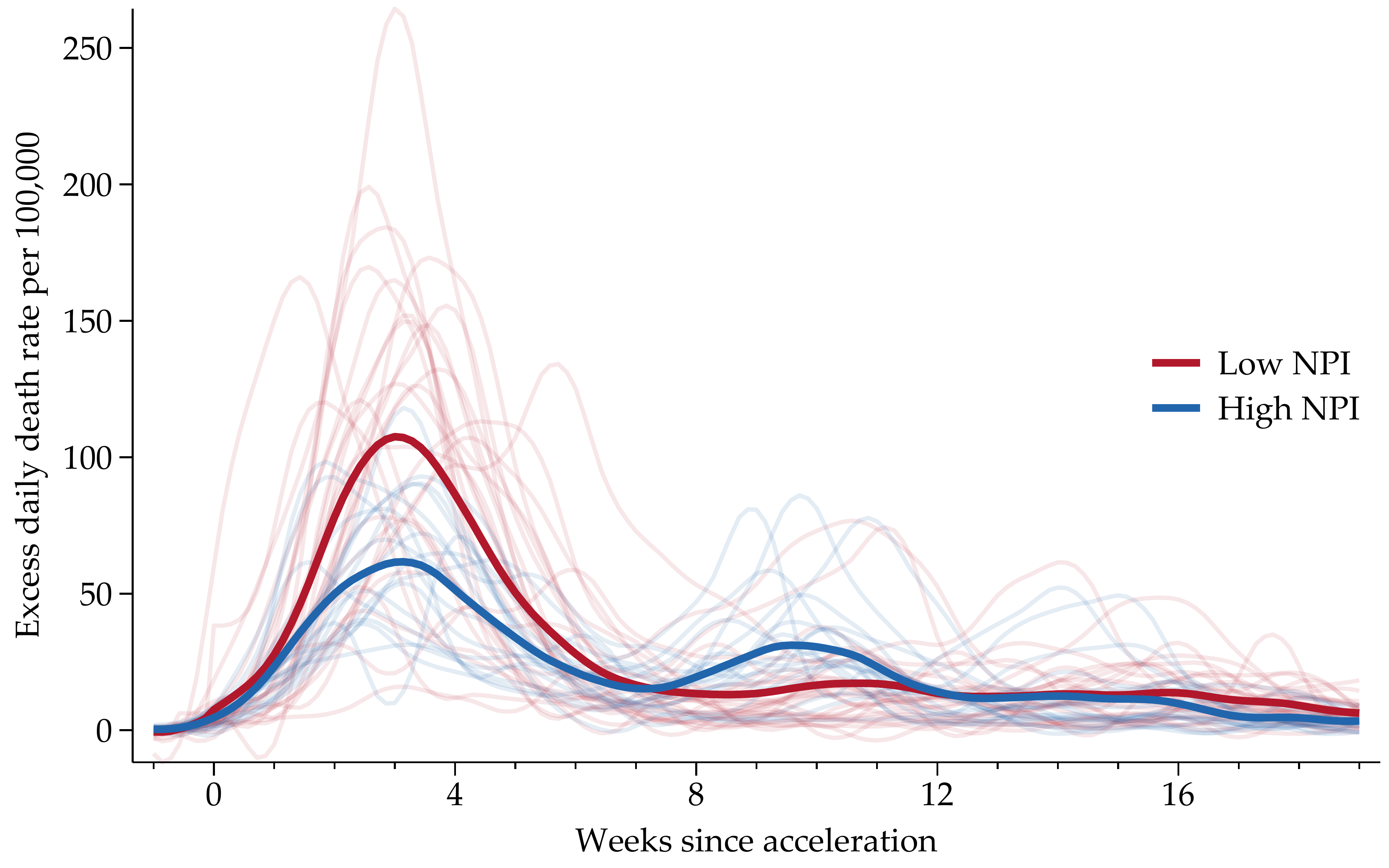}
\caption{\textbf{Excess daily death rates relative to mortality acceleration rate.} This figure shows the average excess daily death rate due to influenza and pneumonia for low and high NPI cities. The thick solid lines represent averages across time, while the thin semitransparent lines represent the individual paths of each city, colored accordingly. High NPI cities are defined as those with above-median NPI intensity and speed. The origin of the x-axis corresponds to the date where mortality accelerated in a city, as defined by \citet{Markel2007}. To prevent missing values, the x-axis stops at 19 weeks, as cities hit later in the pandemic do not have 24 weeks of data \emph{after} their mortality acceleration date.}
\label{fig:city-funkymain}
\end{figure}

This section examines the relation between NPIs and mortality.
We begin by classifying cities according to the \HighNPI indicator variable, equal to one for cities with above-median NPI speed and intensity.
\Cref{fig:city-funkymain} compares the average excess daily death rates from influenza and pneumonia for low NPI cities (thick red line) and high NPI cities (thick blue line), relative to the mortality acceleration dates of each city.
Mortality tended to peak 3 weeks after it accelerated, but remained elevated for most of the 19 weeks. Comparing both lines suggests that cities with stricter NPIs had lower peak mortality and a flatter mortality curve, although they were more likely to experience a second peak. However, the area under the curve is smaller for high NPI cities, indicating that cities with stricter NPIs had lower cumulative excess mortality over the course of the pandemic, why may have been achieved by mitigating epidemic overshoot \citep{Bootsmaa2007}.


\subsection{Empirical framework}\label{sec:framework}

To more formally investigate the pattern revealed in \Cref{fig:city-funkymain}, we estimate city-level regressions of the form
\begin{align}
    Mortality_c = \alpha + \beta \, NPI_c + X_c \, \delta + u_c, \label{eq:mort}
\end{align}
where $Mortality_c$ is a city-level measure of either influenza and pneumonia mortality or all-cause mortality per 100,000 inhabitants, $NPI_c$ is one of the three NPI measures, and $X_c$ is a vector of city-level control variables.

As laid out in \cref{sec:background} above, a key concern with using variation in NPI measures across cities is that NPIs may be endogenous to local health and economic outcomes. Importantly, as also described in \cref{sec:background}, the variation in the decision to implement NPIs can be explained by two main factors. First, differences in the information available to local policymakers at the time of arrival at the flu.\footnote{\Cref{tab:npi_diff} compares the characteristics of cities with high and low NPIs.}  Second, differences in the policy preferences, beliefs, knowledge, and incentives of local policymakers, as well as local political economy factors. The variation from the former is a particular concern for being able to identify the causal effects of NPIs. Hence, we control for several city-level characteristics in our regressions that capture the differences between cities that had more or less time to prepare for the arrival of the flu.

As a set of baseline controls, we include the log city population in 1900 and 1910 to account for city size and past growth. Given that the flu was more likely to spread in crowded areas, we control for city density in 1910. To account for the structure of the local economy we include manufacturing employment in 1914 to 1910 population. To proxy for the quality of the local health care system, we also control for public health expenditure in 1917 relative to 1910 population.

Given the modest sample size of 46 cities, we select a relatively small set of baseline control variables. Nonetheless, we also provide several other specifications in which we add additional controls. For instance, to capture baseline differences in influenza exposure, we control for lagged influenza and pneumonia mortality in 1917. Furthermore, to account for fact that the  cities further west were affected later than those in the east, we  control for a city's longitude. As discussed above, failing to account for the timing of the arrival of the flu may generate a spurious correlation between NPIs and mortality if the severity of the virus weakened over time. A later arrival of the flu may also have led to more voluntary distancing, as people read about the virus in the east, which would also overstate the impact of NPIs.\footnote{For example, \citet{Barry2004} (p. 339) notes that, before the virus arrived in the west, people there read Red Cross newspaper advertisements for people with nursing experience to place themselves at the disposal of the government.} We argue that variation conditional on city longitude reflects local policy preferences, beliefs, and political economy factors discussed in section \ref{sec:background}. In additional robustness checks, instead of just controlling for a city's longitude, we use alternative measures of the timing of the flu's arrival and also control for the week of the arrival of the flu as indicated by \citet{Sydenstricker1918} or the number of days between the first case in Camp Devens in Boston and the acceleration of all-cause mortality in a given city.

Another concern is that NPIs may be correlated with local poverty, air pollution, or the quality of local institutions, which may independently affect mortality. For example, existing work has shown that mortality in the pandemic was higher in cities with higher measures of poverty (as captured, for instance, by the illiteracy rate) and in cities with higher air pollution due to a greater prevalence of coal-fired plants \citep{Clay1918,clay2018}.\footnote{\citet{Clay1918} calculate the coal-fired capacity for electricity generation within a 30-mile radius of each city-centroid. Following \citet{Clay1918}, we use indicators for the city's tercile of this variable.} We therefore  also include both the illiteracy rate and local coal-fired capacity as control variables. In additional tests, we control for infant mortality per capita, another proxy for poverty suggested by \citet{Clay1918}.

WWI and the Armistice is another potentially important confounder. To control for the exposure to WWI arms production, we use data from \citet{Garrett2009}, who constructs a binary variable for states that were involved in WWI production. War production during WWI was concentrated in the east and mid-west of the U.S. 
We also present robustness tests controlling for state-level WWI casualties per capita, which captures mortality from the war and also proxies for the extent of recruitment of the local population.\footnote{An implication of WWI was that many doctors and nurses were in the army and thus not able to treat civilians, making mortality potentially worse among civilians if an area had more recruits or casualties.} Finally, various narratives of the pandemic argue that the second wave was amplified via military camps \citep{Crosby2003,Barry2004}.\footnote{\citet{clay2018} show, however, that there is only limited evidence in favor of this narrative.} We therefore also provide specifications in which we control for the log of a cities distance to the closest army camp.


\subsection{Results}

\Cref{tab:npi_mort_pi} presents estimates of equation \eqref{eq:mort} for mortality from influenza and pneumonia as the outcome variable.
Panel A reports the estimates for weekly excess peak mortality. 
Results for \NPIIntensity, \NPISpeed, and \HighNPI are presented in columns (1)-(3), (4)-(6), and (7)-(9), respectively.
Columns (1), (4), and (7) present regressions without controls. The estimates are negative and statistically significant for \NPIIntensity and \HighNPI, indicating that cities with more stringent NPIs saw lower peak mortality from influenza and pneumonia. For \NPISpeed, the point estimate is negative but not statistically significant, and the $R^2$ is substantially lower. Columns (2), (5), and (8) show that the estimates are similar when we include our baseline controls as well as lagged influenza mortality. Further, columns (3), (6), and (9) reveal that the estimates are essentially unchanged when including additional controls such as longitude, the prevalence of illiteracy, and the reliance on coal-fired power plants. In terms of magnitudes, the estimate in column (9) implies that high NPI cities experienced a 50\% reduction in peak mortality relative to the mean. These estimates suggest that NPIs in the fall of 1918 were successful in flattening the curve.

\begin{table}[htpb]
\centering
\begin{threeparttable}
\caption{\textbf{Non-pharmaceutical interventions and influenza and pneumonia mortality.}}\label{tab:npi_mort_pi}
\scriptsize
    \begin{tabular}{l*{9}{c}}   
    \multicolumn{10}{c}{\textbf{Panel A: Peak excess mortality}} \\ \toprule
                        &         (1)   &         (2)   &         (3)   &         (4)   &         (5)   &         (6)   &         (7)   &         (8)   &         (9)   \\
\cmidrule{1-10} \NPIIntensity&        -0.6***&        -0.5** &        -0.5** &               &               &               &               &               &               \\
                    &       (0.1)   &       (0.2)   &       (0.2)   &               &               &               &               &               &               \\
\NPISpeed           &               &               &               &        -1.2   &        -1.0   &        -1.0   &               &               &               \\
                    &               &               &               &       (1.0)   &       (1.1)   &       (1.1)   &               &               &               \\
\HighNPI            &               &               &               &               &               &               &       -54.5***&       -48.7***&       -50.1***\\
                    &               &               &               &               &               &               &      (10.8)   &      (15.9)   &      (17.7)   \\
1917 I\&P mortality (per 100k)&               &         0.1   &         0.1   &               &         0.2** &         0.0   &               &         0.1   &         0.1   \\
                    &               &       (0.1)   &       (0.1)   &               &       (0.1)   &       (0.1)   &               &       (0.1)   &       (0.1)   \\
Longitude           &               &               &        -0.5   &               &               &         0.1   &               &               &        -0.5   \\
                    &               &               &       (0.7)   &               &               &       (0.9)   &               &               &       (0.6)   \\
Illiteracy          &               &               &         1.3   &               &               &         5.6*  &               &               &         2.5   \\
                    &               &               &       (4.0)   &               &               &       (3.2)   &               &               &       (3.4)   \\
Medium coal-fired capacity&               &               &        14.6   &               &               &        10.9   &               &               &        26.0   \\
                    &               &               &      (36.6)   &               &               &      (34.7)   &               &               &      (35.5)   \\
High coal-fired capacity&               &               &        27.3   &               &               &        25.0   &               &               &        33.9   \\
                    &               &               &      (31.9)   &               &               &      (30.9)   &               &               &      (30.7)   \\
\cmidrule{1-10}
\RSq                &        0.30   &        0.36   &        0.39   &        0.03   &        0.26   &        0.32   &        0.30   &        0.41   &        0.45   \\
Number of cities    &          46   &          46   &          46   &          46   &          46   &          46   &          46   &          46   &          46   \\
\Oster              &           .   &        -.25   &        -.37   &           .   &        -.97   &        -.91   &           .   &         -43   &         -46   \\
Baseline controls   &           -   &         Yes   &         Yes   &           -   &         Yes   &         Yes   &           -   &         Yes   &         Yes   \\
   \bottomrule
 \\ 
    \multicolumn{10}{c}{\textbf{Panel B: Cumulative excess mortality}} \\ \midrule
                        &         (1)   &         (2)   &         (3)   &         (4)   &         (5)   &         (6)   &         (7)   &         (8)   &         (9)   \\
\cmidrule{1-10} \NPIIntensity&        -1.0** &        -0.6   &        -1.3** &               &               &               &               &               &               \\
                    &       (0.4)   &       (0.5)   &       (0.6)   &               &               &               &               &               &               \\
\NPISpeed           &               &               &               &        -4.1   &        -4.8*  &        -5.1*  &               &               &               \\
                    &               &               &               &       (2.8)   &       (2.7)   &       (2.8)   &               &               &               \\
\HighNPI            &               &               &               &               &               &               &      -135.6***&      -122.4***&      -172.6***\\
                    &               &               &               &               &               &               &      (34.8)   &      (43.1)   &      (38.3)   \\
1917 I\&P mortality (per 100k)&               &         1.1***&         1.5***&               &         1.3***&         1.3***&               &         1.0** &         1.4***\\
                    &               &       (0.4)   &       (0.3)   &               &       (0.3)   &       (0.3)   &               &       (0.4)   &       (0.4)   \\
Longitude           &               &               &        -3.1   &               &               &        -2.1   &               &               &        -3.5*  \\
                    &               &               &       (2.3)   &               &               &       (2.7)   &               &               &       (2.1)   \\
Illiteracy          &               &               &        -8.9   &               &               &         3.1   &               &               &        -8.1   \\
                    &               &               &      (11.1)   &               &               &       (9.6)   &               &               &       (9.6)   \\
Medium coal-fired capacity&               &               &        74.1   &               &               &        76.5   &               &               &       119.9   \\
                    &               &               &      (93.5)   &               &               &      (88.9)   &               &               &      (93.2)   \\
High coal-fired capacity&               &               &       122.0   &               &               &       116.9   &               &               &       146.8*  \\
                    &               &               &      (80.2)   &               &               &      (76.6)   &               &               &      (81.2)   \\
\cmidrule{1-10}
\RSq                &        0.11   &        0.30   &        0.40   &        0.05   &        0.32   &        0.39   &        0.22   &        0.40   &        0.53   \\
Number of cities    &          46   &          46   &          46   &          46   &          46   &          46   &          46   &          46   &          46   \\
\Oster              &           .   &        -.46   &        -1.4   &           .   &          -5   &        -5.5   &           .   &        -114   &        -192   \\
Baseline controls   &           -   &         Yes   &         Yes   &           -   &         Yes   &         Yes   &           -   &         Yes   &         Yes   \\
   \bottomrule
 \\ 
 \end{tabular}
	\footnotesize
		\begin{tablenotes} \item
	    Notes: This table presents city-level regressions of peak mortality (panel A) and cumulative excess mortality (panel B) on NPI measures.
	    Mortality refers to influenza and pneumonia mortality, and is based on data from \citet{Collins1930}, \citet{MortalityStatistics}, \citet{InfluenzaArchive}, \citet{PublicHealthReports}, and \citet{PopCensus1920}; see  \cref{appendix:data-mortality} for details.
	    NPI data is based on \citet{Markel2007}, \citet{Berkes2022}, as well as our own data collection; see \cref{appendix:data-npi} for details.
	    Peak mortality is the weekly excess death rate per 100,000 in the first peak of the fall wave of the 1918 pandemic
	    Cumulative excess mortality is the total excess death rate from September 8, 1918 to February 22, 1919.
	    Baseline controls not reported in the table are log of 1900 and 1910 city population, city 1914 manufacturing employment to 1910 population, city density in 1910 and per capita city-level health spending as of 1917.
	    Robust standard errors in parentheses. 
	    *, **, and *** indicate significance at the 10\%, 5\%, and 1\% level, respectively.
\end{tablenotes}
	\end{threeparttable}
\end{table}

\begin{table}[htpb]
\centering
  \begin{threeparttable}
  \caption{\textbf{Non-pharmaceutical interventions and all-cause mortality.} } \label{tab:npi_mort_all_cause}
    \scriptsize
    \begin{tabular}{l*{9}{c}}   
    \multicolumn{10}{c}{\textbf{Panel A: Peak excess mortality}} \\ \midrule
                        &         (1)   &         (2)   &         (3)   &         (4)   &         (5)   &         (6)   &         (7)   &         (8)   &         (9)   \\
\cmidrule{1-10} \NPIIntensity&        -0.6***&        -0.4** &        -0.4*  &               &               &               &               &               &               \\
                    &       (0.1)   &       (0.2)   &       (0.2)   &               &               &               &               &               &               \\
\NPISpeed           &               &               &               &        -1.2   &        -1.1   &        -1.0   &               &               &               \\
                    &               &               &               &       (1.0)   &       (1.1)   &       (1.2)   &               &               &               \\
\HighNPI            &               &               &               &               &               &               &       -54.8***&       -47.8***&       -50.9***\\
                    &               &               &               &               &               &               &      (11.2)   &      (15.9)   &      (17.8)   \\
1917 I\&P mortality (per 100k)&               &         0.2   &         0.2   &               &         0.3***&         0.1   &               &         0.1   &         0.1   \\
                    &               &       (0.1)   &       (0.1)   &               &       (0.1)   &       (0.1)   &               &       (0.1)   &       (0.1)   \\
Longitude           &               &               &        -0.3   &               &               &         0.2   &               &               &        -0.3   \\
                    &               &               &       (0.7)   &               &               &       (0.8)   &               &               &       (0.7)   \\
Illiteracy          &               &               &        -0.4   &               &               &         3.4   &               &               &         0.3   \\
                    &               &               &       (4.3)   &               &               &       (3.4)   &               &               &       (3.6)   \\
Medium coal-fired capacity&               &               &        -1.0   &               &               &        -4.0   &               &               &        11.5   \\
                    &               &               &      (34.6)   &               &               &      (33.1)   &               &               &      (33.8)   \\
High coal-fired capacity&               &               &        15.5   &               &               &        13.5   &               &               &        22.5   \\
                    &               &               &      (29.8)   &               &               &      (29.3)   &               &               &      (29.0)   \\
\cmidrule{1-10}
\RSq                &        0.25   &        0.35   &        0.37   &        0.03   &        0.29   &        0.32   &        0.28   &        0.43   &        0.44   \\
Number of cities    &          46   &          46   &          46   &          46   &          46   &          46   &          46   &          46   &          46   \\
\Oster              &           .   &        -.19   &        -.32   &           .   &          -1   &        -.87   &           .   &         -42   &         -48   \\
Baseline controls   &           -   &         Yes   &         Yes   &           -   &         Yes   &         Yes   &           -   &         Yes   &         Yes   \\
   \bottomrule
 \\ 
    \multicolumn{10}{c}{\textbf{Panel B: Cumulative excess mortality}} \\ \midrule
                        &         (1)   &         (2)   &         (3)   &         (4)   &         (5)   &         (6)   &         (7)   &         (8)   &         (9)   \\
\cmidrule{1-10} \NPIIntensity&        -0.9** &        -0.4   &        -1.2*  &               &               &               &               &               &               \\
                    &       (0.4)   &       (0.6)   &       (0.6)   &               &               &               &               &               &               \\
\NPISpeed           &               &               &               &        -4.3   &        -5.2*  &        -5.8*  &               &               &               \\
                    &               &               &               &       (3.3)   &       (2.9)   &       (3.0)   &               &               &               \\
\HighNPI            &               &               &               &               &               &               &      -120.6***&      -106.5** &      -167.4***\\
                    &               &               &               &               &               &               &      (38.9)   &      (45.2)   &      (41.8)   \\
1917 I\&P mortality (per 100k)&               &         1.0** &         1.4***&               &         1.1** &         1.3***&               &         0.9   &         1.3***\\
                    &               &       (0.5)   &       (0.4)   &               &       (0.5)   &       (0.5)   &               &       (0.5)   &       (0.4)   \\
Longitude           &               &               &        -3.6   &               &               &        -3.0   &               &               &        -4.2   \\
                    &               &               &       (2.5)   &               &               &       (2.8)   &               &               &       (2.5)   \\
Illiteracy          &               &               &        -9.3   &               &               &         2.0   &               &               &        -9.1   \\
                    &               &               &      (15.2)   &               &               &      (13.6)   &               &               &      (13.3)   \\
Medium coal-fired capacity&               &               &       116.9   &               &               &       124.3   &               &               &       162.4*  \\
                    &               &               &      (93.6)   &               &               &      (91.4)   &               &               &      (93.3)   \\
High coal-fired capacity&               &               &       176.2** &               &               &       171.8** &               &               &       200.5** \\
                    &               &               &      (80.9)   &               &               &      (79.4)   &               &               &      (80.3)   \\
\cmidrule{1-10}
\RSq                &        0.07   &        0.23   &        0.37   &        0.05   &        0.28   &        0.38   &        0.15   &        0.31   &        0.48   \\
Number of cities    &          46   &          46   &          46   &          46   &          46   &          46   &          46   &          46   &          46   \\
\Oster              &           .   &        -.21   &        -1.3   &           .   &        -5.5   &        -6.3   &           .   &         -98   &        -188   \\
Baseline controls   &           -   &         Yes   &         Yes   &           -   &         Yes   &         Yes   &           -   &         Yes   &         Yes   \\
   \bottomrule
 \\ 
 \end{tabular}
	\footnotesize
		\begin{tablenotes} \item
	    Notes: This table presents city-level regressions of peak mortality (panel A) and cumulative excess mortality (panel B) on NPI measures.
	    Mortality refers to all-cause mortality, and is bas\textbf{}ed on data from \citet{MortalityStatistics}, \citet{PublicHealthReports}, and \citet{PopCensus1920}; see  \cref{appendix:data-mortality} for details.
	    NPI data is based on \citet{Markel2007}, \citet{Berkes2022}, as well as our own data collection; see \cref{appendix:data-npi} for details.
	    Peak mortality is the weekly excess death rate per 100,000 in the first peak of the fall wave of the 1918 pandemic
	    Cumulative excess mortality is the total excess death rate from September 8, 1918 to February 22, 1919.
	    Baseline controls not reported in table are log of 1900 and 1910 city population, city 1914 manufacturing employment to 1910 population, city density in 1910 and per capita city-level health spending as of 1917.
	    Robust standard errors in parentheses. 
	    *, **, and *** indicate significance at the 10\%, 5\%, and 1\% level, respectively.
\end{tablenotes}
	\end{threeparttable}
\end{table}

In Panel B of \cref{tab:npi_mort_pi}, we examine the relation between NPIs and cumulative excess mortality from influenza and pneumonia over the 24-week period from September 8, 1918 to February 22, 1919. Columns (1) and (7) show that \NPIIntensity and \HighNPI are associated with statistically significantly lower cumulative excess mortality in a regression without controls (see also \cref{fig:excess_mort_NPI}). As in the regressions for peak mortality in Panel A, the estimate on \NPISpeed in column (4) is negative but not statistically significant. Subsequent columns show that the estimates are similar but slightly lower when including additional controls for city characteristics.
However, the estimate on \HighNPI in columns (8)-(9) remains significant at the 1\% level. This estimate implies a reduction in cumulative mortality of 24-34\% relative to the mean, a magnitude similar to \citet{Hatchett2007} (20\% reduction) and \citet{Bootsmaa2007} (10-30\% reduction).

We next study the effect of NPIs on all-cause as opposed to influenza and pneumonia related mortality. Reported data on influenza and pneumonia mortality may contain measurement error due to challenges in establishing cause of death, especially during a period of extremely high mortality. For instance, precise data collection may become difficult when the health care system is overwhelmed, leading to an under-count of influenza and pneumonia mortality in cities that were severely affected by the pandemic. 



The results for all-cause mortality are reported in \cref{tab:npi_mort_all_cause}. There is a high correlation between excess influenza and pneumonia mortality and excess all-cause mortality during the second wave. Hence, the findings in \cref{tab:npi_mort_all_cause} closely align with those from \cref{tab:npi_mort_pi}. As before, cities with high \NPIIntensity and  high \NPISpeed (and thus \HighNPI) experienced significantly lower all-cause mortality. 


In both \cref{tab:npi_mort_pi} and \cref{tab:npi_mort_all_cause}, we also report the coefficients for a selected set of control variables. Our findings confirm some of the relationships between local characteristics and mortality discussed above. For instance, cities with a higher influenza pneumonia mortality in the past tend to also experience higher mortality during the second wave. Further, cities further west tended to have lower mortality. This may be driven by a weakening of the virus strain, a higher degree of voluntary distancing, or other factors correlated with western status. In line with findings in \citet{Clay1918,clay2018}, mortality was higher in cities with a greater reliance on coal-fired energy and a higher illiteracy rate. Finally, areas with a higher past exposure to influenza and pneumonia mortality also experienced higher mortality throughout the second wave of the 1918 pandemic. However, irrespective of the controls, the effect of $High \, NPI$ is essentially unchanged.\footnote{To gauge the potential importance of omitted variable bias, Tables \ref{tab:npi_mort_pi} and \ref{tab:npi_mort_all_cause} also report the bounding value from \citet{Oster2019}. The bounding value provides a sense of the degree of selection on unobservables inferred by the change in the coefficient estimates and $R^2$ following the inclusion of control variables, under the assumption that selection on unobservables is proportional to selection on observables. We follow \citet{Oster2019} and calibrate the maximum $R^2$ to the minimum of 1 or 1.3 times the estimated $R^2$ from the regression with controls. The bound is negative across all specifications, suggesting that our results are not driven by omitted variable bias. }

We provide a range of additional robustness checks in the appendix. For instance, we show that the result that $High \,NPI$ cities had lower mortality is robust to controlling for the timing of the arrival of the flu, measures of city exposure to WWI production and mortality, and infant mortality. It is also robust to excluding the western-most cities in the sample (see \cref{tab:mortality_robustness_pi} and \cref{tab:mortality_robustness_pi_drop_west}).

Taken together, we find a robust effect of NPIs on mortality. Importantly, our analysis reveals that the efficacy of NPIs depends crucially on both the speed at which they are implemented and their intensity. In our analysis, this is reflected by the consistently negative and significant effect of the  $High \ NPI$ measure on all four mortality measures. 


These findings are an important advancement over the existing literature.  \citet{Markel2007} find that NPIs had negative effects on mortality. However, other work either finds no  discernible effects \citep{Kellogg}\footnote{\citet{Kellogg} provided an  early evaluation in 1919 of the effectiveness of NPIs in reducing mortality. He informally compared mortality curves across cities and concluded that NPIs had no discernible effect on mortality \citep{Kellogg}. \citet{Crosby2003}, and \citet{Brainerd2003} argue that NPIs were unlikely to have significantly reduced mortality, while \citet{Barry2004} suggests that NPIs were effective (based on evidence in \citet{Markel2007}). However, these studies do not themselves systematically analyze the relation between NPIs and mortality.} or weak or imprecisely estimated effects  \citep{Hatchett2007,Clay1918,BarroNPI2020}.\footnote{For a smaller sample of 32 cities, \citet{Clay1918} find negative, but not statistically significant, effects of NPIs on 1918 mortality. \citet{BarroNPI2020} finds that NPIs measured by \citet{Markel2007} led to a significant reduction in peak mortality, but finds that NPIs did not reduce cumulative mortality. \citet{BarroNPI2020} suggests this may be because they were not in place long enough.} Our findings support and complement the original findings in \citet{Markel2007}. While we do find that the effect of $NPI \, Intensity$ and $NPI \, Speed$ is sensitive to the exact specification chosen, we find a robust negative effect of NPIs that were \textit{both} fast and aggressive. The effect is precisely estimated, and the magnitudes are meaningful. It is robust to including a wide range of control variables, using four different measures of mortality, and expanding the sample used in \citet{Markel2007}.

The view that NPIs lowered mortality is also in line with several case studies, as depicted in \cref{fig:cities}. As discussed above, the Twin Cities of Minneapolis and St.\ Paul pursued different strategies in response to the pandemic. Minneapolis, which implemented NPIs relatively quickly and aggressively, saw considerably lower peak weekly influenza and pneumonia mortality compared to St.\ Paul (37.6 vs. 55.5 per 100,000). Minneapolis also experienced lower cumulative excess mortality (267.1 vs. 413.2). 

Another interesting comparison is Oakland and San Francisco. Oakland saw significantly lower mortality than San Francisco (506.2 vs. 672.7 cumulative excess mortality). As discussed by \citet{InfluenzaArchive}, ``The reasons for the striking differences between two cities only miles apart are uncertain, but may lie in the relative quickness in each city’s response as well as the length of which epidemic control measures were kept in place: Oakland was twice as quick to respond with closure orders in its epidemic as was San Francisco, and kept those measures in place for nearly twice as long.''


\section{Non-pharmaceutical interventions and economic activity}

\label{sec:economy}
\subsection{Economic activity in the short run}

Were NPIs that flattened the mortality curve associated with a worse downturn in fall 1918? We next examine the impact of the pandemic and NPIs on city-level business disruptions during the pandemic. For this, we rely on a monthly index of business disruptions constructed from \textit{Bradstreet's} trade conditions reports. 

\begin{figure}
    \centering
    \includegraphics[width=1\textwidth]{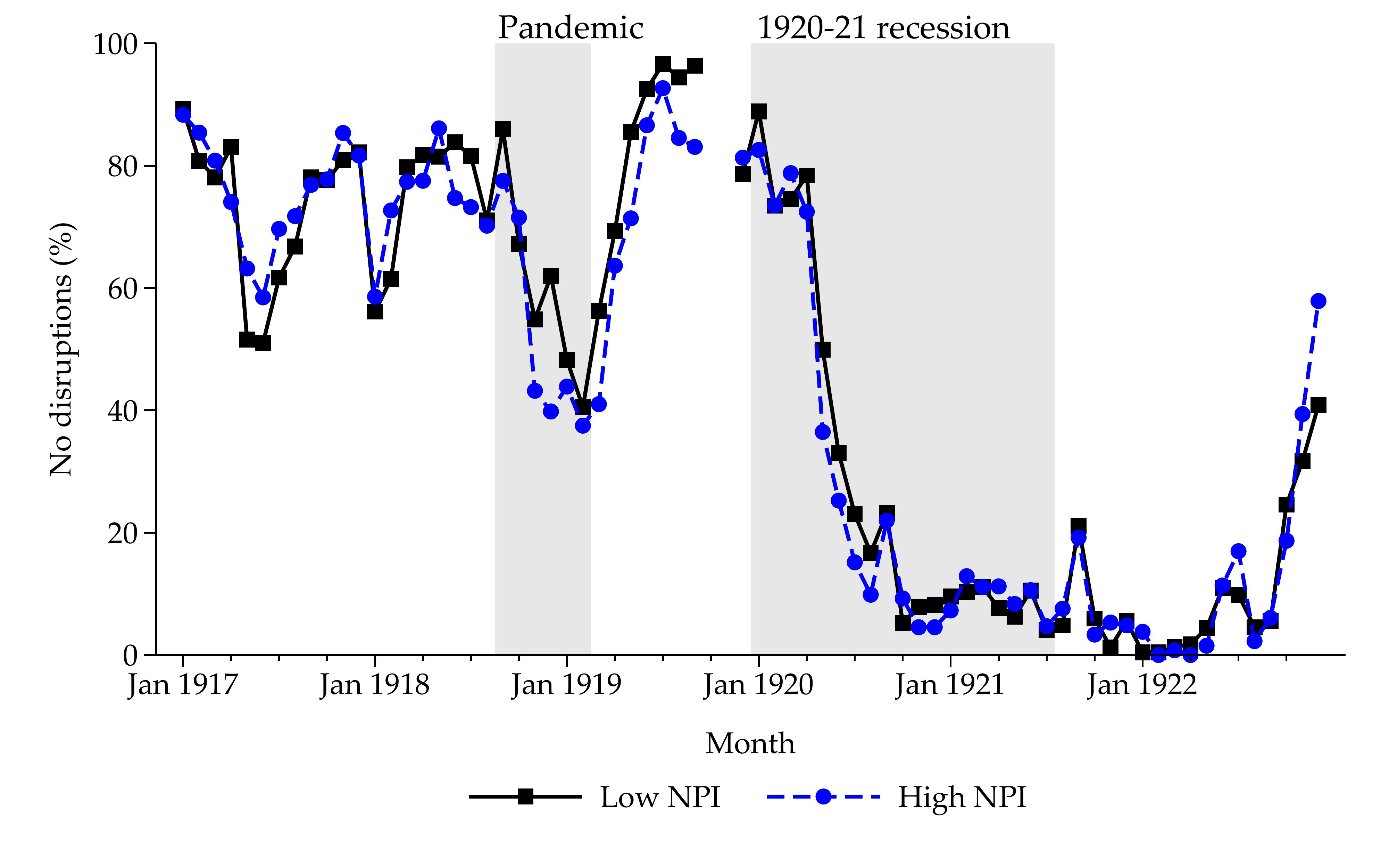}
    \caption{\textbf{Non-pharmaceutical interventions and short-run economic disruptions.} \footnotesize This figure plots the average of an index of economic conditions, for both low and high NPI cities. High NPI cities are defined as cities with above median \NPIIntensity and \NPISpeed. The index is based on the ``Trade at Glance'' tables from \textit{Bradstreet's} weekly magazine, which reported the conditions of three sectors (wholesale, retail, and manufacturing) in brief text snippets (e.g., ``good'', ``poor''). To compute the index, we first convert the snippets into an indicator variable equal to 100 for when there were no disruptions and 0 when there were disruptions. We then average this indicator variable across the three sectors and further aggregate it from a weekly to a monthly frequency. The shaded regions correspond to the 1918-19 Pandemic (September 1918 to February 1918) and to the 1920-21 recession (January 1920 to July 1921). No data is available for October and November 1919 as the magazine was not published due to the New York printing press strikes.}
    \label{fig:bradstreet}
\end{figure}

\Cref{fig:bradstreet} plots the average of our ``No disruptions'' variable across $High \, NPI$ cities -- cities with above-median $NPI \, Intensity$ and $NPI \, Speed$. The figure is based on the sample of 27 cities for which both the index and the information on NPIs is available. ``No disruptions'' are assigned a value of 100; ``Disruptions'' are assigned a value of 0. \Cref{fig:bradstreet} plots the combined index for Wholesale Trade, Retail Trade, and Manufacturing sectors. Appendix \Cref{fig:bradstreet_all} plots the index for each sector separately.

The first takeaway from \cref{fig:bradstreet} is that the pandemic itself was associated with disruptions in economic activity. \Cref{fig:bradstreet} shows that from September 1918 to February 1919 there is a decline in the share of cities with no disruptions.\footnote{Further, studying the disruptions by industry \cref{fig:bradstreet_all} in the appendix shows that the disruptions were most widespread in manufacturing, followed by wholesale trade. The decline in retail trade was more modest, and retail trade saw a rebound already in December 1918.} The business disruptions index then displays a gradual recovery through spring 1919. By early 1920, however, the U.S. economy entered a severe recession. Trade disruption became widespread and even more severe than the repercussions of the 1918 pandemic, in line with the evidence in \citet{Velde2020}.

The second takeaway from \cref{fig:bradstreet} is that the decline in activity during the pandemic was similar in high and low NPI cities. In particular, \cref{fig:bradstreet} shows that high and low NPI cities see approximately equal declines in the combined business disruptions index. For example, from September 1918 to February 1919, high NPI cities saw a 41 point decline in the index, while low NPI cities saw a 52 point decline.  



To  examine the patterns in \cref{fig:bradstreet} formally, \cref{tab:bradstreets} presents results from estimating difference-in-differences models of the form
\begin{align}
    TradeDisruptions_{ct}=\alpha_c + \tau_t + \beta (NPI_c \times Post_t) + (X_c \times Post_t) \Gamma + \epsilon_{ct}, \label{DD_bradstreet}
\end{align}
where $TradeDisruptions_{ct}$ is one of the four trade disruptions indexes from Bradstreet's, $NPI_c$ is one of the three NPI measures, and $X_{c}$ contains a set of city-level controls. The estimation period is January 1918 to March 1919, and $Post_t$ is a dummy that equals one from August 1918 onward. 

Across all three NPI measures, higher NPIs are generally not associated with significant reductions in the combined index, as shown in \cref{tab:bradstreets}. Our preferred specification using the $High \, NPI$ measure with controls (column 6), implies that $High \, NPI$ cities saw a 5.0 point relative decline in economic activity (increase in disruptions). However, the effect is statistically insignificant across all specifications, indicating that there is little evidence supporting the view that NPIs had a substantially negative effect on economic activity. 

\begin{table}
\centering
\begin{threeparttable}
\small
\caption{\textbf{Non-pharmaceutical interventions and local economic activity in Bradstreet's Trade Conditions.}}
\label{tab:bradstreets}
\begin{tabular}{lcccccc}
    \toprule \multicolumn{7}{c}{\textbf{Dependent Variable: Combined Bradstreet Index}} \\ \midrule 
                        &         (1)   &         (2)   &         (3)   &         (4)   &         (5)   &         (6)   \\
 \cmidrule(lr){1-7}  $NPI \ Intensity_c \times Post_{t}$&      -0.036   &      -0.011   &               &               &               &               \\
                    &     (0.065)   &     (0.053)   &               &               &               &               \\
$NPI \ Speed_c \times  Post_{t}$&               &               &      -0.405   &       0.022   &               &               \\
                    &               &               &     (0.356)   &     (0.391)   &               &               \\
 $  High \ NPI_c \times  Post_{t}$&               &               &               &               &      -4.990   &      -5.039   \\
                    &               &               &               &               &     (6.171)   &     (5.657)   \\
\cmidrule{1-7}
\cmidrule(lr){1-7} Within \RSq&        0.00   &        0.06   &        0.01   &        0.06   &        0.00   &        0.06   \\
Observations        &         399   &         399   &         399   &         399   &         399   &         399   \\
Number of cities    &          27   &          27   &          27   &          27   &          27   &          27   \\
\Oster              &           .   &      -.0034   &           .   &         .17   &           .   &        -5.1   \\
Baseline controls   &           -   &         Yes   &           -   &         Yes   &           -   &         Yes   \\
    \bottomrule

\end{tabular}
  \footnotesize
  \begin{tablenotes} \item
     Notes: This table presents estimates of equation \eqref{DD_bradstreet}. The dependent variable is a monthly city-level index of economic disruptions that take a value of 100 for ``No disruptions'' and 0 for ``Disruptions'' (see \cref{appendix:data-bradstreets} for details). Controls interacted with $Post_t$ are the log of 1900 and 1910 city population, 1910 city density, 1917 city health spending per capita, and manufacturing employment in 1914 to 1910 population. Robust standard errors clustered by city in parenthesis.
     *, **, and *** indicate significance at the 10\%, 5\%, and 1\% level, respectively.
  \end{tablenotes}
\end{threeparttable}
\end{table}

Further, in the appendix in \cref{tab:bradstreets_by_index} we also report the effects by the different underlying components of the main index: wholesale trade index (Panel A), retail trade (Panel C), or manufacturing (Panel C). We find both negative and positive estimates, but none is statistically significant. 
Further, in \cref{tab:bradstreet_robustness} in the appendix, we show that the above findings are also robust to including a range of additional controls discussed in section \ref{sec:framework}. Accounting for the timing of the arrival of the flu, WWI exposure, poverty, and air pollution do not alter the conclusion that NPIs did not significantly exacerbate the local economic downturn during the pandemic. The coefficients are small and  insignificant across all specifications except one specification that controls for longitude in conjunction with the baseline controls.\footnote{Controlling only for longitude without our baseline controls does not result in a statistically significant estimate.}


Taken together, monthly information on business disruptions indicates that the cities that were able to flatten the curve through NPIs did not experience  larger disruptions in local business activity as a consequence to their NPI measures. Thus, while the pandemic itself was disruptive for the economy, there is no evidence supporting a view that these public health interventions  exacerbated the disruptions of economic activity.

A specific concern with this interpretation is that the Bradstreet's ``Trade at Glance'' sample is relatively small (27 cities), so there may not be enough power to detect modest negative effects.
It is therefore useful to consider what effect sizes we can reject.
Looking into the \HighNPI estimate for the combined index in Panel A column (6), its 90\% confidence interval is $(-14.7, 4.6)$. At this level, we can reject negative effects below -14.7, which is one-third of the peak-to-trough decline during the pandemic, and about one-sixth of its decline during the 1920-21 recession.\footnote{To define the peak-to-trough decline, we first collapse the data to a monthly frequency. The peak is the maximum across the quarter preceding the decline, and the trough is the minimum across the decline period.}
With the industry-specific indexes, the standard errors are larger, so there is more uncertainty about these point estimates. Nonetheless, across all specifications, there is no clear evidence of large negative effects of NPIs on economic activity in the short run. 
\subsection{Economic activity in the medium run}

Our findings in the previous section indicate that cities with stricter NPIs did not experience more severe short-term business disruptions. We now examine how NPIs affected economic activity in the medium run after the pandemic. To do so, we use city-level data from the Census of Manufactures on employment and output. The Census data have several advantages relative to the \textit{Bradstreet's} data. First, the Census data are measures of actual economic outcomes instead of qualitative reports, which may be subjective. Second, the Census data cover all 46 to 54 cities for which we can obtain NPI data, which doubles the sample compared to the \textit{Bradstreet's} analysis.\footnote{The Census of Manufactures' city-level data covers 54 cities for which we have data on $NPI \ Intensity$, as this measure does not rely on weekly information about precisely when mortality accelerated.} However, a drawback of the Census of Manufactures is that it was only collected every five years until 1919 and every two years from 1919 onwards. As a result, the Census is not informative about pre-trends between 1915 and 1918 or the immediate effects of the pandemic. Instead, the Census allows us to analyze whether NPIs had beneficial or damaging effects on the economy in the medium run.

To  study the medium-term impact of NPIs around the 1918 Flu Pandemic and to control for other observable characteristics and longer pre-trends, we estimate a dynamic difference-in-differences equation of the form
\begin{align}
    Y_{ct} =\alpha_c  + \tau_t + \sum_{j\neq 1914}\beta_j \, NPI_{c} \, \mathbf{1}_{j=t} + \sum_{j\neq 1914} X_s \, \gamma_j \, \mathbf{1}_{j=t}  + \varepsilon_{ct}, \label{eq:spec2}
\end{align}
where $Y_{ct}$ is a measure of economic activity in city $c$, such as the log of manufacturing employment,  $NPI_{c}$ is one of the NPI measures, $\alpha_c$ is a city fixed effect,  $\tau_t$ is a time fixed effect, and $X_{s}$ is a set of control variables that are interacted with time indicator variables to allow for changes in the relationship between outcome variables and controls. The set of coefficients ${\beta_j}$ capture the relative dynamics of cities with strict versus lenient NPIs.

\begin{figure}[htpb]
    \centering
    \subfloat[Impact of $NPI \ Intensity$ on log manufacturing employment. \label{fig:coefplot_intensity}]{
   \includegraphics[width=0.47\textwidth]{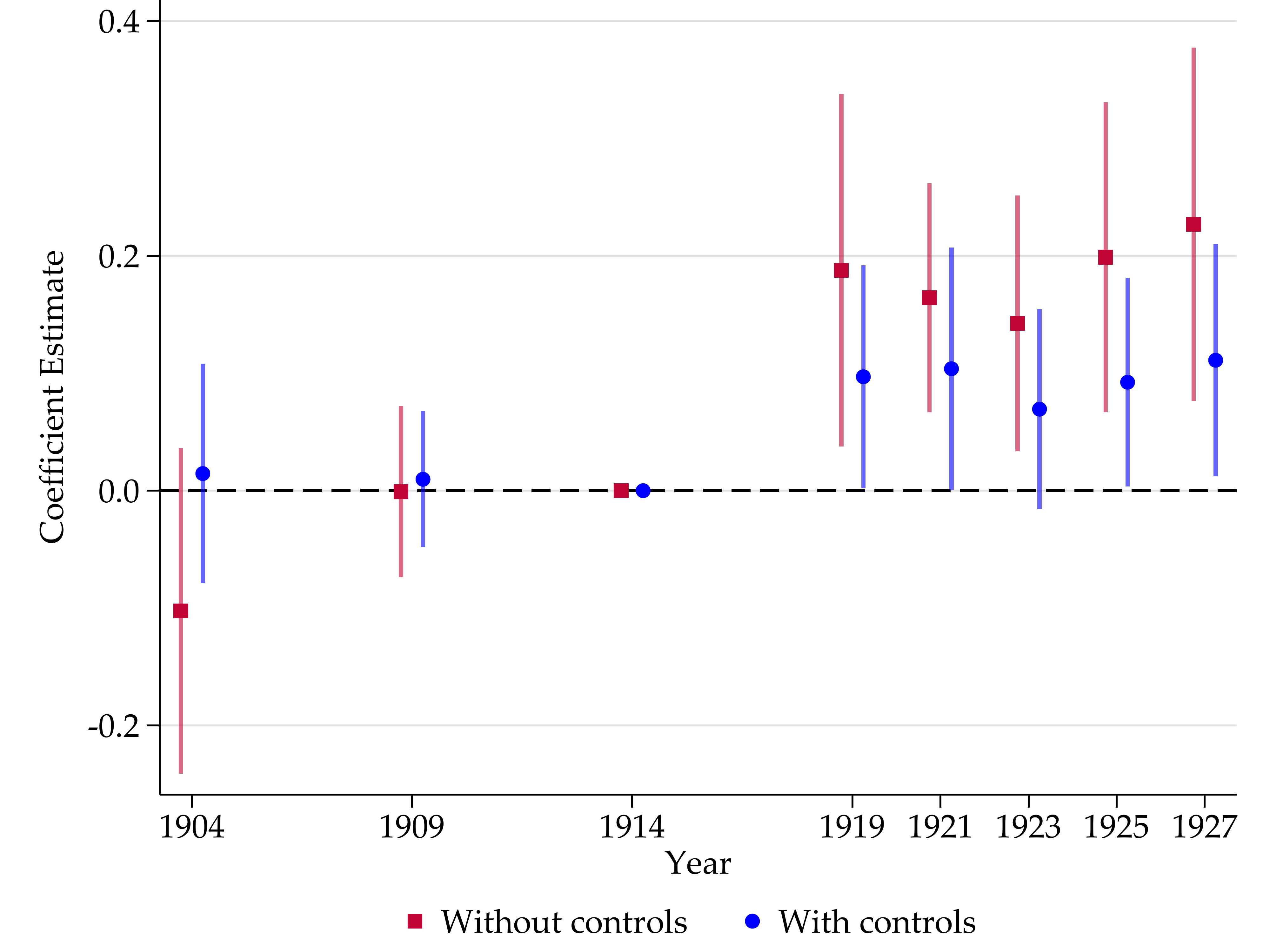}}
     \hfill
   \subfloat[Impact of  $High \ NPI$ on log manufacturing employment. \label{fig:coefplot_highnpi}]{
   \includegraphics[width=0.47\textwidth]{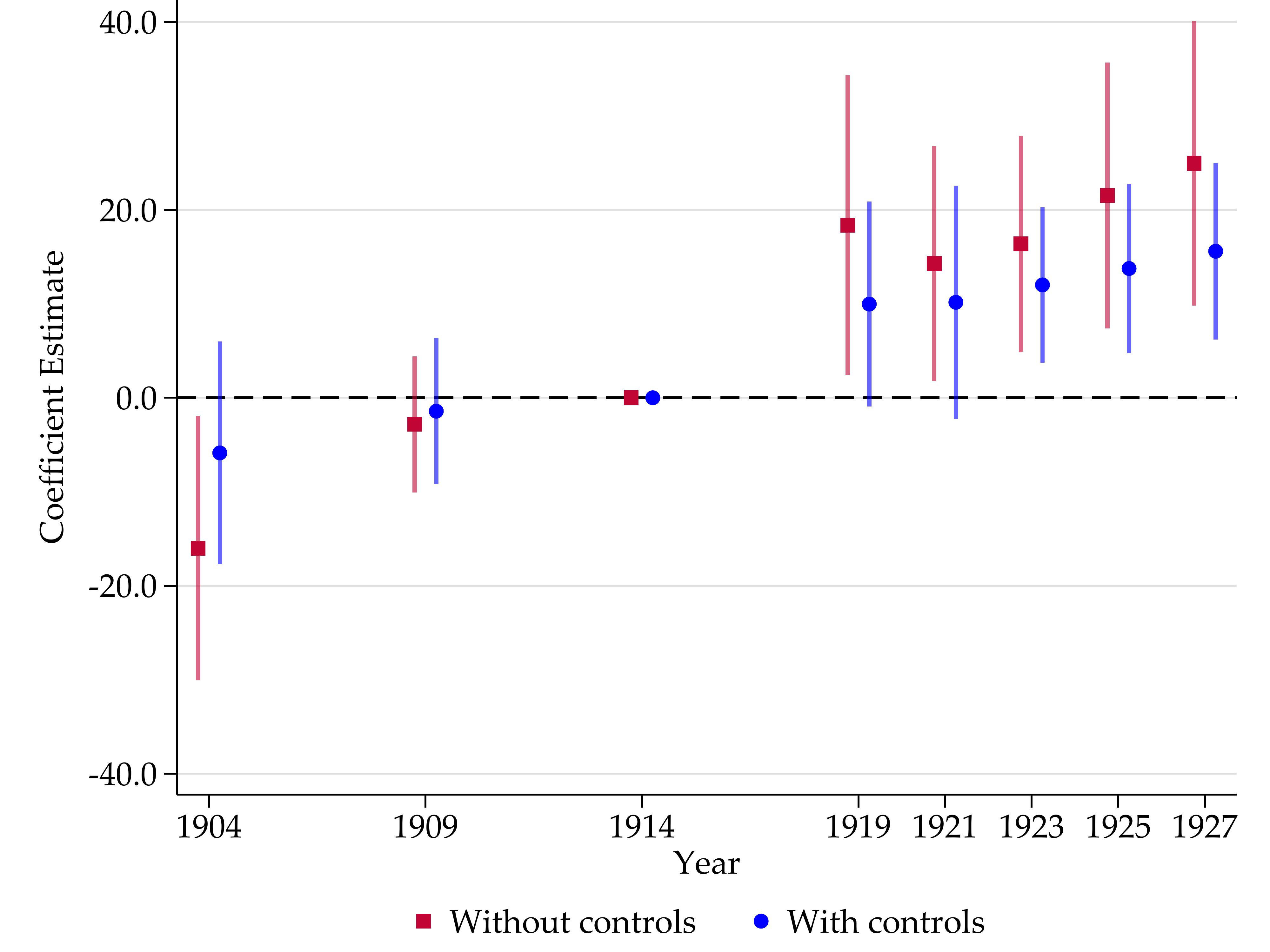}}
   \caption{\textbf{Non-pharmaceutical interventions in fall 1918 and manufacturing employment growth across U.S. cities.} \footnotesize This figure presents results from estimating equation \eqref{eq:spec2} on log manufacturing employment with and without baseline controls. Baseline controls are city log 1900 and 1910 population, city 1914 manufacturing employment to 1910 population, city density in 1910, and per capita city-level health spending as of 1917. Panels (a) and (b) use $NPI\; Intensity$ and $High\; NPI$ as the NPI measures, respectively. Error bands denote 95\% confidence intervals with robust standard errors clustered at the city level.}
    \label{fig:NPI_coefplot}
\end{figure}

\Cref{fig:NPI_coefplot} presents the results from estimating \eqref{eq:spec2}
for manufacturing employment using the $NPI \, Intensity$ and $High \, NPI$ measures as regressors. The estimates without controls show that, relative to 1914, cities with stricter NPIs had a higher level of employment from 1919 onward than those with more lenient NPIs. For instance, the estimate for 1919 implies that $High \ NPI$ cities experienced 18\% higher employment growth from 1914 to 1919. Further, the confidence bands indicate that growth lower than 2\% can be rejected at the 95\% level.  

However, the estimates without controls in panel (b) of \cref{fig:NPI_coefplot} also show that cities with stricter NPIs  grew faster between 1904-1909, indicating a pre-trend from 14 to 9 years before the pandemic. This raises the concern that the results may be driven by more general city-growth patterns. This is not entirely surprising given that most cities with strict NPIs were located further west, and as the structure of the U.S. economy changed quickly at the turn of the 20th century, the westernmost cities such as Los Angeles and Seattle experienced some of the most rapid expansion.\footnote{For a detailed discussion of pre-trends in this context, see \citet{llr}, \citet{clv}, and \citet{santanna}. \citet{llr} raise a specific concern that the city-level population growth from 1910 to 1917 is a confounding factor that explains the positive correlation between NPIs and employment growth from 1914 to 1919. However, this 1917 population value is primarily based on a linear extrapolation of population growth between 1900 and 1910 censuses. Hence, that variable reflects population growth from 1900 to 1910, not 1910 to 1917, which leads to a large and systematic measurement error of population growth from 1910 to 1917 \citep[for additional details, see the discussion in][]{clv}.}

One approach to addressing this concern is to control for observable differences across cities with strict and lenient NPIs.
Once we include our baseline controls, the estimates after 1918 remain positive but are sometimes only significant at a 90\% level.\footnote{The baseline controls are listed in the note in \cref{fig:NPI_coefplot}.}  For instance, the point estimates in panel (b) indicate that cities with high NPIs had 10\% higher manufacturing employment in 1919 compared with low NPI cities. The 95\% confidence interval is $(-1\%,20\%)$, ruling out substantial negative effects. In terms of economic significance, the average growth of manufacturing employment between 1914 and 1919 was 33 percent, with a standard deviation of 22 percent. Thus, the point estimate corresponds to one-third of the average growth and one-half of the cross-city standard deviation in growth. 

To confirm this visual pattern, \cref{tab:reg_NPI} compares the pre- and post-period average in manufacturing employment and output for cities with strict and lenient NPIs, controlling for city observables. The estimates for both employment and output are generally positive across all three NPI measures.  The estimates are not always significant, but the point estimates suggest moderate positive effects. Our preferred specification suggests $High \ NPI$ cities see around 17\% higher manufacturing employment and 12\% higher manufacturing output after the pandemic (column 6). The confidence intervals reject a large negative effect of NPIs on both measures of economic activity. For example, based on the estimates in column 6, we can reject at the 95\% level that the effect of $High \, NPI_c$ on employment and output are below $6.2\%$ and $-5.0\%$, respectively. 


\begin{table}
\centering
\begin{threeparttable}
\caption{\textbf{ Non-pharmaceutical interventions and local manufacturing employment and output.}}\label{tab:reg_NPI}
\small
\begin{tabular}{lcccccc}
    \multicolumn{7}{c}{\textbf{Panel A: Manufacturing employment}} \\ \toprule
                        &         (1)   &         (2)   &         (3)   &         (4)   &         (5)   &         (6)   \\
 \cmidrule(lr){1-7}  $NPI \ Intensity_c \times Post_{t}$&       0.218***&       0.087** &               &               &               &               \\
                    &     (0.076)   &     (0.036)   &               &               &               &               \\
$NPI \ Speed_c \times  Post_{t}$&               &               &       0.753*  &       0.334   &               &               \\
                    &               &               &     (0.390)   &     (0.250)   &               &               \\
 $  High \ NPI_c \times  Post_{t}$&               &               &               &               &      25.386***&      14.728***\\
                    &               &               &               &               &     (8.033)   &     (4.356)   \\
\cmidrule{1-7}
\cmidrule(lr){1-7} Within \RSq&        0.10   &        0.40   &        0.03   &        0.40   &        0.14   &        0.44   \\
Observations        &         428   &         428   &         368   &         368   &         368   &         368   \\
Number of cities    &          54   &          54   &          46   &          46   &          46   &          46   \\
\Oster              &           .   &        .034   &           .   &          .2   &           .   &          10   \\
Controls            &           -   &         Yes   &           -   &         Yes   &           -   &         Yes   \\
    \bottomrule
 \\
    \multicolumn{7}{c}{\textbf{Panel B: Manufacturing output}} \\ \midrule 
                        &         (1)   &         (2)   &         (3)   &         (4)   &         (5)   &         (6)   \\
 \cmidrule(lr){1-7}  $NPI \ Intensity_c \times Post_{t}$&       0.113   &      -0.005   &               &               &               &               \\
                    &     (0.078)   &     (0.053)   &               &               &               &               \\
$NPI \ Speed_c \times  Post_{t}$&               &               &       0.742*  &       0.351   &               &               \\
                    &               &               &     (0.392)   &     (0.347)   &               &               \\
 $  High \ NPI_c \times  Post_{t}$&               &               &               &               &      18.980** &       8.749   \\
                    &               &               &               &               &     (8.862)   &     (6.997)   \\
\cmidrule{1-7}
\cmidrule(lr){1-7} Within \RSq&        0.02   &        0.22   &        0.02   &        0.23   &        0.06   &        0.24   \\
Observations        &         428   &         428   &         368   &         368   &         368   &         368   \\
Number of cities    &          54   &          54   &          46   &          46   &          46   &          46   \\
\Oster              &           .   &       -.043   &           .   &         .22   &           .   &         4.6   \\
Controls            &           -   &         Yes   &           -   &         Yes   &           -   &         Yes   \\
    \bottomrule
 \\
\end{tabular}
\footnotesize
\begin{tablenotes} \item
Notes: This table presents estimates of equation \eqref{DD_bradstreet}.
The dependent variables are the log of manufacturing employment (Panel A) and log of manufacturing output (Panel B), using data from the 1904, 1909, 1914, 1919, 1921, 1923, 1925, and 1927 census.
Controls interacted with $Post_t$ are the share of manufacturing employment in 1914, log of population in 1900 and 1910, city density in 1910, and per capita city health spending in 1917.
Robust standard errors clustered by city in parenthesis. *, **, and *** indicate significance at the 10\%, 5\%, and 1\% level, respectively.
\end{tablenotes}
\end{threeparttable}
\end{table}

In Appendix \cref{tab:manufacturing_robustness}, we provide additional robustness checks and show that the above results are robust to include a wide set of additional controls. We show that the results are robust to controlling for longitude, the timing of the flu's arrival, WWI exposure, poverty, and air pollution, as discussed in section \ref{sec:framework}. Furthermore, in \cref{tab:manufacturing_robustness_west} we show that the same holds true even when excluding the westernmost cities. Fast and stringent NPIs are never associated with negative effects on local economic activity in the aftermath of the flu. The estimates are positive across all specifications and, in some cases, statistically significant at the 5\% level.

As mentioned above, a key drawback of using the Census of Manufacturers is that data are not available from 1915 through 1918. This raises the concern that cities with strict and lenient NPIs could have been on different trajectories in this time period.\footnote{However, the results using the monthly \textit{Bradstreet's} trade conditions in \cref{fig:bradstreet} do not point to a differential pre-trend in local economic activity.} To partly address this concern, we use annual data on total national bank assets as a proxy of local economic activity.\footnote{The reporting date for the OCC data on national bank assets is either August or September of a given year. We normalize coefficients to August 31, 1918. In order to account for Liberty Bond issuance impacting the local banking system \citep[see, e.g.,][]{Hilt2020}, we control for the ratio of the amount subscribed to the Third Liberty Bond to total bank assets in 1918. Moreover, to account for the 1913 founding of the Federal Reserve System as well as for heterogeneity across Federal Reserve Districts, we include reserve district-year fixed effects.}  With our standard controls, there is no indication of a pre-trend in national bank assets for cities with stricter NPIs between 1910 an 1917 (see \cref{fig:NPI_coefplot-national-banks}). This is reassuring because if cities with stricter NPIs were growing at a faster pace before 1918, this should arguably be reflected in the size of the local banking system.\footnote{Further, there is an uptick in bank assets using both the  $High \, NPI$ and  $NPI \ Intensity$ measure after August 1918. This suggests that there was a slight increase in the size of the local banking system in the medium term for cities with strict NPIs. These patterns can be confirmed by comparing the pre- and post-period average in bank assets in \cref{tab:reg_NPI-national-banks}, which suggest that national bank assets growth tended to be higher in cities with stricter NPIs after 1918.} 


\subsection{Newspaper evidence of effects of the pandemic and NPIs on economic activity}

The economic disruption from the pandemic and the effect of NPIs were extensively discussed in contemporary newspaper accounts. In this section, we present narrative evidence that the pandemic led to significant economic disruptions, fear of the virus, and voluntary distancing. While NPIs directly reduced the revenues of businesses subject to closures, newspapers also report that these businesses saw a large decline in sales even in the absence of NPIs due to voluntary distancing and the spike in illness and mortality.\footnote{Appendix \ref{appendix:narrative} provides further evidence from contemporary newspaper accounts of impact of the pandemic and NPIs on economic activity. \citet{Garrett2008} also provides narrative evidence from local newspaper reports that the pandemic caused severe disruption to businesses in many sectors of the economy, but he does not discuss the impact of NPIs on economic or social activity.} These narrative accounts thus provide clues for why NPIs reduced mortality without significantly reducing economic activity. 


Newspaper accounts indicate that the pandemic depressed the economy through both supply and demand-side channels in the form of productivity reduction, labor shortages, and falling demand for retail goods. Output declines were seen across many sectors of the economy, including coal and copper mining, shipbuilding, textile production, retail and wholesale trade, and entertainment.
For example, on October 24, 1918, the \textit{Wall Street Journal} reported:
\begin{quote}
 \emph{In some parts of the country [the influenza epidemic] has caused a decrease in production of approximately 50\% and almost everywhere it has occasioned more or less falling off. The loss of trade which the retail merchants throughout the country have met with has been very large. The impairment of efficiency has also been noticeable. There never has been in this country, so the experts say, so complete domination by an epidemic as has been the case with this one.} (WSJ, Oct. 24, 1918.)
\end{quote}

\paragraph{Supply-side effects.} Contemporary accounts indicate that the pandemic depressed the economy through supply-side channels in the form of productivity reduction and labor shortages (see Appendix \ref{sec:newspapers_direct}). Businesses in many sectors reported labor shortages due to illness and death. The Bell Telephone Company called on citizens of Philadelphia to ``use the telephones as little as possible during the emergency as its service is already being rushed to the breaking point. The company has more than 850 operators or more than twenty-six percent of the entire force, out owing to illness from the epidemic.''\footnote{``Influenza Leaps As 1,480 New Cases Are Listed Today.'' {\normalfont \emph{Philadelphia Evening Bulletin}, October 4, 1918.}} Meat packing plants in Omaha were ``crippled'' by the epidemic; one large meat packing plant reported 30\% of employees were absent due to influenza.\footnote{``Influenza Cripples Omaha Packing Plants. One of the Larger Concerns Reports Thirty Per Cent Ill With Malady. All Four of Big Ones Show Large Number of Employees Not at Work.'' {\normalfont \emph{ The Omaha World-Herald}, October 18, 1918.}} Garbage collection in Oakland came to standstill as 75 out of 90 garbage collectors were sick with influenza.\footnote{``Arrest All Maskless As Cases Grow.'' {\normalfont \emph{Oakland Tribune}, November 1, 1918, p. 18.}}

\paragraph{Demand-side effects and voluntary distancing.} Fear of influenza depressed demand and social activity in many cities. On October 25, 1918, the \textit{Wall Street Journal} reported:
\begin{quote}
\emph{Widespread epidemic of influenza has caused serious inroads on the retail merchandise trade during the current month. Heads of large organizations report that not only has sickness cut down the shopping crowds, but in many cities the health authorities have shut down the stores. The chain store companies have felt the effect of the sickness not a little, for in addition to the smaller business done a number of their employees are sick.} (WSJ, Oct. 25, 1918.)
\end{quote}
Cafes, restaurants, and saloons in Oakland closed early as a result of ``a lack of business due to the Spanish influenza epidemic.''\footnote{``Bars, Cafes in Dark; Patrons Are Missing'' {\normalfont \emph{Oakland Tribune}, October 24, 1918.} See Appendix \ref{sec:newspapers_voluntary} for additional examples.} In many cities, the streets were silent. For example, a doctor who attended medical school in Philadelphia during the 1918 pandemic recounted that he did not see a single other car on a late-night drive home from the emergency hospital where he worked \citep{Starr}.

Theater attendance was depressed in many cities, even before NPIs closed theaters. In New York, the \emph{New York Times} reported that ``an unprecedented theatrical depression, which managers attribute in large part to the influenza scare, resulted in sudden decisions yesterday to close five playhouses tonight.''\footnote{``5 Theatres Close Tonight: Theatrical Depression Attributed in Large Part to Influenza Scare.'' {\normalfont\emph{New York Times}, October 12, 1918, p. 13.}} St. Paul was slower to close movie theaters than Minneapolis. But movie theaters in St. Paul still experienced a significant loss of revenues. The \emph{St. Paul Pioneer Press} reported that ``fear of influenza contagion in crowded places has reduced the patronage of St. Paul motion picture theaters by nearly half.''\footnote{``See Less Influenza'' {\normalfont \emph{St. Paul Pioneer Press,} October 15, 1918, p. 8.} On the other hand, the announcement of closures in Minneapolis led citizens to rush to activities that would be closed for the coming weeks (see Appendix \ref{sec:newspapers_NPIbehavioral}).} As discussed in section \ref{sec:background}, some theater owners supported NPIs because they believed that revenues would fall even in the absence of NPIs.

While the pandemic reduced demand for in-person activities, it also led to a increase in spending on toys, games, books, and magazines, as people searched for home entertainment while places of public amusement were closed. A department store clerk in Spokane said that customers ``come in and say `Well, give us a magazine. We can’t do anything else so we may as well read.'"\footnote{``Influenza Edict Aids Home Life. Demand for Parlor Games and Books Resembles Annual Yuletide Rush. Streets Look Deserted.'' {\normalfont\emph{The Spokesman-Review} (Spokane, WA), October 11, 1918.}}

When schools were not closed, school absenteeism was high in cities severely affected by influenza, both because children were ill and because many parents kept their children home for fear that they would become infected. A survey of Cincinnati schools from December 2, 1918, found that 32\% of students were absent, 12\% of which were ill with influenza and the remaining 20\% kept home as a precautionary measure by their parents. Pittsburgh, which only closed schools three weeks after implementing other closure orders, saw attendance fall 30\% from normal levels before schools were closed. In an order to the Pittburgh school superintendent, the Director of the Department of Health Dr. W. H. Davis noted  that the rise in absenteeism was ``very largely due to the fears of the parents'' and that ``it is reasonable to expect that this absenteeism will increase until the feeling of fear has grown less.''\footnote{``City Schools Are Closed by Grip Fighters'' {\normalfont \emph{The Pittsburgh Gazette Times}, October 24, 1918.}}

Chicago, New Haven, and New York were the only cities in our sample that did not close schools. In Chicago, absentee rates reached 50\% \citep{InfluenzaArchive}. In New York, the Health Commissioner reported that ``careful survey shows that about half of the absences are due to the fear of parents and not to the illness of children.''\footnote{``Copeland Refuses to Close Schools.''\footnote{\normalfont \emph{New York Times}, October 19, 1918, p. 24.}} The \emph{New Haven Evening Register} reported that in three New Haven schools, ``the attendance in the rooms has fallen from 40 to 50, which is a normal attendance, to seven or eight with a maximum of fourteen.'' Influenza also disrupted teaching in New Haven, where teachers absences reached 80 or 90 percent and the school board had difficulty in finding substitutes.\footnote{``155 New Cases And 4 Deaths In Flu Wave--But Health Officials Think Worst of Epidemic Is Over--Not Vaccinating School Children Yet. Teachers for Nurses.'' {\normalfont \emph{New Haven Evening Register}, October 22, 1918.}} 

Fear of influenza also had other effects on social activity. For example, it affected the 1918 election in some cities. The \textit{Oakland Tribune} reported that many polling stations were understaffed, as many election officials were ill and ``substitutes have been difficult to obtain by reason of fear of the epidemic.''\footnote{``Early Vote is Light; Officials Lacking--County Handles Election Under Difficulties\dots'' {\normalfont \emph{Oakland Tribune}, November 5, 1918.}} Influenza also reportedly lowered turnout in some precincts in the Bay Area.

\paragraph{Direct cost of NPIs on businesses.} While the pandemic itself clearly depressed economic activity, NPIs also directly impacted some businesses. The businesses directly affected by closures in the entertainment, restaurant and retail, and hospitality sectors reported significant revenues losses. In Salt Lake City, the labor union reported 1,000 employees, including 300 musicians and several hundred theater workers, were out of employment for nine weeks during the closing order.\footnote{``Influenza Rules Will Be Rigidly Enforced.'' {\normalfont \emph{Salt Lake Tribune}, December 8, 1918.} See Appendix \ref{sec:newspapers_costs} for additional examples.} 
In Rochester ``400 persons directly connected with the theaters have been thrown into idleness and the majority of them deprived of their salaries'' due to the closure orders.\footnote{``Theatrical People Hit By Epidemic. Fully 300 at Gayety Theater Affected by Closing Order. Shows Forced To Lay Off. Proves Especially Tough on Chorus Girls.'' {\normalfont \emph{Rochester Times-Union}, October 19, 1918.}} \emph{The Birmingham News} news listed the estimated costs to theaters, movie theaters, and department stores, concluding that ``undertakers are the only ones who have profited from the epidemic'' from the financial perspective.\footnote{However, the newspaper went on to say: ``the suffering that [undertakers] have seen for the last four weeks, the grieving of relatives for their loved ones and the burying of some of their own friends have cost them in heart aches.'' See ``Influenza Cost City More Than Half A Million—Four Hundred of 10,000 Died, But Plague Is Over and Town Is Reopened.'' {\normalfont \emph{The Birmingham News}, October 31, 1918.}} 


There are also accounts of people circumventing NPIs, illustrating frustration with closures and restrictions. In some place, the closures led to a reallocation of activity toward counties that were not closed. The \textit{Toledo News - Bee} newspaper reported that outside Toledo, OH, ``bars in the county, and adjacent to the city are `getting a big play' from pedestrians and auto parties.''\footnote{``May Lift Order In 10 Days.'' {\normalfont \emph{Toledo News - Bee}, October 15, 1918, p. 1, 2.}} After weeks of the pandemic and associated restrictions, pandemic fatigue set in. For instance, the mayor of Cincinnati noted that ``the people are tired of hearing of influenza and want to forget it. The psychological time for raising the restrictions has arrived. You can no more control the people’s enthusiasm nor regulate their actions on the street than you can control the Ohio River.''\footnote{``It’s Off! Influenza Lid Lifted,’’ \emph{The Cincinnati Enquirer,} November 12, 1918.} 

Reopenings in November 1918 thus led to short-run booms in spending on activities that had been closed in some cities. For example, in Omaha, the reopening was met with excitement, as stores and theaters planned for strong demand.\footnote{See Appendix \ref{sec:newspapers_reopening}.} In Syracuse, NY, ``theaters filled to capacity and church services attended by entire congregations gave definite proof\dots that the full confidence of the public, once shattered by the epidemic, has been restored.''\footnote{``Theaters and Churches Filled to Capacity Show Fear of Epidemic Is Over. Record Audiences Flock to Playhouses and Pastors Greet Unusual Congregations.'' {\normalfont \emph{The Post-Standard} (Syracuse, NY), October 28, 1918.}}  However, in other cities, such as Fall River, attendances at theaters and saloons remained low as residents remained concerned about influenza.\footnote{``Epidemic Still Affects Business. People Evidently Are Not Taking Chances in Assemblages. 26 Cases and Nine Deaths Today’s Report.'' {\normalfont \emph{ Fall River Evening Herald} October 25, 1918.}}

\paragraph{Relative costs of NPIs and mortality from the pandemic.} We close this section by noting that some contemporaries compared the direct costs of NPIs to the costs of deaths from influenza based on the value of a statistical life. In Philadelphia, movie theaters, theaters, liquor stores, and the Transit Company estimated total losses of \$2,000,000 during the period when the Pennsylvania's closures order was in effect in Philadelphia. When these costs came to the attention of Acting State Commissioner of Health, Dr. B. Franklin Royer, he estimated that that the cost in terms of the 11,000 estimated deaths was an order of magnitude larger: ``The loss of life in the city of Philadelphia alone during the present epidemic of influenza, if figured in dollars and cents according to estimates on the value of human life fixed by medical and insurance statisticians and court decisions amounts to approximately fifty-five million dollars since October 1''.\footnote{``Ban Raised Here As Epidemic Ends.'' {\normalfont \emph{Philadelphia Inquirer}, October 31, 1918.}}

In January of 1919, following a resurgence in cases, a pastor of a church in Portland called for a renewed 30-day closure to stamp out the epidemic. Citing an estimate of the statistical value of life computed by economist Irving Fisher, he argued that failing to prevent preventable deaths was more costly to society than the financial losses from a shut-down, which were often cited by those opposed to closures.\footnote{See ``Portland Churches Ask Epidemic Ban,'' \textit{The Oregonian}, January 20, 1919 and Appendix \ref{sec:newspapers_notradeoff}.} 

\section{Conclusion}

\label{sec:conclusion}

Using both narrative evidence from contemporary newspapers and statistical analysis, this paper examines the impact of non-pharmaceutical interventions during the 1918 Flu Pandemic on mortality and economic activity. We find that while NPIs flattened the curve of disease transmission, they were not associated with worse economic performance during or after the pandemic. Instead, our findings suggest that the main source of economic disruption was the pandemic itself.

There are several important caveats to our results. First, our empirical analysis is limited to 27 to 46 cities, resulting in non-trivial uncertainty around some point estimates. Second, we cannot examine pre-trends for manufacturing outcomes in the years 1915 through 1917, as the data is not available at an annual frequency. Third, the economic environment toward the end of 1918 was unusual due to the end of WWI. Fourth, our cross-regional analysis does not allow us to capture aggregate equilibrium effects of NPIs. Nevertheless, the collection of evidence from various data sources and contemporary newspaper accounts paint a picture of significant health benefits of NPIs at limited economic cost, if at all. 

Our findings raise the question: Why were NPIs not economically harmful during the pandemic, and possibly even beneficial in the medium-term? It is challenging to shed light on the exact mechanisms through which NPIs affected the economy with the limited data available for 1918, but we discuss some potential channels that are supported by the narrative evidence.

The direct effect of NPIs such as theater closures and public gathering bans is contractionary, as these policies necessarily restrict economic activity. Newspaper accounts of lobbying by theater and store owners against closures reflect this direct negative effect. However, the pandemic itself can be highly disruptive for the economy. Many activities that NPIs restrict would likely not have occurred even in the absence of NPIs. Narrative accounts reveal that there was significant voluntary distancing. For example, attendance in theaters, schools, and churches was significantly depressed even when NPIs were not implemented. Moreover, the pandemic resulted in significant disruptions in production due to precautionary reductions in labor supply, illness, and mortality. As a result, the counterfactual without NPIs would still involve a downturn.

Moreover, NPIs may have indirect economic benefits by addressing the root of the economic disruption---the pandemic itself---in a coordinated fashion. Mitigating the pandemic may have prevented an ultimately worse economic downturn. Our narrative evidence suggests some contemporaries had this view. For example, shipbuilding companies lobbied for stricter NPIs to reduce the spread of the disease and thereby limit labor shortages and associated disruptions to production. 
Further, by reducing cumulative infection rates, NPIs may have medium-term economic benefits by directly reducing illness and mortality and by reducing the costs associated with increased morbidity.


More specific historical details also shed light on why NPIs in 1918 did not worsen the economic downturn. NPIs implemented in 1918 were milder than the measures adopted in some countries during COVID-19. On the spectrum of costs and benefits of specific NPIs, many measures implemented in 1918---such as public gathering bans---had relatively modest economic costs. More severe measures such as widespread closures of businesses likely increase the cost of NPIs. School closures were less costly in 1918, as female labor force participation was lower. Estimates suggest that 1918 Flu was more deadly than COVID-19, especially for prime-age workers, which also suggests more severe economic impacts of the 1918 Flu and greater medium-run benefits of NPIs. The 1918 H1N1 virus also had a shorter incubation period than COVID-19, which facilitated identifying and isolating suspected cases. As a result, we stress the limits of external validity of lessons from the 1918 Flu Pandemic. Despite these important differences, ongoing research  finds that NPIs implemented in 2020 have reduced disease transmission \citep{Allcott2020}, while accounting for a relatively small fraction of the overall decline in economic activity \citep[see, e.g.,][]{Sheridan2020,Baek2020,DemirgucKunt2020,Goolsbee2020}. 

\clearpage
\singlespacing
\bibliographystyle{chicago}
\bibliography{literature.bib} 
\clearpage
\doublespacing

\begin{appendices}

\crefalias{section}{appendix}
\crefalias{subsection}{appendix}

    \begin{itemize}
    \tightlist
    \item \Cref{appendix:supplementary}\hyperref[appendix:supplementary]{: Supplementary Tables and Figures}
    \item \Cref{appendix:narrative}\hyperref[appendix:narrative]{: Narrative Evidence from Historical Newspaper Archives}
    \item \Cref{sec:data-appendix}\hyperref[sec:data-appendix]{: Data Appendix}
      \begin{itemize}
      \tightlist
      \item \Cref{appendix:data-mortality}\hyperref[appendix:data-mortality]{: Mortality statistics}
      \item \Cref{appendix:data-npi}\hyperref[appendix:data-npi]{: Non-pharmaceutical interventions}
      \item \Cref{appendix:data-bradstreets}\hyperref[appendix:data-bradstreets]{: Bradstreet's Trade Conditions}
      \item \Cref{appendix:data-manufacturing}\hyperref[appendix:data-manufacturing]{: Census of Manufactures}
      \item \Cref{appendix:data-military}\hyperref[appendix:data-military]{: Military camps}
      \item \Cref{appendix:city-mortality-npi}\hyperref[appendix:city-mortality-npi]{: City-level mortality and NPIs}
      \end{itemize}
    \end{itemize}

\section{Supplementary Figures and Tables}\label{appendix:supplementary}
\setcounter{figure}{0}
\renewcommand{\thefigure}{A\arabic{figure}}

\begin{figure}[ht]
    \centering
    \subfloat[NPI speed and I\&P mortality.]{\includegraphics[width=0.45\textwidth]{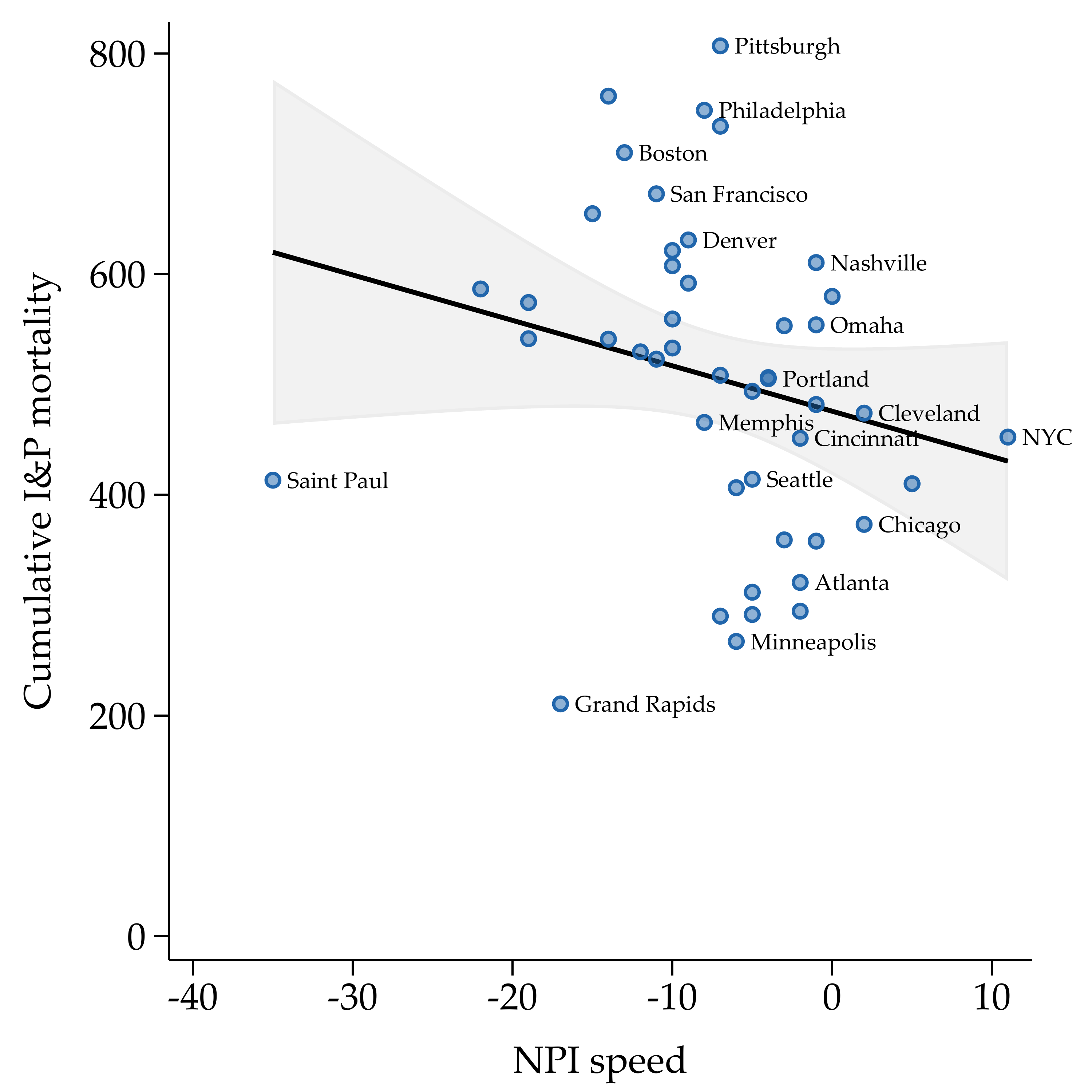}}\hfill
    \subfloat[NPI intensity and I\&P mortality.]{\includegraphics[width=0.45\textwidth]{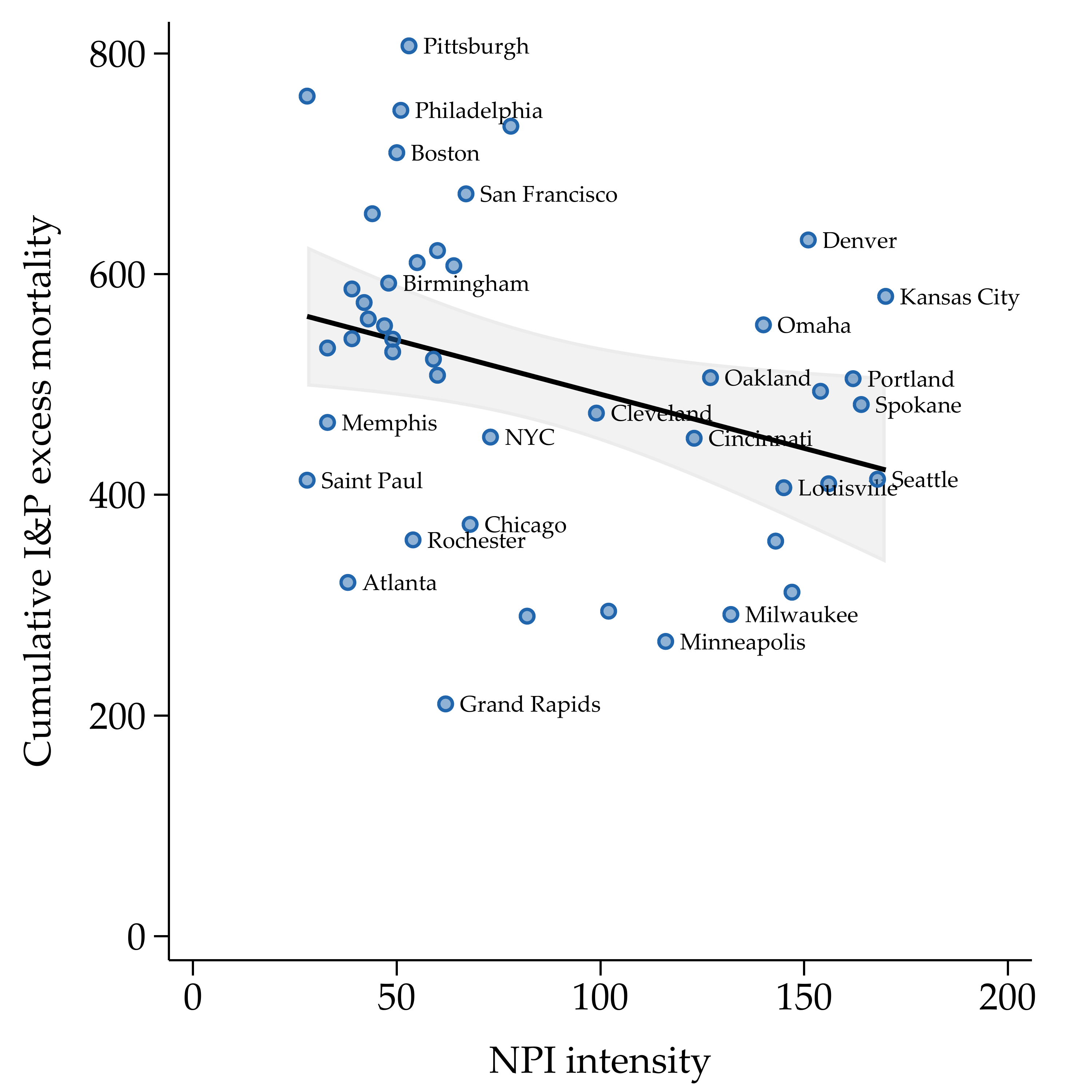}}

    \subfloat[NPI speed and all-cause mortality.]{\includegraphics[width=0.45\textwidth]{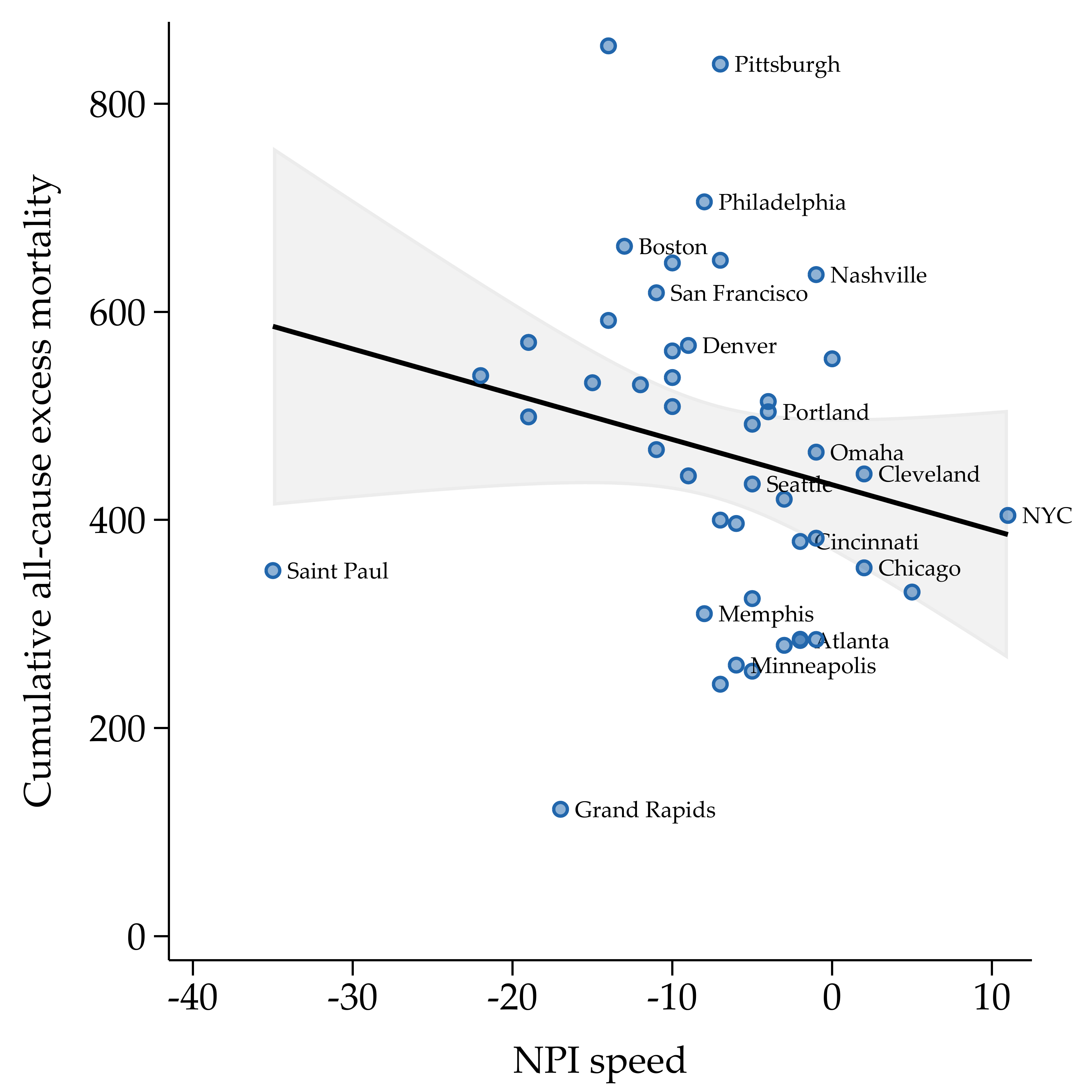}}\hfill
    \subfloat[NPI intensity and all-cause mortality.]{\includegraphics[width=0.45\textwidth]{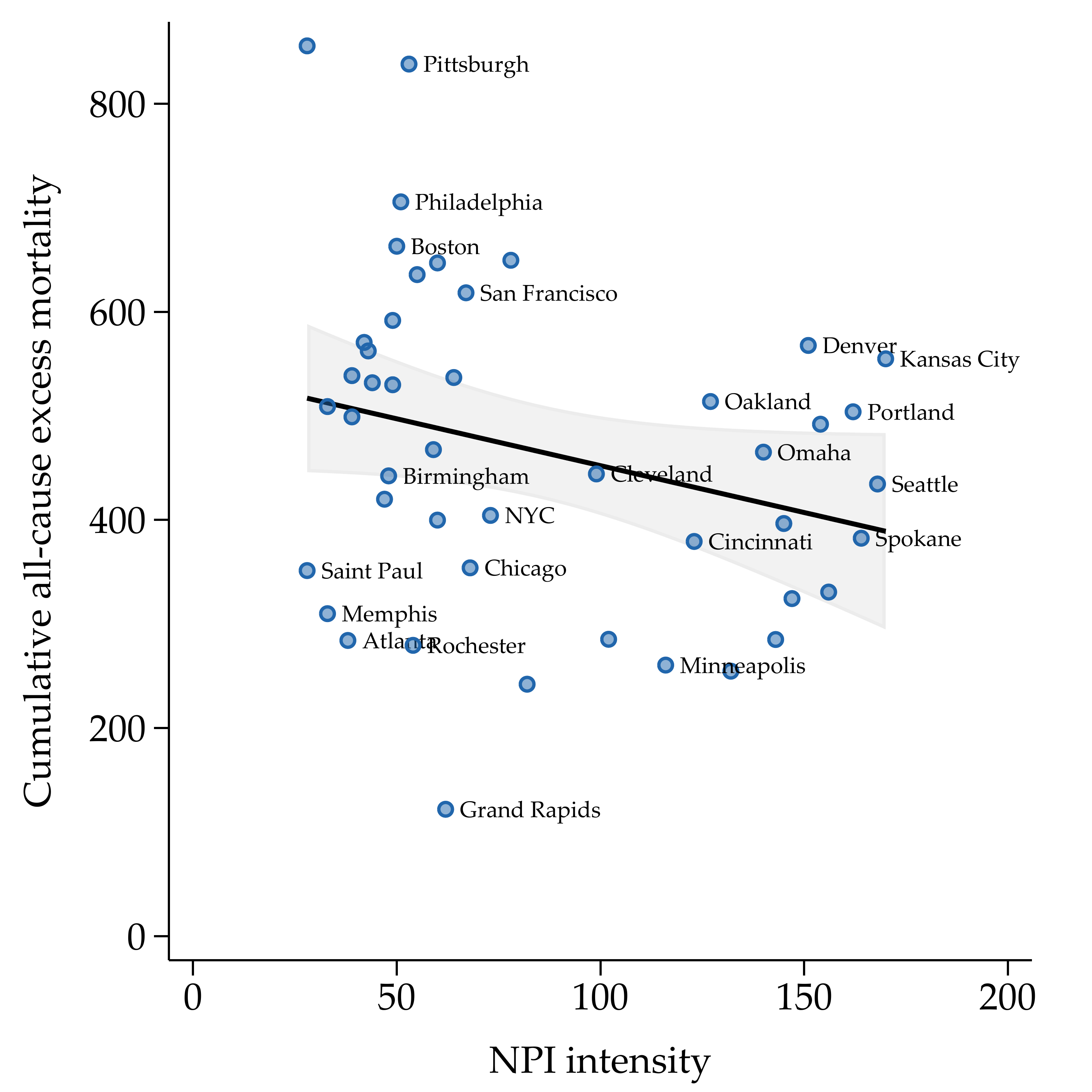}}
    
    \caption{\textbf{Non-pharmaceutical interventions and city-level excess mortality.} This figure correlates the excess the speed and intensity of city-level NPI implementations during fall 1918 with mortality due to influenza and pneumonia as well as all-cause mortality. Sample consists of 46 cities with mortality data, as listed in \cref{tab:samplecities}. See also \cref{tab:npi-sources} for the sources of NPI data by city.}
    \label{fig:excess_mort_NPI}
    \end{figure}

\begin{landscape}
\begin{figure}
    \centering
    \subfloat[Combined index (wholesale, retail, and manufacturing)]{
    \includegraphics[width=0.6\textwidth]{./figures/Bradstreet-WRM-Ok-joint.pdf}}
    \hfill
    \subfloat[Wholesale]{
    \includegraphics[width=0.6\textwidth]{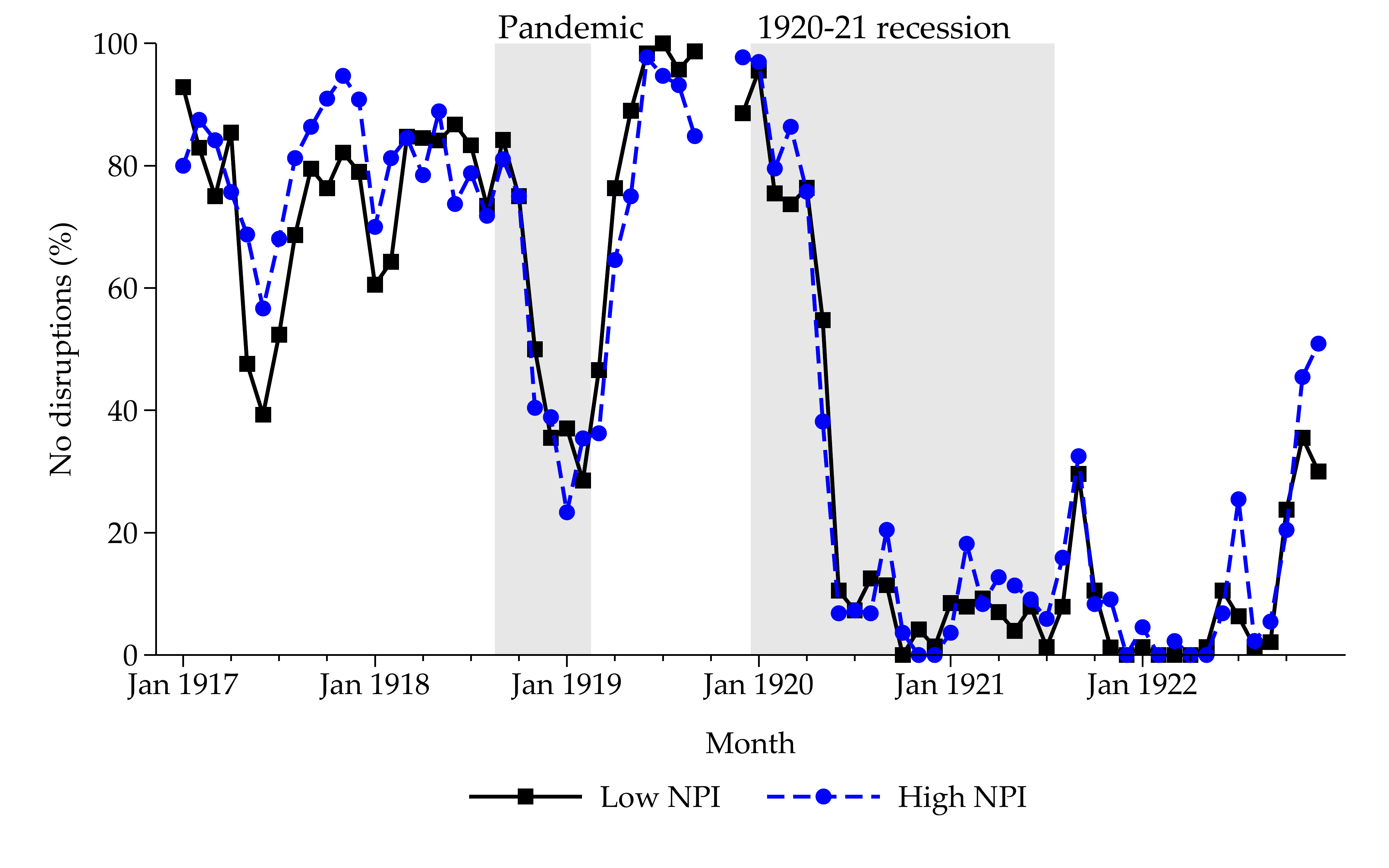}}
    
    \subfloat[Retail]{
    \includegraphics[width=0.6\textwidth]{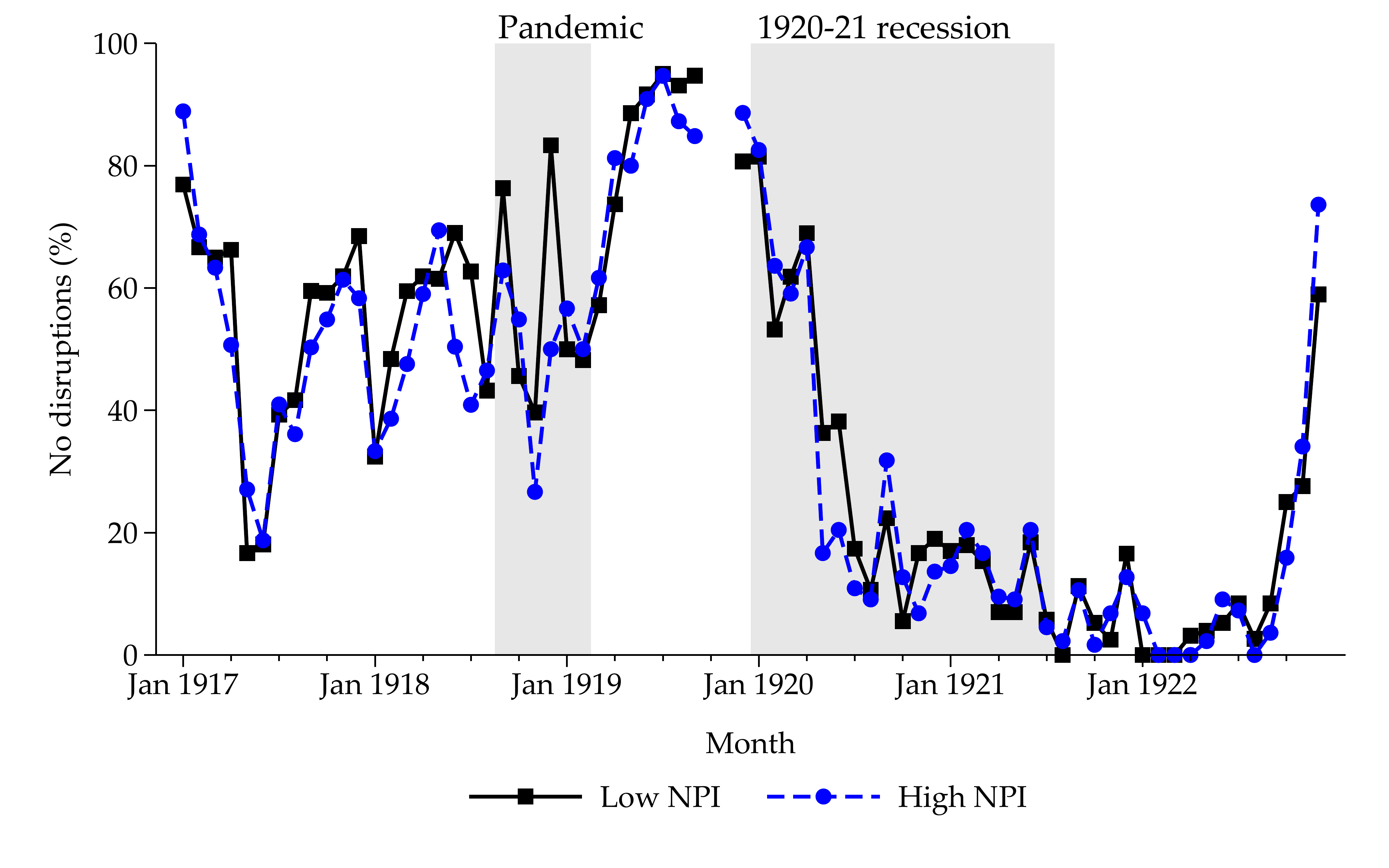}}
    \hfill
    \subfloat[Manufacturing]{
    \includegraphics[width=0.6\textwidth]{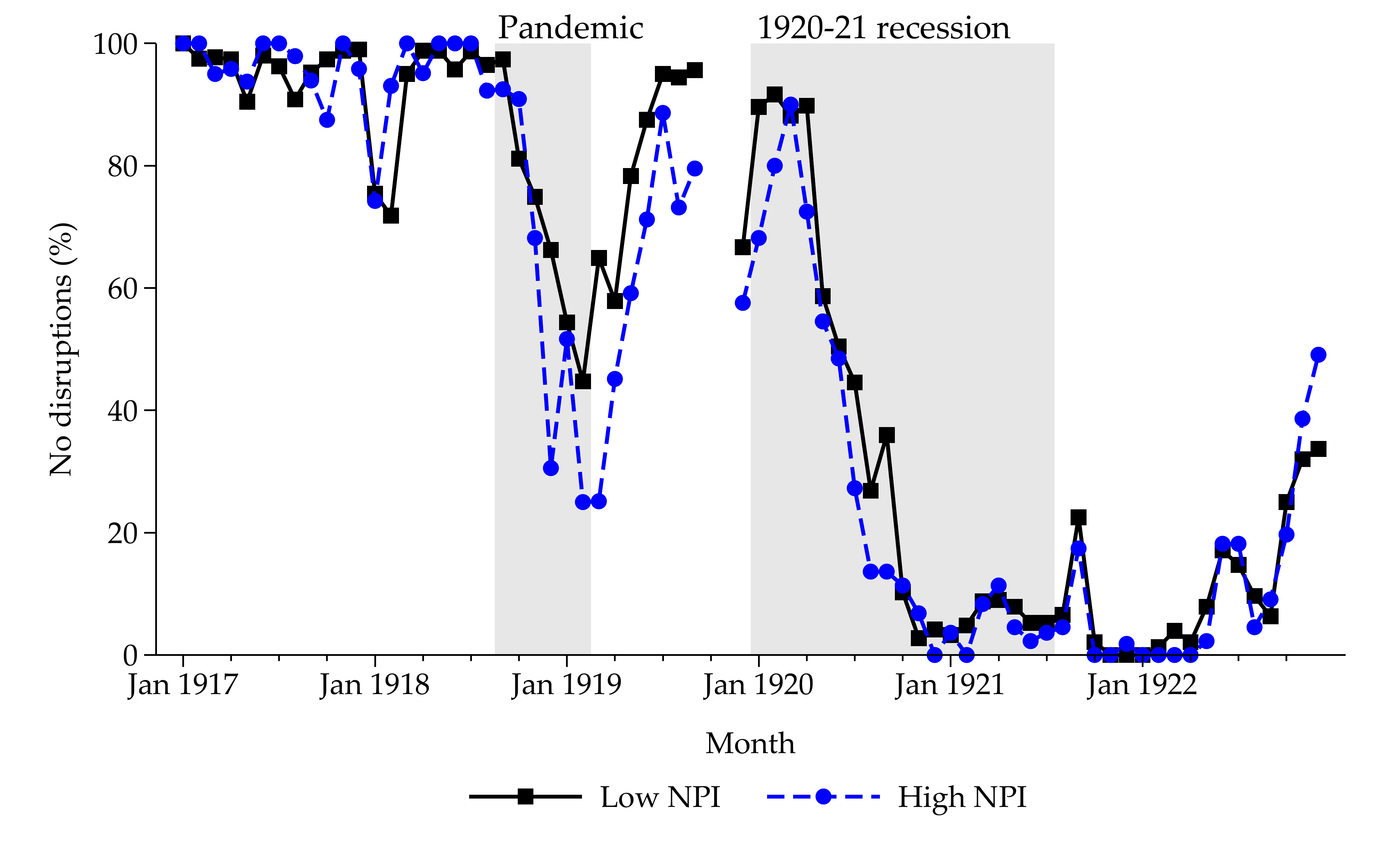}}
    \caption{\textbf{Non-pharmaceutical interventions and short-run economic disruptions.} This figure plots the average across high and low NPI cities of an indicator variable for whether the Bradstreet Trade conditions suggest ``disruptions'' in specific sectors. High NPI cities are defined as cities with above median $NPI \, Intensity$ and $NPI \, Speed$.}
    \label{fig:bradstreet_all}
\end{figure}
\end{landscape}

\begin{landscape}
\begin{figure}
    \centering
       \subfloat[Combined index (wholesale, retail, and manufacturing)]{
    \includegraphics[width=0.6\textwidth]{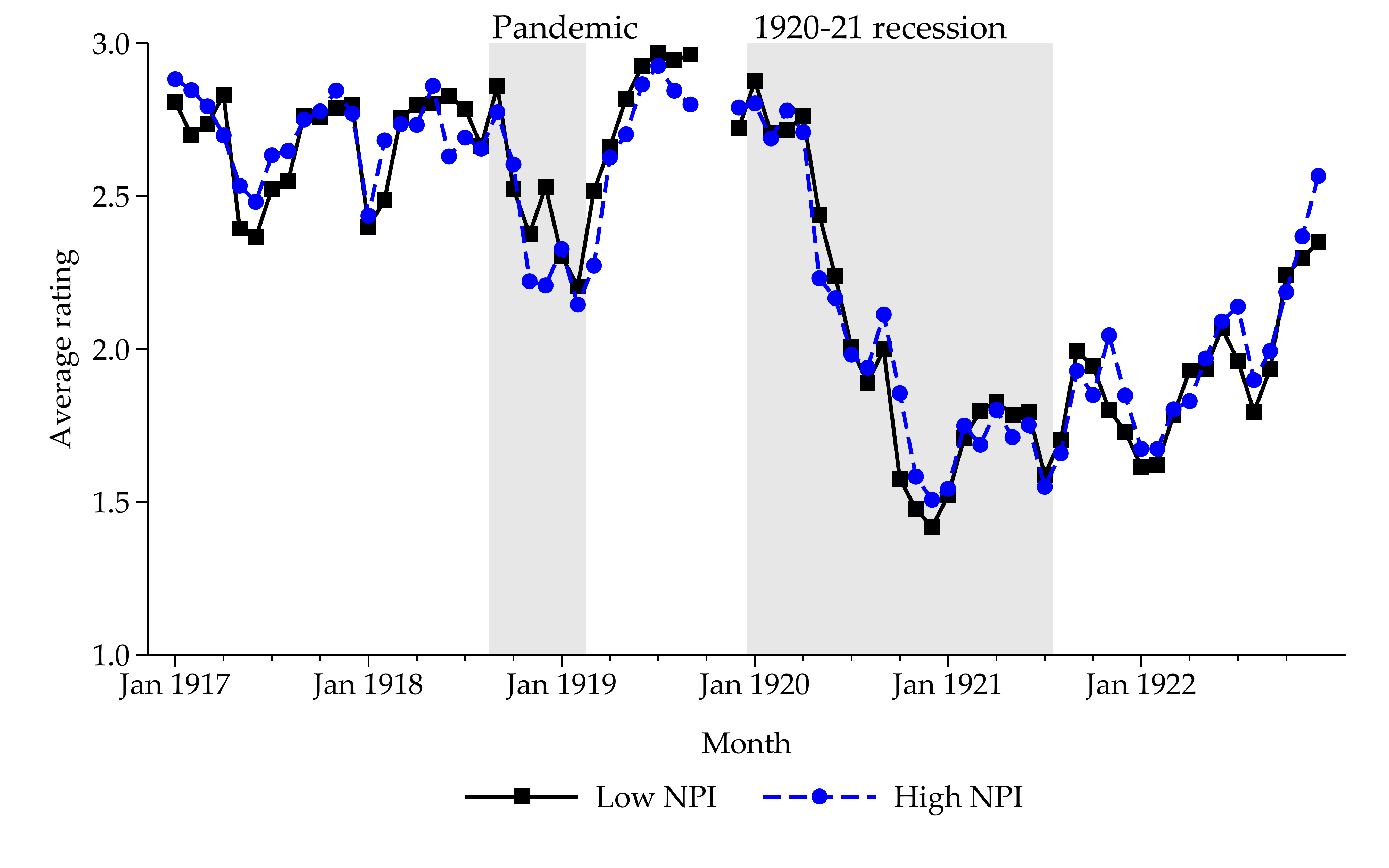}}
    \hfill
    \subfloat[Wholesale]{
    \includegraphics[width=0.6\textwidth]{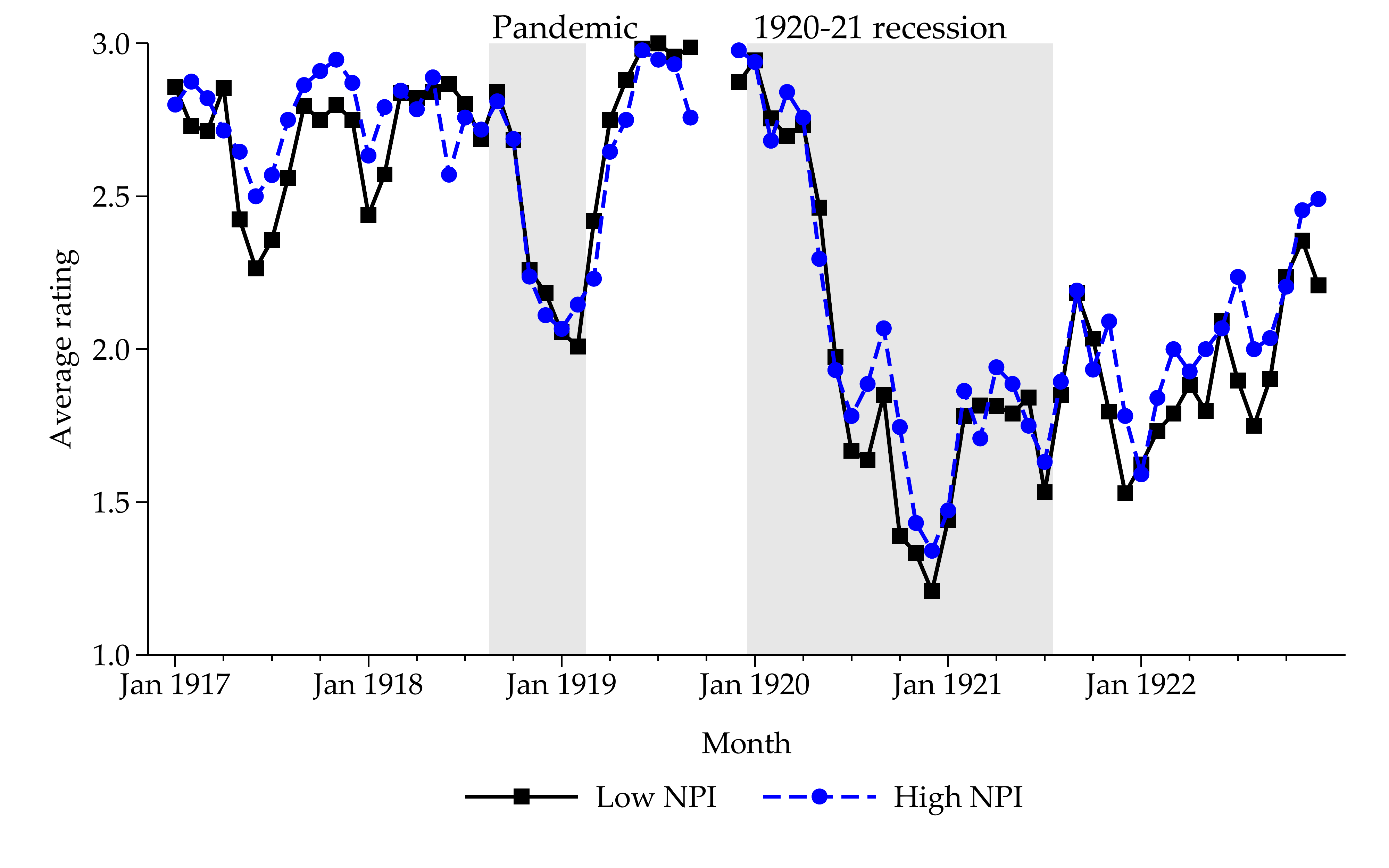}}
    
    \subfloat[Retail]{
    \includegraphics[width=0.6\textwidth]{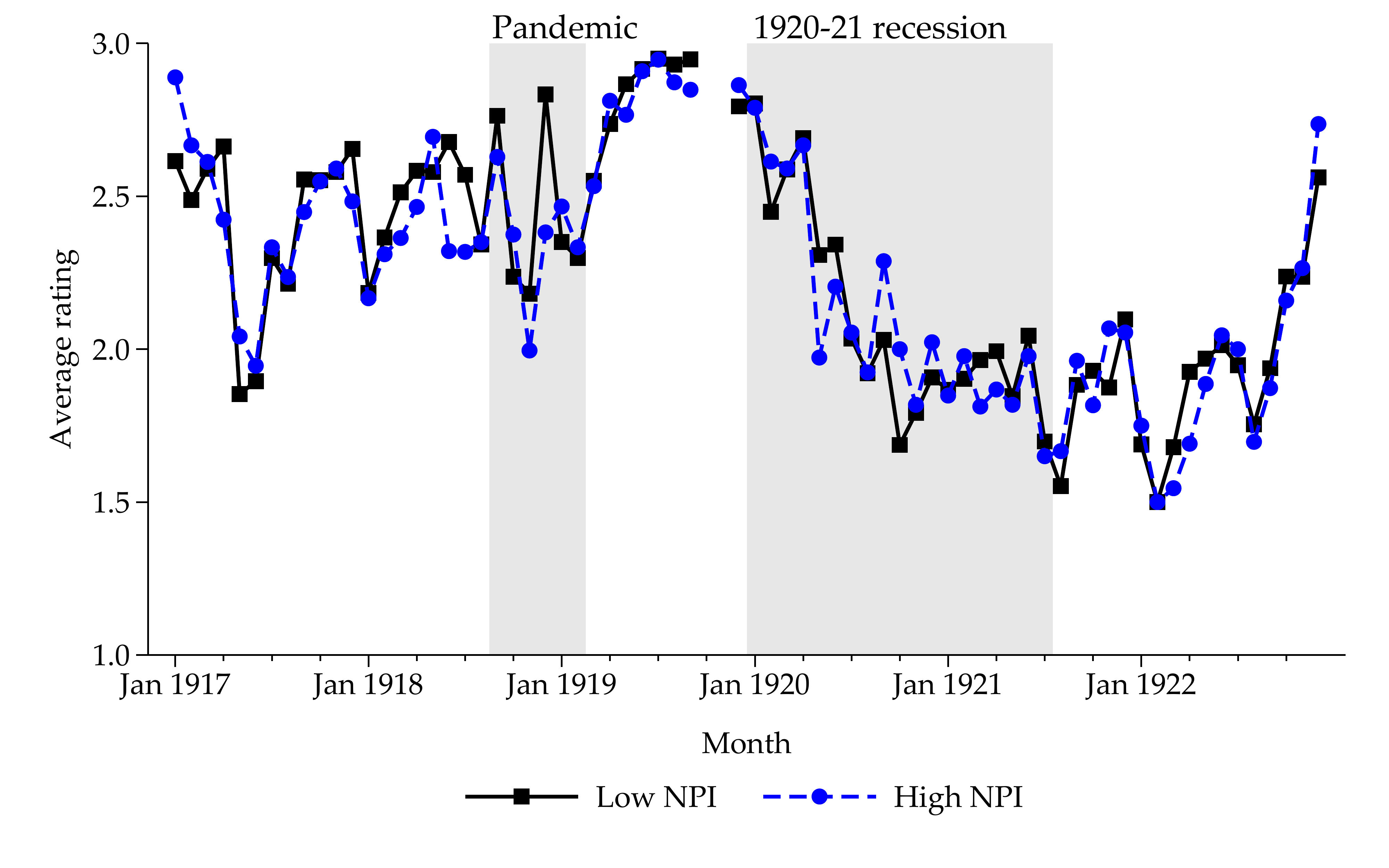}}
    \hfill
    \subfloat[Manufacturing]{
    \includegraphics[width=0.6\textwidth]{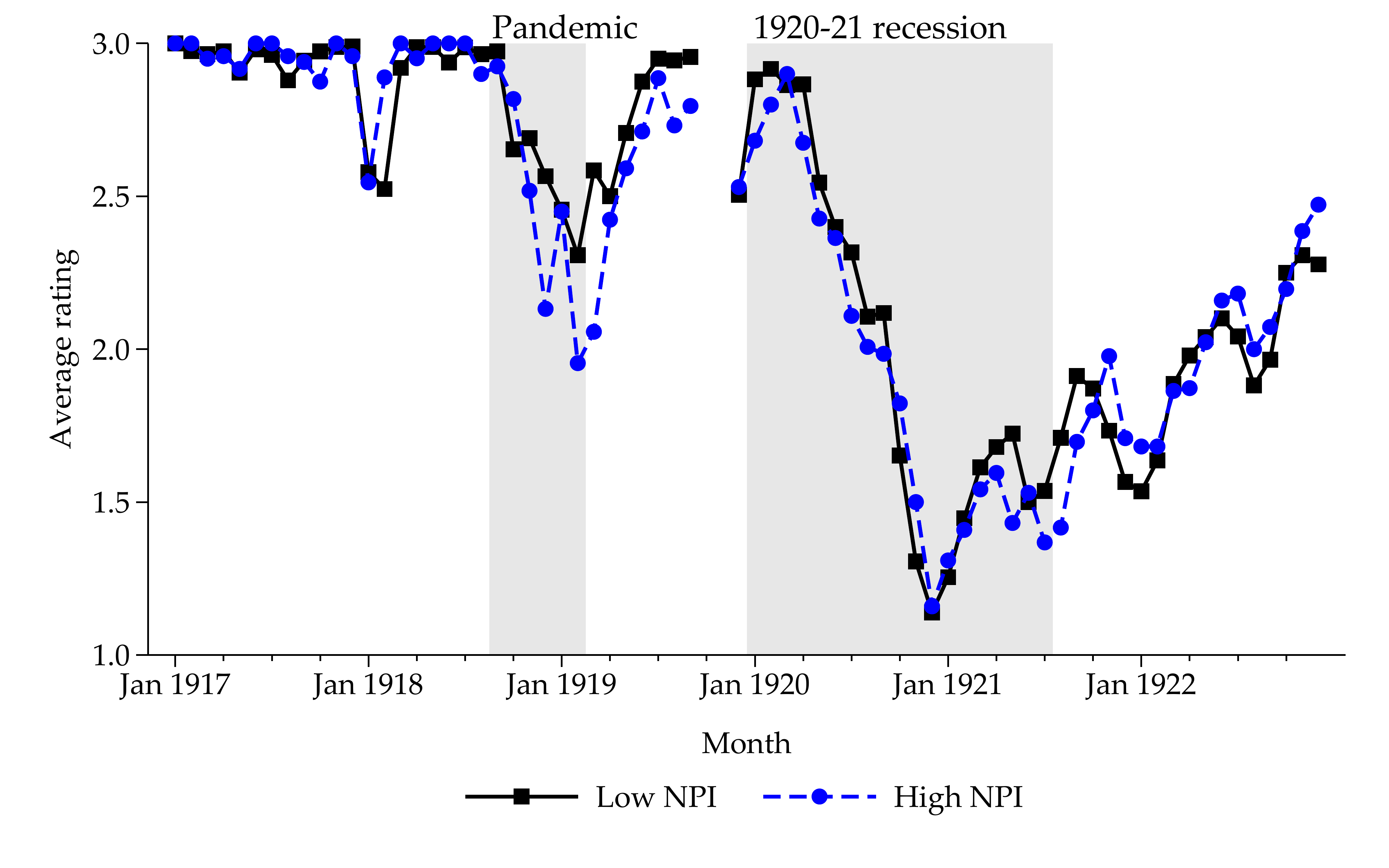}}
   
    \caption{\textbf{Non-pharmaceutical interventions and short-term economic disruptions. Robustness to a three-step index of trade conditions.} High NPI cities are defined as cities with above median $NPI \, Intensity$ and $NPI \, Speed$. The three-step index of trade conditions runs from 1 to 3, with 1=``Bad'', 2=``Fair'', and 3=``Good''.}
    \label{fig:bradstreet_cat}
\end{figure}
\end{landscape}

\begin{figure}[htpb]
    \centering
    \subfloat[\NPIIntensity and log National Bank Assets. \label{fig:coefplot_assets_day_npi}]{\includegraphics[width=0.49\textwidth]{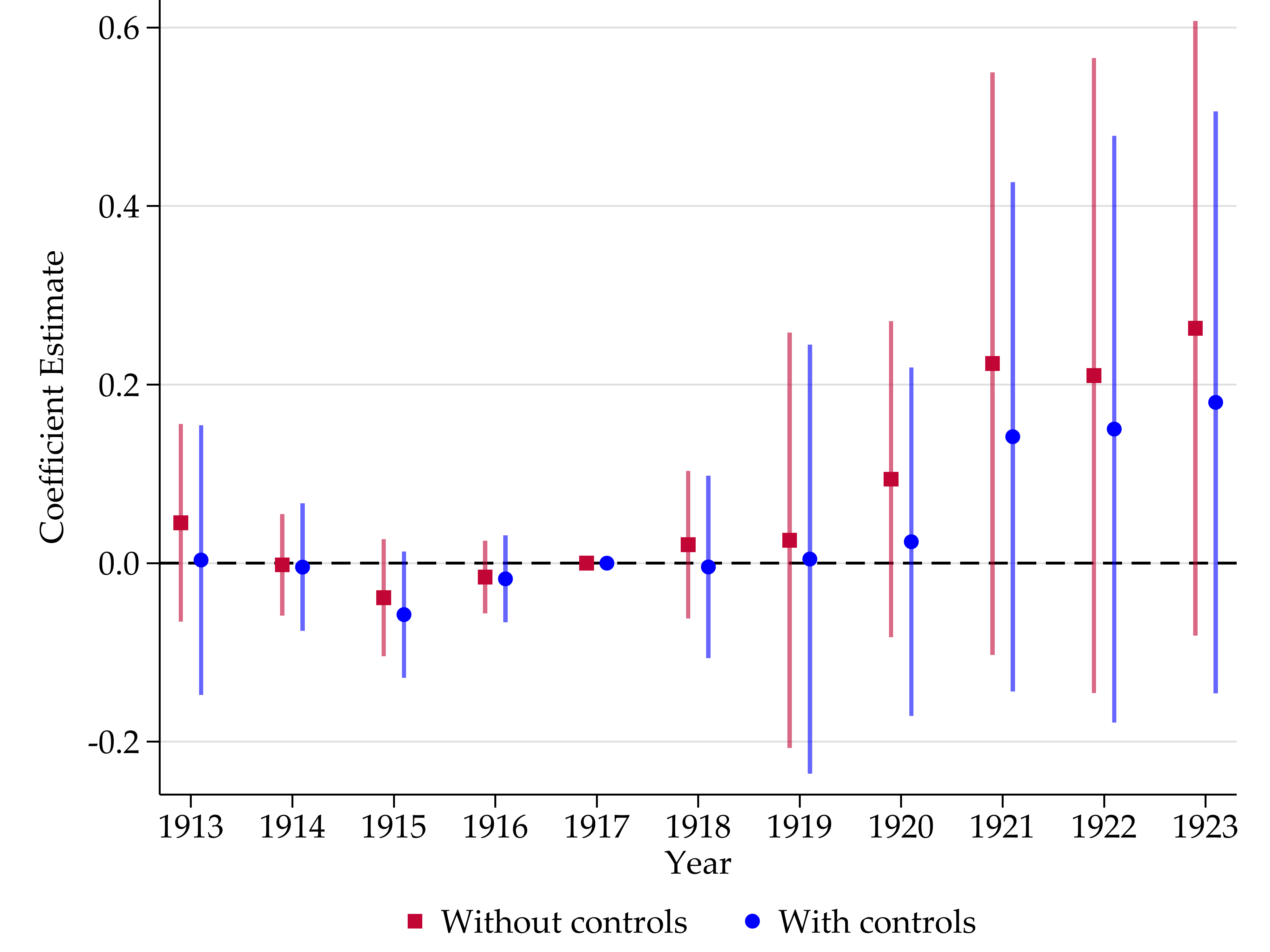}}
    \hfill
    \subfloat[\HighNPI and log National Bank Assets. \label{fig:coefplot_assets_highnpi}]{\includegraphics[width=0.49\textwidth]{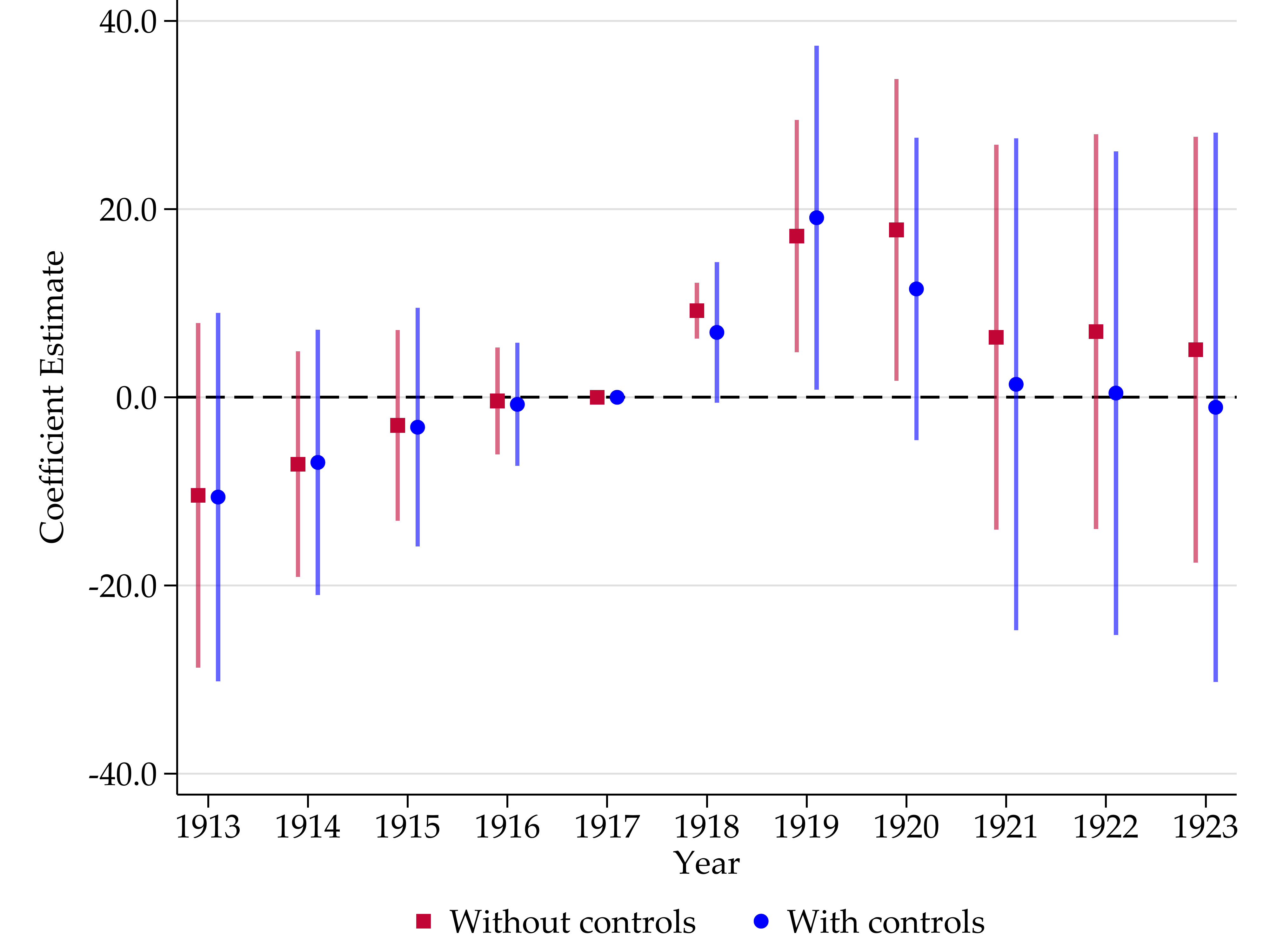}}
   \caption{\textbf{Non-pharmaceutical interventions in Fall 1918 and medium-run economic outcomes.}
   This figure presents estimates of equation \eqref{eq:spec2} on the log of national bank assets, with \NPIIntensity (panel a) and \HighNPI (panel b) as regressors.
   Controls interacted with year are the log of 1900 and 1910 city population, 1910 city density, 1917 city health spending per capita, manufacturing employment in 1914 to 1910 population, and Third Liberty Loan quotas required for each city (obtained from \href{https://books.google.com/books?id=0WUWAAAAIAAJ&pg=PA20}{1918 hearing before the House of Representatives Committee on Ways and Means}).
   Error bands denote 95\% confidence bands with robust standard errors clustered at the city level.}
    \label{fig:NPI_coefplot-national-banks}
\end{figure}

\clearpage

\setcounter{table}{0}
\renewcommand{\thetable}{A\arabic{table}}

\begin{table}[htpb]
\centering
\caption{\textbf{Sample of cities.}}
\label{tab:samplecities}
{\tiny
    \begin{tabular}{ll | ccc | ccc | cccc}
    \toprule
    \multirow{8}{*}{City} & \multirow{8}{*}{State} & \multicolumn{3}{c}{Regressions} & \multicolumn{3}{c}{Other research} & \multicolumn{4}{c}{Sources} \\
    \cmidrule(lr){3-5} \cmidrule(lr){6-8} \cmidrule(lr){9-12}        
    & & \rotatebox[origin=c]{90}{Mortality} & \rotatebox[origin=c]{90}{Bradstreet} & \rotatebox[origin=c]{90}{Manuf.} & \rotatebox[origin=c]{90}{Markel et al.} & \rotatebox[origin=c]{90}{Berkes et al.} & \rotatebox[origin=c]{90}{Collins et al.} & \rotatebox[origin=c]{90}{WHI} & \rotatebox[origin=c]{90}{AMS} & \rotatebox[origin=c]{90}{Bradstreet's} & \rotatebox[origin=c]{90}{CoM} \\
    \cmidrule{1-12}
    Albany&NY&\CheckYes&&\CheckYes&\CheckYes&\CheckYes&\CheckYes&\CheckYes&\CheckYes&&\CheckYes\\
Atlanta&GA&\CheckYes&\CheckYes&\CheckYes&&\CheckYes&\CheckYes&\CheckYes&\CheckYes&\CheckYes&\CheckYes\\
Baltimore&MD&\CheckYes&\CheckYes&\CheckYes&\CheckYes&\CheckYes&\CheckYes&\CheckYes&\CheckYes&\CheckYes&\CheckYes\\
Birmingham&AL&\CheckYes&&\CheckYes&\CheckYes&\CheckYes&\CheckYes&\CheckYes&\CheckYes&&\CheckYes\\
Boston&MA&\CheckYes&\CheckYes&\CheckYes&\CheckYes&\CheckYes&\CheckYes&\CheckYes&\CheckYes&\CheckYes&\CheckYes\\
Buffalo&NY&\CheckYes&\CheckYes&\CheckYes&\CheckYes&\CheckYes&\CheckYes&\CheckYes&\CheckYes&\CheckYes&\CheckYes\\
Cambridge&MA&\CheckYes&&\CheckYes&\CheckYes&\CheckYes&\CheckYes&\CheckYes&\CheckYes&&\CheckYes\\
Charleston&SC&&&\CheckYes&&\CheckYes&&&&\CheckYes&\CheckYes\\
Chicago&IL&\CheckYes&\CheckYes&\CheckYes&\CheckYes&\CheckYes&\CheckYes&\CheckYes&\CheckYes&\CheckYes&\CheckYes\\
Cincinnati&OH&\CheckYes&\CheckYes&\CheckYes&\CheckYes&\CheckYes&\CheckYes&\CheckYes&\CheckYes&\CheckYes&\CheckYes\\
Cleveland&OH&\CheckYes&\CheckYes&\CheckYes&\CheckYes&\CheckYes&\CheckYes&\CheckYes&\CheckYes&\CheckYes&\CheckYes\\
Columbus&OH&\CheckYes&&\CheckYes&\CheckYes&\CheckYes&\CheckYes&\CheckYes&\CheckYes&&\CheckYes\\
Dallas&TX&&&\CheckYes&&\CheckYes&&&&\CheckYes&\CheckYes\\
Dayton&OH&\CheckYes&&\CheckYes&\CheckYes&\CheckYes&\CheckYes&\CheckYes&\CheckYes&&\CheckYes\\
Denver&CO&\CheckYes&&\CheckYes&\CheckYes&\CheckYes&\CheckYes&\CheckYes&\CheckYes&&\CheckYes\\
Des Moines&IA&&&\CheckYes&&\CheckYes&&&\CheckYes&\CheckYes&\CheckYes\\
Detroit&MI&&&\CheckYes&&\CheckYes&\CheckYes&&\CheckYes&\CheckYes&\CheckYes\\
Fall River&MA&\CheckYes&&\CheckYes&\CheckYes&\CheckYes&\CheckYes&\CheckYes&\CheckYes&&\CheckYes\\
Grand Rapids&MI&\CheckYes&&\CheckYes&\CheckYes&\CheckYes&\CheckYes&\CheckYes&\CheckYes&&\CheckYes\\
Indianapolis&IN&\CheckYes&\CheckYes&\CheckYes&\CheckYes&\CheckYes&\CheckYes&\CheckYes&\CheckYes&\CheckYes&\CheckYes\\
Jersey City&NJ&\CheckYes&&\CheckYes&&&\CheckYes&\CheckYes&\CheckYes&&\CheckYes\\
Kansas City&MO&\CheckYes&\CheckYes&\CheckYes&\CheckYes&\CheckYes&\CheckYes&\CheckYes&\CheckYes&\CheckYes&\CheckYes\\
Los Angeles&CA&\CheckYes&\CheckYes&\CheckYes&\CheckYes&\CheckYes&\CheckYes&\CheckYes&\CheckYes&\CheckYes&\CheckYes\\
Louisville&KY&\CheckYes&\CheckYes&\CheckYes&\CheckYes&\CheckYes&\CheckYes&\CheckYes&\CheckYes&\CheckYes&\CheckYes\\
Lowell&MA&\CheckYes&&\CheckYes&\CheckYes&\CheckYes&\CheckYes&\CheckYes&\CheckYes&&\CheckYes\\
Memphis&TN&\CheckYes&\CheckYes&\CheckYes&&&\CheckYes&\CheckYes&\CheckYes&\CheckYes&\CheckYes\\
Milwaukee&WI&\CheckYes&\CheckYes&\CheckYes&\CheckYes&\CheckYes&\CheckYes&\CheckYes&\CheckYes&\CheckYes&\CheckYes\\
Minneapolis&MN&\CheckYes&\CheckYes&\CheckYes&\CheckYes&\CheckYes&\CheckYes&\CheckYes&\CheckYes&\CheckYes&\CheckYes\\
Nashville&TN&\CheckYes&\CheckYes&\CheckYes&\CheckYes&\CheckYes&\CheckYes&\CheckYes&\CheckYes&\CheckYes&\CheckYes\\
New Haven&CT&\CheckYes&&\CheckYes&\CheckYes&\CheckYes&\CheckYes&\CheckYes&\CheckYes&&\CheckYes\\
New Orleans&LA&\CheckYes&\CheckYes&\CheckYes&\CheckYes&\CheckYes&\CheckYes&\CheckYes&\CheckYes&\CheckYes&\CheckYes\\
New York City&NY&\CheckYes&\CheckYes&\CheckYes&\CheckYes&\CheckYes&\CheckYes&\CheckYes&\CheckYes&\CheckYes&\CheckYes\\
Newark&NJ&\CheckYes&&\CheckYes&\CheckYes&\CheckYes&\CheckYes&\CheckYes&\CheckYes&&\CheckYes\\
Oakland&CA&\CheckYes&&\CheckYes&\CheckYes&\CheckYes&\CheckYes&\CheckYes&\CheckYes&&\CheckYes\\
Omaha&NE&\CheckYes&\CheckYes&\CheckYes&\CheckYes&\CheckYes&\CheckYes&\CheckYes&\CheckYes&\CheckYes&\CheckYes\\
Paterson&NJ&&&\CheckYes&&&\CheckYes&&\CheckYes&&\CheckYes\\
Philadelphia&PA&\CheckYes&\CheckYes&\CheckYes&\CheckYes&\CheckYes&\CheckYes&\CheckYes&\CheckYes&\CheckYes&\CheckYes\\
Pittsburgh&PA&\CheckYes&\CheckYes&\CheckYes&\CheckYes&\CheckYes&\CheckYes&\CheckYes&\CheckYes&\CheckYes&\CheckYes\\
Portland&OR&\CheckYes&\CheckYes&\CheckYes&\CheckYes&\CheckYes&\CheckYes&\CheckYes&\CheckYes&\CheckYes&\CheckYes\\
Providence&RI&\CheckYes&&\CheckYes&\CheckYes&\CheckYes&\CheckYes&\CheckYes&\CheckYes&&\CheckYes\\
Richmond&VA&\CheckYes&\CheckYes&\CheckYes&\CheckYes&\CheckYes&\CheckYes&\CheckYes&\CheckYes&\CheckYes&\CheckYes\\
Rochester&NY&\CheckYes&&\CheckYes&\CheckYes&\CheckYes&\CheckYes&\CheckYes&\CheckYes&&\CheckYes\\
Saint Louis&MO&\CheckYes&\CheckYes&\CheckYes&\CheckYes&\CheckYes&\CheckYes&\CheckYes&\CheckYes&\CheckYes&\CheckYes\\
Saint Paul&MN&\CheckYes&\CheckYes&\CheckYes&\CheckYes&\CheckYes&\CheckYes&\CheckYes&\CheckYes&\CheckYes&\CheckYes\\
Salt Lake City&UT&&&\CheckYes&&\CheckYes&&&&\CheckYes&\CheckYes\\
San Antonio&TX&&&\CheckYes&&\CheckYes&&&&&\CheckYes\\
San Francisco&CA&\CheckYes&\CheckYes&\CheckYes&\CheckYes&\CheckYes&\CheckYes&\CheckYes&\CheckYes&\CheckYes&\CheckYes\\
Scranton&PA&&&\CheckYes&&&\CheckYes&&\CheckYes&&\CheckYes\\
Seattle&WA&\CheckYes&\CheckYes&\CheckYes&\CheckYes&\CheckYes&\CheckYes&\CheckYes&\CheckYes&\CheckYes&\CheckYes\\
Spokane&WA&\CheckYes&\CheckYes&\CheckYes&\CheckYes&\CheckYes&\CheckYes&\CheckYes&\CheckYes&\CheckYes&\CheckYes\\
Syracuse&NY&\CheckYes&&\CheckYes&\CheckYes&\CheckYes&\CheckYes&\CheckYes&\CheckYes&&\CheckYes\\
Toledo&OH&\CheckYes&&\CheckYes&\CheckYes&\CheckYes&\CheckYes&\CheckYes&\CheckYes&&\CheckYes\\
Washington&DC&\CheckYes&&\CheckYes&\CheckYes&\CheckYes&\CheckYes&\CheckYes&\CheckYes&&\CheckYes\\
Worcester&MA&\CheckYes&&\CheckYes&\CheckYes&\CheckYes&\CheckYes&\CheckYes&\CheckYes&&\CheckYes\\
\cmidrule{1-12} Total&&46&27&54&43&50&49&46&50&32&54\\
  \bottomrule
    \end{tabular}
}
\parbox{\textwidth}{\small
\vspace{1eX} 
Notes: This table lists the sample of cities employed throughout the paper, in three related papers, and whether they were available in the primary sources.
Columns (1)-(3) indicate whether a city was used in each of the three types of regressions employed in the paper.
Columns (4)-(6) indicate whether the cities were in the sample of  \citet{Markel2007}, \citet{Berkes2022}, and \citet{Collins1930} respectively.
Column (7) indicates whether pandemic mortality information was available from the Weekly Health Index (1918-1919).
Column (8) corresponds to the Annual Mortality Statistics (1910-1916), which provides monthly death rates prior to the pandemic, used to compute baseline mortality rates
Column (9) corresponds to Bradstreet's Trade at Glance (1917-1922).
Lastly, column (10) indicates whether manufacturing data was available from either the Census of Manufactures (1904-1927) or the Statistical Abstract of the United States (1931).}
\end{table}

\begin{table}[htpb]
\centering
\caption{\textbf{Non-pharmaceutical interventions in 54 cities during Fall 1918}.}
\label{tab:npi-sources}
{\tiny
\begin{tabular}{ll|rrr|rrc}
    \toprule
    City &  State & First case & Acceleration date & Response date & \NPISpeed & \NPIIntensity & Source \\
    \cmidrule{1-8}
    Albany&NY&27 Sep&6 Oct&9 Oct&-3&47&Markel et al.\\
Atlanta&GA&-&5 Oct&-&-2&38&Markel raw data\\
Baltimore&MD&18 Sep&29 Sep&9 Oct&-10&43&Markel et al.\\
Birmingham&AL&24 Sep&30 Sep&9 Oct&-9&48&Markel et al.\\
Boston&MA&4 Sep&12 Sep&25 Sep&-13&50&Markel et al.\\
Buffalo&NY&24 Sep&28 Sep&10 Oct&-12&49&Markel et al.\\
Cambridge&MA&4 Sep&11 Sep&25 Sep&-14&49&Markel et al.\\
Charleston&SC&-&-&-&&69&Berkes et al.\\
Chicago&IL&17 Sep&28 Sep&26 Sep&2&68&Markel et al.\\
Cincinnati&OH&24 Sep&4 Oct&6 Oct&-2&123&Markel et al.\\
Cleveland&OH&20 Sep&7 Oct&5 Oct&2&99&Markel et al.\\
Columbus&OH&20 Sep&6 Oct&11 Oct&-5&147&Markel et al.\\
Dallas&TX&-&-&-&&41&Berkes et al.\\
Dayton&OH&20 Sep&5 Oct&30 Sep&5&156&Markel et al.\\
Denver&CO&17 Sep&27 Sep&6 Oct&-9&151&Markel et al.\\
Des Moines&IA&-&-&-&&56&Berkes et al.\\
Detroit&MI&-&-&-&&29&Berkes et al.\\
Fall River&MA&9 Sep&16 Sep&26 Sep&-10&60&Markel et al.\\
Grand Rapids&MI&23 Sep&2 Oct&19 Oct&-17&62&Markel et al.\\
Indianapolis&IN&22 Sep&30 Sep&7 Oct&-7&82&Markel et al.\\
Jersey City&NJ&-&25 Sep&-&-14&28&Authors\\
Kansas City&MO&20 Sep&26 Sep&26 Sep&0&170&Markel et al.\\
Los Angeles&CA&27 Sep&6 Oct&11 Oct&-5&154&Markel et al.\\
Louisville&KY&13 Sep&1 Oct&7 Oct&-6&145&Markel et al.\\
Lowell&MA&9 Sep&16 Sep&27 Sep&-11&59&Markel et al.\\
Memphis&TN&-&5 Oct&-&-8&33&Authors\\
Milwaukee&WI&14 Sep&6 Oct&11 Oct&-5&132&Markel et al.\\
Minneapolis&MN&21 Sep&6 Oct&12 Oct&-6&116&Markel et al.\\
Nashville&TN&21 Sep&6 Oct&7 Oct&-1&55&Markel et al.\\
New Haven&CT&14 Sep&23 Sep&15 Oct&-22&39&Markel et al.\\
New Orleans&LA&10 Sep&1 Oct&8 Oct&-7&78&Markel et al.\\
New York City&NY&5 Sep&29 Sep&18 Sep&11&73&Markel et al.\\
Newark&NJ&6 Sep&30 Sep&10 Oct&-10&33&Markel et al.\\
Oakland&CA&1 Oct&8 Oct&12 Oct&-4&127&Markel et al.\\
Omaha&NE&18 Sep&4 Oct&5 Oct&-1&140&Markel et al.\\
Paterson&NJ&-&-&-&&172&Authors\\
Philadelphia&PA&27 Aug&25 Sep&3 Oct&-8&51&Markel et al.\\
Pittsburgh&PA&4 Sep&27 Sep&4 Oct&-7&53&Markel et al.\\
Portland&OR&2 Oct&7 Oct&11 Oct&-4&162&Markel et al.\\
Providence&RI&8 Sep&17 Sep&6 Oct&-19&42&Markel et al.\\
Richmond&VA&21 Sep&29 Sep&6 Oct&-7&60&Markel et al.\\
Rochester&NY&22 Sep&6 Oct&9 Oct&-3&54&Markel et al.\\
Saint Louis&MO&23 Sep&7 Oct&8 Oct&-1&143&Markel et al.\\
Saint Paul&MN&21 Sep&2 Oct&6 Nov&-35&28&Markel et al.\\
Salt Lake City&UT&-&-&-&&141&Berkes et al.\\
San Antonio&TX&-&-&-&&164&Authors\\
San Francisco&CA&24 Sep&7 Oct&18 Oct&-11&67&Markel et al.\\
Scranton&PA&-&-&-&&69&Authors\\
Seattle&WA&24 Sep&1 Oct&6 Oct&-5&168&Markel et al.\\
Spokane&WA&28 Sep&9 Oct&10 Oct&-1&164&Markel et al.\\
Syracuse&NY&12 Sep&18 Sep&7 Oct&-19&39&Markel et al.\\
Toledo&OH&21 Sep&13 Oct&15 Oct&-2&102&Markel et al.\\
Washington&DC&11 Sep&23 Sep&3 Oct&-10&64&Markel et al.\\
Worcester&MA&9 Sep&12 Sep&27 Sep&-15&44&Markel et al.\\
  \bottomrule
 \\ 
\end{tabular}
}
\parbox{\textwidth}{\small
\vspace{1eX} 
    Notes: This table lists the values of the NPI variables for each of the 54 cities studied in the paper, as well as related variables.
    \emph{First case} is the date where influenza cases related to the pandemic were first reported in the city.
    \emph{Acceleration date} is the date where the mortality due to influenza and pneumonia first exceeded twice the baseline mortality rate.
    \emph{Response time} corresponds to the date where NPIs were first implemented in the city.
    Following \citet{Markel2007}, we consider three categories of NPIs: public gathering bans, school closures, and other measures including quarantine/isolation.
    \NPISpeed is the difference between the response date and the mortality acceleration date.
    \NPIIntensity is the sum of the number of days where each the three categories of NPIs was active.
    \emph{Source} indicates whether the information on NPIs was taken from \citet{Markel2007}, from the raw data collected by \citet{Markel2007}, from \citet{Berkes2022}, or was collected directly by the authors.
}
\end{table}

\begin{landscape}
\setlength{\tabcolsep}{.1in}
\begin{table}
\centering
\begin{threeparttable}
    \caption{\textbf{Summary statistics and comparison of cities with low and high NPIs.}}
    \small
    \begin{tabular}{l cc cc cc cc}
        \toprule & \multicolumn{2}{c}{All cities} & \multicolumn{2}{c}{$Low \ NPI$} & \multicolumn{2}{c}{$High \ NPI$} & \multicolumn{2\textbf{}}{c}{Difference} \\
        \cmidrule(lr){2-3} \cmidrule(lr){4-5} \cmidrule(lr){6-7} \cmidrule(lr){8-9}
        & \multicolumn{1}{c}{Mean} & \multicolumn{1}{c}{Std} 
         & \multicolumn{1}{c}{Mean} & \multicolumn{1}{c}{Std} & \multicolumn{1}{c}{Mean} & \multicolumn{1}{c}{Std} & \multicolumn{1}{c}{Diff} & \multicolumn{1}{c}{t-stat} \\  \midrule  
        I\&P peak mort. rate (Markel)&98.8&49.4&120.1&50.9&65.6&21.1&-54.5&-5.02\\
I\&P peak (replication)&101.96&47.15&122.32&48.52&70.27&19.99&-52.05&-5.04\\
All-cause peak mortality rate&106.1&51.2&127.6&53.9&72.8&19.8&-54.8&-4.88\\
I\&P cumulative mort. rate (Markel)&506.1&141.7&559.2&143.6&423.6&92.6&-135.6&-3.89\\
I\&P cumulative (replication)&487.77&133.81&535.49&135.23&413.53&93.60&-121.96&-3.61\\
All-cause cumulative mortality rate&466.0&155.9&513.2&171.2&392.5&91.7&-120.7&-3.10\\
\cmidrule{1-9} \NPIIntensity&84.7&46.9&53.8&23.3&132.7&30.9&78.9&9.31\\
\NPISpeed&-7.39&7.68&-11.18&6.94&-1.50&4.40&9.68&5.79\\
Longitude&-87.0&15.4&-81.1&11.6&-96.1&16.5&-14.9&-3.36\\
I\&P mort. (1917, per 1,000)&1.81&0.60&1.95&0.60&1.58&0.53&-0.38&-2.22\\
All-cause mort. (1917, per 1,000)&15.50&2.78&16.66&2.33&13.69&2.48&-2.97&-4.07\\
Infant mort. (1916, per capita)&0.26&0.09&0.28&0.09&0.22&0.09&-0.06&-2.36\\
Illiteracy rate (Clay 2019)&4.70&2.79&5.68&2.99&3.17&1.53&-2.51&-3.74\\
Density (city pop. per sq.mi.)&8,864&4,710&9,271&4,995&8,232&4,290&-1,039&-0.75\\
Health spending per capita (1917)&0.54&0.24&0.56&0.25&0.52&0.22&-0.04&-0.57\\
Per capita manuf. output (1914)&0.13&0.06&0.13&0.07&0.11&0.05&-0.02&-1.25\\
Log Pop in 1910&12.45&0.83&12.31&0.69&12.66&1.00&0.36&1.33\\
Log Pop in 1900&12.10&0.92&12.03&0.76&12.20&1.14&0.16&0.53\\
\cmidrule{1-9} Observations & \multicolumn{2}{c}{46} & \multicolumn{2}{c}{28} & \multicolumn{2}{c}{18} & \multicolumn{2}{c}{-} \\ \bottomrule
    \end{tabular}
	\begin{tablenotes}
    	\footnotesize
    	\item Notes:  This table reports summary statistics and differences in city characteristics for the 46 cities with data on NPI intensity and NPI speed. \HighNPI cities are those with above-median \NPISpeed and above-median \NPIIntensity. Peak mortality rates are the maximum excess weekly death rates per 100,000 people on the first wave of the pandemic. Cumulative mortality rates are the sum of excess daily death rates per 100,000 people over the 24 weeks of the pandemic. See \cref{appendix:data-mortality} for more details.
    \end{tablenotes}
    \label{tab:npi_diff}
\end{threeparttable}
\end{table}
\end{landscape}

\begin{table}[htpb]
\centering
\begin{threeparttable}
\caption{\textbf{Non-pharmaceutical interventions and influenza and pneumonia mortality. Robustness to additional controls.}} \label{tab:mortality_robustness_pi}
\tiny
\begin{tabular}{l*{9}{c}}   
\multicolumn{10}{c}{\textbf{Panel A: Peak Mortality}} \\ \toprule
                    &         (1)   &         (2)   &         (3)   &         (4)   &         (5)   &         (6)   &         (7)   &         (8)   &         (9)   \\
\cmidrule{1-10} \HighNPI&       -49.7***&       -44.6** &       -48.1** &       -51.1***&       -50.9***&       -50.6***&       -54.1***&       -55.0***&       -50.8** \\
                    &      (17.6)   &      (19.1)   &      (18.1)   &      (18.2)   &      (17.6)   &      (18.0)   &      (18.7)   &      (16.9)   &      (18.8)   \\
Sep. 14--Sep. 21    &               &         5.8   &               &               &               &               &               &               &        -8.2   \\
                    &               &      (24.6)   &               &               &               &               &               &               &      (29.4)   \\
Sep. 21--Sep. 28    &               &        11.2   &               &               &               &               &               &               &         8.4   \\
                    &               &      (31.5)   &               &               &               &               &               &               &      (39.0)   \\
Sep. 28--Oct. 5     &               &       -13.5   &               &               &               &               &               &               &       -31.9   \\
                    &               &      (21.6)   &               &               &               &               &               &               &      (28.2)   \\
After Oct. 5        &               &        51.1   &               &               &               &               &               &               &        49.0   \\
                    &               &      (47.9)   &               &               &               &               &               &               &      (58.8)   \\
Mortality acceleration date&               &               &        -1.0   &               &               &               &               &        -2.5** &               \\
                    &               &               &       (0.9)   &               &               &               &               &       (1.2)   &               \\
WW1 production      &               &               &               &        17.3   &               &               &               &        55.1***&        41.8** \\
                    &               &               &               &      (16.0)   &               &               &               &      (19.3)   &      (16.7)   \\
WW1 state death rate&               &               &               &         0.2   &               &               &               &         0.1   &         0.2   \\
                    &               &               &               &       (0.6)   &               &               &               &       (0.6)   &       (0.6)   \\
Distance to military camp (log))&               &               &               &               &        -5.0   &               &               &        -5.0   &        -9.8   \\
                    &               &               &               &               &       (7.3)   &               &               &       (6.5)   &       (7.9)   \\
Infant mortality 1916 (per capita)&               &               &               &               &               &       241.1** &               &       420.5***&       360.3***\\
                    &               &               &               &               &               &     (118.3)   &               &     (117.8)   &     (115.0)   \\
Infant mort. (Clay et al.)&               &               &               &               &               &               &         6.4*  &               &               \\
                    &               &               &               &               &               &               &       (3.6)   &               &               \\
\cmidrule{1-10}
\RSq                &        0.45   &        0.49   &        0.46   &        0.47   &        0.45   &        0.49   &        0.54   &        0.63   &        0.66   \\
Observations        &          46   &          46   &          46   &          46   &          46   &          44   &          38   &          44   &          44   \\
\Oster              &         -42   &         -32   &         -39   &         -47   &         -46   &         -46   &         -54   &         -56   &         -47   \\
Baseline controls   &         Yes   &         Yes   &         Yes   &         Yes   &         Yes   &         Yes   &         Yes   &         Yes   &         Yes   \\
   \bottomrule
 \\ \multicolumn{10}{c}{\textbf{Panel B: Cumulative Excess Mortality}} \\ \toprule
                    &         (1)   &         (2)   &         (3)   &         (4)   &         (5)   &         (6)   &         (7)   &         (8)   &         (9)   \\
\cmidrule{1-10} \HighNPI&      -162.4***&      -162.7***&      -156.3***&      -161.1***&      -161.5***&      -161.7***&      -169.6***&      -157.9***&      -166.8***\\
                    &      (45.9)   &      (52.7)   &      (49.0)   &      (51.1)   &      (46.8)   &      (45.9)   &      (52.8)   &      (49.2)   &      (54.2)   \\
Sep. 14--Sep. 21    &               &       -27.0   &               &               &               &               &               &               &       -63.8   \\
                    &               &      (64.5)   &               &               &               &               &               &               &      (72.8)   \\
Sep. 21--Sep. 28    &               &        64.8   &               &               &               &               &               &               &        58.4   \\
                    &               &      (82.1)   &               &               &               &               &               &               &     (103.8)   \\
Sep. 28--Oct. 5     &               &       -18.7   &               &               &               &               &               &               &       -45.4   \\
                    &               &      (73.1)   &               &               &               &               &               &               &      (91.6)   \\
After Oct. 5        &               &       172.4   &               &               &               &               &               &               &       146.1   \\
                    &               &     (107.8)   &               &               &               &               &               &               &     (139.0)   \\
Mortality acceleration date&               &               &        -4.0   &               &               &               &               &        -8.2   &               \\
                    &               &               &       (3.0)   &               &               &               &               &       (5.0)   &               \\
WW1 production      &               &               &               &         2.4   &               &               &               &       119.4   &        53.8   \\
                    &               &               &               &      (50.0)   &               &               &               &      (76.4)   &      (65.7)   \\
WW1 state death rate&               &               &               &         0.6   &               &               &               &         0.4   &         1.7   \\
                    &               &               &               &       (2.1)   &               &               &               &       (2.0)   &       (2.3)   \\
Distance to military camp (log))&               &               &               &               &         3.6   &               &               &        16.8   &       -10.8   \\
                    &               &               &               &               &      (22.5)   &               &               &      (22.0)   &      (23.6)   \\
Infant mortality 1916 (per capita)&               &               &               &               &               &       806.9** &               &      1107.6***&       974.9** \\
                    &               &               &               &               &               &     (358.2)   &               &     (361.8)   &     (378.2)   \\
Infant mort. (Clay et al.)&               &               &               &               &               &               &        10.0   &               &               \\
                    &               &               &               &               &               &               &      (14.0)   &               &               \\
\cmidrule{1-10}
\RSq                &        0.40   &        0.47   &        0.42   &        0.40   &        0.40   &        0.48   &        0.56   &        0.57   &        0.60   \\
Observations        &          46   &          46   &          46   &          46   &          46   &          44   &          38   &          44   &          44   \\
\Oster              &        -193   &        -188   &        -178   &        -190   &        -191   &        -186   &        -198   &        -176   &        -192   \\
Baseline controls   &         Yes   &         Yes   &         Yes   &         Yes   &         Yes   &         Yes   &         Yes   &         Yes   &         Yes   \\
   \bottomrule
 \\ 
\end{tabular}
	\footnotesize
		\begin{tablenotes} \item
	    Notes: This table presents city-level regressions of peak mortality (panel A) and cumulative excess mortality (panel B). Mortality refers to influenza and pneumonia mortality. Peak mortality is the weekly excess death rate per 100,000 in the first peak of the fall 1918 pandemic. Cumulative excess mortality is the total excess death rate from September 8, 1918 to February 22, 1919. Baseline controls not reported in the table are the log of city population in 1900 and 1910, the city 1914 manufacturing employment to 1910 population, city public health spending per capita in 1917, city density in 1910, longitude, share of illiteracy, and reliance on coal-fired power plants. Additional controls as indicated. \emph{WW1 state death rate} is per 100,000 inhabitants and using the 1910 population counts. \emph{Infant mortality 1916} is per capita of the population in 1910. \emph{Infant mort. (Clay et al.)} is infant mortality in 1918 normalized by 1921 birth rates, as reported by \citet{clay2018}.
	    Robust standard errors in parentheses.
	    *, **, and *** indicate significance at the 10\%, 5\%, and 1\% level, respectively.
\end{tablenotes}
\end{threeparttable}
\end{table}

\begin{table}[htpb]
\centering
  \begin{threeparttable}
  \caption{\textbf{Non-pharmaceutical interventions and all-cause mortality. Robustness to additional controls and to dropping western cities.}} \label{tab:mortality_robustness_pi_drop_west}
\tiny
\begin{tabular}{l*{9}{c}}   
    \multicolumn{10}{c}{\textbf{Panel A: Peak Mortality}} \\ \toprule 
                        &         (1)   &         (2)   &         (3)   &         (4)   &         (5)   &         (6)   &         (7)   &         (8)   &         (9)   \\
\cmidrule{1-10} \NPIIntensity&        -0.6***&        -0.6***&        -0.2   &               &               &               &               &               &               \\
                    &       (0.1)   &       (0.2)   &       (0.2)   &               &               &               &               &               &               \\
\NPISpeed           &               &               &               &        -1.0   &        -1.1   &        -0.4   &               &               &               \\
                    &               &               &               &       (1.0)   &       (1.0)   &       (0.8)   &               &               &               \\
\HighNPI            &               &               &               &               &               &               &       -57.7***&       -59.7***&       -37.3*  \\
                    &               &               &               &               &               &               &      (12.1)   &      (15.4)   &      (19.6)   \\
1917 I\&P mortality (per 100k)&               &         0.4** &         0.4** &               &         0.5***&         0.4** &               &         0.4** &         0.4*  \\
                    &               &       (0.2)   &       (0.2)   &               &       (0.2)   &       (0.2)   &               &       (0.2)   &       (0.2)   \\
Longitude           &               &               &         3.4***&               &               &         3.9***&               &               &         2.5** \\
                    &               &               &       (1.0)   &               &               &       (0.9)   &               &               &       (1.1)   \\
Illiteracy          &               &               &         1.8   &               &               &         3.4   &               &               &         1.6   \\
                    &               &               &       (3.6)   &               &               &       (3.1)   &               &               &       (3.2)   \\
Medium coal-fired capacity&               &               &        -9.9   &               &               &       -10.3   &               &               &         5.2   \\
                    &               &               &      (20.8)   &               &               &      (19.9)   &               &               &      (23.2)   \\
High coal-fired capacity&               &               &       -29.0   &               &               &       -31.9   &               &               &       -21.0   \\
                    &               &               &      (20.1)   &               &               &      (20.0)   &               &               &      (20.2)   \\
\cmidrule{1-10}
\RSq                &        0.21   &        0.55   &        0.65   &        0.02   &        0.41   &        0.64   &        0.26   &        0.61   &        0.68   \\
Observations        &          39   &          39   &          39   &          39   &          39   &          39   &          39   &          39   &          39   \\
\Oster              &           .   &        -.55   &       -.047   &           .   &        -1.2   &        -.26   &           .   &         -61   &         -27   \\
Baseline controls   &           -   &         Yes   &         Yes   &           -   &         Yes   &         Yes   &           -   &         Yes   &         Yes   \\
   \bottomrule
 \\
    \multicolumn{10}{c}{\textbf{Panel B: Cumulative Excess Mortality}} \\ \toprule 
                        &         (1)   &         (2)   &         (3)   &         (4)   &         (5)   &         (6)   &         (7)   &         (8)   &         (9)   \\
\cmidrule{1-10} \NPIIntensity&        -1.4** &        -1.1*  &        -0.6   &               &               &               &               &               &               \\
                    &       (0.6)   &       (0.6)   &       (0.6)   &               &               &               &               &               &               \\
\NPISpeed           &               &               &               &        -4.2   &        -4.8** &        -3.6   &               &               &               \\
                    &               &               &               &       (3.2)   &       (2.1)   &       (2.3)   &               &               &               \\
\HighNPI            &               &               &               &               &               &               &      -142.5***&      -115.0***&       -98.0** \\
                    &               &               &               &               &               &               &      (42.4)   &      (41.6)   &      (46.7)   \\
1917 I\&P mortality (per 100k)&               &         2.1***&         2.0***&               &         2.2***&         1.9***&               &         1.9***&         1.9***\\
                    &               &       (0.3)   &       (0.4)   &               &       (0.3)   &       (0.4)   &               &       (0.4)   &       (0.4)   \\
Longitude           &               &               &         7.6*  &               &               &         8.1*  &               &               &         5.4   \\
                    &               &               &       (4.1)   &               &               &       (3.9)   &               &               &       (4.2)   \\
Illiteracy          &               &               &        -5.8   &               &               &        -0.2   &               &               &        -6.2   \\
                    &               &               &      (13.6)   &               &               &      (14.3)   &               &               &      (13.3)   \\
Medium coal-fired capacity&               &               &        80.7   &               &               &        93.4   &               &               &       119.8*  \\
                    &               &               &      (59.9)   &               &               &      (55.0)   &               &               &      (68.4)   \\
High coal-fired capacity&               &               &        81.4   &               &               &        74.9   &               &               &       101.8   \\
                    &               &               &      (55.3)   &               &               &      (53.5)   &               &               &      (60.8)   \\
\cmidrule{1-10}
\RSq                &        0.13   &        0.55   &        0.61   &        0.04   &        0.53   &        0.62   &        0.17   &        0.58   &        0.63   \\
Observations        &          39   &          39   &          39   &          39   &          39   &          39   &          39   &          39   &          39   \\
\Oster              &           .   &         -.9   &        -.28   &           .   &          -5   &        -3.4   &           .   &        -103   &         -80   \\
Baseline controls   &           -   &         Yes   &         Yes   &           -   &         Yes   &         Yes   &           -   &         Yes   &         Yes   \\
   \bottomrule
 \\
 \end{tabular}
	\footnotesize
		\begin{tablenotes} \item
	    Notes: This table presents city-level regressions of peak mortality (panel A) and cumulative excess mortality (panel B). Mortality refers to all-cause mortality, and is based on data from \citet{MortalityStatistics}, \citet{InfluenzaArchive}, \citet{PublicHealthReports}, and \citet{PopCensus1920}; see  \cref{appendix:data-mortality} for details.
	    NPI data is based on \citet{Markel2007}, \citet{Berkes2022}, as well as our own data collection; see \cref{appendix:data-npi} for details. Peak mortality is the weekly excess death rate per 100,000 in the first peak of the fall 1918 pandemic. Cumulative excess mortality is the total excess death rate from September 8, 1918 to February 22, 1919. ``Baseline Controls" note reported in the table are city log 1900 and 1910 population, city 1914 manufacturing employment to 1910 population, city public health spending per capita, city density. Additional controls as indicated. The analysis drops cities with a longitude of less than -100 (Denver, Los Angeles, Oakland, San Francisco, Seattle, and Spokane).
	    Robust standard errors in parentheses. 
	    *, **, and *** indicate significance at the 10\%, 5\%, and 1\% level, respectively.
\end{tablenotes}
\end{threeparttable}
\end{table}

\newgeometry{top=.8in}
\begin{table}
\scriptsize
\centering
  \begin{threeparttable}
  \caption{\textbf{Non-pharmaceutical interventions and local economic activity in Bradstreet's Trade Conditions. Individual sectors.}}\label{tab:bradstreets_by_index}
\begin{tabular}{lcccccc}
\multicolumn{7}{c}{\textbf{Panel A: Wholesale}} \\ \toprule
                    &         (1)   &         (2)   &         (3)   &         (4)   &         (5)   &         (6)   \\
 \cmidrule(lr){1-7}  $NPI \ Intensity_c \times Post_{t}$&       0.022   &      -0.035   &               &               &               &               \\
                    &     (0.091)   &     (0.092)   &               &               &               &               \\
$NPI \ Speed_c \times  Post_{t}$&               &               &      -0.599   &      -0.343   &               &               \\
                    &               &               &     (0.476)   &     (0.587)   &               &               \\
 $  High \ NPI_c \times  Post_{t}$&               &               &               &               &       1.155   &      -6.943   \\
                    &               &               &               &               &     (9.020)   &     (9.460)   \\
\cmidrule{1-7}
\cmidrule(lr){1-7} Within \RSq&        0.00   &        0.03   &        0.01   &        0.03   &        0.00   &        0.03   \\
Observations        &         398   &         398   &         398   &         398   &         398   &         398   \\
Number of cities    &          27   &          27   &          27   &          27   &          27   &          27   \\
\Oster              &           .   &       -.053   &           .   &        -.24   &           .   &        -9.4   \\
Baseline controls   &           -   &         Yes   &           -   &         Yes   &           -   &         Yes   \\
    \bottomrule
 \\
\multicolumn{7}{c}{\textbf{Panel B: Retail}} \\ \toprule
                    &         (1)   &         (2)   &         (3)   &         (4)   &         (5)   &         (6)   \\
 \cmidrule(lr){1-7}  $NPI \ Intensity_c \times Post_{t}$&       0.037   &       0.065   &               &               &               &               \\
                    &     (0.100)   &     (0.102)   &               &               &               &               \\
$NPI \ Speed_c \times  Post_{t}$&               &               &      -0.144   &       0.143   &               &               \\
                    &               &               &     (0.694)   &     (0.755)   &               &               \\
 $  High \ NPI_c \times  Post_{t}$&               &               &               &               &       2.812   &       4.359   \\
                    &               &               &               &               &     (9.252)   &    (11.209)   \\
\cmidrule{1-7}
\cmidrule(lr){1-7} Within \RSq&        0.00   &        0.02   &        0.00   &        0.02   &        0.00   &        0.02   \\
Observations        &         397   &         397   &         397   &         397   &         397   &         397   \\
Number of cities    &          27   &          27   &          27   &          27   &          27   &          27   \\
\Oster              &           .   &        .074   &           .   &         .23   &           .   &         4.8   \\
Baseline controls   &           -   &         Yes   &           -   &         Yes   &           -   &         Yes   \\
    \bottomrule
 \\  
\multicolumn{7}{c}{\textbf{Panel C: Manufacturing}} \\ \toprule
                    &         (1)   &         (2)   &         (3)   &         (4)   &         (5)   &         (6)   \\
 \cmidrule(lr){1-7}  $NPI \ Intensity_c \times Post_{t}$&      -0.130   &      -0.059   &               &               &               &               \\
                    &     (0.097)   &     (0.097)   &               &               &               &               \\
$NPI \ Speed_c \times  Post_{t}$&               &               &      -0.432   &       0.202   &               &               \\
                    &               &               &     (0.364)   &     (0.299)   &               &               \\
 $  High \ NPI_c \times  Post_{t}$&               &               &               &               &     -15.658   &     -11.841   \\
                    &               &               &               &               &     (9.381)   &     (9.603)   \\
\cmidrule{1-7}
\cmidrule(lr){1-7} Within \RSq&        0.02   &        0.09   &        0.00   &        0.09   &        0.02   &        0.09   \\
Observations        &         395   &         395   &         395   &         395   &         395   &         395   \\
Number of cities    &          27   &          27   &          27   &          27   &          27   &          27   \\
\Oster              &           .   &       -.033   &           .   &          .4   &           .   &         -10   \\
Baseline controls   &           -   &         Yes   &           -   &         Yes   &           -   &         Yes   \\
    \bottomrule
 \\ 
\end{tabular}
  \footnotesize
    \begin{tablenotes} \item
     Notes: This table presents estimates of equation \eqref{DD_bradstreet}. The dependent variables are monthly city-level indexes of economic disruptions that take a value of 100 for ``No disruptions'' and 0 for ``Disruptions''  (see \cref{appendix:data-bradstreets} for details). Controls interacted with $Post_t$ are log 1900 and 1910 city population, 1910 city density, 1917 health spending per capita, and manufacturing employment in 1914 to 1910 population. Standard errors are clustered at the city level.
\end{tablenotes}
  \end{threeparttable}
\end{table}

 \begin{landscape}
\begin{table}
\scriptsize
\centering
  \begin{threeparttable}
  \caption{\textbf{Non-pharmaceutical interventions and short-term economic disruptions in Bradstreet's Trade Conditions. Controlling for Western status, WW1 exposure, poverty, and air pollution.}}\label{tab:bradstreet_robustness}
\begin{tabular}{lccccccccccc}
\toprule
  \multicolumn{12}{c}{\textbf{Combined Index of Trade Disruptions}} \\ \midrule 
                        &         (1)   &         (2)   &         (3)   &         (4)   &         (5)   &         (6)   &         (7)   &         (8)   &         (9)   &        (10)   &        (11)   \\
\cmidrule(lr){1-12} $  High \ NPI_c \times  Post_{t}$&        -5.0   &       -12.2*  &         1.0   &        -7.9   &        -4.1   &        -5.0   &        -4.6   &        -4.8   &        -8.1   &        -8.4   &        -6.4   \\
                    &       (5.7)   &       (6.0)   &       (7.4)   &       (5.8)   &       (4.9)   &       (5.9)   &       (6.2)   &       (6.8)   &       (7.3)   &       (5.3)   &       (5.4)   \\
\cmidrule{1-7}
\cmidrule(lr){1-12} Within \RSq&        0.06   &        0.08   &        0.08   &        0.06   &        0.09   &        0.06   &        0.06   &        0.06   &        0.06   &        0.07   &        0.15   \\
Observations        &         399   &         399   &         399   &         399   &         399   &         399   &         399   &         399   &         399   &         399   &         399   \\
Number of cities    &          27   &          27   &          27   &          27   &          27   &          27   &          27   &          27   &          27   &          27   &          27   \\
\Oster              &        -5.1   &         -14   &         2.9   &        -8.8   &        -3.8   &          -5   &        -4.5   &        -4.8   &          -9   &        -9.5   &        -6.8   \\
Baseline controls   &         Yes   &         Yes   &         Yes   &         Yes   &         Yes   &         Yes   &         Yes   &         Yes   &         Yes   &         Yes   &         Yes   \\
Additional controls &           -   &        Lon.   &Sydenstricker   &Mort. accel.   &   WW1 prod.   &  Casualties   &  Camp dist.   &Infant mort.   &  Illiteracy   &        Coal   &         All   \\
    \bottomrule
 \\
\end{tabular}
	\footnotesize
		\begin{tablenotes} \item
Notes: This table presents estimates of equation \eqref{DD_bradstreet}.
The dependent variables are monthly city-level indexes of economic disruptions that take a value of 100 for ``No disruptions'' and 0 for ``Disruptions'' (see \cref{appendix:data-bradstreets} for details).
Time and city fixed effects included.
Baseline controls interacted with $Post_t$ are the share of manufacturing employment in 1914, log of population in 1900 and 1910, city density in 1910, and per capita city health spending in 1917.
Columns (2)-(10) add alternating controls interacted with the $\text{Post}_{t}$  indicator: city longitude, flu arrival dates according to \citet{Sydenstricker1918}, the mortality acceleration date, an indicator of heavy involvement in WW1 production according to \citet{Garrett2009}, state-level WW1 casualties, distance to the closest military camp, infant mortality in 1916, the illiteracy share, and measures of reliance to coal-fired plants according to \citet{clay2018}.
Column (11) includes all baseline and additional controls.
Robust standard errors clustered at the city level in parentheses. 
\end{tablenotes}
	\end{threeparttable}
\end{table}
\end{landscape}

\begin{landscape}
\begin{table}
\scriptsize
\centering
 \begin{threeparttable}
 \caption{\textbf{Non-pharmaceutical interventions and local manufacturing employment and output. Controlling for Western status, WW1 exposure, poverty, and air pollution.}}\label{tab:manufacturing_robustness}
\begin{tabular}{lccccccccccc}
  \multicolumn{12}{c}{\textbf{Panel A: Manufacturing Employment}} \\ \toprule
                      &         (1)   &         (2)   &         (3)   &         (4)   &         (5)   &         (6)   &         (7)   &         (8)   &         (9)   &        (10)   &        (11)   \\
\cmidrule(lr){1-12} $  High \ NPI_c \times  Post_{t}$&        14.7***&        11.3** &        16.8***&        14.0***&        14.1***&        14.9***&        14.6***&        12.8** &        11.0** &        15.4***&        13.3** \\
                    &       (4.4)   &       (4.6)   &       (4.7)   &       (4.2)   &       (4.2)   &       (4.3)   &       (4.4)   &       (5.4)   &       (4.7)   &       (4.6)   &       (5.0)   \\
\cmidrule{1-7}
\cmidrule(lr){1-12} Within \RSq&        0.44   &        0.45   &        0.47   &        0.44   &        0.44   &        0.44   &        0.44   &        0.44   &        0.45   &        0.45   &        0.51   \\
Num. obs.           &         368   &         368   &         368   &         368   &         368   &         368   &         368   &         368   &         368   &         368   &         368   \\
Number of cities    &          46   &          46   &          46   &          46   &          46   &          46   &          46   &          46   &          46   &          46   &          46   \\
\Oster              &          10   &         5.2   &          13   &           9   &         9.1   &          10   &         9.9   &         7.2   &         4.7   &          11   &         8.4   \\
Baseline controls   &         Yes   &         Yes   &         Yes   &         Yes   &         Yes   &         Yes   &         Yes   &         Yes   &         Yes   &         Yes   &         Yes   \\
Additional control  &           -   &        Lon.   &Sydenstricker   &Mort. accel.   &   WW1 prod.   &  Casualties   &  Camp dist.   &Infant mort.   &  Illiteracy   &        Coal   &         All   \\
    \bottomrule
 \\
  \multicolumn{12}{c}{\textbf{Panel B: Manufacturing Output}} \\ \toprule
                      &         (1)   &         (2)   &         (3)   &         (4)   &         (5)   &         (6)   &         (7)   &         (8)   &         (9)   &        (10)   &        (11)   \\
\cmidrule(lr){1-12} $  High \ NPI_c \times  Post_{t}$&         8.7   &         8.7   &        10.6   &         7.2   &         7.4   &         8.7   &         9.0   &         9.9   &         9.1   &        10.2   &        12.2*  \\
                    &       (7.0)   &       (7.2)   &       (7.3)   &       (6.5)   &       (6.3)   &       (6.9)   &       (7.1)   &       (8.1)   &       (7.4)   &       (7.3)   &       (7.3)   \\
\cmidrule{1-7}
\cmidrule(lr){1-12} Within \RSq&        0.24   &        0.24   &        0.31   &        0.25   &        0.25   &        0.24   &        0.24   &        0.24   &        0.24   &        0.26   &        0.34   \\
Num. obs.           &         368   &         368   &         368   &         368   &         368   &         368   &         368   &         368   &         368   &         368   &         368   \\
Number of cities    &          46   &          46   &          46   &          46   &          46   &          46   &          46   &          46   &          46   &          46   &          46   \\
\Oster              &         4.6   &         4.5   &         7.5   &         2.4   &         2.8   &         4.5   &           5   &         6.3   &         5.1   &         6.7   &         9.8   \\
Baseline controls   &         Yes   &         Yes   &         Yes   &         Yes   &         Yes   &         Yes   &         Yes   &         Yes   &         Yes   &         Yes   &         Yes   \\
Additional control  &           -   &        Lon.   &Sydenstricker   &Mort. accel.   &   WW1 prod.   &  Casualties   &  Camp dist.   &Infant mort.   &  Illiteracy   &        Coal   &         All   \\
    \bottomrule
 \\
\end{tabular}
	\footnotesize
		\begin{tablenotes} \item
Notes: This table presents estimates of equation \eqref{DD_bradstreet}.
The dependent variable is average manufacturing employment for Panel A and manufacturing output for Panel B, using data from the 1904, 1909, 1914, 1919, 1921, 1923, 1925, and 1927 manufacturing census.
Time and city fixed effects included.
Baseline controls interacted with $Post_{1919}$ are the share of manufacturing employment in 1914, log of population in 1900 and 1910, city density in 1910, and per capita city health spending in 1917.
Columns (2)-(10) add alternating controls interacted with the $\text{Post}_{1919}$  indicator: city longitude, flu arrival dates according to \citet{Sydenstricker1918}, the mortality acceleration date, an indicator of heavy involvement in WW1 production according to \citet{Garrett2009}, state-level WW1 casualties, distance to the closest military camp, infant mortality in 1916, the illiteracy share, and measures of reliance to coal-fired plants according to \citet{clay2018}.
Column (11) includes all baseline and additional controls.
Robust standard errors clustered at the city level in parentheses. 
\end{tablenotes}
\end{threeparttable}
\end{table}
\end{landscape}

\begin{landscape}
\begin{table}

\scriptsize
\centering
  \begin{threeparttable}
  \caption{\textbf{Non-pharmaceutical interventions and local manufacturing employment and output. Controlling for Western status, WW1 exposure, poverty, and air pollution. Robustness to dropping western cities.}}\label{tab:manufacturing_robustness_west}
\begin{tabular}{lccccccccccc}
  \multicolumn{12}{c}{\textbf{Panel A: Manufacturing Employment}} \\ \toprule
                    &         (1)   &         (2)   &         (3)   &         (4)   &         (5)   &         (6)   &         (7)   &         (8)   &         (9)   &        (10)   &        (11)   \\
\cmidrule(lr){1-12} $  High \ NPI_c \times  Post_{t}$&        14.0***&        11.0*  &        12.4***&        12.9***&        12.9***&        14.0***&        13.2***&        12.7** &         9.7** &        13.9***&         8.0   \\
                    &       (4.2)   &       (6.1)   &       (4.3)   &       (4.0)   &       (3.8)   &       (4.2)   &       (4.3)   &       (5.7)   &       (4.5)   &       (4.1)   &       (6.5)   \\
\cmidrule{1-7}
\cmidrule(lr){1-12} Within \RSq&        0.28   &        0.29   &        0.30   &        0.29   &        0.30   &        0.28   &        0.30   &        0.29   &        0.31   &        0.29   &        0.36   \\
Num. obs.           &         312   &         312   &         312   &         312   &         312   &         312   &         312   &         312   &         312   &         312   &         312   \\
Number of cities    &          39   &          39   &          39   &          39   &          39   &          39   &          39   &          39   &          39   &          39   &          39   \\
\Oster              &          15   &          11   &          13   &          13   &          14   &          15   &          14   &          13   &         9.1   &          15   &         6.8   \\
Baseline controls   &         Yes   &         Yes   &         Yes   &         Yes   &         Yes   &         Yes   &         Yes   &         Yes   &         Yes   &         Yes   &         Yes   \\
Additional control  &           -   &        Lon.   &Sydenstricker   &Mort. accel.   &   WW1 prod.   &  Casualties   &  Camp dist.   &Infant mort.   &  Illiteracy   &        Coal   &         All   \\
    \bottomrule
 \\
\multicolumn{12}{c}{\textbf{Panel B: Manufacturing Output}} \\ \toprule
                    &         (1)   &         (2)   &         (3)   &         (4)   &         (5)   &         (6)   &         (7)   &         (8)   &         (9)   &        (10)   &        (11)   \\
\cmidrule(lr){1-12} $  High \ NPI_c \times  Post_{t}$&         9.1   &        11.6   &         5.4   &         6.8   &         7.4   &         9.1   &         8.7   &         8.6   &         6.6   &         9.3   &        12.3   \\
                    &       (7.4)   &       (8.9)   &       (7.2)   &       (6.9)   &       (6.4)   &       (7.3)   &       (7.2)   &       (8.6)   &       (7.1)   &       (6.9)   &       (9.7)   \\
\cmidrule{1-7}
\cmidrule(lr){1-12} Within \RSq&        0.17   &        0.17   &        0.19   &        0.18   &        0.19   &        0.17   &        0.17   &        0.17   &        0.17   &        0.19   &        0.22   \\
Num. obs.           &         312   &         312   &         312   &         312   &         312   &         312   &         312   &         312   &         312   &         312   &         312   \\
Number of cities    &          39   &          39   &          39   &          39   &          39   &          39   &          39   &          39   &          39   &          39   &          39   \\
\Oster              &         9.7   &          13   &         4.8   &         6.7   &         7.4   &         9.7   &         9.1   &         9.1   &         6.4   &          10   &          14   \\
Baseline controls   &         Yes   &         Yes   &         Yes   &         Yes   &         Yes   &         Yes   &         Yes   &         Yes   &         Yes   &         Yes   &         Yes   \\
Additional control  &           -   &        Lon.   &Sydenstricker   &Mort. accel.   &   WW1 prod.   &  Casualties   &  Camp dist.   &Infant mort.   &  Illiteracy   &        Coal   &         All   \\
    \bottomrule
 \\
\end{tabular}
	\footnotesize
		\begin{tablenotes} \item
Notes: This table presents estimates of equation \eqref{DD_bradstreet}.
The dependent variable is average manufacturing employment for Panel A and manufacturing output for Panel B, using data from the 1904, 1909, 1914, 1919, 1921, 1923, 1925, and 1927 manufacturing census.
Time and city fixed effects included.
Baseline controls interacted with $Post_{1919}$ are the share of manufacturing employment in 1914, log of population in 1900 and 1910, city density in 1910, and per capita city health spending in 1917.
Columns (2)-(10) add alternating controls interacted with the $\text{Post}_{1919}$  indicator: city longitude, flu arrival dates according to \citet{Sydenstricker1918}, the mortality acceleration date, an indicator of heavy involvement in WW1 production according to \citet{Garrett2009}, state-level WW1 casualties, distance to the closest military camp, infant mortality in 1916, the illiteracy share, and measures of reliance to coal-fired plants according to \citet{clay2018}.
Column (11) includes all baseline and additional controls.
We exclude the westernmost cities from the sample, i.e. those with a longitude of less than -100 (Denver, Los Angeles, Oakland, San Francisco, Seattle, and Spokane).
Robust standard errors clustered at the city level in parentheses. 
\end{tablenotes}    
	\end{threeparttable}
\end{table}
\end{landscape}

\newgeometry{top=.8in}
\begin{table}
\centering
  \begin{threeparttable}
  \caption{\textbf{Non-pharmaceutical interventions and national bank assets.}}\label{tab:reg_NPI-national-banks}
\begin{tabular}{lcccccc}
\toprule \multicolumn{7}{c}{\textbf{Dependent Variable: National Bank Assets}} \\ \midrule 
                    &         (1)   &         (2)   &         (3)   &         (4)   &         (5)   &         (6)   \\
 \cmidrule(lr){1-7}  $NPI \ Intensity_c \times Post_{t}$&       0.161   &       0.136   &               &               &               &               \\
                    &     (0.144)   &     (0.115)   &               &               &               &               \\
$NPI \ Speed_c \times  Post_{t}$&               &               &       1.056   &       1.143*  &               &               \\
                    &               &               &     (0.644)   &     (0.612)   &               &               \\
 $  High \ NPI_c \times  Post_{t}$&               &               &               &               &      12.621   &      10.057   \\
                    &               &               &               &               &     (9.109)   &     (9.366)   \\
\cmidrule{1-7}
\cmidrule(lr){1-7} Within \RSq&        .028   &         .15   &        .047   &         .13   &         .02   &          .1   \\
Observations        &         589   &         545   &         490   &         446   &         490   &         446   \\
Number of cities    &          54   &          50   &          45   &          41   &          45   &          41   \\
\Oster              &           .   &        0.13   &           .   &        1.18   &           .   &        9.10   \\
Controls            &           -   &         Yes   &           -   &         Yes   &           -   &         Yes   \\
    \bottomrule
 \\
\end{tabular}
\footnotesize
\begin{tablenotes} \item
Notes: This table presents estimates of equation \eqref{DD_bradstreet}.
The dependent variable is the log of total assets of national banks per city, at annual frequency from 1910 to 1925.
City and Federal Reserve District times year fixed effects are included.
Columns (2), (4), and (6) add the following controls interacted with $Post_{1919}$: log of 1900 and 1910 city population, 1910 city density, 1917 city health spending per capita, manufacturing employment in 1914 to 1910 population, and the city-specific quota for the Third Liberty Loan program (May 1918), retrieved from a \href{https://books.google.com/books?id=0WUWAAAAIAAJ&pg=PA20}{1918 hearing before the House of Representatives Committee on Ways and Means}.
Robust standard errors clustered by city in parenthesis.
 
\end{tablenotes}
	\end{threeparttable}
\end{table}

\restoregeometry

    \clearpage
\newgeometry{left=1.0in,right=1.0in}
\section{Narrative Evidence from Historical Newspaper Articles and Other Sources}
\label{appendix:narrative}
\singlespacing
This appendix contains excerpts of newspaper articles during 1918 flu pandemic. The articles are organized by theme. We first present articles chronicling the the timeline, debates, and impact of non-pharmaceutical public interventions. We then present articles that document the real effects of the pandemic economic activity, including on production and trade.



\subsection{Non-pharmaceutical public health interventions}


\subsubsection{Examples of introduction of NPIs} 
\label{sec:newspapers_NPIgeneral}

\paragraph{``Drastic Steps Taken To Fight Influenza Here: Health Board Issues 4 P.M. Closing Orders for All Stores Except Food and Drug Shops. Hours for Factories Fixed. Plan, in Effect Today, to Reduce Crowding in Transportation Lines in Rush Periods. Time Table for Theatres. Radical Regulations Necessary to Prevent Shutting City Up Tight, Says Dr. Copeland.'' {\normalfont\emph{New York Times}, October 5, 1918, p. 1.}}

In order to prevent the complete shutdown of industry and amusement in this city to check the spread of Spanish influenza, Health Commissioner Copeland, by proclamation, yesterday ordered a change in the hours for opening stores, theatres and other places of business.

The Department is of the opinion that the greatest sources of spread of the disease are crowded subway and elevated trains and cars on the surface lines and the purpose of the order is to diminish the ``peak'' load in the evenings and mornings on these lines by distributing the travelers over a greater space of time.
This will reduce crowding to a minimum.

Dr. Copeland's action was taken after a statement made by Surgeon General Blue, Chief of the Public Health Service in Washington, was called to his attention, in which Dr. Blue advocated the closing of churches, schools, theatres and public institutions in every community where the epidemic has been developed. Dr Blue said:

``There is no way to put a nationwide closing order into effect, as this is a matter which is up to the individual communities. In some States the State Board of Health has this power, but in many others it is a matter of municipal regulation.
I hope that those having the proper authority will close all public gathering places if their community is threatened with the epidemic. This will do much toward checking the spread of the disease''

\dots  One of the decisions reached is to close all stores other than those dealing exclusively in food or drugs at 4 o'clock in the afternoon. \dots

All moving picture houses and theatres outside of a certain district are considered community houses and are held to draw their patronage from within walking distance.
There was debate on the proposition to close the schools and churches and other places of assemblage, but it was decided against it at this time. \dots

 
\paragraph{``The Spanish Influenza.'' {\normalfont\emph{New York Times}, October 7, 1918, p. 12.}}

Under adverse conditions the health authorities of American communities are now grappling with an epidemic that they do not understand very well.
But they understand it well enough to know that it spreads rapidly where people are crowded together in railway trains, in theatres and places of amusement, in stores and factories and schools.
In some cities and towns where the influenza seems to be malignant the schools and many places of amusement have been closed.
Pennsylvania, taking a serious view of the hazards of the disease, because it is raging in the shipyards and increasing ominously elsewhere, has taken drastic measures to protect the public health.
The sale of liquor has been generally prohibited in Philadelphia, the courts stand adjourned, Liberty Loan meetings have been abandoned, public assemblies of all kinds have been forbidden, the theatres are not allowed to give performances, and it is recommended that the churches hold no services.
In some other parts of Pennsylvania the authorities have gone further, closing churches and Sunday schools.
Football games have been canceled.
In localities in New Jersey the public schools have been closed.
This is the case in Omaha and other Western cities. In Oswego, where about 15 per cent of the population is down with influenza, the Health Board has acted vigorously. \dots

New York City has thus far escaped lightly compared with Boston, which has had 100,000 cases, and with Philadelphia, where the total two days ago was 20,000. Up to yesterday only 8,000 cases had been reported in this city of about 6,000,000 people, according to the Health Department, although there are perhaps many cases still unreported.
It seems providential that there have been so few cases in our congested districts, and generally in a population that packs the transportation lines.
But unless our health authorities are vigilant and practical, there may soon be another story to tell.
The precautionary and restrictive regulations adopted by the Department of Health are at best tentative.
It is a question whether the schools should not be temporarily closed, as in other places.
As business must go on, if not as usual, it was advisable to vary the opening and closing hours of business establishments to regulate the ``rush hours'' on transportation lines.
The opening time of theatres has been changed with a similar purpose.
It is evident that the Health Department hesitates to be strenuous, because, as Dr. Copeland says, ``this community is not striken with the epidemic''.

But it may be if only half measures at taken. A stitch in time saves nine.
The closing of the schools is a debatable question.
Dr. Copeland's reasons for keeping them open are not altogether convincing. \dots

\paragraph{``Delays In Reports Swell Grip Figures: 1,450 New Cases Recorded, Largest Number for a Single Day Since Epidemic Began. Newark Officials Clash. Mayor Raises Closing Bank OVer Head of the State Board of Health.'' {\normalfont\emph{New York Times}, October 24, 1918, p. 12.}}

For the twenty-four hours ended at 10 o'clock yesterday morning, 1,450 new cases of Spanish influenza were reported to the Board of Health. This is the largest number of new cases reported in a single day since the disease became epidemic in New York.

\dots Major Gillen of Newark, and the New Jersey State Board of Health yesterday began a controversy over the authority of the city officials in ordering the raising of the closing order on schools, theatres, saloons, soda fountains and churches after the State Board had ruled that all should be closed until it lifted the ban.
A meeting of the State Board will be held in Trenton today to consider measures compelling the Newark City Government to enforce the rule.
The Newark City Commission also will hold a meeting to discuss whether it has jurisdiction upon health superior to that of the State Board.

\dots After being held twenty-four hours in Quarantine for examination and fumigation the Holland-America liner Nieuw Amsterdam was permitted to leave for the pier to land her 900 passengers yesterday. The health officers at Quarantine said there had been fifty cases of Spanish influenza on the voyage from Holland, but only twelve passengers in the second cabin were still confined to their berths when the steamship reached port on Tuesday. \dots

\paragraph{``Mayor Closes Theatres, Schools and Churches. Sudden Spread of Spanish Influenza Forces City Officials to Take Drastic Steps. 25 Flu Cases in Seattle Reported.'' {\normalfont\emph{The Seattle Star}, October 5, 1918, p. 1.}}
 
All churches, schools, theatres and places of assemblage were ordered closed by proclamation of Mayor Hanson at noon Saturday, to check the spread of the Spanish influenza.

Police officers were immediately sent to the motion picture houses to enforce the order.

At 2 p.m. policemen had served notice on all the downtown theatres, including movie houses, and the had close their doors.

While latitude was given to officers in orders to close all other assemblages in buildings.

No church services will be permitted Sunday.

``We will enforce the order to the letter,'' Mayor Hanson declared. ``The chief of police has been given orders. Dance halls were ordered closed last night. No private dances must be held. Persons spitting on sidewalks or in street cars are to be immediately places under arrest.''

His order followed consultation with Health Commissioner McBride, who reported that there were 25 civilian cases on record at noon.

New cases are being reported every few minutes.

There has been one civilian death. \dots

\paragraph{``Not Ready to Lift the Influenza Ban.'' {\normalfont\emph{The Seattle star}, October 23, 1918, p. 3.}}
 
Twelve influenza and pneumonia cases have been reported in Seattle to the health department within the last 24 hours, while 194 new cases were reported Wednesday morning.
Five deaths occurred late Tuesday night and Wednesday morning. \dots

Wednesday, Dr. J. S. McBride, city health commissioner, announces that the crest of the epidemic has been passed, but that great caution must be observed by every individual for some time yet.
He has not announced when the ban will be lifted.

\subsubsection{Examples of sources of NPI variation across cities: Learning from cities in the east, individual public health officials, local political economy, patriotism and WWI, and other drivers of the variation}

\label{sec:newspapers_NPIvariation}

\paragraph{``Health Officer Explains Order—Best of Reasons for Exempting the Liberty Loan—Playgrounds Are Closed—All Local Funerals Must Be Held in Cemeteries, Not in Church or Home.'' {\normalfont\emph{Charleston News and Courier}, October 9, 1918, p. 8.}} The head of the local board of health has listened to so many queries sent the fourth liberty loan meetings being held this week, although meetings of almost every other sort are temporarily prohibited on account of the epidemic of Spanish ``flu,'' that he explains in the following manner: 

``It is quite true that it is just as dangerous for a large number of individuals to congregate in the interests of the fourth liberty loan as it is for said persons to assemble for social or other reasons \dots But that is just the point which all America, Charleston included, is trying to make. It is this way: part of the American people are having to face the bullets of the enemy. We can’t all do that. Some of us are unable to go to war \dots But that portion of us who are thus incapacitated must make martyrs of ourselves, if it can be called that, by facing a possible illness in order to help our country in time of need.\dots That is how we can explain the meetings of the fourth liberty loan being held at the time. We may contract influenza. But if we do, we shall have done so in a splendid cause. \dots''

\paragraph{``No Quarantine!'' {\normalfont\emph{The Cincinnati Enquirer}, October 7, 1918, p. 14.}} Reports have gained currency in various cities, particularly Indianapolis, that a quarantine has been established in Cincinnati, preventing arrival and departure, Mayor Galvin was informed last night.

The reports are absolutely without foundation.

``Cincinnati is endeavoring to prevent an epidemic of Spanish influenza,’’ said Mayor Galvin last night. ``There is no epidemic here. We are doing what other cities should have done—we are preventing. \dots''  

\paragraph{``Theaters, Churches Are Closed By Mayor.'' {\normalfont\emph{Oakland Tribune}, October 18, 1918.}}

Mayor John L. Davie today issued a proclamation designed to assist in abatement of the Spanish influenza, which prohibits until further notice public gatherings of every sort. The order is of effect at midnight tonight. Authority is given the chief of police to enforce the terms of the order.

Theaters, motion picture playhouses, and all other places of amusement are included. Schools will not reopen next Monday morning. Church services are suspended. Lodges are forbidden to convene and social gatherings must be indefinitely postponed. Poolrooms are included in the taboo. \dots

The passing of the resolution authorizing the mayor to issue the above proclamation followed an address by Dr. Daniel Crosby, newly appointed health officer, before the council this morning. Dr. Crosby urged that immediate action be taken in order to make the preventative measures adopted by San Francisco and other bay cities more effective, arguing that if Oakland were to remain open it would be a menace to the entire section. \dots

\paragraph{``Weather Cause of Deaths’'{\normalfont \emph{The Milwaukee Journal} October 26, 1918, p. 2.}} \dots That Milwaukee is fighting the epidemic of Spanish influenza with better results than other cities is borne out by a statement from Surgeon General Rupert Blue received Friday by Health Commissioner Ruhland. The prompt closing orders are believed to have resulted in the favorable condition here. \dots Warned Against Overconfidence. The public is warned not to be overconfident that the epidemic is abating when reports show a decline in the number of new cases. Some eastern cities made this mistake and later found the decrease due to physicians being too busy to turn in their reports. Physicians here are also overworked and a sudden decrease in cases reported may in reality mean a big increase.

\paragraph{``No Epidemic, But Omaha is Near Closed.'' {\normalfont \emph{The Omaha World-Herald}, October 5, 1918, p. 1, 2.}} \dots City Health Commissioner E. T. Manning, in issuing an order yesterday prohibiting all public gatherings in Omaha until danger of an influenza epidemic is passed, took care to point out that the present situation is not serious and that people should not be unduly alarmed. 

``The condition in Omaha is by no means as serious as in eastern cities,'' he said. ``We are taking this drastic step to keep it from becoming so. I would rather be blamed for being over cautious than be responsible for a single death. Prohibition of public gatherings is the only way known to medical science for checking the spread of the disease, and I believe we are justified in order that to prevent a more serious situation. Such action was approved yesterday in a statement issued by Dr. Rupert Blue, head of the federal public health service.''

\dots ``The thing for Omaha people to remember,'' he said, ``is that hysteria helps no one. I confidently expect that in a few days the semiquarantine may be raised and Omaha may escape such epidemics as are sweeping easter cities. Most of these cities permitted the disease to get a firm hold before shutting down on public gatherings.'' 

\paragraph{``Halls and Churches to be Flu Hospitals.'' {\normalfont\emph{The Seattle Star}, October 07, 1918, p. 1.}}
Don't Be A Grumbler. Don't grumble because you can't see a movie or play a game of billiards---or because the schools and churches closed.

The health of the city is more important than all else. An ounce of prevention now is worth a thousand cures. In Boston, influenza has taken a toll of thousands. We do not  want to court that situation here.

Preparations were under way Monday by Mayor Hanson and municipal health authorities to transform Seattle's big public dance halls, and churches if necessary, into emergency hospitals to care for Spanish influenza cases if the epidemic is not checked.

This action was decided upon as a preparatory measure, supplementing the order of Saturday that closed schools, theatres, motion picture houses, pool halls, and all indoor assemblages. \dots

``We don't know how long it will be necessary to enforce the general closing order,'' said Mayor Hanson Monday. ``I have not made any predictions, and cannot make any. We have received citywide co-operation with practically everyone affected except school authorities, who objected.''

\paragraph{``Hospital Full, Police Worried.'' {\normalfont\emph{Worcester Evening Post}, September 27, 1918, p. 15.}}

If there ever was a time in which the man of the hour was needed in Worcester it is today. And from the talk on the streets and all around the city no one appears on the scene large enough to grapple with it. Men, women and children are dying in their homes because of this condition. The police are ready and anxious to remove sick people to the City Hospital, but that institution has been closed to all applicants. 

Today the police took the matter up with the Board of Health, but up to early afternoon nothing has been done. \dots 

As one well-known citizen said today: ``Isn't there a man in Worcester big enough to take the situation in hand and help save the men, women and children who are dying in Worcester for want of care.'' 


\subsubsection{Debates about NPIs among policymakers, health officials, and commentators}

\label{sec:newspapers_debate}

\paragraph{``Wants Boston Schools Closed — Councilor to Appeal to the School Board.'' {\normalfont\emph{Boston Globe}, September 20, 1918, p. 7.}}

Executive Councilor Lewis R. Sullivan announced at the State House today that he will ask the Boston School Committee to close the schools until the epidemic of influenza abased. Four of his children are afflicted with the malady. 

``I believe the Boston schools should be closed,’’ Councillor Sullivan said today. ``Health authorities agree that there is no disease so contagious as influenza, and the number of deaths that have recently occurred seem to indicate that there are few more dangerous. Last year the School Committee closed the schools in order to save coal, and it seems to me that they should now be closed in order to protect the health of the little pupils that don’t know how to protect their own health. \dots ''  


\paragraph{``Be Calm, Cool; Check Disease—Public Given Advice on Influenza at Meeting of Doctors and Laymen. City Not To Be Closed.'' {\normalfont \emph{Detroit News} October 18, 1918, p. 1, 2.}} Keep cool; prevent hysteria; use common sense, and by so doing Detroit will escape the ravages of an epidemic of influenza. This was the opinion of 50 or more representatives of the Red Cross, large business and manufacturing interests and medical men held at the Board of Health Building today. \dots

It was decided that nothing would be accomplished--save increasing hysteria--by closing the schools, amusements and places of public gatherings. This decision was endorsed by Dr. M. R. Donovan, health officer of Lynn, Mass., which city has had an epidemic. The closing had little effect, Dr. Donovan said.

\paragraph{``Spanish Influenza and the Fear of It.'' {\normalfont \emph{Philadelphia Inquirer}, October 5, 1918, p. 12.}} \dots The so-called Spanish influenza and the fear of it have got a grip upon a goodly number of people. The authorities have seen fit to close public schools, theaters, churches and many other places. Since crowds gather in congested eating houses and press into elevators and hang to the straps of illy-ventilated streetcars, it is a little difficult to understand what is to be gained by shutting up well ventilated churches and theaters. The authorities seem to be going daft. What are they trying to do, scare everybody to death? 

What we do wish to impress upon the public is the desirability of keeping a clear head. So much has been printed about the influenza, and there has been so much discussion of it, that abject fear has been implanted in the minds of thousands of citizens. From fear to panic is only a step\dots Panic is the worst thing that can happen to an individual or a community. \dots The fear of influenza is creating a panic, an unreasonable panic that will be promoted, we suspect, by the the drastic commands of the authorities. \dots What, then, should a man do? A very sensible physician tells us to sleep enough, eat enough and carefully, and take outdoor exercise. \dots  

\paragraph{``The Influenza Edicts.'' {\normalfont \emph{Philadelphia Evening Bulletin}, October 5, 1918, p.6.}} The orders which have been issued by the health authorities in the State and the city for safeguarding the public against the influenza are unprecedentedly severe, as regards the strictness and the extent of such edicts on any past occasion of pestilence or of epidemic. For the present, however, or until the full effects of the malady shall be measured, there will be little or no disposition to raise objections to even those which seem to be far fetched or ill considered. \dots

It is held that the extremely rapid progress of the disease in and about Philadelphia during the past few days is due, as its like progress has been elsewhere, to the assembling of people in mass, and that no immediate headway can be made in checking it until all such gatherings shall have been reduced to the smallest number consistent with the maintenance of the necessary business and functions of the community. \dots 

It is a common duty, therefore, to co-operate with the authorities at every reasonable point in combating the pest. \dots 

\paragraph{``Ban Gatherings In An Effort To Halt Influenza.'' {\normalfont \emph{Seattle Post-Intelligencer}, October 6, 1918, p. 1, 10, 13.}}

Aroused at last to the imminence of the danger facing Seattle on account of the prevalence of Spanish influenza, City Health Commissioner J. S. McBride, with the approval of Mayor Hanson, yesterday placed in effect the most drastic regulations to which the city has ever been subjected. In its efforts to prevent a further spread of the malady, which has already taken its toll of death in the city, the city health department forbids every form of public assemblage. 

Superintendent of Schools Frank B. Cooper last night said that it did not seem to him that the situation at present warranted the drastic action taken. ``The state health authorities having requested that the schools be closed, they will be closed, although, personally, I consider the step unwise,'' said Supt. Cooper. 

Mr. Cooper, who said that there is no influenza in the schools are present, said that he considered that one of the worse things that could be done at the present time would be the turning into the streets of Seattle’s schools children. ``I consider it more dangerous to have children running around the streets loose than to have them in school, where they will be under strict medical supervision,'' said Supt. Cooper. ``Since the schools are to be closed I must urge that all school children be kept at home by their parents until the schools reopen.''


\paragraph{``Mayor Voted Down in Effort to Take Off Influenza Ban. Medical Men Show St. Louis' Precautions Kept Disease from Spreading.'' {\normalfont \emph{St. Louis Globe Democrat}, October 25, 1918, p. 9.}}
A proposal by Mayor Kiel to lift the influenza ban for one or two days to notice the effect on the epidemic was voted down by a majority of forty medical advisors and others who had assembled for a conference in Commissioner of Health Starkloff's Office in the municipal courts building yesterday. 

Mayor Kiel said it was not fair to the business men and the public to keep the ban on any longer than is absolutely necessary, but a dozen men rapidly advised that it would not be lifted. \dots 

Maj. Hahrenburg read a telegram from Surgeon General Blue urging that the order be continued until all danger is past, and from a memorandum which recited that 75,000 persons were stricken in Massachusetts in one day, and that 709 died in Boston alone in one day. 

Fears Wildfire Spread. Dr. Smith said he believed St. Louis had escaped the misfortune which befell Boston, Philadelphia, and New York only by the timely closing of theaters, churches and schools and prohibiting public gatherings. He said that lifting the ban now would ``spread the disease like wildfire.'' \dots 

Frank Tate assured the meeting of the support of the moving picture people and other theater proprietors, despite their losses, but said he thought the order should have been more extended. 


\paragraph{``Ridiculous To Reopen Public Places, Says Nurse Jacobus.'' {\normalfont \emph{Worcester Daily Telegram}, October 18, 1918, p. 18.}}

``It is absolutely ridiculous to think of yet lifting the ban on the closing of public places,'' said Miss Rosabelle Jacobus, supervisor of the Worcester society for district nursing, who with her capable staff of nurses has been working night and day to relieve the sick and suffering affected by the epidemic.

``Things are clearing up, but it has not yet cleared off; things are better, but it is not over,'' said Miss Jacobus in speaking of conditions at the present time. ``If schools, churches, theaters and other public places reopen Monday, it may mean another outbreak of influenza.'' 

Things are now being handled quite well, but with nurses, doctors, canteen workers and all who have come to the call in this emergency worn out, it might be impossible to handle another such trying situation. \dots  

\paragraph{``Influenza Ban Abrogated by City Officials. Mayor Babcock Advises Disregard of State Health Authorities’ Ruling. Churches May Open. Schools to Resume and Other Activities Expected to Become Normal Again.''  {\normalfont \emph{The Pittsburgh Gazette Times}, November 2, 1918, p. 1, 5}} Mayor E. V. Babcock last night lifted the influenza ban on Pittsburgh, as far as the city authorities are concerned. \dots

``We have been closed for one month now, and the state is acting as if we are being punished for something,'' said the Mayor. ``No one would act more vigorously to carry out the orders of the state health department than I would or than I did when they seemed necessary. Now, in the opinion of the men most competent to judge, the keeping on of the ban for another week would simply have an effect directly opposite to that desired. While the state is, of course, supreme in this respect, the city will refuse to act as its police agent\dots”  

In regard to saloons and theatrical enterprises, the opening is dependent upon the judgment of the proprietors. These places were closed by the state authorities. If they open in violation of this order, they may come in conflict with the state health commissioner. The city, however, will refused to act as prosecuting agent \dots  


\paragraph{``New `Flu' Cases Drop to Fifty-Nine.'' {\normalfont \emph{Salt Lake Tribune}, December 10, 1918, , p. 12.}} \dots  The board of county commissioners addressed a letter to Governor Bamberger yesterday, protesting against the raising of the influenza ban in Salt Lake City. The letter, which was signed by the three commissioners, stated that the officials regarded the action as unfair to Salt Lake county both as to the health of the people and the business interests. It admitted the impracticability of quarantining against the city. 


\subsubsection{Arguments that NPIs reduce mortality and benefit economy; cost-benefit arguments based on statistical value of life}

\label{sec:newspapers_notradeoff}

\paragraph{``Denver Closes Churches And Theaters—\dots—All Civic and Business Interests Unite to Safeguard Lives of People and Halt Plague.'' {\normalfont\emph{The Denver Post}, October 6, 1918, p. 1, 2.}} All places of public assembly in Denver will close Sunday morning at 6 o’clock and remain closed for such time as the bureau of health deems necessary in stamping out and preventing the spread of Spanish influenza, or grip, in this community. 

This drastic and speedy action was taken Saturday [illegible] at a meeting attended by representatives of the city council, the bureau of health, the medical association, the ministry, the Civic and Commercial association, the school board, the theaters and various other commercial organizations held in the office of Mayor Mills. Opinion among those present was unanimous as to the action that should be taken.\dots 

From an economic viewpoint, the doctors were agreed that one or two or three weeks closing of public assemblies now would save many dollars in the long run, for they confidently predicted that 40 per cent of the population would be stricken if strict measures were not taken to prevent the spread of the contagion and that another 40 percent of the population would be caring for the great mass afflicted\dots Homer E. Ellison, manager of the Rialto theater, speaking for the moving picture interests—commercial institutions affected more than any other by the closing order—sounded the keynote of the meeting when he declared in earnestness: ``I shall sacrifice gladly all that I have and hope to have, if by so doing I can be the means of saving one life.'' 


\paragraph{``Nov. 16 Day Fixed For Raising Ban Against Crowds'' {\normalfont \emph{New Orleans Times-Picayune}, November 7, 1918.}} \dots Dr. Corput’s Warning [local U.S. Public Health Service representative Dr. Gustave M. Corput] Dr. Corput implied that the situation is now in the hands of the public. He warns that crowding is just as dangerous as ever, and that the danger is not over. His statement is as follows:

`In recommending to the State Board of Health and representatives of the various cities at the conference in the rooms of the State Board of Health, the removal of restrictions, I have assumed that the public will continue to co-operate with the authorities in stamping out one of the most severe epidemics which has visited the state of Louisiana within the past forty years. The fact that the restrictions against churches, schools, moving pictures, theaters and other places where people congregate is removed does not mean that all danger has passed. \dots Patience and compliance on the part of the public for the next few weeks means the wiping out of the epidemic in the state. Failure to do these things undoubtedly means the loss of many lives and an inestimable damage to business conditions.'  

\paragraph{``Drastic Rules to Combat Influenza.'' {\normalfont \emph{The Oregonian}, November 3, 1918, p 22.}} All downtown stores to close at 3:30 P.M. and offices to close at 4 o’clock — this is the gist of an order issued yesterday by Mayor Baker as a means of avoiding crowds on streetcars, checking the influenza epidemic and shortening the period of ban on public gatherings\dots 

In explaining the further order, Mayor Baker made the following statement: ``We have not called this conference because of any marked increase in the number of reported cases of Spanish Influenza, but because of our belief that more stringent regulations during the next few days will have a direct tendency to shorten the period during which regulations of any sort will be needed. Preventive measures promulgated at the beginning of the epidemic resulted in Portland suffering less from the illness than any other city of like population in the Nation, and the present step is taken with the view that the sacrifices and loss entailed will be more than compensated by the early ending of the epidemic. In other words, it will save lives, prevent suffering and lessen economy hardships if all of us for a short time do our utmost to stamp out this epidemic than to use only halfway measures over a long period of time.''

\paragraph{``Portland Churches Ask Epidemic Ban.'' {\normalfont \emph{The Oregonian}, January 20, 1919, p. 8.}} \dots Declaring himself in favor of a rigid quarantine of existing cases of the influenza, the closing of all public gatherings for 30 days and a rigid enforcement of the use of masks by all persons entering public buildings or conveyances, Dr. Edward H. Pence, pastor of the Westminster Presbyterian Church, yesterday morning asserted from his pulpit that the economic losses in Portland thus far have run into millions. 

This, he said, is a cold-blooded way of considering the epidemic, but as many are opposed to a shut-down as a remedy because of a financial loss, he thought he would work out some figures to prove that the financial side of the situation is now running on the wrong side of the ledger. 

Dr. Pence pointed out that large numbers of supposedly preventable deaths have taken place in the city and that the plea of the dollar must not be permitted to stand before ``1000 needless graves.'' 

Analysis Is Cold-blooded. ``It is objected that the prohibition of public gatherings in order to reduce the number of infections involves large financial loss,'' said Dr. Pence. ``The objection provokes us to an analysis—a decidedly cold-blooded one, of the economic loss due to the death of a few over a thousand human beings since October 1, 1918.''

``The fact that strong restrictive measures in the past, here and elsewhere, though sporadically prosecuted, have sharply reduced the number of infections proves that influenza belongs to the class of preventable diseases, and, equally, that its deaths belong to the class of preventable deaths. 

``Professor Irving Fisher of Yale, has evolved some startlingly significant facts concerning the value of the individual to society. In a report made to the United States Senate in 1910, he submits the following conclusions: 

``Value of human beings to society: At birth, \$90; at five years \$950; at 10 years \$2000; at 20 years \$4000; at 30 years, \$4100; at 50 years, \$2100; at 80 years, \$700. 

``He deduces the average economic value of lives now sacrificed by preventable deaths to be \$1700. Applying these figures to the Portland situation, we have \$1700 multiplied by the number of deaths (alleged to have been preventable), since October 1, 1918, which gives us \$1,700,000. \dots

``Over against these naked figures, coldly, sordidly calculated, it is argued that restrictive measures will reduce the profits upon certain businesses. I challenge any statistician to match the sheer economic losses stated in the foregoing by the total reduction of profits due to a 30 days' tight ban. \dots

``Let us urge a reasonable maximum program involving drastic measures where needed—anything to bring the scourge to an end. The church of Portland, fully realizing the loss to itself, both bookable and in imponderables, insists that a 30-day closed period be proposed.''

\subsubsection{Opposition and support for NPIs from businesses, churches, and other interest groups}

\label{sec:newspapers_business}

\paragraph{`` `Flu’ Cases Gain In Denver, Relaxed Care New Danger—Health Authorities, While Making Best Fight Against Epidemic of Any City in Nation, Warn Against Letting Down of Rules by People.'' {\normalfont\emph{The Denver Post}, October 11, 1918, p. 1, 3.}} Panicky excitement gone and relaxation taking the place of strict observance of health rules in the crisis have been responsible for a jump in the new cases of Spanish influenza here in the last twenty-four hours. \dots 

Vigorous complaints were lodged Friday against open-air assemblies by persons who are desirous that normal business may be opened as soon as possible and that the schools and other institutions may resume work. The open-air assembly is not such a dangerous place as the herding of the people on street cars, it is held. In most of the larger cities where the death rate has more than trebled, open-air assemblies have been stopped. \dots

Business interests in Denver affected by the closing ban are losing thousands of dollars daily and many persons are out of employment. Such affected businesses contend that the plague cannot be stamped out successfully without stopping outdoor assemblies and the crowding of street cars.

\paragraph{``Theater Men Protest Ban — Equal Chance With Stores Asked and Offer to Furnish Masks Made.''  {\normalfont\emph{Rocky Mountain News} November 23, 1918, p. 7.}}  Managers of all the Denver theaters and moving picture houses last night asked The News to put them on record as voicing a strong protest against the health departments new order closing their places because of the influenza epidemic. The managers of the various houses declared the action was discriminatory, and they will go to the city hall this morning to make their formal protests to the mayor and manager of health. The theater and moving picture men said they were anxious to join with all business men in the city and follow any rule to stamp out the disease if the rule applied to all alike. They said, however, that the order which allowed the stores and other places to remain open, and forced them to close, even if their patrons did agree to wear masks, was showing discrimination. \dots

\paragraph{``Health Board May Act Today—Postpones Matter of Raising Ban on Theaters, Saloons, Etc.—Special Meeting Called for 3 o’clock This Afternoon—Results Dependent on Conditions at That Time.'' {\normalfont \emph{Fall River Evening News} October 22, 1918, p. 1.}} Contrary to the opinions of many who believed they were on the inside track, the Board of Health at its meeting Monday afternoon did not remove the restrictions on public gatherings adopted as the most effective means of combating the ravages of Spanish influenza. \dots 

Agent Morriss informed the board that he had received numerous protests from different people, asking that the restrictions now in force be kept up for a week or so longer. Included in these communications was one from the Chamber of Commerce and one from W. Frank Shove, representing the Cotton Manufacturers Association.

Mr. Morriss said that he had assured all who spoke to him on the matter that the board would be careful and would handle the situation to the best of their ability, even in the absence of any suggestion from outsiders. He said also that many tradesmen affected by the rulings of the board were anxious to know when they could resume their businesses.

Dr. Peckham said he thought it was the part of wisdom to continue the present restrictions for at least three or four days longer. \dots

Mr. Borden thought that, compared with the interests involved the danger was slight. \dots He said the influenza epidemics run their course as a general thing, no matter what is done to prevent them. \dots


\paragraph{``Urge Tighter Lid'' {\normalfont \emph{Los Angeles Evening Herald} November 7, 1918, p. 9.}} Frank A. McDonald, president of the Theater Owners’ association, and 25 members of the association, appeared in masks. McDonald, for the association, appealed not to open the theaters, but rather to close up everything else as tight as the theaters, and make the fight against influenza more radical. He deplored the fact that department stores, cafeterias, parks and other places where people still gather were allowed to be open, while churches, schools and theaters were closed. 

McDonald said his people were willing to co-operate in everything the city council and the health authorities demanded. 

\paragraph{ ``Influenza On Steady Drop. \dots Theater Men are to Petition the City Council. {\normalfont \emph{Los Angeles Times} November 10, 1918}} \dots A committee representing the Theater Owners’ Association called at the health office yesterday and informed Dr. George L. Codle, chairman of the advisory board, that the committee had presented a written communication to the City Council to be read before that body Monday morning. The theater men said that they will show that it is just as necessary to close the entire city as it is to shut down the theaters. They will advocate, the say, closing stores and placing a quarantine on every case of influenza. \dots

\paragraph{``Theaters Closed by Authorities.'' {\normalfont \emph{Nashville Banner},  October 7, 1918, p. 8.}} \dots Although every theater and movie show in the city had arranged special programs of unusual merit and expense for this week in honor of the fall opening season, the managers cheerfully acquiesced in the order putting them out of business for the time being. There will, of course, be a sense of disappointment on the part of the amusement-loving public, but all recognize that human health and life are considerations too valuable to be jeopardized in any degree for the gratification of a mere temporary pleasure. \dots


\paragraph{``Copeland Refuses to Close Schools--\dots Staten Island Shipbuilders Report 40 Per Cent of Men Ill and Ask the Mayor to Act.'' {\normalfont \emph{New York Times}, October 19, 1918, p. 24.}} \dots There is still much discussion as to whether schools, theaters, and places of public gatherings should be closed, but Dr. Copeland, the Heath Commissioner, said emphatically in reply to all criticisms, that he saw no reason for closing such places or varying the department’s program. One thing thing that induced Dr. Copeland to defend his program was a report that Borough President Van Name of Richmond had sent to Mayor Hylan a letter from three shipbuilding firms in that borough asking that public places be closed, with the endorsement from President Van Name that it be done. The letter came from the Standard Shipbuilding Corporation, the State Island Shipbuilding Company, and the Downey Shipbuilding Corporation. It said that influenza had decreased efficiency at the plants approximately 40 per cent and was seriously hindering work for the Government. It pointed out that about 15,000 men were employed in the years and the disease had been epidemic for two weeks. The private hospitals maintained in the yards are overtaxed and the nurses and attendants are overworked.

\paragraph{``No Epidemic, But Omaha is Near Closed.'' {\normalfont \emph{The Omaha World-Herald}, October 5, 1918, p. 1,2.}} \dots Church pastors and theater managers joined yesterday in expressing doubt as to the necessity of the health commissioner’s order. 

``Only as a last recourse should the churches of the city be ordered closed,'' said Secretary Mayer of the Omaha Church federation. ``Since the order is made, however, we will abide by it.''\dots 

Theatrical managers, meeting at the Boyd theater, said they would obey the order, although with regret. ``All that I can learn is that there is one case in Omaha and six at Fort Omaha,'' said Joy Sutphen, manager of the Brandeis theater. ``If they closed business houses every time there were six cases of one kind of lease at once, we would be closed all the time. It looks to me as though the health commissioner is scared at nothing. Such a move will throw from thirty to eighty people in every theater out of work and will work a big hardship on show people, traveling in and out of Omaha. If there was an excuse for such a drastic step, we would not object, but no condition exists to make such a step necessary.'' \dots 

R. H. Manley, commissioner of the Chamber of Commerce, appeared at the physicians' meeting and pledged co-operation of the business interests, while urging in their behalf that no drastic action be taken unless deemed certainly necessary for the preservation of public health. The physicians told him that the action was necessary. 

\paragraph{``Plan To Put End To Epidemic Here. Business Men and Others Discuss Methods of Combating Influenza.'' {\normalfont  \emph{Salt Lake Tribune}, November 20, 1918, , p. 11.}} \dots  Following reports that indicate a general setback in the influenza situation throughout the state, a meeting of physicians and business men was held last night at the Salt Lake Commercial club to discuss ways and means of putting an end to the epidemic. The gatherings resulted in the appointment of a committee of five business men, which will cooperate with the state and city boards of health in devising measures to bring about the desired results. James P. Casey, M. H. Hannuer, Walter C. Lewis, M.F. Lipman, and Edward P. Levy constitute the committee, which agreed that isolation and prevention of crowds in stores and other places were the two most effective weapons to combat the spread of the disease. \dots 

\paragraph{``Minutes, Special Meeting of the San Francisco Board of Health.'' October 17, 1918.} \dots Mr. Don. W. Bingham, of the U.S. Shipping Board, stated: ``The influenza epidemic in the East has seriously hampered the ship building program. According to the figures we have at the present time, approximately 25\% of the building has been decrease during the past month. Here on the Coast, especially in the San Francisco Bay District, the effect has not been very serious so far; from 8 to 10 cases per yard per day being reported for the past week. In regard to the proposed plan of closing all theaters and other places where people congregate, I consider it will limit the length of the epidemic, altho it has been the history in large epidemics of this nature that probably 70\% of the people will be affected.'' 

Mr. Frey of the U.S. Shipping Board, reiterated the statements made by Mr. Bingham and added that the East is now looking to California and the western states to carry on the work of ship building. He also stated that we should close all public places of amusement and by so doing reduce the infection. \dots 

Dr. Fred Rothganger, of the 12th Naval District, stated: ``I am in harmony with the ideas of the Public Health Officer. I do not think the closing of these activities will be successful in controlling the epidemic. We have adopted the closure system in the Navy and it is only in places where the man can be entirely shut off, such as Yerba Buena Island, that any success has been obtained. I think forcible closing would do more harm than good. I think people who have the disease should be instructed to keep away from others. The U.S. Public Health Service has prepared circulars relative to the disease and the same have been printed by the Red Cross for distribution among the people of the community.'' 

\dots Mr. Arnstein stated that judging from the discussion had within the last few days the consensus of opinion of doctors seems to be that we should avoid crowds and he is of the opinion that if a brief closing order were enforced at this time it might obviate the necessity of closing various activities for from six to eight weeks later. \dots

Mr. Sprouls stated: ``It occurs to me that there can hardly be any debate about the closing of the schools and I think such action should be taken for a limited time at least, and by so doing we would be giving a warning to the entire city. Further than that I have some doubts as to the wisdom of taking any drastic action. The panic created by unnecessary actions promotes the very thing it is intended to prevent \dots''

The President called for an opinion from the various motion picture houses represented: Mr. Roth of the California Theater stated: ``In the early stages of the epidemic, a committee from the moving picture houses called upon the Health Officer and offered the use of their screens for public education and very good service was rendered in this regard. The situation in the city at the present time is somewhat trying and as the movie houses are now suffering from lack of patronage it might be of vital interest to the city to order the houses entirely closed for a period of whatever time may be necessary to clear up the situation.''

Mr. Sheehan, of the Rialto Theater, stated: ``I have interviewed the managers and owners of many theaters and show houses and they have all stated that as the people appear to be voluntarily staying away anyhow, that the amusement managers would not suffer much more if all theaters were ordered closed. Even tho the public would still continue to patronize use, we are public spirited enough to want to do anything and everything possible to protect the public health.''

Mr. George R. Puckett, representing the dancing hall interests, stated that his people are anxious to help and eager to abide by whatever decision is reached by this Board and are perfectly willing to discontinue their dances for the time being. \dots 

Mr. Del Lawrence, of the Majestic Theater, stated that receipts have fallen off 40

Mr. Irving Ackermann, of the Hippodrome stated that if the contemplated action is to be productive of any real good, he cannot see the wisdom of delaying the date of closing and action should be taken immediately in this regard. If prompt closing of all the show houses of this city will result in saving but one life, it is certainly advisable to close down everything immediately.  

\paragraph{``Plan to Head Off Influenza.'' {\normalfont  \emph{The Spokesman-Review} (Spokane, WA), October 8, 1918, p. 10.}} 

Chamber Committee to Launch Educational Campaign. 

At its weekly luncheon yesterday the health and sanitation committee of the chamber of commerce pledged its moral support to Dr. J. B. Anderson, city health officer, in an educational campaign to head off any epidemic of Spanish influenza in Spokane.  \dots 

The committee passed a resolution in which it expressed its willingness to back up the health office in the necessary steps to head off the disease in Spokane. The committee urged that drastic measures be adopted by the health office.

\paragraph{``Order Fixing 9:30 To 4:30 As Business Hours For Downtown Stores Rescinded.'' {\normalfont  \emph{St. Louis Post-Dispatch}, October 23, 1918, p. 3.}} 

Health Commissioner Starkloff today rescinded his order establishing 9:30 a.m. to 4:30 p.m. as the daily business hours of retail stores in a defined downtown district. Insofar as it affected small retailers, including saloons, it became effective immediately. The order was designed to relieve congestion on street cars during the influenza situation. 

He said that he had requested the department stores to voluntarily continue to follow the restricted hours. \dots 

The Health Commissioner said that he had been convinced that the order was working a hardship on small businesses, such as cigar dealers, and that it was accomplishing little in preventing gathering of persons, because there were no congregations of size in small stores.

\paragraph{``Theater Men Protest On New Order. Claim Other Amusement Places Should Also Be Closed.'' {\normalfont  \emph{Worcester Evening Post}, October 4, 1918, p. 1, 10.}} 

At a meeting this noon of the members of the Worcester Theatre Managers' Association, keen disappointment was shown at the order this morning by the board of health prohibiting them from opening their houses as anticipated on next Monday. 

After the meeting, when the situation was gone over very carefully, members of the association said: ``Having closed our houses 10 days ago at the request of the authorities, alone of all the industries in Worcester, we have made gala preparations of re-opening Monday, the Board of Health having assured us as late as last night that the opening would be allowed on that date. 

``It is a peculiar predicament that we find ourselves in. Compelled by law to have the most sanitary of establishments and with a wonder record of not one of our employes affected by the disease, our only request today is that the board make an order covering other places of assembly and amusement.'' 

The managers contend that the crowds of men gathering at saloons, bowling alleys, poolrooms, etc., are constantly increasing the local danger far greater than sanitary theaters and they ask for the closing of these places so that the disease may be quickly arrested. While admitting the dangers of some of these places, the health board refused to close any other industry than the theater, say the managers. 

\paragraph{``Nurses Bureau Is Opened,'' {\normalfont\emph{The Milwaukee Journal,} October 20, 1918.}} \dots 

`Why are saloons permitted to remain open while churches must be closed? Why is our church constituency not called into consultation when all other interests are recognized?' These are the questions that will be asked of Health Commissioner Ruhland by about 100 members of the Milwaukee federation of churches, if they comply with a request of the executive committee of the federation.


\subsubsection{Debates about mask ordinances}

\label{sec:newspapers_masks}

\paragraph{``2779 Cases of Influenza Now on Hand.'' {\normalfont \emph{Oakland Tribune}, October 26, 1918, p. 7.}} ``Cigar Stores Are Miffed at Masks \dots ``Cigar stores will be hard hit from now until the ``flu'' epidemic is in check and the gauze masks relegated to the ashcan,'' was the statement made today by a prominent cigar dealer. ``The gauze masks does not permit of the use of cigars, cigarettes or pipes and even the habitual tobacco chewer would be forced to forego his usual `chew of terbaky,' for with the `flu' mask covering his visage he cannot very well engage in the pastime.''  

\paragraph{``Flu Masking Ordinance Is Turned Down.'' {\normalfont \emph{Oakland Tribune}, January 21, 1919, p. 1.}} The influenza mask ordinance, which came up before the city council today for final passage, did not pass. 

The council, following a vigorous protest from Christian Scientists, labor representatives and others, who packed the lobby, voted to lay the ordinance on the table, to be called up in case of emergency. 

Mayor John L. Davie was the first to have a fling at the flu mask. He recounted with a good deal of emotion how he was arrested in Sacramento for not wearing a mask and how he had to await at the jail until the police arrested someone else who had bail money so they could get change for the mayor’s \$20. 

Then he read from a report of the State Board of Health which said that records from masking and non-masking communities indicated ``conclusively that wearing masks does not effect the progress of the epidemic.'' ``The mask is just what it is name — a mask — a camouflage,'' he concluded. \dots

Peter V. Ross, representing the Christian Scientists, said that the people were not demanding that their liberties be interfered with, but on the contrary are opposed to the ordinance. There was nothing, he said, to prevent any one wearing a mask who wanted to.\dots 

Dr. Crosby said that he was and always had been [in favor of masks]. The state board of health, said Dr. Crosby, had failed to include in its report that the mask had been worn voluntarily and with success in combatting the epidemic in San Diego. \dots 

Evidence Aplenty in Favor of Masks. Dr. Ewer, who appeared masked, referred sarcastically to this expression ``lunatics'' saying that the members of the Alameda County Medical Association who recently unanimously approved wearing masks, much be ``lunatics.'' ``There is plenty of evidence in favor of the mask,'' he declared. ``If we had been masked for the last four weeks we would not have had this recent wave of the epidemic.'' 


\subsubsection{Behavioral effects of NPIs}
\label{sec:newspapers_NPIbehavioral}
\paragraph{``Influenza Lid Clamped Tight All Over City--No Church Services to Be Held Today, Decision. Schools Will Remain Closed, Also Places of Amusement.'' {\normalfont \emph{Minneapolis Morning Tribune}, October 13, 1918, p. 1, 10.}} 
 
The influenza lid went on in Minneapolis at midnight last night. \dots Downtown theaters were packed last night with patrons who took advantage of their last chance to see a performance until the ban is lifted. Long lines of men and women waited in front of the motion picture and vaudeville theaters during the early hours of the evening.

\paragraph{``Arrests for Drunkenness Show Marked Decrease Since Saloons Are Closed--Police Records Disclose Benefits Attendant Upon Stopping Sale of Liquor. Homes are Happier.'' {\normalfont \emph{The Pittsburgh Gazette Times}, October 27, 1918, p. 6.}} \dots Facing the worst scourge in the history of the nation, the State Commissioner of Health, believing that the epidemic of Spanish influenza might thus be the more readily combated, closed the saloons of the state, and a few days later ordered the closing of the wholesale liquor establishments. 

Chief among those who have benefited from the closing order, beside the wives and children of the tipplers, are the police, whose duties have materially lessened. \dots

since the week of October 12 it [Pittsburgh] has seen a most remarkable decrease in the crime of drunkenness. \dots 

In the week ending October 25, 1917, there were 899 such arrests [for drunkenness] in Pittsburgh. For the week beginning September 22, 1918, there were 807 arrests for drunkenness in Pittsburgh. Then came a dry period. For the week beginning October 18, there were 272 arrests for drunkenness in Pittsburgh. In the five days ending October 25, 1918, there were but 178 such arrests. There was a falling off, too, in crimes directly associated with booze, such as street soliciting, visiting disorderly houses and the like.


\subsubsection{Evidence on compliance and avoidance of NPIs}

\label{sec:newspapers_noncompliance}
\paragraph{``Fewer Influenza Cases In Lowell Saturday and Sunday'' {\normalfont \emph{Lowell Courier-Citizen} October 14, 1918, p. 3.}} \dots Catholic Churches Open. Notwithstanding the fact that the board of health voted to order all churches to suspend sessions yesterday, the Roman Catholic Churches were open and low masses were celebrated. The attendance was not as large as usual, but nevertheless many went to the masses. \dots 

\paragraph{``Everyone Is Compelled to Wear Masks by City Resolution; Great Variety in Styles of Face Adornment in Evidence.'' {\normalfont \emph{San Francisco Chronicle}, October 25, 1918, }} The Board of Supervisors yesterday passed a resolution whereby every person who appears on the streets of San Francisco until a rescinding order is made, is compelled to wear a mask as a protection against influenza. \dots

Meanwhile, since the first order recommending the wearing of a mask was promulgated three days ago, the number of persons who accepted that advised increased so materially that yesterday about four out of every five persons in San Francisco were wearing masks.

\paragraph{``Gloomy Sunday Is Result Of the Influenza Ban On All Places of Amusement.'' {\normalfont \emph{Seattle Post-Intelligencer}, October 7, 1918.}} \dots Autoists Brave Rain. 

Most persons who owned machines had them out. Although it rained for a short time in the afternoon, and in the morning, the day was not a disagreeable one for autoists. At the gasoline service stations it was said that the sales had been heavy. Many of the stations had sold more gas at noon than the sales of an ordinary Sunday would amount to. There were many out-of-town auto parties. The boulevards and the parks were dotted with machines throughout the day. And to the country folk, who generally spend Sunday at home, anyway, the day was an auto parade for them. \dots 

While the churches were forbidden to hold services, out-of-town meetings were held. These proved to be successful, and the ministers declared that if the closing order is still in effect next Sunday, more elaborate arrangements will be made for the holding of mass meetings in the open. The parks will probably be used for Sunday meetings if weather permits.


\paragraph{``May Lift Order In 10 Days.'' {\normalfont \emph{Toledo News - Bee}, October 15, 1918, p. 1, 2.}}
\dots The public in general took the [closure] order in good spirit. It seemed anxious to do its part to protect the public health in every possible way. All business affected generally obeyed the order. \dots 

Due to the fact that Toledo saloons are closed, bars in the county, and adjacent to the city are ``getting a big play'' from pedestrians and auto parties. Near Calvary Cemetery there is a saloon on one side of the street, in the city, that is closed; across the street, in the country, is one that is open. The county sheriff’s office has rendered its services to the City Health Division to disperse crowds in the county. 


\subsubsection{Schools and influenza: Debates about whether to close schools, parents keeping children home out of fear of influenza in schools, etc.}

\label{sec:newspapers_schools}

\paragraph{``Doctors Advise Parents to Keep Children Out of School.'' {\normalfont\emph{Birmingham Age-Herald}, December 6, 1918, p. 5.}} Thirty-one local doctors yesterday signed a statement wherein they advised parents to keep their children at home. \dots 

\paragraph{``New Ban To Close Many Schools -- Children To Be Barred From Street Cars and Theaters -- Health Board Issues Drastic Rules to Check Malady -- Survey Shows 5,036 Pupils of Cincinnati Are Ill of Influenza -- Nineteen Deaths, Day’s Toll.'' {\normalfont\emph{The Cincinnati Enquirer}, December 3, 1918, p. 16.}} \dots Survey of elementary schools made yesterday under the direction of Superintendent Condon revealed 13,314 pupils out of a total enrollment of 40,780 were absent. Of the absentees, 5,036, or 12 per cent of the total enrollment, were influenza victims. The others were kept home from school as a precautionary measure.

There were 713 high school pupils absent out of a total enrollment of 4,792. 245 of whom, or 5 per cent, being ill of influenza. More than half of the influenza victims are students of Hughes High School. \dots

\paragraph{``Contagious Diseases Increase in Schools--Twenty New Cases, Mostly Influenza, Reported by Director of Hygiene.'' {\normalfont \emph{Minneapolis Morning Tribune}, January 11, 1919, , p. 7.}} \dots Almost half as many absences due to fear of influenza as to actual influenza cases were reported by school principals following a canvass Tuesday. On that day 805 pupils were absent because of suffering from influenza or recovering from the disease, and 486 other pupils were kept out of school because parents feared exposure. During the last week 4,600 pupils were reported absent, the per cent of attendance, on a school population of 51,000, being about 90 per cent. During the last two years the per cent of attendance for this time has been 94. 


\paragraph{``Schools Are Affected--Epidemic Causes Marked Decrease in Attendance'' {\normalfont \emph{New Haven Journal - Courier}, October 29, 1918  p. 9.}} Attendance at the schools of the city has fallen off very considerably because of the epidemic. \dots 

\paragraph{``155 New Cases And 4 Deaths In Flu Wave--But Health Officials Think Worst of Epidemic Is Over--Not Vaccinating School Children Yet. Teachers for Nurses.'' {\normalfont \emph{New Haven Evening Register}, October 22, 1918.}} \dots At Scranton, Bernard, and Worthington Hooker [schools], the attendance in the rooms has fallen from 40 to 50, which is a normal attendance, to seven or eight with a maximum of fourteen. In all, 9,000 children are absent from the schools in the city. 110 teachers are out with sickness and substitutes are sometimes difficult to find. The schools can be closed by the board of health and not by the board of education and Dr. Wright does not think it necessary to close them.


\paragraph{``Lack Substitutes in City Schools.'' {\normalfont \emph{New Haven Journal - Courier}, October 28, 1918, p. 6.}} \dots At the board of education meeting Friday night, Superintendent of Schools F.H. Beede reported that the absence of teachers on account of the epidemic was formerly 80 or 90 per cent, but it has now decreased somewhat. The board is finding difficulty in finding substitutes. 

\paragraph{``Copeland Refuses to Close Schools.'' {\normalfont \emph{New York Times}, October 19, 1918, p. 24.}} \dots Dr. Copeland, when told of Dr. Goldwater’s statement, would not discuss it. Before he learned of it he declared that he stood fast in his determination not to close theaters or schools. ``I have discussed the school situation with those in authority and they agree with me,'' he said. ``The chief difficulty lies in the fact that there is an unusual amount of illness among the teachers. So far as the children are concerned, a careful survey shows that about half of the absences are due to the fear of parents and not to the illness of children.'' 

\paragraph{``City Schools Are Closed by Grip Fighters'' {\normalfont \emph{The Pittsburgh Gazette Times}, October 24, 1918, p. 1, 7.}} The public schools of Pittsburgh will be closed today on account of the epidemic of influenza. \dots 

The pupils and teachers will assemble this morning, but there will be no lessons. The children will be instructed in personal hygiene and preventative measures. 

The principal points of the instructions to be given are contained in the order issued by Director Davis to Superintendent of Schools William M. Davidson after the conference. The order reads as follows:

``After the conference at which the Department of Public Health and the school officials were represented, I have concluded that it would be wise to direct you to close the public schools of the city. I will issue a like order to the management of all parochial and private schools of Pittsburgh. 

``The closest possible analysis was made of the best figures obtainable in reference to conditions in the schools. The has been a very great increase in the number of absentees in the last two days. This appears to be very largely due to the fears of the parents, as the figures do not indicate a larger percentage of sickness among children than would be expected from their proportion of the population. It is reasonable to expect that this absenteeism will increase until the feeling of fear has grown less. 

``Under these circumstances it is not possible to maintain through the schools the medical supervision that would have been possible had the attendance remained at standard. This removes the only reason for keeping the schools open'' \dots  

\paragraph{``10,000 Children Out of School: Increase in Absentees May Necessitate Closing. Many Teachers Also Away. Mayor to Lay Education System’s Situation Before Aldermen Tomorrow. Influenza Continues to Spread, Although the Death Rate Remains Small.'' {\normalfont  \emph{The Providence Daily Journal}, January 5, 1919.}} The seriousness of the school situation in this city, which has been increasing daily for the past week, reached its height Friday when approximately 10,000 children failed to attend either the morning or afternoon sessions. Many teachers are also out because of illness. 

Reports received yesterday by Superintendent of Schools Winslow from the principals showed that in some schools the absentees amounted to practically 50 per cent of the entire enrollment. With one-third of all the school children of the city out, officials are confronted by several serious problems, and if the condition does not improve shortly, it is possible that the schools may again be closed.\dots  

\paragraph{``School Officials Worried Over ``Flu.'' '' 
{\normalfont  \emph{Seattle Post-Intelligencer}, December 12, 1918, p. 13.}}  Second Wave of Epidemic Causing Much Concern--Attendance Below Enrollment. Whether they should close the city schools until the epidemic of influenza shall have passed or leave them open is a question giving school officials much concern according to A. S. Borrows, county superintendent of schools. Reports received by the superintendent show that in many of the schools the attendance is more than 50 per cent below the enrollment, and while they have been hoping that conditions would improve they appear to be growing worse. 

Of the children who are not attending school, reports indicate that the number suffering from influenza is not alarming, but parents are keeping the youngsters at home because of fear of the epidemic. \dots 

\subsubsection{Response to reopening: Caution in some cities, boom in others}

\label{sec:newspapers_reopening}

\paragraph{``Epidemic Still Affects Business. People Evidently Are Not Taking Chances in Assemblages. 26 Cases and Nine Deaths Today’s Report.'' {\normalfont \emph{ Fall River Evening Herald} October 25, 1918, p. 1.}} The epidemic situation showed a further improvement at the noon hour today, according to Agent Morriss of the Board of Health. \dots

This is considered to be most encouraging in view of the fact that all of the restrictions were lifted at midnight Wednesday. Despite the fact that all of the restrictions have been lifted, people are apparently not taking any chances, for the theaters report small attendances and proprietors of drugstores, saloons, etc., are not being rushed to any great extent in the handling of business. 

\paragraph{``Omaha to Bloom Saturday After Month of Gloom. ``Flu'' Ban is to Be Lifted at 12.01 Saturday Morning; Shows and Stores Are Planning for Rush.'' {\normalfont \emph{Omaha Daily Bee}, November 1, 1918, p. 13.}}  Omaha is buzzing with preparations to resume its normal life after more than four weeks closing of public gatherings of all kinds because of the Spanish influenza. 

Theaters, moving picture houses, schools and churches are being put in shape for the grand opening. Most of the theaters will open Saturday, the Gayety being first in the ring with a special performance scheduled to start one minute after the closing order is lifted at midnight Friday night. The rest of the theaters also open Saturday, except the Orpheum, which will start with its new performance Sunday matinee. 

The moving picture films will begin to flicker through the machines by 11 o'clock Saturday morning\dots  

The big stores have been holding back their special sales during the ``flu'' epidemic at the request of the health authorities. They will have some extraordinary bargains for Saturday. \dots  

\paragraph{``Theaters and Churches Filled to Capacity Show Fear of Epidemic Is Over. Record Audiences Flock to Playhouses and Pastors Greet Unusual Congregations.'' {\normalfont \emph{The Post-Standard} (Syracuse, NY), October 28, 1918, p. 6.}}

Theaters filled to capacity and church services attended by entire congregations gave definite proof yesterday that the full confidence of the public, once shattered by the epidemic, has been restored. 

Ideal weather, with the temperature reminding one of a June day, got people out of their homes to enjoy the sunshine and indulge in the much needed recreation they had denied themselves for three weeks and more. The red-lettered placards of the Bureau of Health were the only reminders of the period through which the city has passed. 

Attendances at the playhouses last night exceeded any Sunday night crowds ever recorded in this city. The enforced inactivity of the last three weeks has apparently made the public theater-mad. 

\paragraph{``Influenza Ban Is Lifted. All Business Is Resumed; Schools Open Monday.'' {\normalfont \emph{Toledo News-Bee}, November 7, 1918, p. 2.}}

The resumption of all business on Thursday, thru the lifting of the influenza quarantine ban, was marked by more than usual activity. Moving picture houses and saloons did especially well, tho the happiness of the saloonists was marred a bit by the news that the state has ``gone dry.'' 

\paragraph{``It’s Off! Influenza Lid Lifted -- Health Officials Are Optimistic About Malady Situation -- Cincinnatians Should Be Allowed To Forget the Disease, Mayor Galvin Declares,'' {\normalfont \emph{The Cincinnati Enquirer}, November 12, 1918, p. 8.}}

All anti-influenza closing orders were rescinded yesterday by the Board of Health, effective last midnight. 

Churches, schools, theaters, stores, and saloons may open to-day and resume their accustomed hours of opening and closing.

The action of the Board of Health was taken following an optimistic report of the situation by Health Officer William H. Peters. Mayor Galvin appeared before the board to urge the lifting of the ban.

``The people are tired of hearing of influenza and want to forget it,’’ said Mayor Galvin. ``The psychological time for raising restrictions has arrived. You can no more control the people's enthusiasm nor regulate their actions on the street than you can control the Ohio River.

``The people of Cincinnati have been patient. They have endeavored to live up to the orders of the Board of Health, and now they should be given an opportunity to forget that such a thing as influenza ever threatened our city.

``We should give our people a chance to meet in the house of God and give thanks for the glorious end of the war,’’ concluded the Mayor. \dots

\subsubsection{Direct cost of the epidemic and NPIs on businesses; comparisons with statistical value of life}

\label{sec:newspapers_costs}

\paragraph{``Influenza Cost City More Than Half A Million—Four Hundred of 10,000 Died, But Plague Is Over and Town Is Reopened.'' {\normalfont \emph{The Birmingham News}, October 31, 1918, p. 1.}}

What Plague Has Cost Birmingham. Cost to movies and theaters, closed three and one-half weeks: \$90,000. Cost to department stores depressed business: \$160,000. Cost in doctors’ bills: \$25,000. Cost in hospital bills: \$10,000. Cost in undertaker’s bills: \$90,000. \dots It is estimated that influenza cost Birmingham’s citizens more than a half a million dollars in money. \dots The department stores, it is said, from a financial standpoint have been the hardest hit on account of the epidemic. Loss of traded during the four weeks that the city was closed, it is estimated, has cost the stores about \$160,000. The theaters have also suffered financially. Their loss, it is estimated, is about \$90,000.

Discussing the situation from a financial standpoint, Birmingham’s undertakers are the only ones who have profited from the epidemic. However, the suffering that they have seen for the last four weeks, the grieving of relatives for their loved ones and the burying of some of their own friends have cost them in heart aches. \dots  

\paragraph{``Epidemic Brings Big Money Loss--Five Branches of Business Drop \$52,500 Daily, Is Estimate'' {\normalfont\emph{Cleveland Plain Dealer}, November 11, 1918, p. 7.}} The influenza quarantine, extending twenty-nine days, including five Sundays, cost Cleveland tradesmen hundreds of thousands of dollars, and in the case of those whose business might be classified under the heading of pleasure, more than \$1,250,000. 

Men well versed in five [illegible] furnished the following figures, which they claim are conservative, representing the daily losses in receipts. Theaters: \$11,000. Movies: \$25,000. Restaurants \$[illegible]. Bars \$[illegible]. Soda fountains \$2,000. Estimated [total] daily loss: \$52,500. 

Hotels were also heavily hit by the ban, which forced postponement of conventions and many small daily gatherings. An assistant manager of one of the big hotels said the four week period caused a loss of \$50,000 in revenue to his house and estimated that Cleveland hotels suffered a gross loss of more than \$200,000. 

Theatrical interests were probably hit the hardest as a number of companies which were booked to show remained in the city at a heavy expense. \dots Many of the owners of theaters took advantage of the opportunity of making improvements. Several of the houses were redecorated. \dots

\paragraph{``Closing of Theaters By `Flu' Quarantine Entails Heavy Loss. {\normalfont \emph{Omaha Daily Bee}, October 5, 1918, p. 10.}} Theatrical managers in Omaha are very much depressed over the closing of the theaters on account of the Spanish influenza and are unable to estimate their financial loss. \dots 

Commissioner Manley of the Chamber of Commerce requested that the recommendation be modified if the situation was not too grave, as a quarantine on the city would prove disastrous to the business interests. He said the business men would comply with any order. No action was taken on his request. 

\paragraph{``Ban Raised Here As Epidemic Ends.'' {\normalfont \emph{Philadelphia Inquirer}, October 31, 1918, p. 8}} \dots [The] influenza ban, which had kept many industries closed tight for nearly four weeks, was lifted yesterday and within a few hours virtually every activity returned to normal. As was to be expected, saloons did a lively business during the first few hours after 8 A.M., the official time for their reopening. \dots 

The quarantine which was dissipated yesterday morning cost nearly \$2,000,000 dollars to the proprietors of theaters, motion picture houses, [illegible]\dots 

While the epidemic was at its height the receipts of the Transit Company fell off \$25,000 a day. The total loss to the corporation is fixed at \$250,000. 

Motion picture houses lost at least half a million. \dots

Dr. B. Franklin Rover, acting Commissioner of Health last night issued a statement that the 11,000 lives lost in Philadelphia through influenza had a potential value of fifty-five million dollars. The statement in part follows:

``The loss of life in the city of Philadelphia alone during the present epidemic of influenza, if figured in dollars and cents according to estimates on the value of human life fixed by medical and insurance statisticians and court decisions amounts to approximately fifty-five million dollars since October 1. 

``These figures were announced today by Dr. B. Franklin Royer, acting Commissioner of Health, when his attention was called to the fact that the moving picture, theatrical and liquor interests of Philadelphia and the Philadelphia Transit Company had estimated that their losses totaled two million dollars during the time the State Health Department’s closing order was in effect in Philadelphia.'' 


\paragraph{``Theatrical People Hit By Epidemic. Fully 300 at Gayety Theater Affected by Closing Order. Shows Forced To Lay Off. Proves Especially Tough on Chorus Girls.'' {\normalfont \emph{Rochester Times-Union}, October 19, 1918, p. 8.}}  While Rochester’s theatergoing public has been deprived of amusement as a result of the order of Commissioner for Public Safety Hamilton closing the theaters because of the prevalency of Spanish Influenza, more than 400 persons directly connected with the theaters have been thrown into idleness and the majority of them deprived of their salaries. The order closing the theaters has had its strongest effect upon the Gayety Theater where, it was stated this morning, fully 300 persons have been affected. \dots As the performers are working on a ``no play, no pay'' form of contract, the closing order has hit them pretty strong, especially the chorus girls. \dots  

\paragraph{``Influenza Rules Will Be Rigidly Enforced.'' {\normalfont \emph{Salt Lake Tribune}, December 8, 1918, p. 1, 19.}} \dots ``Salt Lake labor unions are rejoicing that they can again meet for discussion of their affairs and also because the lifting of the ban will send many of their members back to the work from which they have been restrained for nine weeks. The proposition certainly looks mighty good to the Salt Lake Federation of Labor, and the health authorities can be assured that every measure of caution will be observed in an effort to prevent undue recurrence of the epidemic,” said Otto E. Asbridge, president of the local labor federation, after heaving been assured by Dr. T. B. Beatty of the state health board that the restrictions against labor meetings had been lifted. 

Mr. Asbridge explained that about 1000 workers of various classes have been out of employment since the issuance of the health department's closing order. The total included about 300 musicians, twenty-five motion picture theater operators, about 150 stage hands, besides janitors and workers on the staffs of several film companies in Salt Lake. \dots 


\subsection{Effect of influenza pandemic on economic activity}

\label{sec:newspapers_direct}

\subsubsection{Effect of influenza on production and supply-side of the economy}

\paragraph{``Holland's Letter: Effect Of Influenza on Loan and Output---Reasons For and Against Imposing a Stamp Tax.'' {\normalfont\emph{Wall Street Journal}, Oct 24, 1918, p. 2.}}

At a private and informal meeting last week of some of these who are of important in the world of finance and banking, the suggestion was made that a communication be sent to Secretary of the Treasury McAdoo that he would be justified in extending to another week the campaign for the sale of the Fourth Liberty Loan bonds. \dots

One reason alone influenced those who suggested a recommendation of this kind to Secretary McAdoo. That was the prevalence of the grippe or influenza, which had seriously interfered with the sale of the bonds. \dots

The effect of the influenza epidemic was not exclusively felt, by the loan, however. In some parts of the country it has caused a decrease in production of approximately 50\% and almost everywhere it has occasioned more or less falling off.

The loss of trade which the retail merchants throughout the country have met with has been very large. The impairment of efficiency has also been noticeable. There never has been in this country, so the experts say, so complete domination by an epidemic as has been the case with this one. \dots

\paragraph{``Textile Trade Hit By Spanish Influenza: Many Mills Closed And Others Working Partially---Retail Business Hurt.'' {\normalfont\emph{Wall Street Journal}, Oct 21, 1918, p. 6.}}
 
Both the wholesale and retail trades have been hit badly by the Spanish influenza epidemic.
Mill production is being curtailed, and even Government business is held up.
A great many mills throughout the country have either completely ceased operations or kept only a small fraction of their machinery working.
Consequently, deliveries have been held up in many lines.
Retailers report that the disease has hurt their fall business, but it is hoped particularly among New York merchants that when the epidemic wanes they will quickly catch upon lagging sales. \dots

\paragraph{``Anthracite Output Affected By Influenza: Collieries Shut Down As .'' {\normalfont\emph{Wall Street Journal}, Oct 12, 1918, p. 9.}}
 
Effect of the influenza epidemic in current anthracite production is substantial \dots Around Minersville, Pa., where the ravages of the disease are said to have been probably as severe as in any part of the region, one entire colliery was shut down, but the washery of this particular company resumed working before the close of the week.

\paragraph{``Copper Shortage Is Acute: Influenza At Refineries And Smelters Further Reduces Output Already Curtailed by Labor Scarcity.'' {\normalfont\emph{Wall Street Journal}, Oct 25, 1918, p. 6.}}
 
Scarcity of copper is acute. Even the United States Government is not at present obtaining its full quota of metal, according to interests conversant with the situation.
With Government orders unfilled, there is, of course, no surplus available for the outside trade.

Increased curtailment of production is due largely to influenza at the refineries and smelters.
With the country's output already seriously impaired by labor shortages, a condition which is believed not likely to improve during the war, incapacitation of a large percentage of employees at nearly all the producing plants is resulting in a contraction in the copper supply which is expected to be more severe than was experienced during the worst months of the labor strikes in 1917.

 
\paragraph{``Influenza Impedes Ship Production: About 6,500 Workers Are Ill At Fall River and Hog Island---Other Yards Affected.'' {\normalfont\emph{New York Times}, Oct 3, 1918, Special.}}

The epidemic of Spanish influenza has put 10 per cent of the shipyard workers in the Fall River district and at least 8 per cent of those at Hog Island, Philadelphia, temporarily on the ineffective list and is seriously interfering with rapid ship construction.
Practically all of the yards as far south as Baltimore are affected to some degree, and extraordinary steps are being take in to fight the disease.
At Hog Island and other large plants some of the administration buildings have been converted into hospitals.

\paragraph{``Flu Affects the Cars — Not Enough Platform Men for the Various Lines.'' {\normalfont\emph{Charleston News and Courier}, Oct 9, 1918, p. 7}}
Owing to Spanish ``flu’’ among conductors and motormen, the Consolidated Company has been forced to curtail the service on all city lines and partly on the Navy Yard line during going-to-work and knock-off hours. \dots

\paragraph{``Epidemic Shows Slight Decline'' {\normalfont \emph{New Haven Journal - Courier}, October 29, 1918.}}  \dots H. C. Knight, vice-president and general manager of the Southern New England Telephone Co., returned to his duties yesterday after an attack of Spanish influenza. \dots

In the ranks of the telephone company influenza cut a wide swathe, taking as many as 40 per cent of the entire operating staff at a time and handicapping the company very seriously. The company now seems out of the woods. 

\paragraph{``Influenza on Wane in Manhattan.'' {\normalfont \emph{New York American}, October 23, 1918, p. 11.}} \dots Although there was a marked decrease in new cases of Spanish influenza reported to the Health Board yesterday, scores of big corporations are handicapped by the epidemic\dots The telephone company is crippled because of the influenza. Several of the big hotels and department stores are running short-handed and scores of big companies cannot get help to carry on their business. The Edison Company is tied up to a great extent and the telegraph companies announce they cannot get enough operators to carry out their usual business, much less the added telegrams on account of hundreds of deaths.

\paragraph{``Influenza Leaps As 1,480 New Cases Are Listed Today.'' {\normalfont \emph{Philadelphia Evening Bulletin}, October 4, 1918 p. 1, 2.}} \dots Telephones Overburdened. The Bell Telephone Company has issued another call to the public to use the telephones as little as possible during the emergency as its service is already being rushed to the breaking point. The company has more than 850 operators or more than twenty-six percent of the entire force, out owing to illness from the epidemic. \dots

Hits War Industries. The war industries of the city-have all been hard hit with the disease, and all of them are giving their employees as much protection as they can. 

The J. G. Brill Company estimates about twenty per cent of its employees ill with influenza and colds. About ten per cent off the employees of the Standard Roller Bearing plant are affected. About twenty per cent of those at Baldwin’s, in this city, are out. A few employees are ill at the Eddystone plant. The Link Belt Company also reports that about twenty per cent of its employees are out. The percentage of illness at Hog Island continues where it has stood for days, ten per cent being out. 

The E. G. Budd Company and the Hero Manufacturing Company, both big producers of special war material, are using disinfectants and the employee are being encouraged to spray their noses and throats two or three times a day. The Albefoyle Manufacturing Company, Chester, is closed today while the plant is being fumigated.

\paragraph{``Arrest All Maskless As Cases Grow.'' {\normalfont \emph{Oakland Tribune}, November 1, 1918, p. 18.}} \dots It was reported today to the health office that 75 of the 90 scavengers that gather the garbage in Oakland are sick with the influenza and that garbage collection in many districts was practically at a standstill. Commissioner Morse suggested that prompt measures be taken to relieve the situation and he will call a conference with the scavengers and undertake to make a new arrangement for the handling of the garbed during the present crisis. 

\paragraph{``Influenza Cripples Omaha Packing Plants. One of the Larger Concerns Reports Thirty Per Cent Ill With Malady. All Four of Big Ones Show Large Number of Employees Not at Work.'' {\normalfont \emph{ The Omaha World-Herald}, October 18, 1918.}}  L. Hersey, general manager of Morris \& Co., told the World-Herald that at least 30 percent of the 1,500 employees of the plant were away on account of influenza, that about 500 of his employees were suffering from the disease, and that with the shortage of labor the plant is severely handicapped. 

Michael Murphy, general manager of the Cudahy Packing company said that 240 of their 3,000 employees did not report for duty Thursday and he estimated the production at the plan was reduced fully 8 per cent. General Manager Edwards of Swift \& Co., said 15 per cent of the company’s employees were away from the plan because of influenza being about 400 out of a total of 2,800 persons employed at the establishment.


\subsubsection{Voluntary distancing; fear of influenza curtailing demand and economic and social activity}

\label{sec:newspapers_voluntary}

\paragraph{``5 Theatres Close Tonight: Theatrical Depression Attributed in Large Part to Influenza Scare.'' {\normalfont\emph{New York Times}, October 12, 1918, p. 13.}}
 
An unprecedented theatrical depression, which managers attribute in large part to the influenza scare, resulted in sudden decisions yesterday to close five playhouses tonight. \dots In all, more than a dozen local theatres will be dark next week.

\paragraph{``Influenza Checks Trade: Less Doing In Retail Shops As Illness and Caution Cut Down the Crowds.'' {\normalfont\emph{Wall Street Journal}, Oct 25, 1918, p. 10.}}
 
Widespread epidemic of influenza has caused serious inroads on the retail merchandise trade during the current month.
Heads of large organizations report that not only has sickness cut down the shopping crowds, but in many cities the health authorities have shut down the stores.

The chain store companies have felt the effect of the sickness not a little, for in addition to the smaller business done a number of their employees are sick. \dots

\paragraph{``See Less Influenza'' {\normalfont \emph{St. Paul Pioneer Press,} October 15, 1918, p. 8.}} Change In Hours of Movies Is Suggested. Fear of influenza contagion in crowded places has reduced the patronage of St. Paul motion picture theaters by nearly half, according to reports to Dr. H. M. Bracken, executive officer of the state board of health. \dots 

\paragraph{``Little Change In Local Epidemic,'' {\normalfont \emph{New Haven Journal - Courier}, October 21, 1918, p. 3.}} \dots In many instances means are being taken voluntarily by private interests to combat the spread of the epidemic. At St. Mary’s church the usual Sunday high mass at 10:30 o’clock as been abandoned, so that people may not be obliged to congregate longer than necessary. Rev. John Coyle, pastor of St. John’s church, announced at all the masses yesterday that the parochial school will close this week until further notice, not because of the fear of influenza, but because attendance is so poor that it has been deemed inexpedient to keep the school open for the few who attend.

Funerals Are Limited. Father Coyle also announced that funerals would be limited in attendance to the family and immediate relatives and he advised against wakes. It is understood that public funerals will be discouraged at St. Mary’s church, and that only the bearers will be allowed at the services at the church, although this report could not be verified last night. \dots  An increasing shortage of caskets is reported by local undertakers. 

\paragraph{``Bars, Cafes in Dark; Patrons Are Missing'' {\normalfont \emph{Oakland Tribune}, October 24, 1918, p. 7.}} Cafes, restaurants and saloons closed early last night without a police order for the first time on record in Oakland. The early closing was the result of a lack of business due to the Spanish influenza epidemic. Captain J. F. Lynch of the police department reported that many of the saloons had closed at 10 o’clock, while almost all the cafes and restaurants were closed by 10:30 o’clock. Many of the hotels and restaurants found it necessary to reduce their hours because of lack of cooks, waiters and other help. One restaurant has already reduced its business to that of a bakery and confectionary shop because the chef and cooks are down with Spanish influenza. 

\paragraph{``Health Officer Visits Theaters'' {\normalfont \emph{The Spokesman-Review} (Spokane, WA), December 8, 1918,  p. 6.}}

Dr. J.B. Anderson Insists Alternate Rows of Seats Must Not Be Occupied. \dots ``Reserved by the city health office'' will be the sign displayed on alternate rows of seats in theaters. \dots Many of the theater managers saw disaster in the new rule vacating half of the seating capacity of the houses. Others thought it made no difference, as the influenza had reduced the attendance to the extent that the vacant seats could be spared. \dots 

\paragraph{``Plan To Put End To Epidemic Here.'' {\normalfont \emph{Salt Lake Tribune}, November 20, 1918, , p. 11.}}  \dots A Frightened Community. Referring to the conditions as he found them on a recent visit to Tremonton, Dr. Beatty said, ``Not very long ago I read an account of conditions as they existed in London during the Black Plague in the sixteenth century, and I did not fully appreciate the description given until I arrived in Tremonton shortly after this epidemic started. Everyone seemed scared out of his wits, and regarded everyone else as if he expected to meet an assassin. Crape was visible everywhere, and even the dogs in the street seemed to sense that danger was impending from some invisible enemy. While the situation is critical enough, there is, of course, no sense in being frightened out of one’s wits.''

\paragraph{``Early Vote is Light; Officials Lacking--County Handles Election Under Difficulties\dots'' {\normalfont \emph{Oakland Tribune}, November 5, 1918,  p. 1.}} Oakland and virtually all of Alameda County is handling today’s election under difficulties. Many of those selected as election booth officials have been made ill by Spanish influenza and substitutes have been difficult to obtain by reason of fear of the epidemic. 

All of the polling places are open, however, although many of them are being run shorthanded. It is the purpose of County Clerk Gross to fill the vacancies by drafting persons who come to work at the polls for candidates, or from those who come to vote. 

Early morning voting was very light, according to the reports that have been received by the county clerks. \dots  
The vote throughout the county was light up to noon. In San Leandro less than 25 per cent of the normal vote was cast in the morning according to the city clerk’s office, although it was expected that the night vote of workers returning from nearby industrial plants would raise this somewhat. \dots 

Members of the health department and physicians generally have announced that if every one wears a mask there is no danger of contracting the influenza in polling places. \dots 
The influenza epidemic scared Berkeley voters away from the polls, despite a rigid enforcement of the mask ordinance, and an unusually light vote was recorded at noon today. \dots


\paragraph{``City Continues Vigorous Efforts To Stamp Out Epidemic Influenza. Plague's Grip Is Laid Upon Every Section. Use of City Hospital Found Necessary to Care for Sick Persons. Keen Suffering Among Poor. Nearly 3000 Cases Reported by Physicians--Urgent Need of More Nurses.'' {\normalfont \emph{The Post-Standard} (Syracuse, NY), October 10, 1918,  p. 6, 12.}}

\dots Stop Crowding Cars. 

Acting upon reports that street cars were being overcrowded, Mayor Stone yesterday ordered the New York State Railways to see that this practice was ended at once. One report stated that 115 passengers were counted on one car during the rush hour Tuesday night. The railway company officials declared that altho the number of travelers on its various lines had dropped off considerably in the past few days, the same service is being maintained and there should be no crowding on any cars. 


\subsubsection{Effects of influenza on consumer behavior and society}

\label{sec:newspapers_behavior}

 
\paragraph{``Drug Markets Affected By Spanish Influenza: Big Demand For Camphor Causes Advance in Wholesale and Retail Prices---Aspirin, Rhinitis and Quinine Taken in Big Quantities.'' {\normalfont\emph{Wall Street Journal}, Oct 21, 1918, p. 6.}}

The countrywide epidemic of Spanish influenza has had considerable influence on the drug markets and the demand for camphor, aspirin, quinine and many disinfectants has been unprecedented. \dots

\paragraph{``Influenza Edict Aids Home Life. Demand for Parlor Games and Books Resembles Annual Yuletide Rush. Streets Look Deserted.'' {\normalfont\emph{The Spokesman-Review} (Spokane, WA), October 11, 1918, p. 5.}}
\dots With all its supplies of canned amusements shut down, Spokane has had to turn to the home grown variety because of the Spanish influenza epidemic. It is an ill wind that blows nobody good, so the toy and game counters of the book and department stores have flourished for the last two days. The sales almost equaled the Christmas rush. 

\dots The book counters and the magazine counters are crowded. As one clerk said: ``They come in and say `Well, give us a magazine. We can’t do anything else so we may as well read.'''

Streets Look Deserted. Wednesday evening the streets looked as sparsely filled as in a blizzard. A few brave ones walked by whistling, trying to be cheerful, but there were no swooping automobiles full of young bloods, or sidewalks filled with short skirted damsels. \dots 

The mothers are in despair over the schools closing. As one said: ``The children are much happier to be out of school than I am to have them.'' \dots 

Families with children and cottages at the lakes are going to the country for the next two weeks. \dots  

\paragraph{``Fewer Flu Cases, But More Deaths.'' {\normalfont\emph{The Spokesman-Review} (Spokane, WA), November 12, 1918, p. 6.}}

\dots More Poor People Need Help. 

The social service bureau reported a marked increase in calls from destitute families with influenza. Three visiting nurses and two household field workers are giving all of their time to this work. \dots 

The present influenza epidemic has been the most trying time in the history of the social service bureau, the office reported yesterday. 

\subsubsection{Effect of influenza on financial markets}

\paragraph{``Corporation Bonds Comparatively Low: Present Average Price Over Eleven Points Under High Price Reached Since Stock Exchange Reopened.'' {\normalfont\emph{Wall Street Journal}, Jan 22, 1919, p. 5.}}
 
High Point Recorded January 18, 1917, and Low Since September 28, 1918---Influenza Epidemic an Influence in Decline of Railroad Bonds Which Are Usually Bought Heavily by Life Insurance Companies

\dots Several other factors which have tended to unsettle the bond market will be removed in the near future.
The influenza epidemic, which caused heavy claims on life insurance companies, thus temporarily putting them out of the market for high-grade railroad bonds, is an example.

    \clearpage
    \hypertarget{data-appendix}{%
\section{Data Appendix}\label{sec:data-appendix}}

\hypertarget{appendix:data-mortality}{%
\subsection{Mortality statistics}\label{appendix:data-mortality}}

We use information on mortality to construct several of the key
variables used throughout the paper. For instance,
\cref{tab:npi_mort_pi} studies the effect of NPIs on \emph{peak
mortality} and \emph{cumulative excess mortality}. In the case of
regressors, the NPI speed variable is defined for each city as the
difference between the NPI implementation date and the \emph{mortality
acceleration date}. Moreover, besides studying mortality due to
influenza and pneumonia we also look at all-cause mortality, which for
COVID-19 has been found to be a more accurate and less systematically
biased measure of mortality \citep{WhyAllCause}.

In this appendix, we describe in detail the sources and methods used to
calculate these mortality variables.

\hypertarget{procedures}{%
\subsubsection{\texorpdfstring{The procedures by \citet{Collins1930} and
\citet{Markel2007}}{The procedures by Collins et al (1930) and Markel et al (2007)}}\label{procedures}}

The first step of our work involves understanding and replicating the
work of \citet{Collins1930} and \citet{Markel2007}. With this, we should
be able to extend their approach to additional cities and from mortality
due to influenza and pneumonia to all-cause mortality, which they did
not study.

In this respect, \citet{Markel2007} states:

\begin{quote}
From the census reports, we extracted the weekly pneumonia and influenza
mortality data covering the 24 weeks from September 8, 1918 through
February 22, 1919, for the 43 US cities.
\end{quote}

\begin{quote}
A small number of missing values (846 or 0.6\% of 136,563 deaths) was
imputed. Using estimated weekly baseline pneumonia and influenza death
rates generated from the 1910-1916 median monthly values found by
\citet{Collins1930}, weekly excess death rates (EDR) were computed.
\end{quote}

\begin{quote}
We defined\ldots{} the magnitude of the first peak as the first peak
weekly EDR; and (3) the mortality burden as the cumulative EDR during
the entire 24-week study period.
\end{quote}

\begin{quote}
We also defined\ldots{} the date when weekly EDR first exceeds twice the
baseline pneumonia and influenza death rate (2x baseline; i.e., when the
mortality rates begings to accelerate)\ldots{}
\end{quote}

Moreover, \citet{Collins1930} states (table 5, footnote 5):

\begin{quote}
\ldots{} the excesses are computed as deviations from the median death
rate for the corresponding week \ldots{} The series of 52 medians
representing ``normal'' or ``expected'' rates for the different weeks of
the year were smoothed out by a 5-period moving average before
deviations were computed. The median rates were plotted, and a smooth
line passing through each of the 12 monthly medians was drawn to
represent the seasonal curve of mortality from influenza and pneumonia.
From this graph the approximate medians for each week of the year were
read.
\end{quote}

In accordance with the descriptions listed above, below we outline the
steps required to compute the influenza and pneumonia \emph{peak
mortality} and \emph{cumulative excess mortality} variables, as well as
the \emph{mortality acceleration date} required for the NPI days
variable:

\begin{enumerate}
\def\labelenumi{\arabic{enumi}.}
\item
  \textbf{Data collection:} collect monthly influenza and pneumonia
  death rates from 1910 to 1916.
\item
  \textbf{Median:} construct monthly baseline death rates (MBDR) as the
  median death rate of each month (Jan-Dec) between 1910 and 1916.
\item
  \textbf{Interpolate to higher frequency:} interpolate the MBDR series
  to create a weekly base death rate (WBDR) and daily baseline death
  rate (DBDR).
\item
  \textbf{Data collection:} collect weekly death counts (WDC) for the
  duration of the pandemic. Input missing values as needed.
\item
  \textbf{Interpolate to higher frequency:} transform the output of step
  4 to a daily frequency (DDC).
\item
  \textbf{Apply formula:} construct the excess weekly death rate:
  \(\operatorname{EWDR}_t = \operatorname{WDC}_t / \operatorname{Pop}_{1918} \times \allowbreak 10,000 - \operatorname{WBDR}_t\)
\item
  \textbf{Apply formula:} construct the excess daily death rate:
  \(\operatorname{EDDR}_t = \operatorname{DDC}_t / \operatorname{Pop}_{1918} \times \allowbreak 10,000 - \operatorname{DBDR}_t\)
\item
  \textbf{Compute maximum:} construct the peak excess mortality over the
  24 weeks of the pandemic using the daily death rates:
  \(\operatorname{Peak Mortality} = \operatorname{max}(\operatorname{EDDR}_t) \; \blacksquare\)
\item
  \textbf{Compute sum/integral:} construct the cumulative excess
  mortality over the 24 weeks of the pandemic:
  \(\operatorname{Cumulative Excess Mortality} = \operatorname{sum}(\operatorname{EWDR}_t) \; \blacksquare\)
\item
  \textbf{Apply formula:} compute the mortality acceleration date (AD):
  \(\operatorname{min} t \operatorname{st} t > \operatorname{08Sep1918} \, \land \operatorname{EDDR}_t \geq 2 \times \operatorname{DBDR}_t \; \blacksquare\)
\end{enumerate}

The steps above leave a few aspects unresolved, namely i) the exact data
sources used, ii) how the data was interpolated to a higher frequency,
and iii) how the missing values were imputed. These questions are
resolved below.

\hypertarget{data-sources}{%
\subsubsection{Data sources}\label{data-sources}}

\begin{itemize}
\item
  \textbf{Median monthly influenza and pneumonia death rates
  (1910-1916).} Obtained from \citet{Collins1930}, Appendix Table A,
  Pages 28-44.
\item
  \textbf{Monthly all-cause death rates by city (1910-1916).} Because
  \citet{Collins1930} does not list median monthly all-cause mortality
  rates, we obtain the monthly death rates for the years 1910-1916
  directly from the annual Mortality Statistics
  \citep{MortalityStatistics}. The specific tables used were Table 1 on
  page 191 for 1910, table 8 on page 521 for 1911, table 7 on page 311
  for 1912, etc.
\item
  \textbf{Weekly mortality counts (31Aug1918-15Mar1919).} Comprises
  total deaths, deaths from influenza, and deaths from all forms of
  pneumonia. This data was collected by the U.S. Census Bureau and
  published in their \emph{Weekly Health Index}, which we retrieved
  online from \citet{InfluenzaArchive}. Whenever the digitized records
  were illegible or unavailable, we instead used the \emph{Public Health
  Reports} \citep{PublicHealthReports}, which provided the same
  information as part of a broader report.
\item
  \textbf{Pre-pandemic population estimates}. To compute death rates
  from death counts, we need to divide by an estimate of the city
  population right before the onset of the pandemic. We computed this
  estimates via a linear interpolation of the city populations as
  reported in the 1910 and 1920 Decennial Census, interpolated to July
  1918. See Table 51, page 178 of \citet{PopCensus1920}.\footnote{Note
    that using 1920 population counts might be problematic, particularly
    for cities with higher pandemic mortality rates. For these cities,
    an interpolation using 1920 counts would under-estimate the true
    population as of July 1918, in turn leading to an over-estimate of
    the pandemic mortality rate. To assuage these concerns, we adjust
    the 1920 counts by adding to them the number of deaths caused the
    pandemic, to create a counterfactual population count. With this
    adjustment, our population estimates are:
    \(\operatorname{Pop}_{1918} = \operatorname{Pop}_{1910} \times \left( {\frac{\operatorname{Pop}_{1920} + \operatorname{PandemicDeaths}}{\operatorname{Pop}_{1910}} } \right) ^\frac{8.5}{10}\).
    Note, however, that this adjustment is essentially immaterial to the
    results shown throughout the paper.}
\end{itemize}

\hypertarget{interpolation-to-a-higher-frequency}{%
\subsubsection{Interpolation to a higher
frequency}\label{interpolation-to-a-higher-frequency}}

It is not entirely clear how \citet{Collins1930} used the monthly
baseline mortality rates to construct \emph{smoothed} baseline mortality
series at a higher-frequency. In the text, the authors explain that ``a
smooth line passing through each of the 12 monthly medians was drawn to
represent the seasonal curve of mortality from influenza and
pneumonia.'' However, drawing ``a smooth line'' is not a description
sufficient enough for replication, so we instead approximate their
results by applying an iterative rolling-mean algorithm which we
describe in \cref{alg:SmoothMortality}.

\begin{algorithm}
\begin{algorithmic}[0]
\caption{\textbf{SmoothMortality.} Interpolate low-frequency mortality series into a smoothed and mean-preserving high-frequency series}\label{alg:SmoothMortality}
\State \textbf{inputs:}
\Statex \quad variable $x$ \Comment{Series with mortality counts reported at a weekly or monthly frequency}
\Statex \quad frequency $f$ \Comment{Frequency of $x$; either weekly or monthly}
\Statex \quad bandwidth $k$ \Comment{Tuning parameter; $k=3$ (weekly) or $15$ (monthly)}
\State $y = x$
\For{$i=k, \, k-1, \, \dots, \, 1$}
  \State $z \gets \mathrm{RollingMean}(y; \, i)$ \Comment{RollingMean computes a centered mean from $t-i$ to $t+i$}
  \State $w \gets \mathrm{MeanBy}(z; \, f)$ \Comment{MeanBy computes the mean within each week or month}
  \If{$i > 1$} 
      \State $y \gets x + z - w$ \Comment{$Rebalancing: (z-w)$ has mean zero within frequency $f$}
  \EndIf
\EndFor
\State \textbf{output:} y
\end{algorithmic}
\end{algorithm}

This \texttt{SmoothMortality} algorithm has three essential features:

\begin{enumerate}
\def\labelenumi{\arabic{enumi}.}
\item
  It is \emph{smooth}, as it is based on the application of centered
  rolling means.
\item
  To avoid having too little variation, which would occur if we applied
  rolling means only once, we instead rolling means multiple times, at
  an increasingly smaller bandwidth. For instance, with monthly data we
  start with a \(\pm 15\) days bandwidth, iteratively reducing it by one
  day until the bandwidth is as small as possible (\(\pm 1\) day).
\item
  An important property that we wanted to preserve, is that the
  low-frequency input series and the high-frequency smoothed-out output
  series should have the same average the low frequency. For instance,
  Albany had 186 influenza and pneumonia deaths in the week ending on
  Nov 2, 1918. Our goal is that the daily smoothed-out series generated
  by \cref{alg:SmoothMortality} should have about 186 deaths throughout
  the seven days of this week. Otherwise, we risk undercounting or
  overcounting death rates with this transformation.
\end{enumerate}

{
\setstretch{1.0}
  \begin{figure}[htpb]
    \centering
    \includegraphics[width=0.7\textwidth]{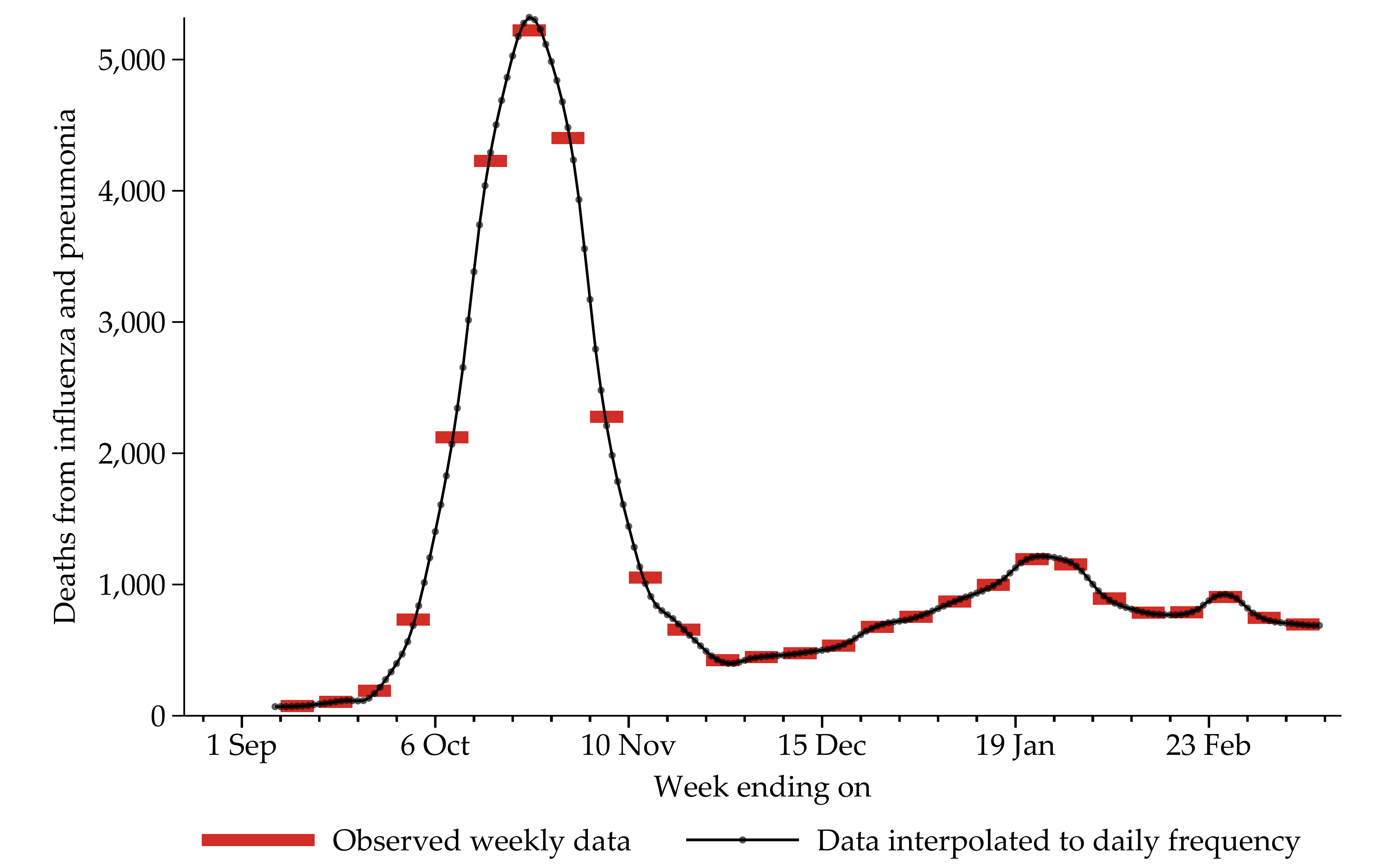}
    \caption{\textbf{\texttt{SmoothMortality} interpolation of weekly influenza and pneumonia
death counts to a daily frequency. }This figures shows the output of applying the \texttt{SmoothMortality}
algorithm to the weekly influenza and pneumonia death counts of New York
City. The bars in gray represent weekly deaths, as reported by the
\emph{Weekly Health Index}. The line in blue represents the data
interpolated to a daily frequency. Notice how its average for each week
is about the same as the gray bar. Note also that to compute daily death
rates this line has to be divided by
\(7 \times \operatorname{Pop}_{1918}\).}
    \label{fig:smoothing1}
  \end{figure}
}

\Cref{fig:smoothing1} illustrates the output of this algorithm for the
case of weekly influenza and pneumonia data for New York City, during
the weeks of the pandemic. As we can see, the daily series is smooth and
averages roughly the same values as the weekly series in gray bars. The
importance of applying the \texttt{SmoothMortality} algorithm, rather
than a simpler rolling means approach, can be seen in
\cref{fig:smoothing2}. Here, we can see that applying rolling means at
different bandwidths yields daily series that are not smooth and are
thus unlikely to represent the true underlying daily data. The
importance of the rebalancing step of the algorithm, which ensures that
the average of the high-frequency series matches the value of the low
frequency series, can be seen in \cref{fig:smoothing3}. This figure
contrasts the \texttt{SmoothMortality} algorithm (in blue) with one in
which the rebalancing step has been turned off (in red). Notice how the
red line underestimates the value at the peak, as the daily death counts
never exceed the \emph{average} daily death counts (in gray). Therefore,
without rebalancing we would be unable to replicate the peak mortality
measure used in this paper without measurement errors.

{
\setstretch{1.0}
  \begin{figure}[htpb]
    \centering
    \includegraphics[width=0.7\textwidth]{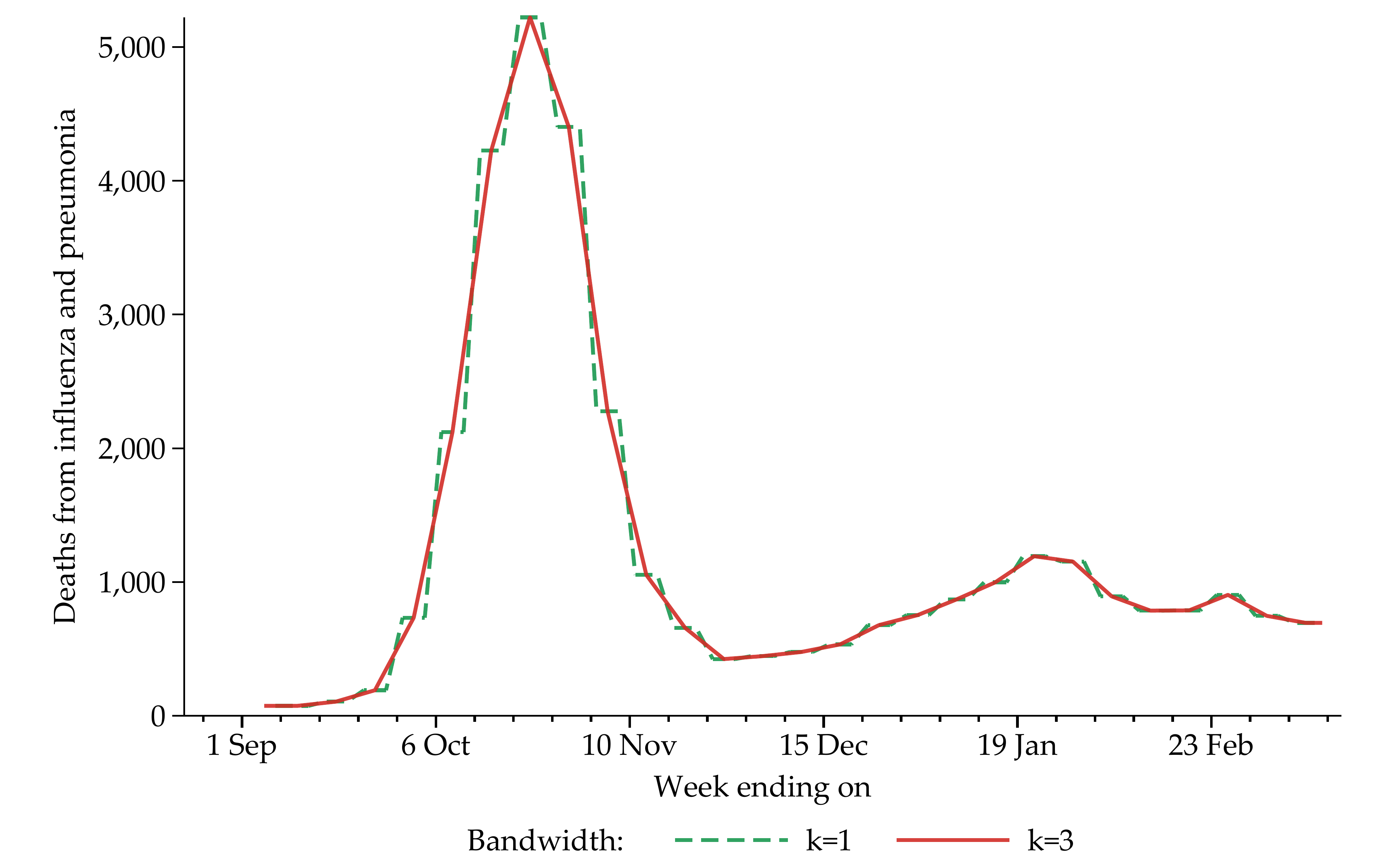}
    \caption{\textbf{Interpolation through rolling means exclusively. }This figures illustrates why centered rolling means by themselves are
unable to obtain smooth series at a higher frequency. The blue line,
based on a rolling mean with a \(\pm 1\) day bandwidth, is discontinuous
and appears similar to a step function. The red line, based on a
\(\pm 3\) day bandwidth, has less abrupt changes but nonetheless is not
smooth, particularly near the peaks.}
    \label{fig:smoothing2}
  \end{figure}
}

{
\setstretch{1.0}
  \begin{figure}[htpb]
    \centering
    \includegraphics[width=0.7\textwidth]{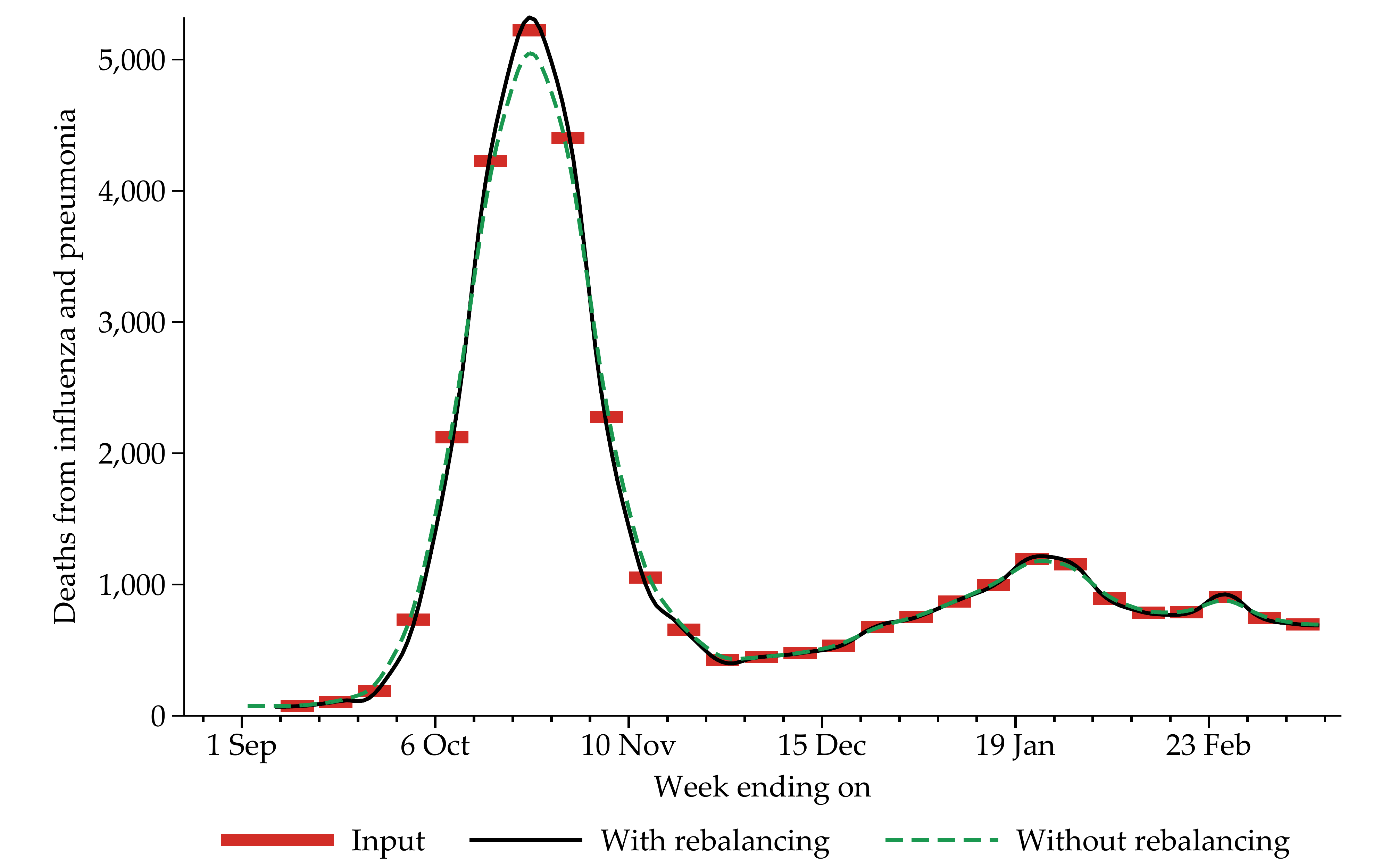}
    \caption{\textbf{\texttt{SmoothMortality} algorithm with and without rebalancing. }This figure shows the importance of rebalancing, i.e.~ensuring that the
average of the high-frequency output series matches the value of the
lower-frequency input series. The first line, in blue, does so, and it
has a peak quite close to the peak of the weekly input series. However,
the line in red has a much lower peak due to its lack of rebalancing,
which leads to potential measurement errors in the \emph{peak mortality}
variable of interest.}
    \label{fig:smoothing3}
  \end{figure}
}

\hypertarget{imputation-of-missing-values}{%
\subsubsection{Imputation of missing
values}\label{imputation-of-missing-values}}

The \emph{Weekly Health Index} compiled by the Census Bureau
occasionally had missing values for certain cities, where it was noted
that ``no report was received''. To address this, \citep{Markel2007}
stated that ``a small number of missing values (846 or 0.6\% of 136,563
deaths) was imputed.''

In our case, we had to input fewer missing values---accounting for 346
deaths rather than 846 as in Markel---because we used the \emph{Public
Health Reports} \citep{PublicHealthReports} to input otherwise illegible
information. We input missing values with a simple linear interpolation,
illustrated in \cref{fig:imputation} for Omaha, NE. This figure shows
weekly reported deaths in black, and two inputted death counts in red.
Note also that no city had more than two missing observations, and that
10 out of the 18 missing observations occurred in the third week of
September, when death counts were zero for most cities. Thus, we believe
missing values to be relatively immaterial for the mortality
calculations.

{
\setstretch{1.0}
  \begin{figure}[htpb]
    \centering
    \includegraphics[width=0.7\textwidth]{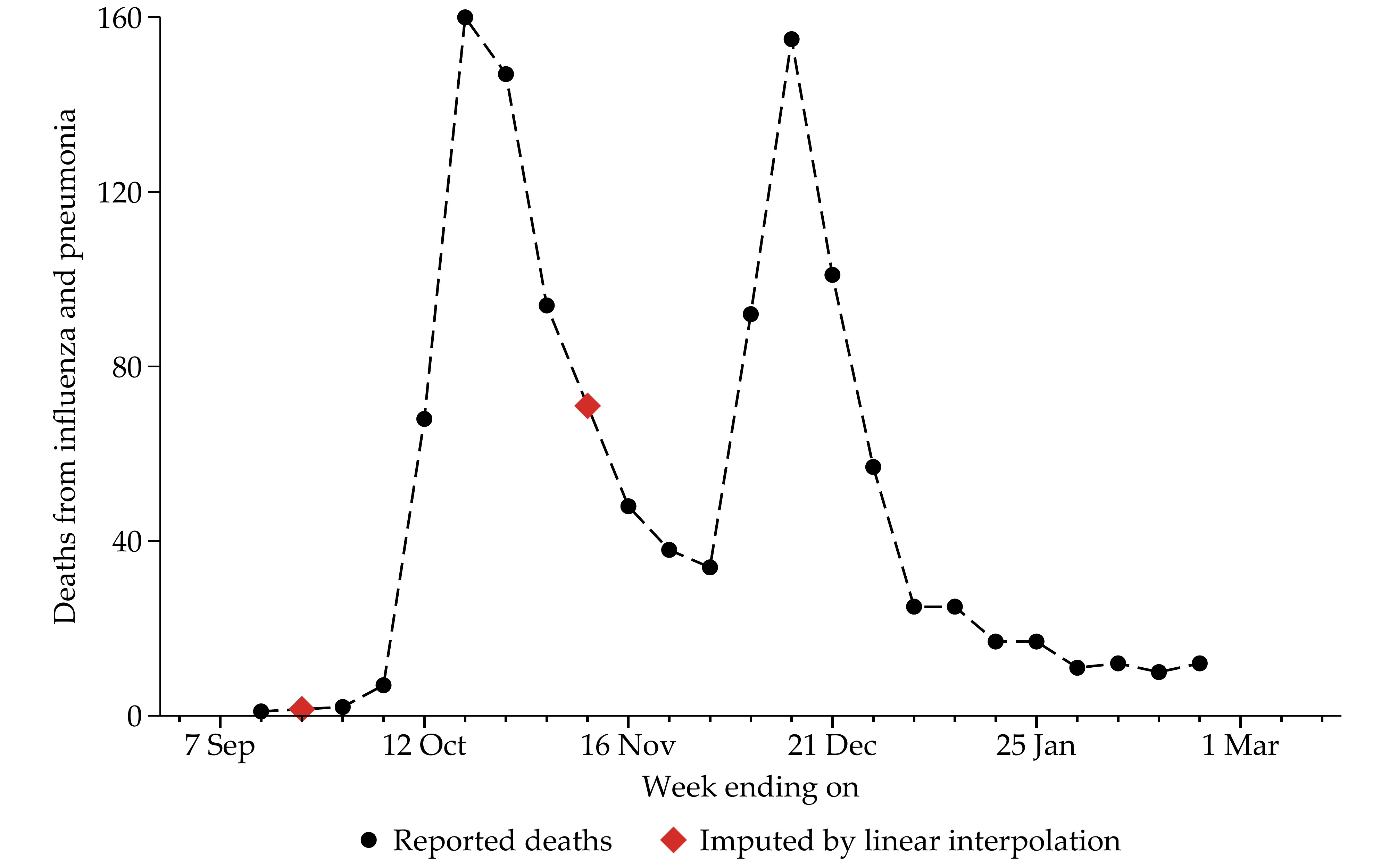}
    \caption{\textbf{Imputation of influenza and pneumonia death counts for the city of
Omaha, Nebraska. }The black markers show deaths by influenza and pneumonia, as reported in
the \emph{Weekly Health Index}. The dashed black line shows a linear
interpolation of those death counts, and the red markers show the two
imputed values, based on the linear interpolation.}
    \label{fig:imputation}
  \end{figure}
}

\hypertarget{validation-of-results}{%
\subsubsection{Validation of results}\label{validation-of-results}}

\Cref{fig:validation-acceleration} compares the mortality acceleration
dates as reported by \citet{Markel2007} with our replication. As we can
see, the two sets of dates are mostly in line, even though we
constructed ours by directly collecting data from primary sources and
without an accurate understanding of the algorithms behind Markel's
results (namely, the smoothing algorithm by Collins et al.~1930 and the
missing data imputation by Markel et al.~2007). This gives us confidence
that the NPI speed variables compiled for the additional cities---which
depend on the mortality acceleration dates---will be in line with those
compiled by Markel et al.

{
\setstretch{1.0}
  \begin{figure}[htpb]
    \centering
    \includegraphics[width=0.5\textwidth]{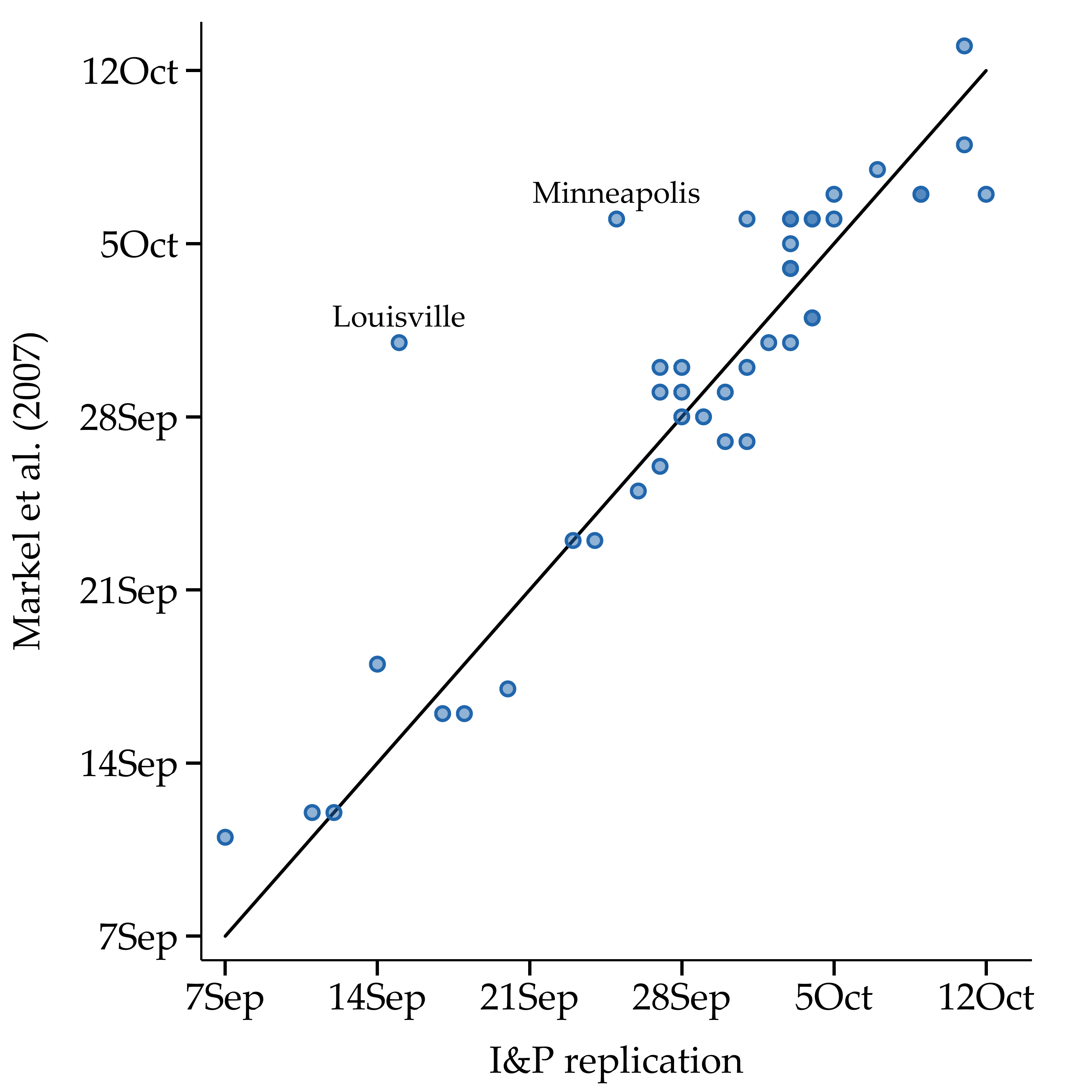}
    \caption{\textbf{Validation of mortality acceleration dates. }This figure compares the Markel et al.~(2007) influenza and pneumonia
acceleration dates against our replication for influenza and pneumonia.
We labelled the two cities where the difference between the two sources
exceeds five days (Louisville and Minneapolis), and tracked the
discrepancy to a small cluster of cases in early september in these
cities, which lead to the excess death rate to slightly exceed twice the
baseline I\&P mortality.}
    \label{fig:validation-acceleration}
  \end{figure}
}

Further, \cref{fig:validation-peak-ip} and
\cref{fig:validation-cumulative} compare both our influenza and
pneumonia replication and our new results on all-cause mortality against
Markel's information, for both peak excess mortality and cumulative
excess mortality. The panels on the left show that our replication
provides results quite similar to those of Markel, with the dots falling
close to the 45-degree line. Moreover, the panels on the right show that
the results for excess all-cause mortality are also quite similar to
Markel's Given the concerns on COVID-19 reported deaths
\citep{WhyAllCause}, we find it quite striking that the results ascribed
to influenza and pneumonia match so closely those obtain from the
all-cause mortality counts.

{
\setstretch{1.0}
\begin{figure}[ht!]
  \centering
  \begin{subfigure}{0.5\textwidth}
    \centering
    \includegraphics[width=0.9\linewidth]{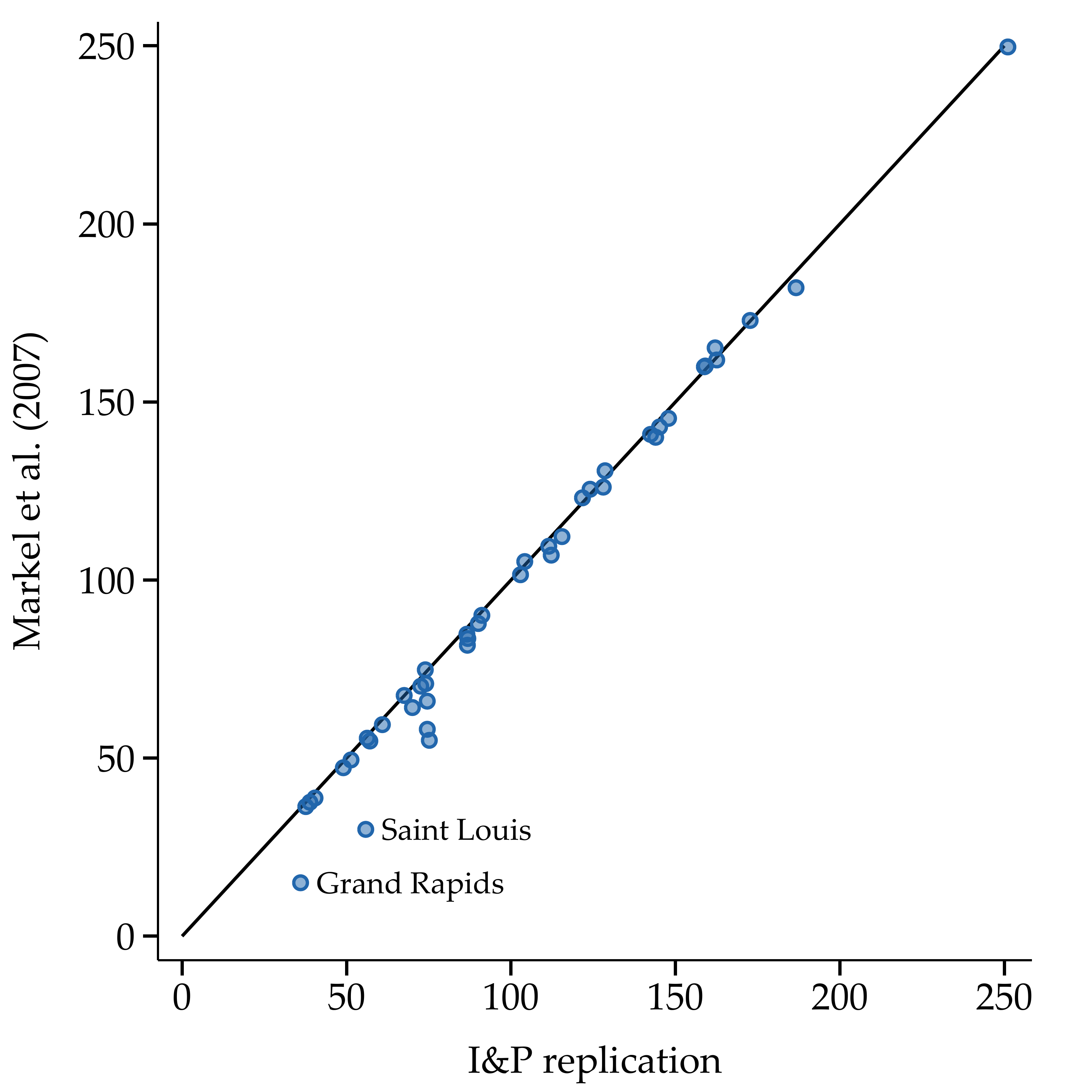}
    \caption{I\&P Replication}
    \label{fig:validation-peak-ip}
  \end{subfigure}
  \begin{subfigure}{0.5\textwidth}
    \centering
    \includegraphics[width=0.9\linewidth]{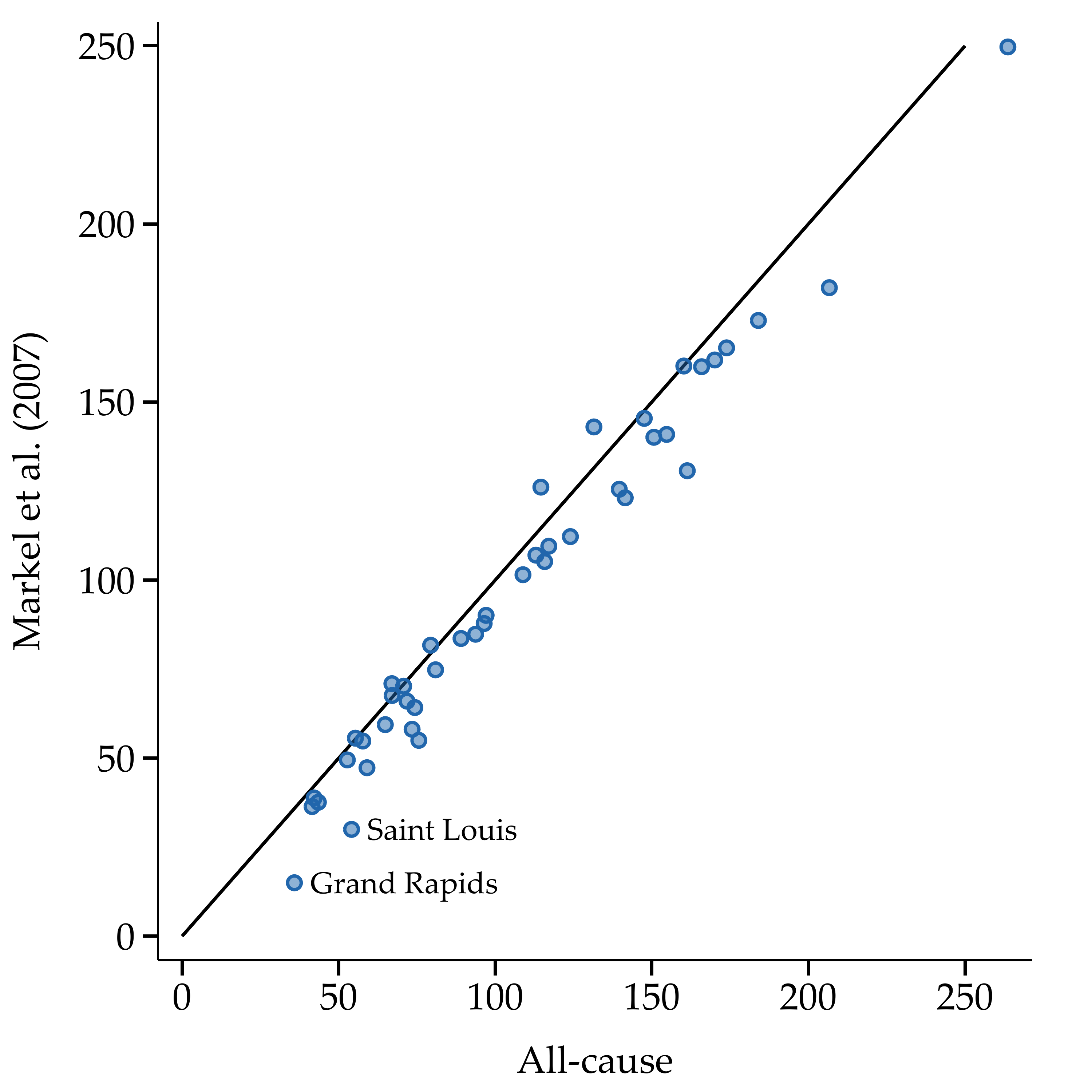}
    \caption{All-cause mortality}
    \label{fig:validation-peak-allcause}
  \end{subfigure}
  \caption{\textbf{Validation of peak excess mortality. }This figure compares the Markel et al.~(2007) influenza and pneumonia
peak mortality values (rate per 100,000 population) against our
replication for influenza and pneumonia (Panel A) and our additional
results for all-cause mortality (Panel B). We labelled the two cities
where the difference between the two measures exceeds thirty percent
(Grand Rapids and St.~Louis). Note that both our influenza and pneumonia
and our all-cause mortality both agree on lower peak mortality rates
than those reported by Markel.}
  \label{fig:validation-peak}
\end{figure}
}

{
\setstretch{1.0}
\begin{figure}[ht!]
  \centering
  \begin{subfigure}{0.5\textwidth}
    \centering
    \includegraphics[width=0.9\linewidth]{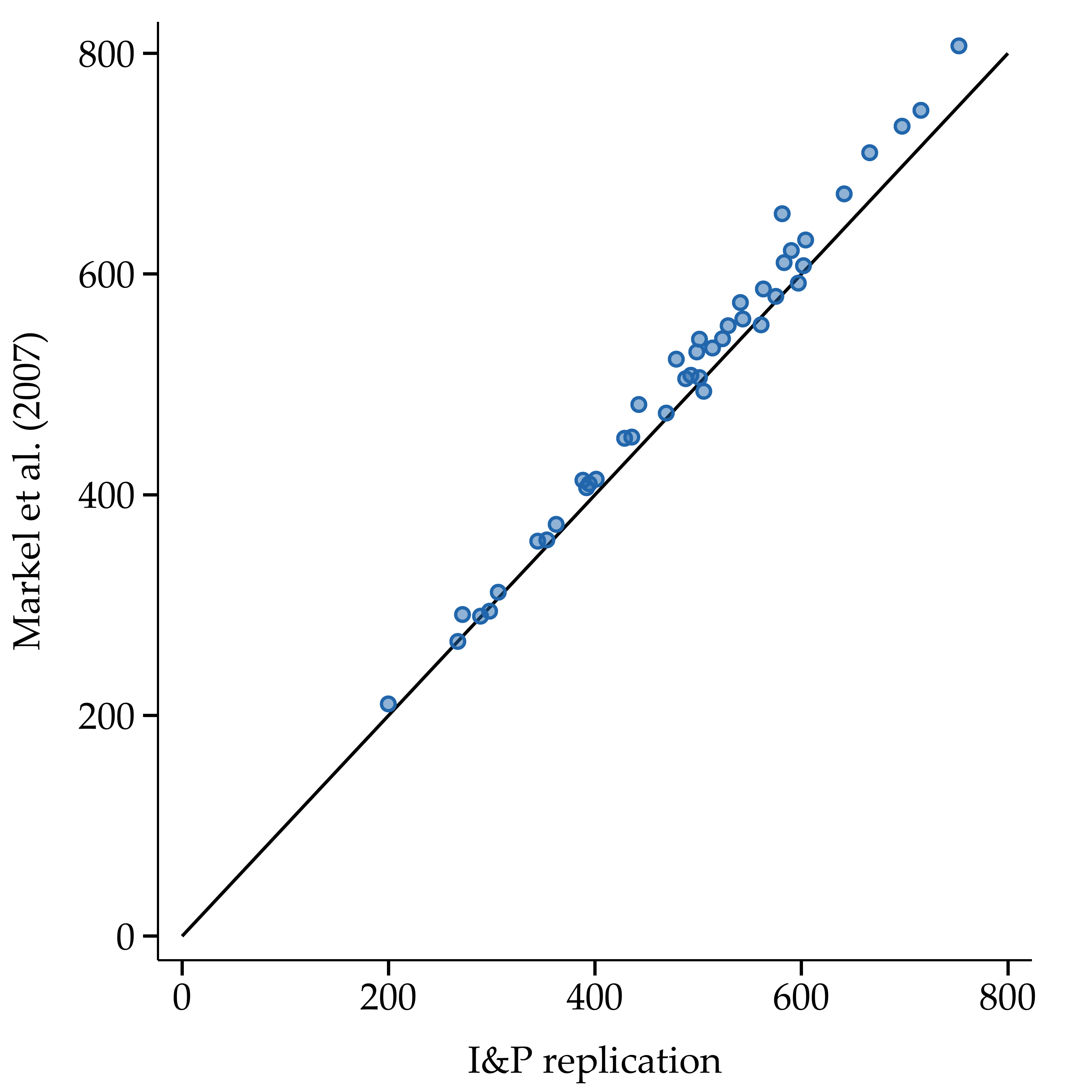}
    \caption{I\&P Replication}
    \label{fig:validation-cumulative-ip}
  \end{subfigure}
  \begin{subfigure}{0.5\textwidth}
    \centering
    \includegraphics[width=0.9\linewidth]{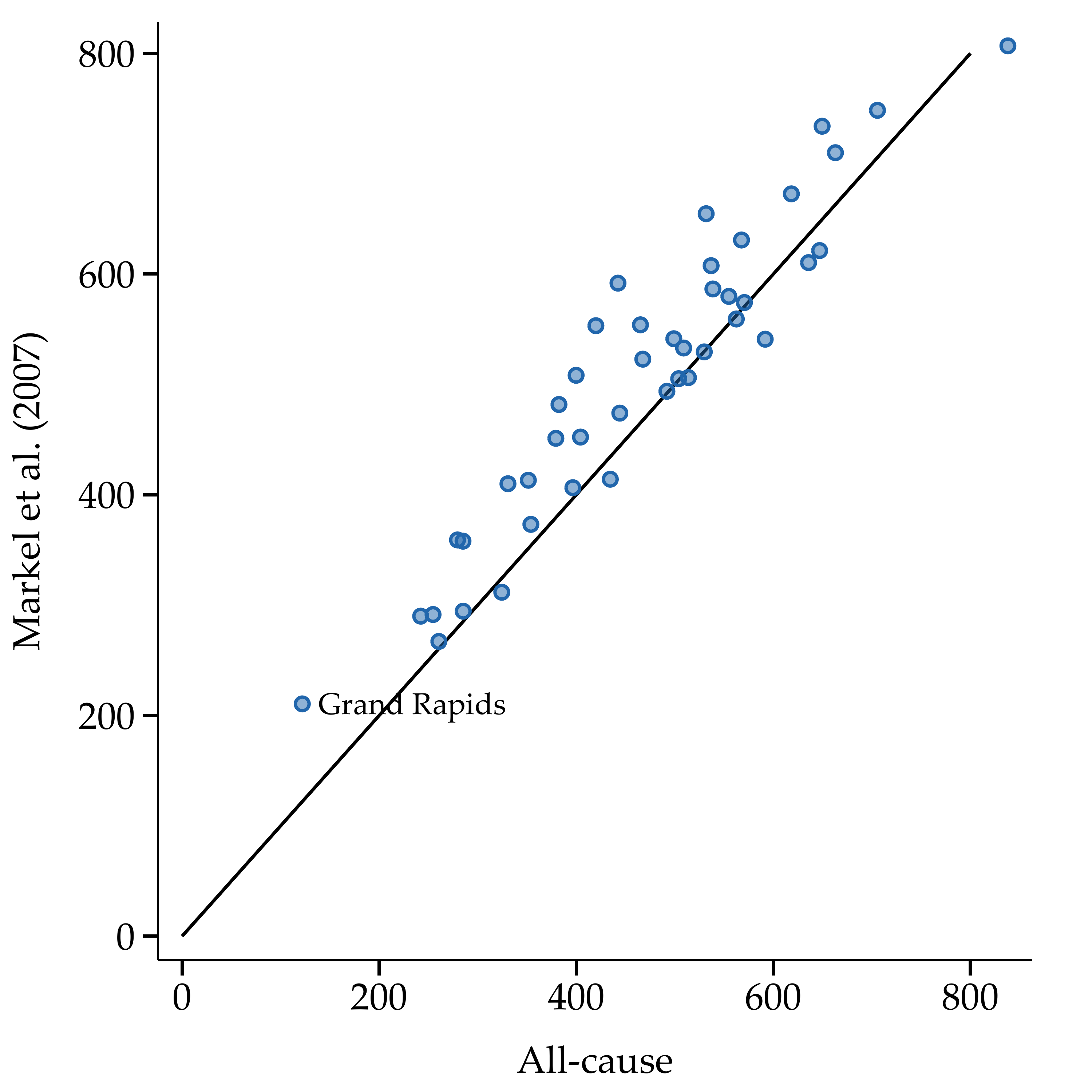}
    \caption{All-cause mortality}
    \label{fig:validation-cumulative-allcause}
  \end{subfigure}
  \caption{\textbf{Validation of cumulative excess mortality. }This figure compares the Markel et al.~(2007) influenza and pneumonia
cumulative mortality (rate per 100,000 population) against our
replication for influenza and pneumonia (Panel A) and our additional
results for all-cause mortality (Panel B). There is only one case (Grand
Rapids, all-cause mortality) where the difference with Markel et
al.~(2007) exceeds 30\%.}
  \label{fig:validation-cumulative}
\end{figure}
}

In terms of differences, we do find that Markel's cumulative excess
mortality is slightly higher than both of our results. We are unsure
about this reason, but believe one possibility might be a slightly
better computation of smoothed baseline mortality rates in our paper
that in the work of \citet{Collins1930}, which had to resort to manual
calculations.

\hypertarget{additional-cities}{%
\subsubsection{Additional cities}\label{additional-cities}}

The mortality regressions have three additional cities compared to
\citet{Markel2007}. These are Atlanta, Jersey City, and Memphis, and all
three were included in both \citet{Collins1930} aand in the Weekly
Health Index, which allows us to compute mortality statistics and NPI
speed for them.\footnote{Note that Jersey City has several missing
  values after the its mortality peak, so its estimates for cumulative
  excess death rate might be noisier than for other cities.}

\hypertarget{appendix:data-npi}{%
\subsection{Non-pharmaceutical interventions}\label{appendix:data-npi}}

In order for us to better understand how to collect and compute NPI
measures for the additional cities in our sample, \citet{Markel2007}
graciously provided the raw information used to construct their NPI days
and NPI speed variables. \Cref{fig:validation-npi} compares their
published NPI measures against those based on our replication based on
the raw information provided. Although this figure has the caveat that
there might be errors in our replication and interpretation of the notes
provided, it is nonetheless reassuring that both panels show strikingly
similar results between the Markel published estimates and our
replication. Note also that the results shown throughout the paper are
robust to choosing our replication NPI measures results instead of the
ones by \citet{Markel2007}.

{
\setstretch{1.0}
\begin{figure}[ht!]
  \centering
  \begin{subfigure}{0.5\textwidth}
    \centering
    \includegraphics[width=0.9\linewidth]{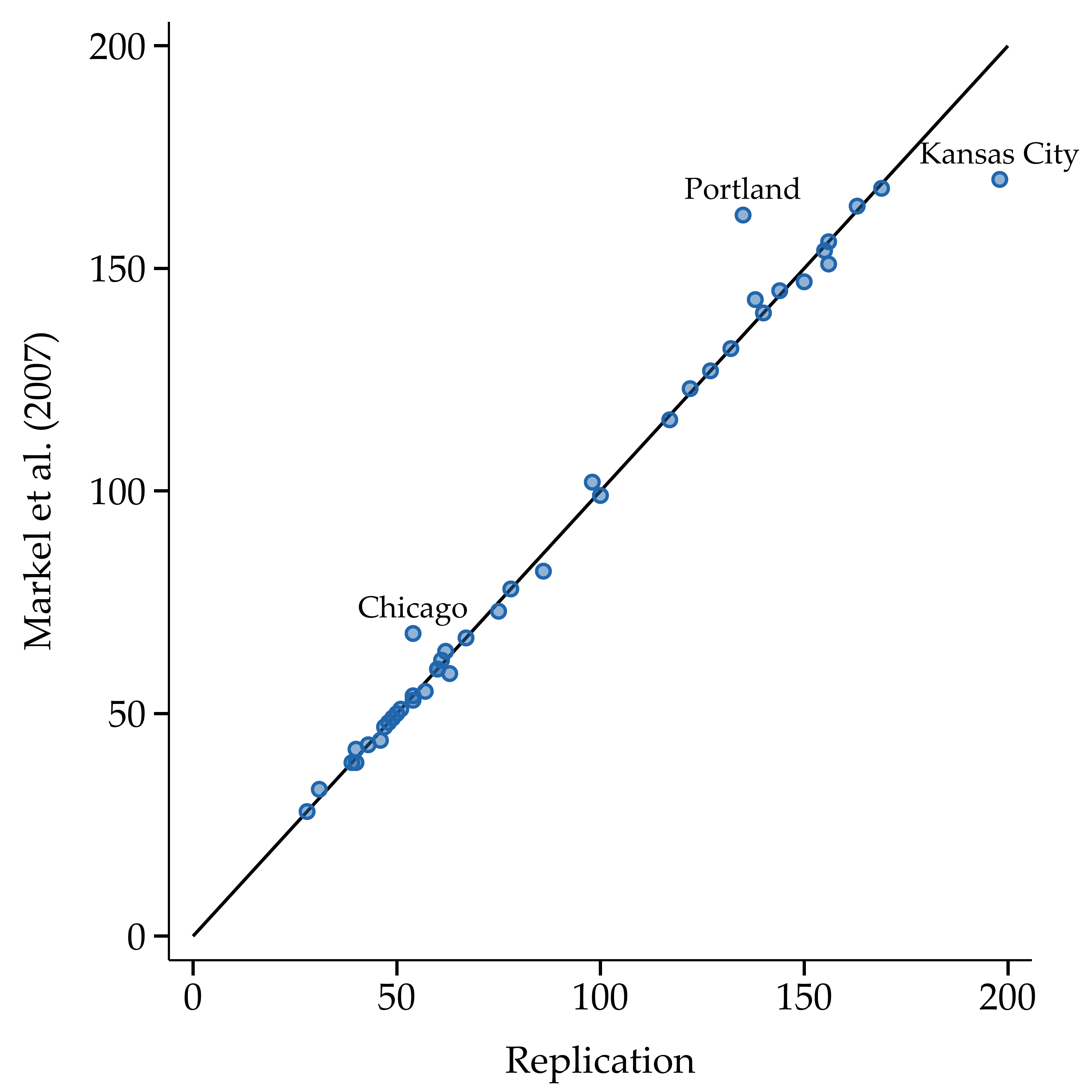}
    \caption{Days of NPI}
    \label{fig:validation-days-npi}
  \end{subfigure}
  \begin{subfigure}{0.5\textwidth}
    \centering
    \includegraphics[width=0.9\linewidth]{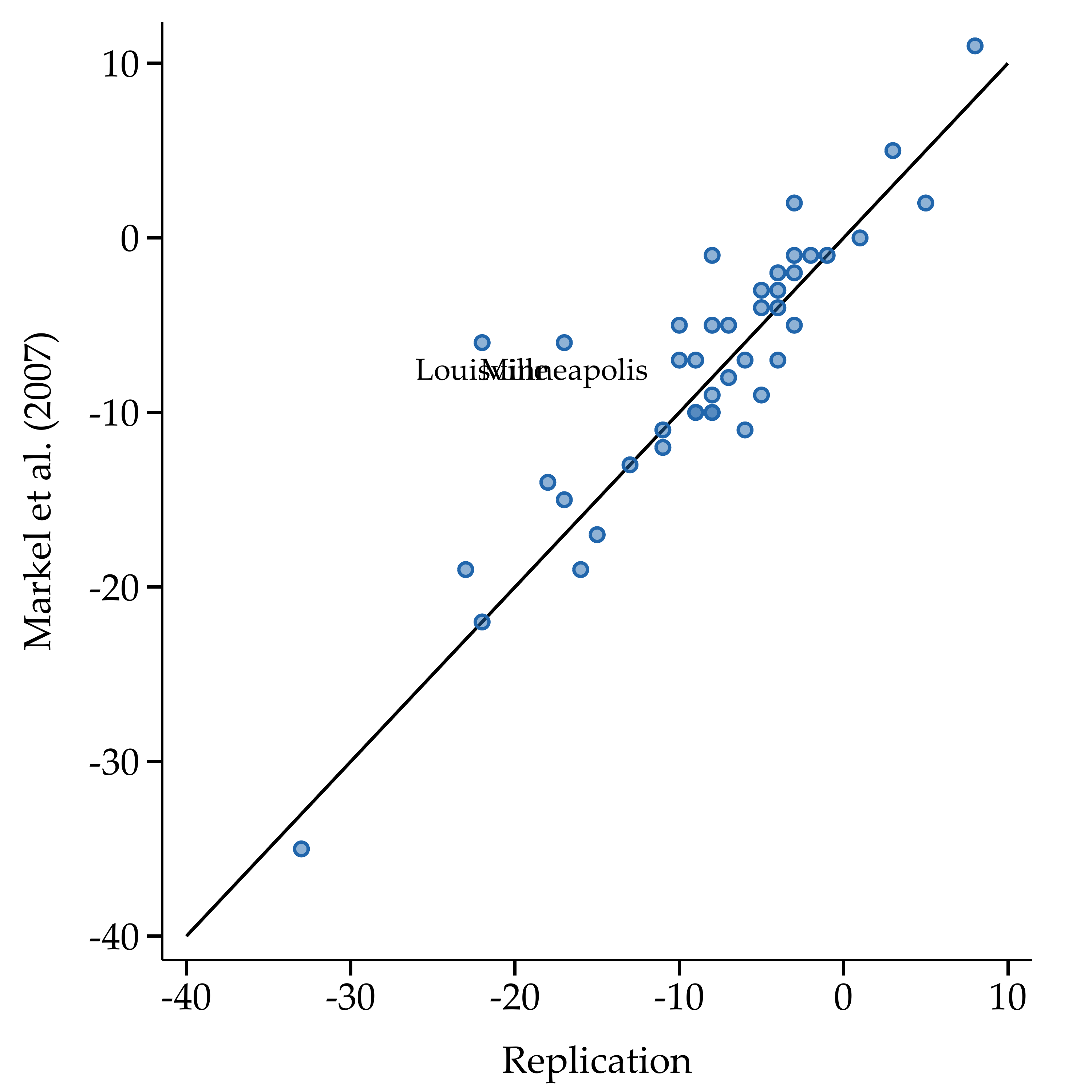}
    \caption{Speed of NPI}
    \label{fig:validation-speed-npi}
  \end{subfigure}
  \caption{\textbf{Replication of Markel et al.~(2007) NPI measures. }This figure compares the Markel et al.~(2007) NPI measures against our
replication based on their provided raw material. Panel (a) corresponds
to the NPI days measure (total days of NPI across the three measures),
and panel (b) corresponds to the NPI speed measure (mortality
acceleration date minus date of first NPI). and our additional results
for all-cause mortality (Panel B). In both panels, we have labelled the
cases where there is a discrepancy above ten days, which corresponds to
three observations for NPI days (Chicago, Portland, Kansas City) and two
observations for NPI speed (Louisville and Minneapolis).}
  \label{fig:validation-npi}
\end{figure}
}

\hypertarget{additional-cities-1}{%
\subsubsection{Additional cities}\label{additional-cities-1}}

Throughout the paper, we use NPI data for eleven cities outside the
\citet{Markel2007} sample (three for the mortality regressions, two for
the Bradstreet's regressions, and eleven for the census manufacturing
regressions). For Atlanta, we rely on Markel's raw information. For
Charleston, Dallas, Des Moines, Detroit, Salt Lake City, and San
Antonio, we rely on \citet{Berkes2022}, and for Jersey City, Memphis,
Paterson, and Scranton we collect the data directly from primary
sources, accessed via
\href{https://www.newspapers.com/}{Newspapers.com},
\href{https://newspaperarchive.com/}{Newspaper Archive},
\href{https://chroniclingamerica.loc.gov/search/titles/}{Chronicling
America} from the Library of Congress.

\hypertarget{appendix:data-bradstreets}{%
\subsection{Bradstreet's ``Trade at a
Glance''}\label{appendix:data-bradstreets}}

To construct the city-level index of economic disruptions we use the
``Trade at a Glance'' report published most Saturdays as part of the
weekly
\textit{Bradstreet's - A Journal of Trade, Finance, and Public Economy}.
In particular, we use Volumes
\href{https://books.google.com/books?id=vno2AQAAMAAJ}{45 (1917)},
\href{https://books.google.com/books?id=jng2AQAAMAAJ}{46 (1918)},
\href{https://books.google.com/books?id=4HM2AQAAMAAJ}{47 (1919)}, 48
(1920), \href{https://books.google.com/books?id=FBkUAQAAMAAJ}{49
(1921)}, and \href{https://books.google.com/books?id=TioUAQAAMAAJ}{50
(1922)} of Bradstreet's annual publication, which comprise most weekly
issues between 1917 and 1922.\footnote{No data is available for October
  and November 1919 as the magazine was not published due to the New
  York printing press strikes, and the Trade at a Glance report was
  often not published on the first weeks of each year.}

As seen on \cref{fig:bradstreets}, Trade at a Glance summarized for the
main cities in the United States the status of wholesale trade, retail
trade, manufacturing and industry, collections, and crops (except on
Winter months). Each of these categories was described with a single
word (``Quiet,'' ``Fair,'' ``Active,'' etc.), with additional
information available for the city as a whole on the ``Remarks''
section. From these five categories, we focus on wholesale trade, retail
trade, and manufacturing, and exclude collections and crops, as the
latter are reported more sparsely and are less related to the day-to-day
economic activity of the city itself.

{
\setstretch{1.0}
  \begin{figure}[htpb]
    \centering
    \includegraphics[width=0.7\textwidth]{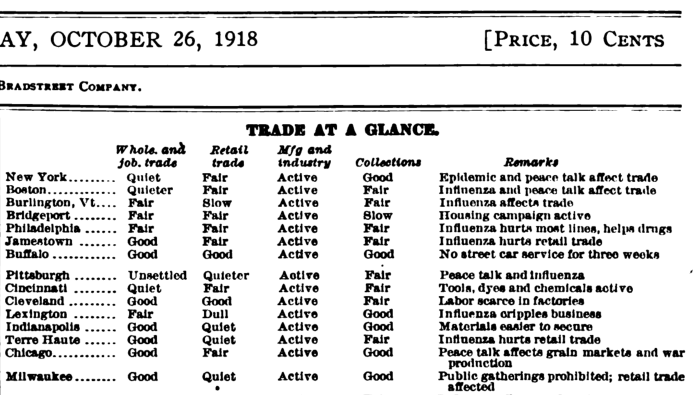}
    \caption{\textbf{Bradstreet's ``Trade at a Glance'', October 26, 1918}}
    \label{fig:bradstreets}
  \end{figure}
}

Because the number of different one-word descriptions we observe in each
category over 1918-1919 is quite
large\footnote{We observe 27 different values for wholesale trade, 26 for retail trade, and 53 for manufacturing.},
we use a series of rules to condense it into a three-valued index that
describe conditions as ``Bad,'' ``Fair,'' or ``Good.'' Further, we also
use this index to construct a binary variable for whether there were
trade disruptions (``Bad'' and ``Fair'') or no trade disruptions
(``Good''). ``No disruptions'' are given a value of 100, while
disruptions are given a value of 0.

The classification rule is as follows:

\begin{enumerate}
\def\labelenumi{\arabic{enumi}.}
\item
  Good: good, brisk, excellent, active, liberal, very active, better,
  record, very good, steady, more active, prompt.
\item
  Fair: fair, moderate, fair to good, satisfactory, close, 3/4 capacity,
  60 percent, 75 percent, 75\% basis, normal, fair activity, fairly
  active, hesitating, hesitation, only fair, slowdown, readjusting, half
  speed, half time, hampered, waiting, slack, uncertain, suspended, many
  strikes, contracted, disturbed, inactive, short time, retarded,
  paralyzed, irregular, unsettled, conservative.
\item
  Bad: quiet, dull, slow, very slow, cautious, interrupted, light,
  restricted, below normal, curtailed, under normal, poor, lagging,
  tardy, delayed, backward, drag.
\end{enumerate}

Further, we occasionally found words that describe conditions
\textit{relative} to previous reports. For instance, ``quieter,''
``improving,'' etc. In these cases we apply the following rule:

\begin{enumerate}
\def\labelenumi{\arabic{enumi}.}
\item
  If the current week has any of the following keywords, we reduce the
  rating by one notch (from Good to Fair, or from Fair to Bad): reduced,
  quieter, slower, slowing down, smaller, less active, receding.
\item
  If the current week has any of the following keywords, we increase the
  rating by one notch (from Fair to Good, or from Bad to Fair):
  improved, improving, slightly better, enlarging, shifting, enlarging,
  improvement, increasing.
\end{enumerate}

Lastly, we exclude observations where remarks mention labor
strikes\footnote{Results are robust to including observations with labor strikes, but are potentially more difficult to interpret, as strikes could signify a booming local economy where labor is in short supply.}
and remove cities with only a few observations (Oakland and Denver with
only one and eight observations, respectively). We then collapse the
data at the city-month level, as information was often sparse and some
cities did not report data every week. To create a combined index of
wholesale trade, retail trade, and manufacturing, we take a simple
average of the monthly disruptions index for the three sectors. This
leaves us with a total of 27 cities, as shown in \cref{tab:samplecities}.

\hypertarget{appendix:data-manufacturing}{%
\subsection{Census of Manufactures}\label{appendix:data-manufacturing}}

Data on city-level manufacturing employment, output (value of products),
and value added is from the Census of Manufactures. Manufacturing
employment is defined as the average number of wage earners over the
twelve monthly values of each census year.\footnote{See
  \citet{Graphical_Book} for details on how annual employment values are
  estimated.} The sources used to obtain the manufacturing data are
listed in table \cref{table:manufsources} below:

\begin{longtable}[]{@{}
  >{\centering\arraybackslash}p{(\columnwidth - 8\tabcolsep) * \real{0.4054}}
  >{\centering\arraybackslash}p{(\columnwidth - 8\tabcolsep) * \real{0.1757}}
  >{\centering\arraybackslash}p{(\columnwidth - 8\tabcolsep) * \real{0.0811}}
  >{\centering\arraybackslash}p{(\columnwidth - 8\tabcolsep) * \real{0.0946}}
  >{\centering\arraybackslash}p{(\columnwidth - 8\tabcolsep) * \real{0.2432}}@{}}
\caption{Manufacturing sources.}\label{table:manufsources}\tabularnewline
\toprule()
\begin{minipage}[b]{\linewidth}\centering
Source
\end{minipage} & \begin{minipage}[b]{\linewidth}\centering
Page in PDF
\end{minipage} & \begin{minipage}[b]{\linewidth}\centering
Page
\end{minipage} & \begin{minipage}[b]{\linewidth}\centering
Table
\end{minipage} & \begin{minipage}[b]{\linewidth}\centering
Years Covered
\end{minipage} \\
\midrule()
\endfirsthead
\toprule()
\begin{minipage}[b]{\linewidth}\centering
Source
\end{minipage} & \begin{minipage}[b]{\linewidth}\centering
Page in PDF
\end{minipage} & \begin{minipage}[b]{\linewidth}\centering
Page
\end{minipage} & \begin{minipage}[b]{\linewidth}\centering
Table
\end{minipage} & \begin{minipage}[b]{\linewidth}\centering
Years Covered
\end{minipage} \\
\midrule()
\endhead
\href{https://www2.census.gov/prod2/decennial/documents/00486469ch05.pdf}{1919
Census of Manufactures} & 10 & 293 & 193 & 1904-1919 \\
\href{https://www2.census.gov/prod2/statcomp/documents/1924-09.pdf}{1924
Statistical Abstract} & 32 & 754 & 692 & 1914-1923 \\
\href{https://www2.census.gov/prod2/statcomp/documents/1926-12.pdf}{1926
Statistical Abstract} & 30 & 774 & 748 & 1914, 1919, 1923, 1925 \\
\href{https://fraser.stlouisfed.org/files/docs/publications/stat_abstract/sa_1931.pdf}{1931
Statistical Abstract} & 860 & 842 & 815 & 1923-1929 \\
\bottomrule()
\end{longtable}

These sources often have overlapping years, which we use to check for
data quality (typos in the original data) as well as to alleviate
potential measurement errors. In particular, we identified and studied
four changes in the canvassing methodology in the years covered in our
analysis (1904-1927):

\begin{enumerate}
\def\labelenumi{\arabic{enumi}.}
\item
  The 1914 and 1919 census canvassed data for ``automobile repairing.''
  This category was excluded in all other censuses, so 1909-1914 growth
  rates will be biased upwards and 1919-1921 growth rates will be biased
  downwards. Both biases are likely to be small for two reasons. First,
  the automobile repairing industry accounted for only 0.18\% of wage
  earners in 1914 (0.61\% in 1919). Second, due to the nature of the
  industry, its output was distributed relatively uniformly across the
  country (see
  \href{https://books.google.com/books?id=LODrAAAAMAAJ\&pg=PA175}{Table
  49, page 175} of Volume 8 of the Fourteenth Census of the United
  States). Moreover, note that this change in classification methodology
  does not involve any bias in the 1914-1919 growth rates.
\item
  The 1904-1919 censuses collected data for all factories with a total
  annual output above \$500. This threshold was increased in 1921 to
  \$5,000, thus creating a downward bias in 1919-1921 growth rates.
  However, output for factories in the \$500-\$5,000 range were
  estimated by the Census Bureau (see general note of Table 685,
  \href{https://www2.census.gov/prod2/statcomp/documents/1924-09.pdf}{page
  723}, 1924 Statistical Abstract) to account for only 0.6\% of
  employment and 0.3\% of output, so potential biases for 1919-1921
  growth rates are likely to be small.
\item
  In contrast to other years, the 1925 census did not canvas data for
  the ``Coffee and spice, roasting and grinding'' industry. Thus,
  naively collecting the data would underestimate 1923-1925 growth rates
  and overestimate 1925-1927 growth rates in cities with a coffee
  roasting industry. To alleviate this potential issue, we exploit the
  fact that the 1926 Statistical Abstract also reported figures for 1923
  that excluded the coffee industry. Thus, we input the 1925
  manufacturing figures by first computing the 1923-1925 growth rate
  from the 1926 Statistical Abstract (which excludes coffee in both
  years) and then multiplying it with the 1923 figures from the 1931
  Statistical Abstract (where the coffee industry was included). Note
  that all our results are robust to not doing any adjustment, and that
  the differences are below 1\% of employment in all cities in our
  sample.
\item
  Occasionally, to preserve the anonymity of figures for specific
  establishments, the Bureau of the Census must report them in a given
  city even though they were ``located elsewhere in the State.'' For
  instance, for the 1921 census the cities of Bridgeport, Cincinnati,
  and Cleveland included one such establishment. More information on
  this practice is available e.g.~on
  \href{https://babel.hathitrust.org/cgi/pt?id=mdp.39015084923757\&view=1up\&seq=29}{section
  26 of chapter 1 (page 10)} of the 1925 Census of Manufacturers.
\end{enumerate}

To address these issues, as well as some changes in city boundaries
(discussed below), we build our manufacturing dataset using the
following steps:

\begin{enumerate}
\def\labelenumi{\arabic{enumi}.}
\item
  We use the 1919 Census of Manufacturers (CoM) as the primary source
  for the years 1904 and 1909, as well as to validate data for 1914 and
  1919.
\item
  We use the 1924 Statistical Abstract (SA) as the primary source for
  the years 1914, 1919, 1921, and to validate data for 1923.
\item
  We use the 1926 SA as the primary source for 1925 (by computing
  1923-1925 growth rates and scaling by the 1923 values).
\item
  We use the 1931 SA as the primary source for 1923 and 1927, and to
  obtain 1925 data corrected for canvassing changes in the coffee
  industry.
\end{enumerate}

\hypertarget{boundary-changes-for-specific-cities}{%
\subsubsection{Boundary changes for specific
cities}\label{boundary-changes-for-specific-cities}}

The Census Bureau computed city-level manufacturing statistics by adding
up establishments within the corporate limits of each city. Annexations
or consolidations have the potential to mechanically bias growth rates
upwards on the census years around an event. The potential magnitude of
these biases depends on the information available, and can be classified
in three types:

\begin{enumerate}
\def\labelenumi{\arabic{enumi}.}
\item
  Cities with statistics that after a boundary change were were
  retabulated by the Census Bureau itself, by ex-post combining the
  microdata of the cities that merged. No adjustments are needed if we
  use the retabulated reports. This is the case of the consolidation of
  Boston and Hyde Park in Massachusetts.
\item
  Cities lacking retabulated statistics, but where enough information is
  available to create combine the information for both cities. This is
  the case of the Omaha annexation of South Omaha in Nebraska, where
  pre-merger statistics of both cities can be added up to create a
  baseline comparable to the post-merger values.
\item
  Cities lacking retabulated statistics, where not enough information is
  available to compute pre-merger values. Here, we can however provide
  upper bounds on how much can the growth rates be affected by the
  annexation. In every notable case, these bounds are fairly small and
  close to the unadjusted values.
\end{enumerate}

Below, we explore all major boundary changes in order to understand how
they might affect our results. We also describe how we adjust the raw
values to account for boundary changes in cases where this is feasible.
We pay particular attention to boundary changes that occurred between
the 1914 and 1919 census years.\footnote{In a discussion of our paper,
  \citet{llr2} state that ``In addition to Omaha, the other cities with
  NPI data and large incorporations between 1914 and 1919---Los Angeles,
  Portland, Richmond, and Toledo---have identical 1909 manufacturing
  data in the 1910 and 1920 Census''. Thus, we place particular
  attention to these cities.}

\hypertarget{boston-ma}{%
\paragraph*{Boston, MA}\label{boston-ma}}
\addcontentsline{toc}{paragraph}{Boston, MA}

As stated in
\href{https://books.google.com/books?id=UBAeCRvtXTkC\&newbks=1\&newbks_redir=0\&dq=Census\%20of\%20manufactures\%201914\%20massachusetts\&pg=PA595}{Volume
I, Page 595, Table 7} of the 1914 Census of Manufactures, information
for 1904 and 1909 already ``includes Hyde Park, consolidated with Boston
Jan.~1, 1912.''

\hypertarget{pittsburgh-pa}{%
\paragraph*{Pittsburgh, PA}\label{pittsburgh-pa}}
\addcontentsline{toc}{paragraph}{Pittsburgh, PA}

As stated in
\href{https://books.google.com/books?id=UBAeCRvtXTkC\&newbks=1\&newbks_redir=0\&dq=Census\%20of\%20manufactures\%201914\%20massachusetts\&pg=PA1280}{Volume
I, Page 1280, Table 7} of the 1914 Census of Manufactures, information
for 1904 ``includes statistics for Allegheny, annexed in 1907.''

\hypertarget{omaha-ne}{%
\paragraph*{Omaha, NE}\label{omaha-ne}}
\addcontentsline{toc}{paragraph}{Omaha, NE}

As stated in Table 193, Page 312 of the 1919 Census of Manufactures,
South Omaha was ``annexed to Omaha in 1915''. Because this document
reports totals for South Omaha for 1904, 1909, and 1914, it is
straightforward to combine the totals of both cities before the
annexation.

\hypertarget{richmond-va}{%
\paragraph*{Richmond, VA}\label{richmond-va}}
\addcontentsline{toc}{paragraph}{Richmond, VA}

On November 5, 1914, Richmond's corporate limits were extended to
include the towns of Barton Heights, Fairmount, and Highland Park. As
shown in \cref{fig:richmond}, this annexation included
industries in the Chamberlayne industrial district, so this case is worth 
a more detailed study.\footnote{\Cref{fig:richmond} is compiled from
  overlaying the 1910 Census Volume 8 Part 2 Page 117 with a map from a
  1923
  \href{http://www.virginiamemory.com/online-exhibitions/items/show/14}{report}
  from Richmond's Department of Public Works.}

{
\setstretch{1.0}
  \begin{figure}[htpb]
    \centering
    \includegraphics[width=0.8\textwidth]{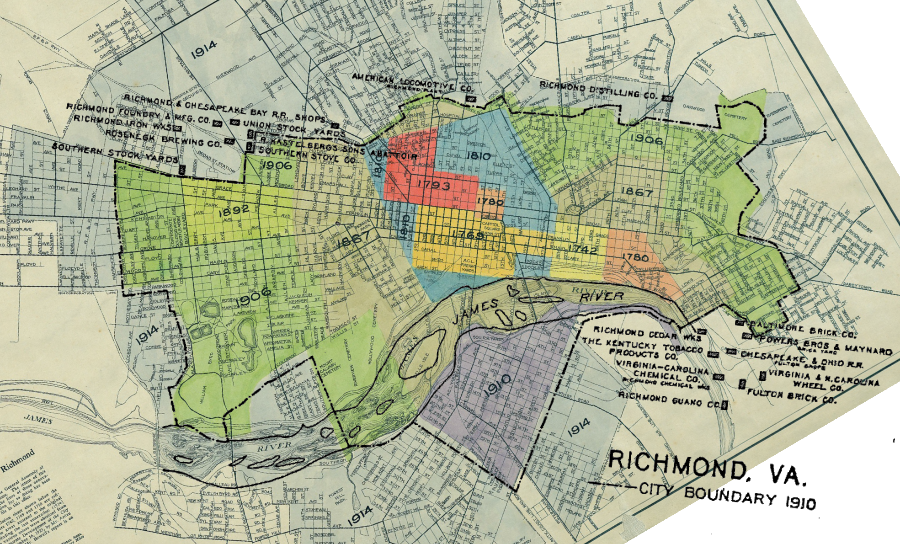}
    \caption{\textbf{Territorial expansion of the City of Richmond. }Map of the manufacturing plants on the outskirts of Richmond, VA as of
the 1909 Census, overlaid against a 1923 map of the territorial
expansion of the City of Richmond (source: Department of Public Works,
City of Richmond). While the southern cluster of plants was not annexed
into the city, the northern cluster was annexed in 1914.}
    \label{fig:richmond}
  \end{figure}
}

For this, the 1914 Census of Manufacturers is particularly useful, as it
discusses the issue in depth (see \cref{fig:richmond2}, extracted from
page 14 of the
\href{https://books.google.com/books?id=LVLOAAAAMAAJ}{Virginia Volume}).
In particular, Table 14 on page 13 reports results not only for the city
pre-annexation but also for the annexed territory and for the enlarged
city. These results show that the annexations were not as substantial,
as manufacturing employment in the annexed territory corresponds to only
5.8\% of the employment in the boundaries pre-annexation. Results are
similar for manufacturing output (5.6\%) and value added (4.5\%).

Nonetheless, to avoid any bias in the 1914-1919 growth figures, we use
the numbers of the expanded city for 1914, and chain values for 1904 and
1909 accordingly (assuming that the expanded city maintains the same
growth rate as the pre-annexation city did). Our results are almost
indistinguishable from those obtained without the adjustment.

{
\setstretch{1.0}
  \begin{figure}[htpb]
    \centering
    \includegraphics[width=0.7\textwidth]{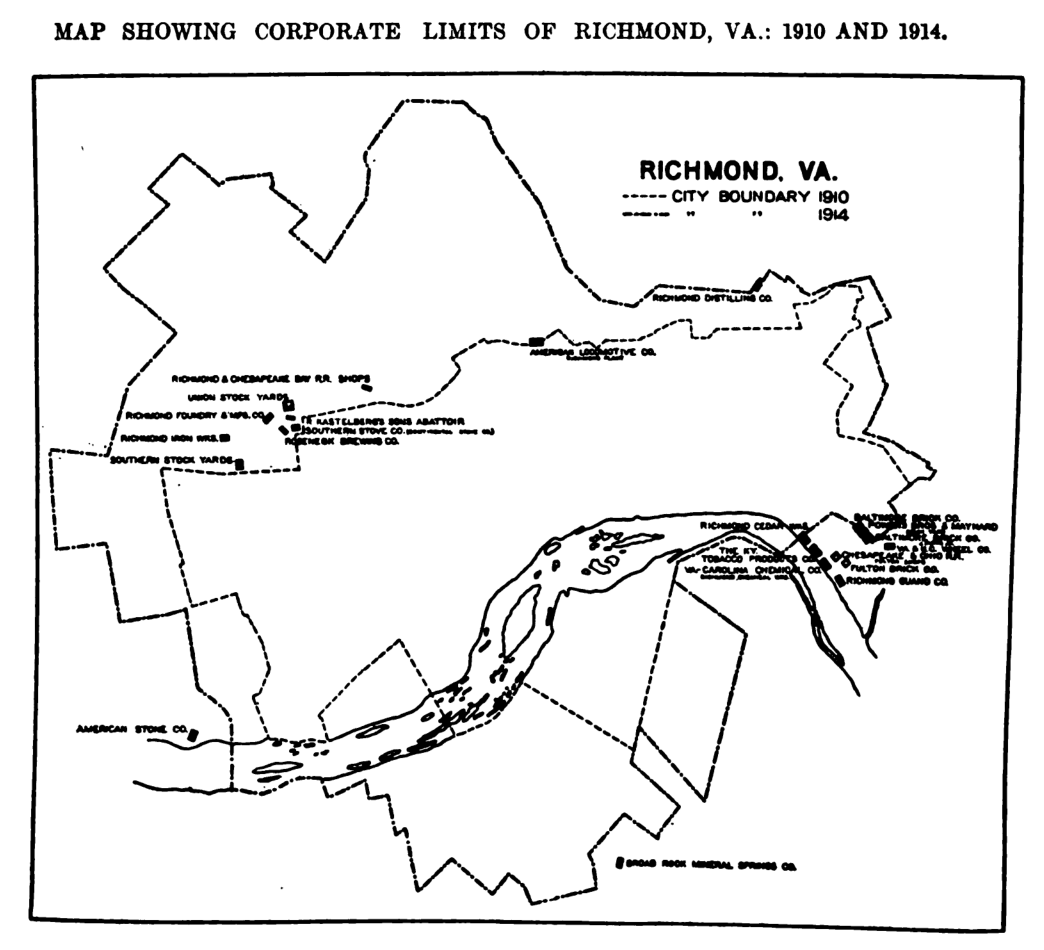}
    \caption{\textbf{Corporate Limits of Richmond, VA in 1910 and 1914}}
    \label{fig:richmond2}
  \end{figure}
}

\hypertarget{los-angeles-ca}{%
\paragraph*{Los Angeles, CA}\label{los-angeles-ca}}
\addcontentsline{toc}{paragraph}{Los Angeles, CA}

The city of Los Angeles has expanded dramatically
\href{https://la.curbed.com/2016/7/5/12097070/los-angeles-growth-map}{through
its history}. Through the 1914-1919 period, most of the drive for
annexation was due to the 1913 opening of the Los Angeles Aqueduct, as
the City of Los Angeles had a surplus of water but was contractually not
allowed to resell it. Particularly important is the annexation of the
San Fernando Valley, which doubled the area of the city. It is still not
clear whether these annexations substantially increased the
manufacturing activity of the city, as no annexations of incorporated
cities happened in these years\footnote{The incorporated cities annexed
  to Los Angeles include Wilmington (1909), San Pedro (1909), Hollywood
  (1910), Sawtelle (1922), Hyde Park (1923), Eagle Rock (1923), Venice
  (1925), Watts (1926), Barnes City (1927), and Tujunga (1932).}
Instead, the annexations between 1914 and 1919 comprised the mostly
rural areas of San Fernando Valley (1915), Palms (1915), Bairdstown
(1915), Westgate (1916), West Coast (1917), Griffith Ranch (1918), and
Hansen Heights (1918).

{
\setstretch{1.0}
  \begin{figure}[htpb]
    \centering
    \includegraphics[width=0.7\textwidth]{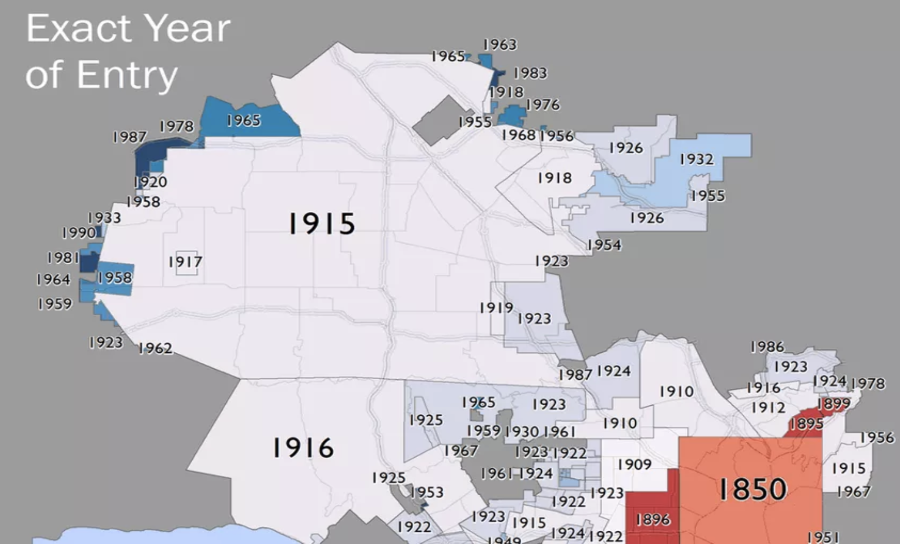}
    \caption{\textbf{Growth of the City of Los Angeles}}
    \label{fig:la1}
  \end{figure}
}

First, note that the San Fernando Valley was only sparsely populated,
most activity was agricultural, and the annexation excluded the areas
with highest population within the valley, such as the City of San
Fernando and Rancho el Escorpion. Thus, as noted by
\citet{jorgensen1982san}:

\begin{quote}
In the Valley, the tabulations showed 681 voting for annexation to Los
Angeles and 25 against.
\end{quote}

In contrast, as of 1910 the city of Los Angeles had a population of
102,479.

Moreover, as an exercise, we can take advantage of a feature of the 1914
Census to compute upper bounds for how much of Los Angeles'
manufacturing could be accounted by the annexed regions. In particular,
in 1914 (but not in 1919) the Census Bureau provided statistics for the
``Los Angeles Metropolitan District'', a region almost four times larger
than the City of Los Angeles itself, encompassing also other neighboring
cities such as Long Beach City, Pasadena, Santa Monica, and Sawtelle.

{
\setstretch{1.0}
  \begin{figure}[htpb]
    \centering
    \includegraphics[width=0.7\textwidth]{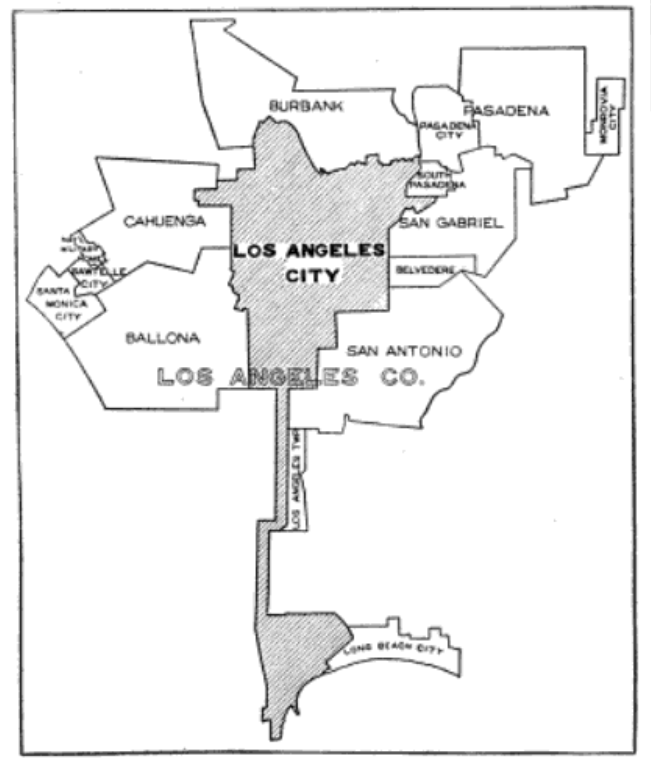}
    \caption{\textbf{Los Angeles Metropolitan District in 1914}}
    \label{fig:la2}
  \end{figure}
}

Using the information for the Metropolitan District, we can do the
following thought exercise: using \cref{fig:la3}, compute the totals for
the Metropolitan District excluding the two other cities listed
separately (Long Beach and Pasadena, which are still not part of Los
Angeles). Assume that in 1914 the city of Los Angeles annexed \emph{the
entirety} of the Metropolitan District bar these two cities (which we
know is false, as e.g.~Santa Monica is still independent). Then, this
simulated annexation would increase manufacturing employment by only
8.8\% (2096/23744). In contrast, growth of manufacturing employment in
Los Angeles was 198\% between 1914 and 1919. Clearly, even at an upper
bound, the annexations are not what drove manufacturing growth in Los
Angeles.

{
\setstretch{1.0}
  \begin{figure}[htpb]
    \centering
    \includegraphics[width=1.0\textwidth]{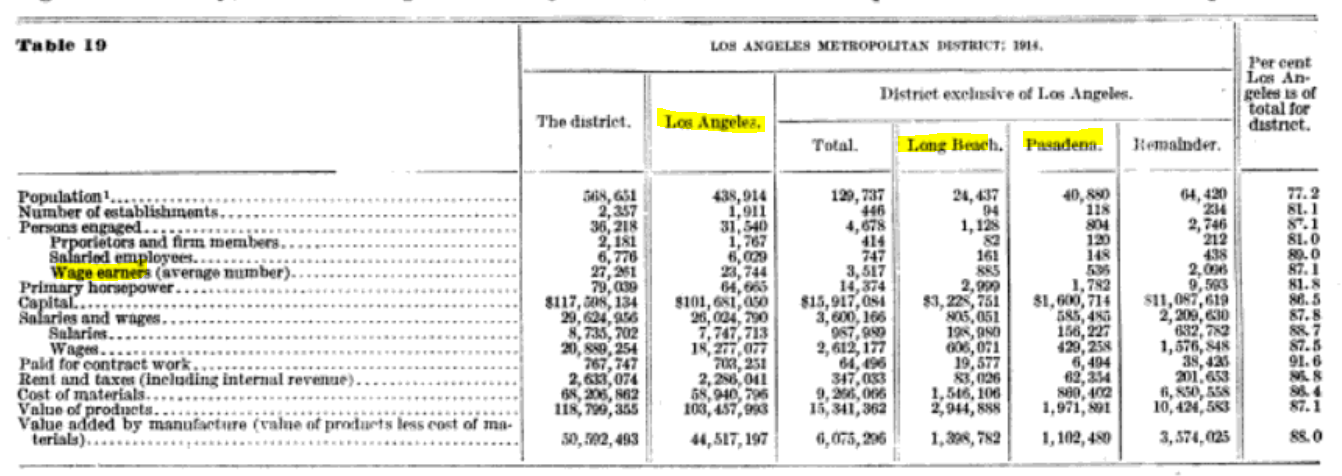}
    \caption{\textbf{Excerpt from the 1914 Manufacturing Census corresponding to ``Los
Angeles Metropolitan District''}}
    \label{fig:la3}
  \end{figure}
}

\hypertarget{toledo-oh}{%
\paragraph*{Toledo, OH}\label{toledo-oh}}
\addcontentsline{toc}{paragraph}{Toledo, OH}

The 1920 census states that ``parts of Adams and Washington townships
{[}were{]} annexed to Toledo city since 1910''
(\href{https://books.google.com/books?id=oei2AAAAIAAJ\&pg=PA565}{Table
53, Page 565, Volume 1, 1920 Census of Population}). We are not aware of
any manufacturing data for these townships, but we can do a similar
exercise to compute upper bounds using population numbers.

In particular, we can assume that not just \emph{parts} but \emph{all}
of these townships were annexed into Toledo, and that both annexations
occurred between 1914 and 1919. Then, as seen in \cref{fig:toledo}, the
annexations would have accounted for at most 7,433 inhabitants.
Considering that as of 1910 the population of Toledo was 168,497, then
the annexations accounted for at most 4\% of Toledo's 1910 population.
This is an order of magnitude lower than Toledo's 1910-1920 reported
population growth, of 44\%.

{
\setstretch{1.0}
  \begin{figure}[ht]
    \centering
    \includegraphics[width=0.8\textwidth]{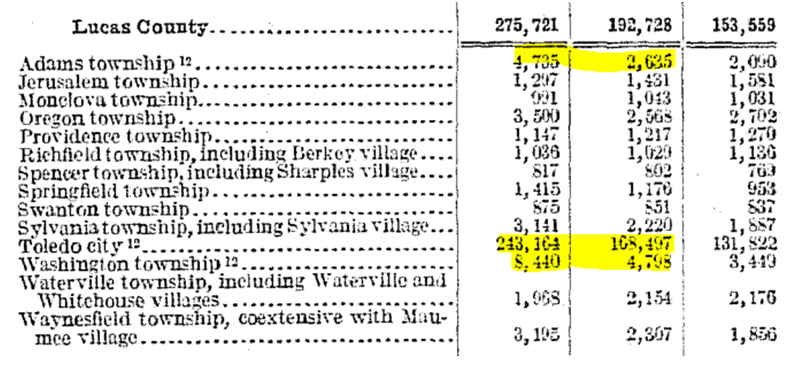}
    \caption{\textbf{Excerpt from the 1920 Decennial Census for Toledo, OH}}
    \label{fig:toledo}
  \end{figure}
}

\hypertarget{portland-or}{%
\paragraph*{Portland, OR}\label{portland-or}}
\addcontentsline{toc}{paragraph}{Portland, OR}

For Portland, we can do a similar exercise as with Toledo. As
\cref{fig:portland} states, between 1910 and 1920, Portland annexed the
town of Linnton as well as St.~Johns city. Linnton was only incorporated
in 1910, so no population figures are available from any census.
However, as of 2000 the neighborhood of Linnton had a population of 541
inhabitants. In the case of St.~Johns, its 1910 population was 4872, two
orders of magnitude lower than Portland's 207,214.

{
\setstretch{1.0}
  \begin{figure}[ht]
    \centering
    \includegraphics[width=0.7\textwidth]{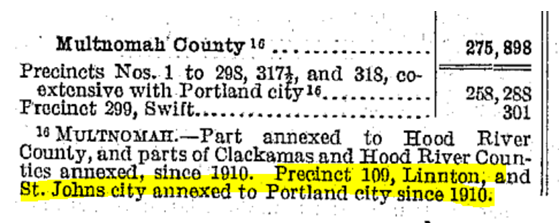}
    \caption{\textbf{Excerpt from the 1920 Decennial Census for Portland, OR}}
    \label{fig:portland}
  \end{figure}
}

\hypertarget{appendix:data-military}{%
\subsection{Military camps}\label{appendix:data-military}}

As part of the robustness checks, we use distance from a city to the closest military camp
as an additional control variable. We use four sources to collect
and validate this data on WWI military camps:

\begin{enumerate}
\def\labelenumi{\arabic{enumi}.}
\item
  \citet{OrderOfBattle},
  \href{https://history.army.mil/html/books/023/23-4/CMH_Pub_23-4.pdf}{part
  2}: this source contains the lists of all Army and National Guard
  training camps, embarkment camps, and Army Forts. Further, it lists
  the strength of each camp, defined as the average number of troops
  located at each camp in a given month (or that transited through a
  given camp, in the case of embarkment camps).
\item
  \citet{setzekorn2017}: this source lists major training camps and
  cantonments on pages 28-29, which we use to validate our first source.
\item
  http://www.fortwiki.com/World\_War\_I: we use this online resource to
  geolocate all camps, as well as to help establish the founding date of
  each camp, in cases where the founding date is missing in the sources
  above.
\item
  \citet{Hilt2020}: we use this source to validate our list of camps.
  This is particularly useful as camps and forts were often located in
  multiple locations (or in the case of embarkment camps, even in
  multiple states).
\end{enumerate}

To compute camp strength, we use the ``Aggregate'' column from the
Average Strength tables available for each camp (see \cref{fig:pike} for
an example). In some cases the information is reported at the fort and
not camp level, in which case we utilize a fort as the unit of analysis.
For instance, Camp Funston was part of Fort Riley. Moreover, we exclude
some specialized training camps with no information on troop counts,
such as Fort Harrison (officer training post), Camp Robison (artillery
training camp), Camp Colt (tank corps training camp), and Camp Crane
(ambulance corps training camp). This leaves us with 40 military camps
in total, composed of 19 Army training camps, 16 National Guard training
camps, and 5 embarkation camps.

{
\setstretch{1.0}
  \begin{figure}[htpb]
    \centering
    \includegraphics[width=0.6\textwidth]{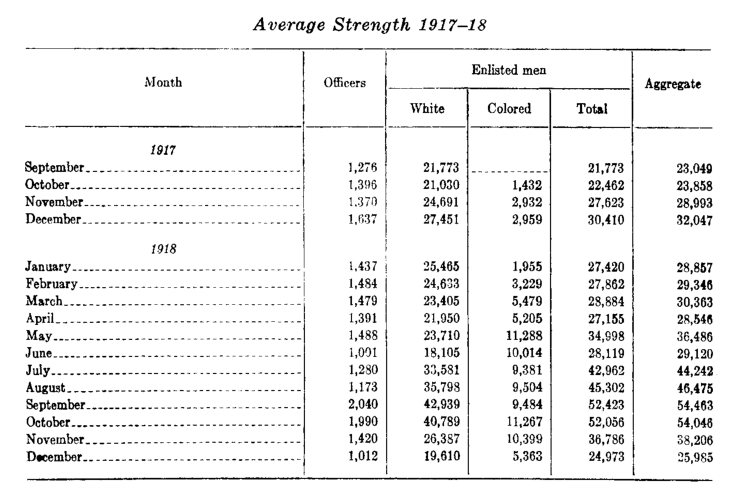}
    \caption{\textbf{Camp Pike - Average strength 1917-18. }This table shows the average strength, in terms of quantity of officers
and enlisted men, for Camp Pike, between September 1917 and December
1918.}
    \label{fig:pike}
  \end{figure}
}

\hypertarget{appendix:city-mortality-npi}{%
\subsection{City-level mortality and NPIs}\label{appendix:city-mortality-npi}}

\begin{figure}[ht]\centering
    \subfloat[Albany]{\includegraphics[width=0.9\textwidth]{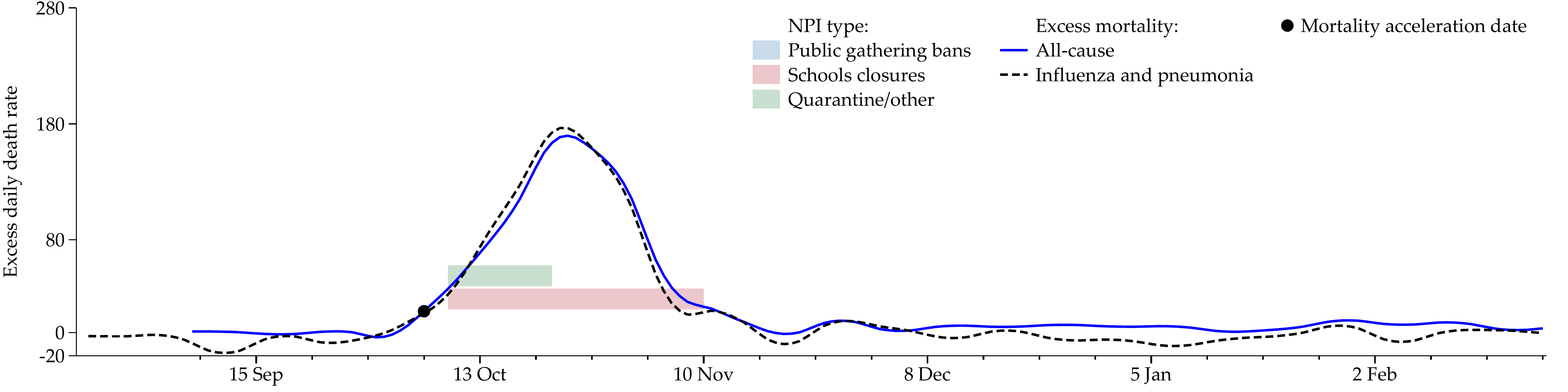}}\hfill
    \subfloat[Atlanta]{\includegraphics[width=0.9\textwidth]{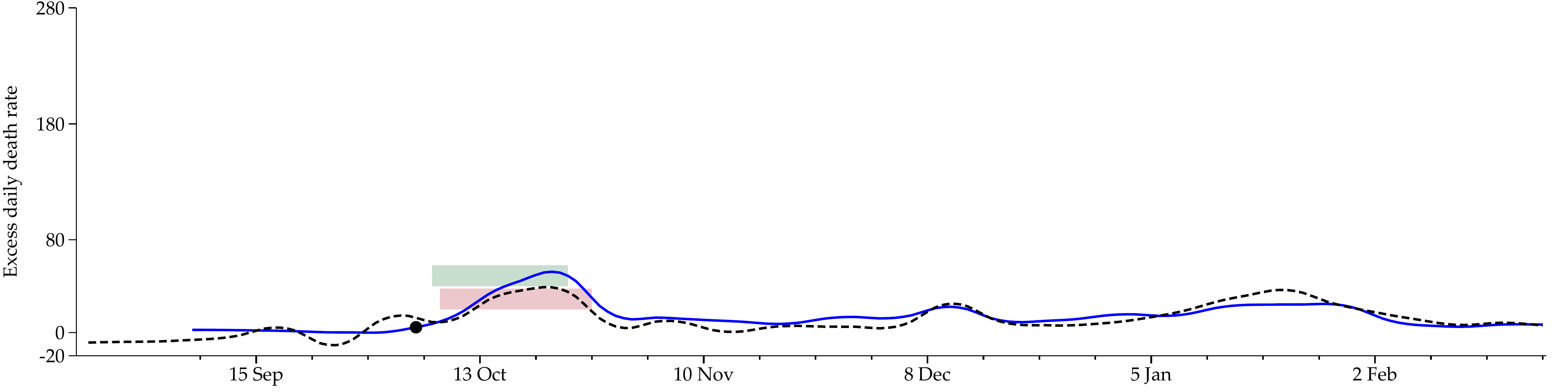}}\hfill
    \subfloat[Baltimore]{\includegraphics[width=0.9\textwidth]{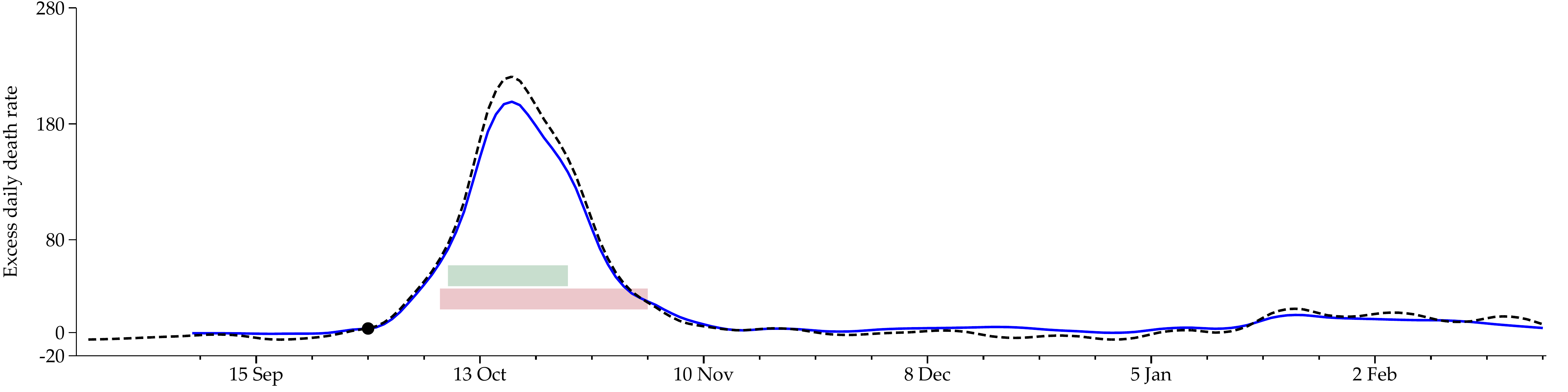}}\hfill
    \subfloat[Birmingham]{\includegraphics[width=0.9\textwidth]{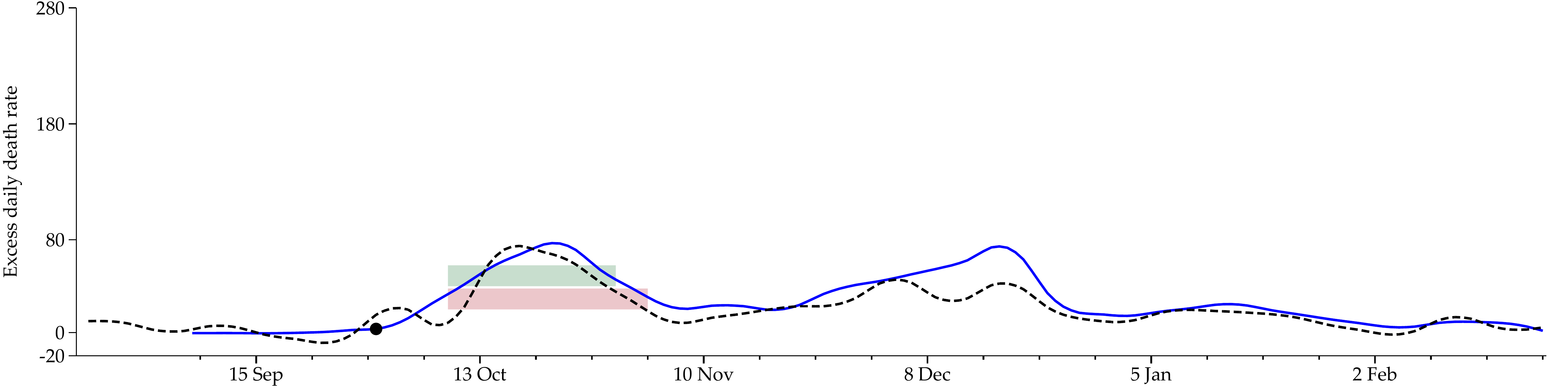}}\hfill
\CitiesCaption\label{fig:city-mortality-npi}
\end{figure}

\begin{figure}[ht]\ContinuedFloat\centering
    \subfloat[Boston]{\includegraphics[width=0.9\textwidth]{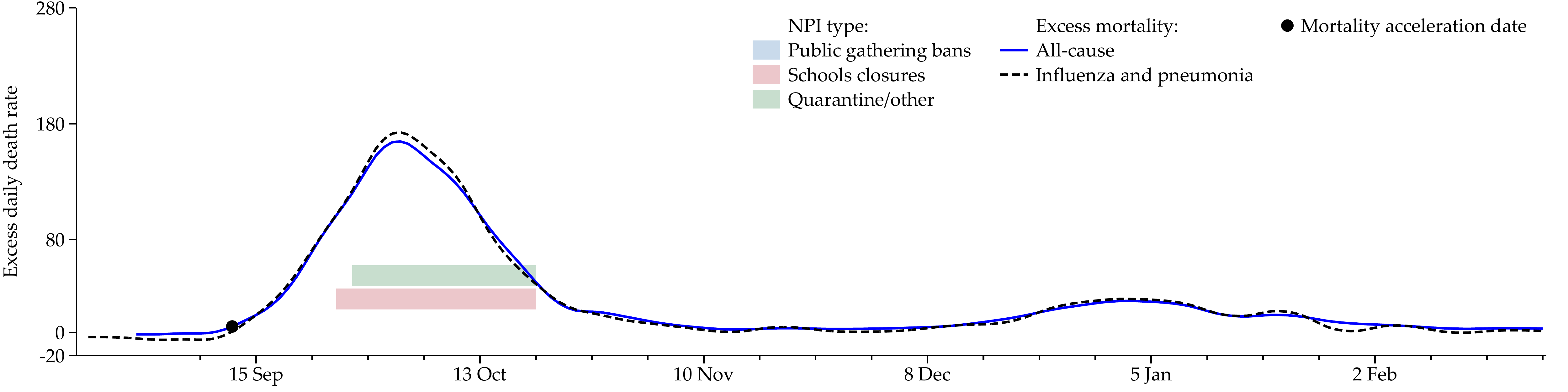}}\hfill
    \subfloat[Buffalo]{\includegraphics[width=0.9\textwidth]{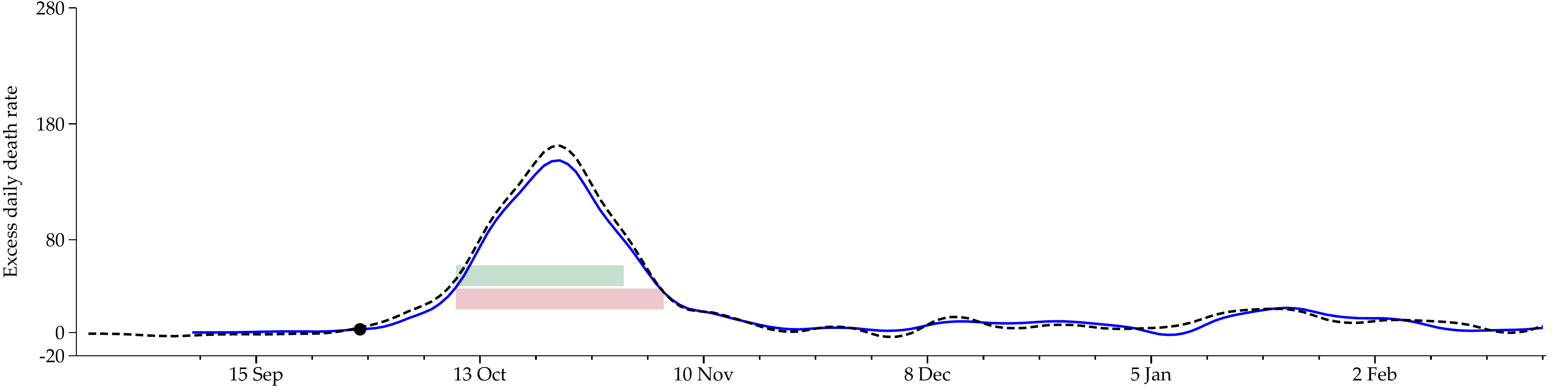}}\hfill
    \subfloat[Cambridge]{\includegraphics[width=0.9\textwidth]{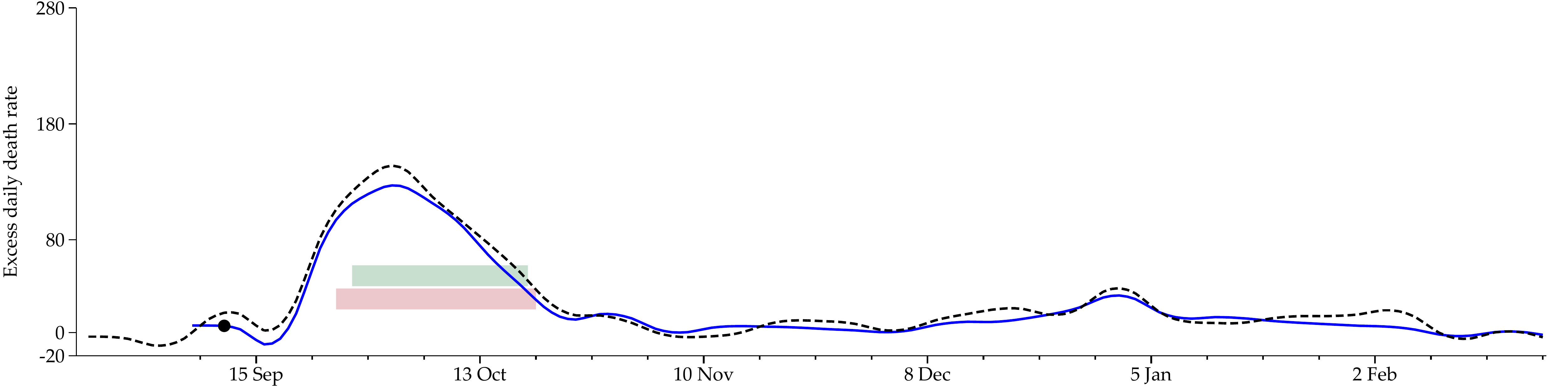}}\hfill
    \subfloat[Chicago]{\includegraphics[width=0.9\textwidth]{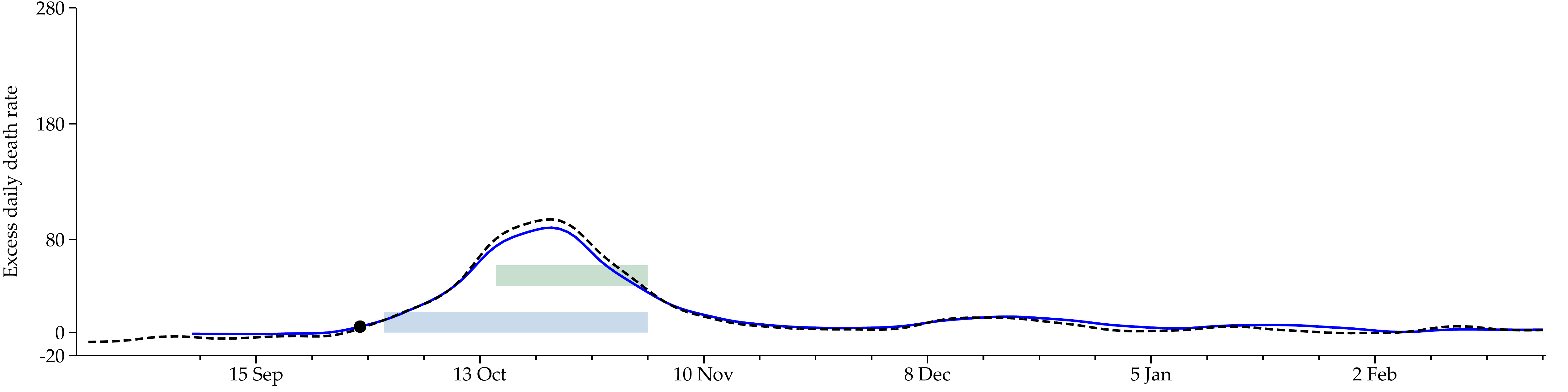}}\hfill
\CitiesCaption
\end{figure}

\begin{figure}[ht]\ContinuedFloat\centering
    \subfloat[Cincinnati]{\includegraphics[width=0.9\textwidth]{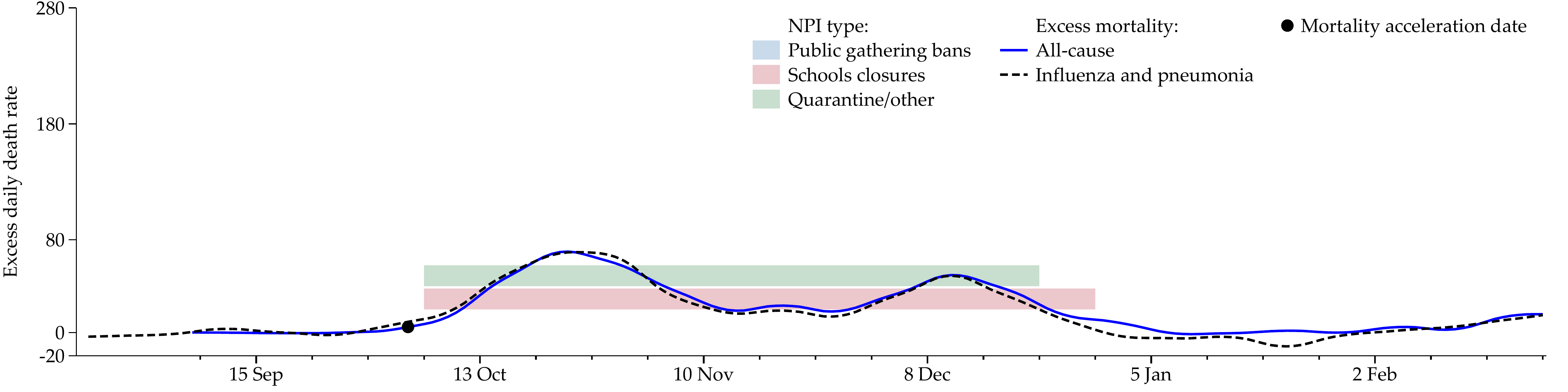}}\hfill
    \subfloat[Cleveland]{\includegraphics[width=0.9\textwidth]{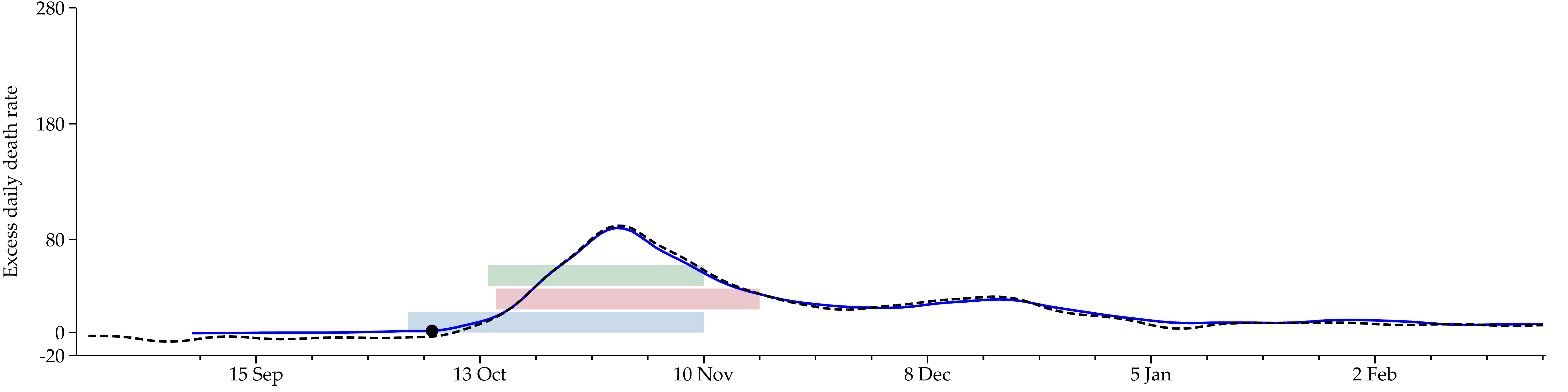}}\hfill
    \subfloat[Columbus]{\includegraphics[width=0.9\textwidth]{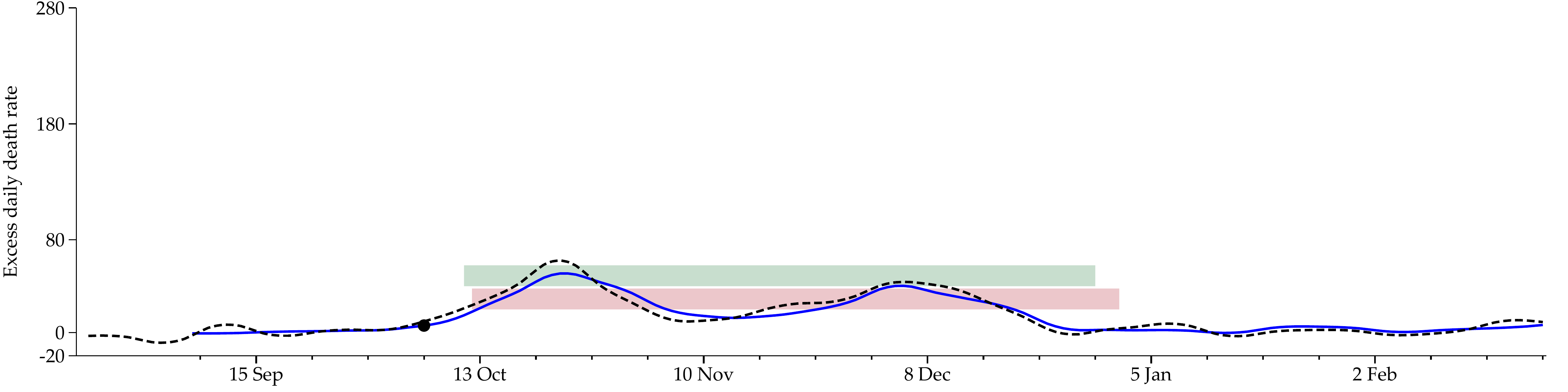}}\hfill
    \subfloat[Dayton]{\includegraphics[width=0.9\textwidth]{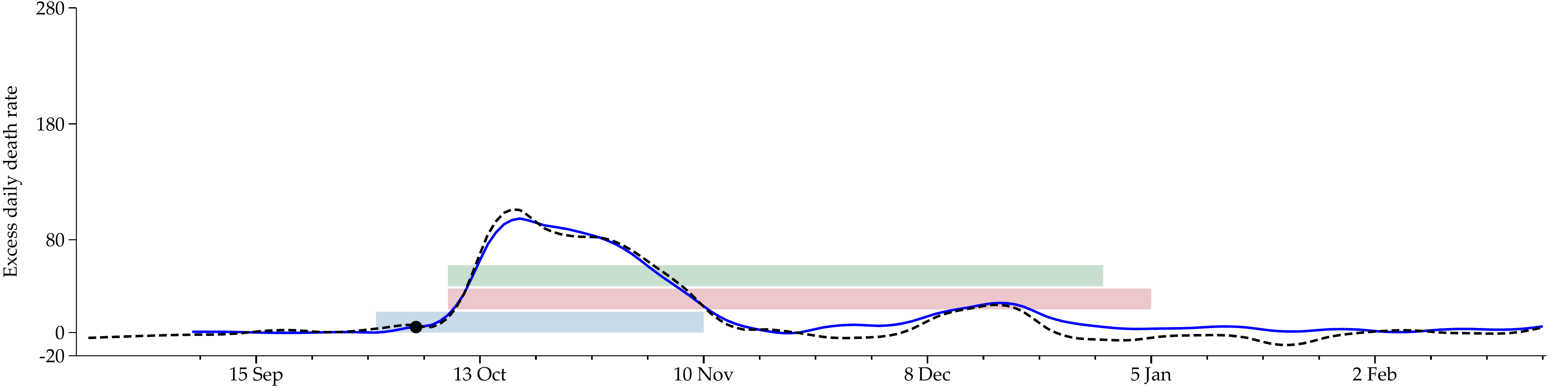}}\hfill
\CitiesCaption
\end{figure}

\begin{figure}[ht]\ContinuedFloat\centering
    \subfloat[Denver]{\includegraphics[width=0.9\textwidth]{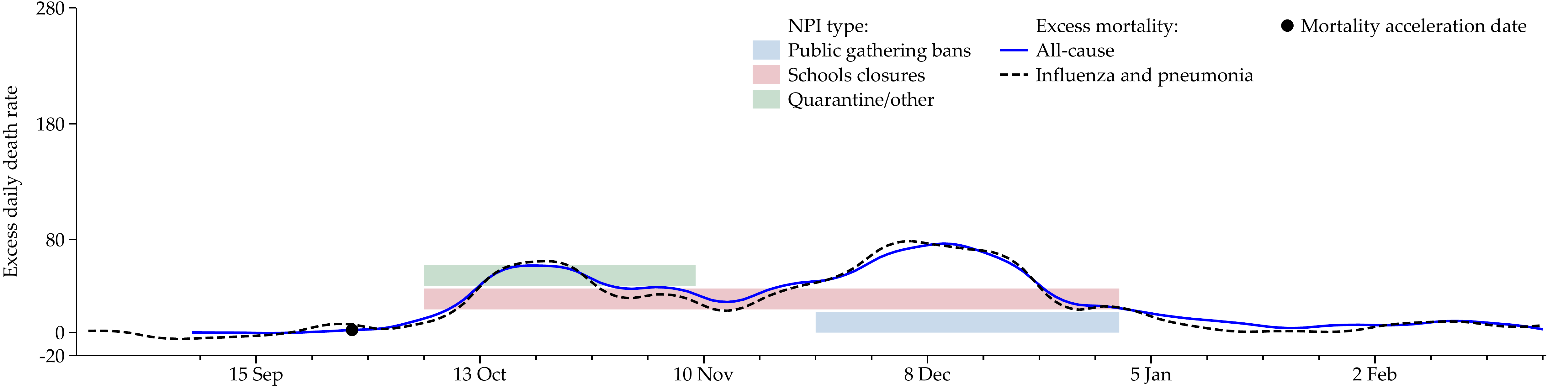}}\hfill
    \subfloat[Fall River]{\includegraphics[width=0.9\textwidth]{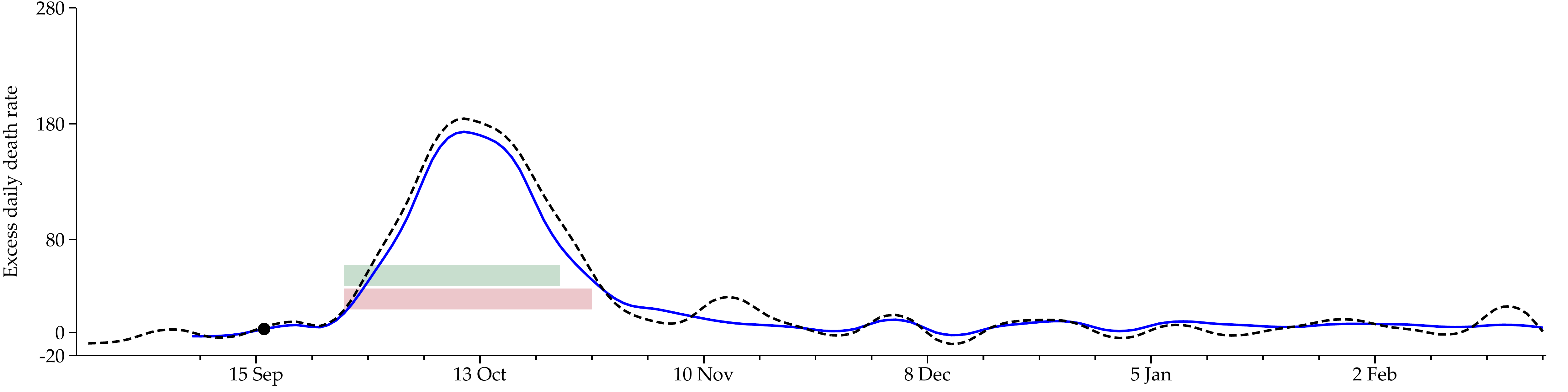}}\hfill
    \subfloat[Grand Rapids]{\includegraphics[width=0.9\textwidth]{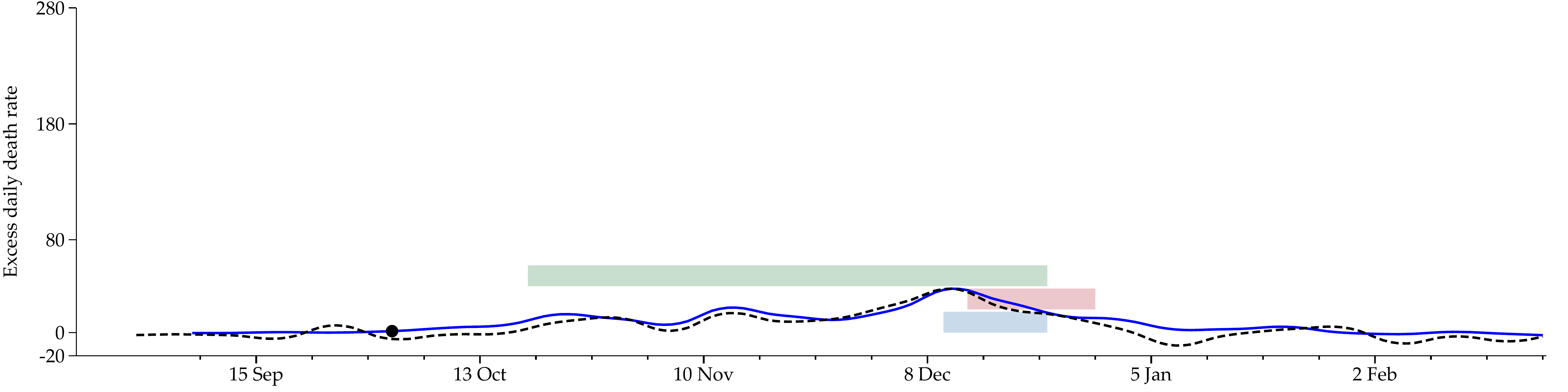}}\hfill
    \subfloat[Indianapolis]{\includegraphics[width=0.9\textwidth]{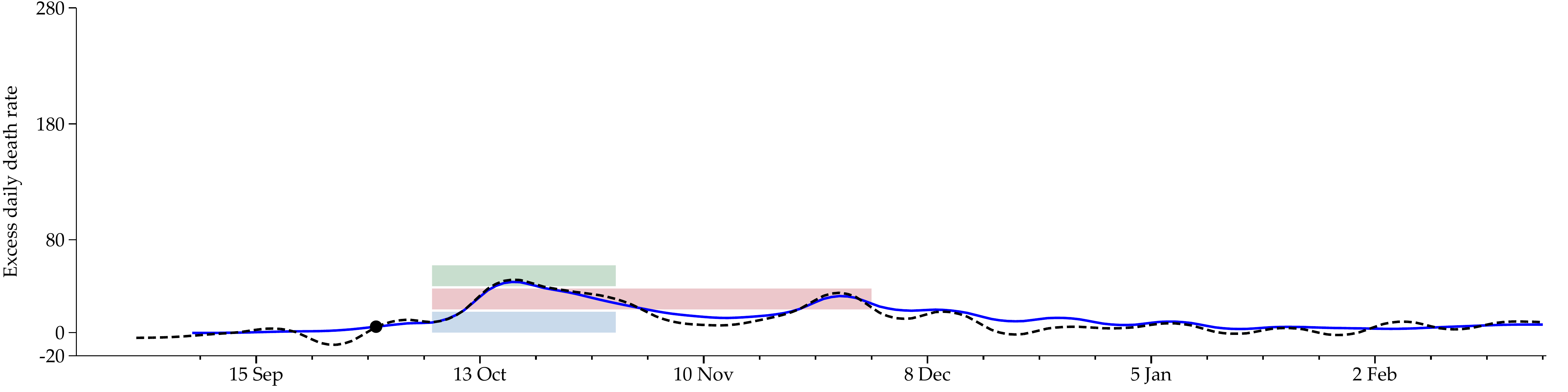}}\hfill
\CitiesCaption
\end{figure}

\begin{figure}[ht]\ContinuedFloat\centering
    \subfloat[Jersey City]{\includegraphics[width=0.9\textwidth]{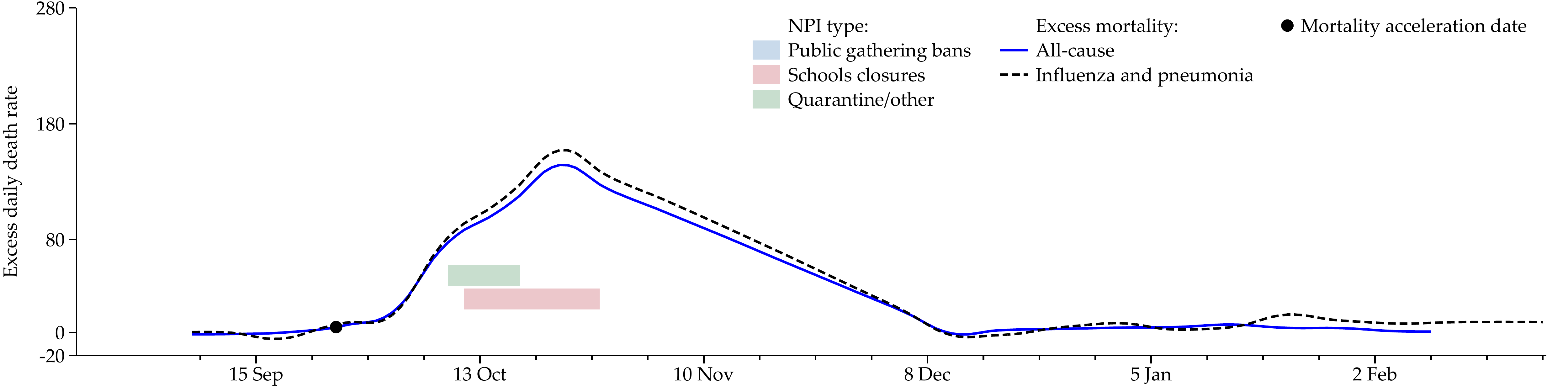}}\hfill
    \subfloat[Kansas City]{\includegraphics[width=0.9\textwidth]{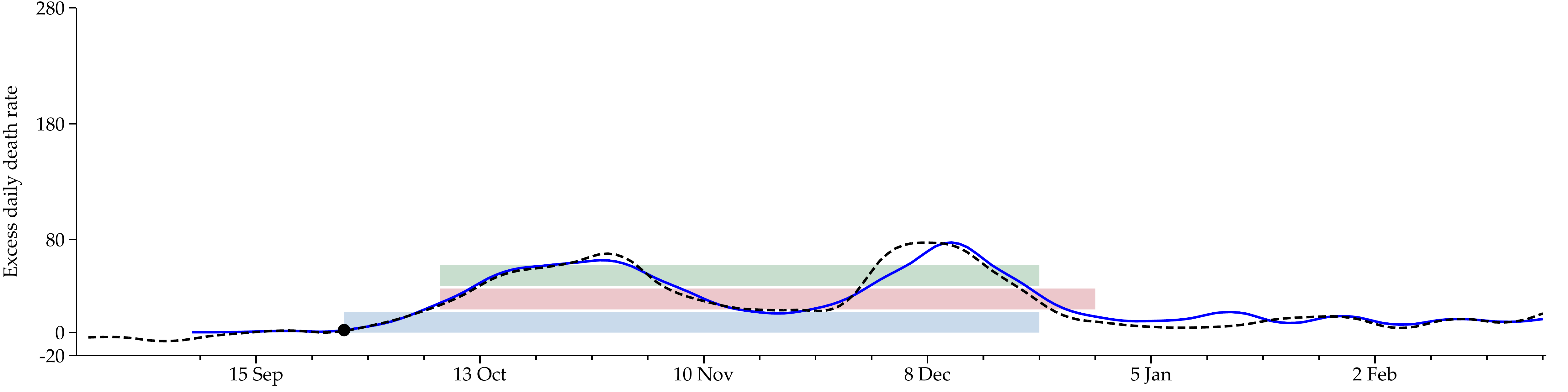}}\hfill
    \subfloat[Los Angeles]{\includegraphics[width=0.9\textwidth]{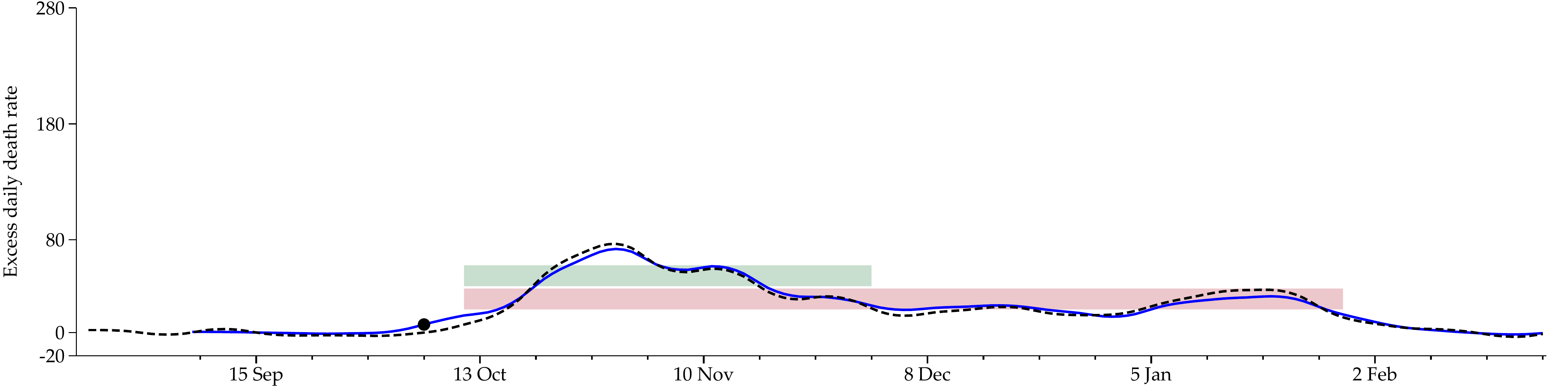}}\hfill
    \subfloat[Louisville]{\includegraphics[width=0.9\textwidth]{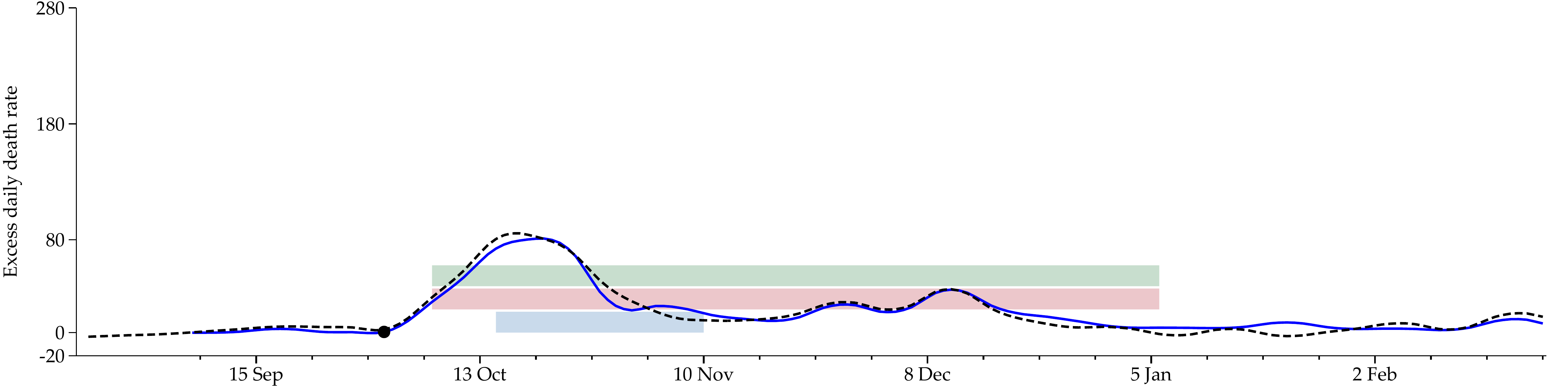}}\hfill
\CitiesCaption
\end{figure}

\begin{figure}[ht]\ContinuedFloat\centering
    \subfloat[Lowell]{\includegraphics[width=0.9\textwidth]{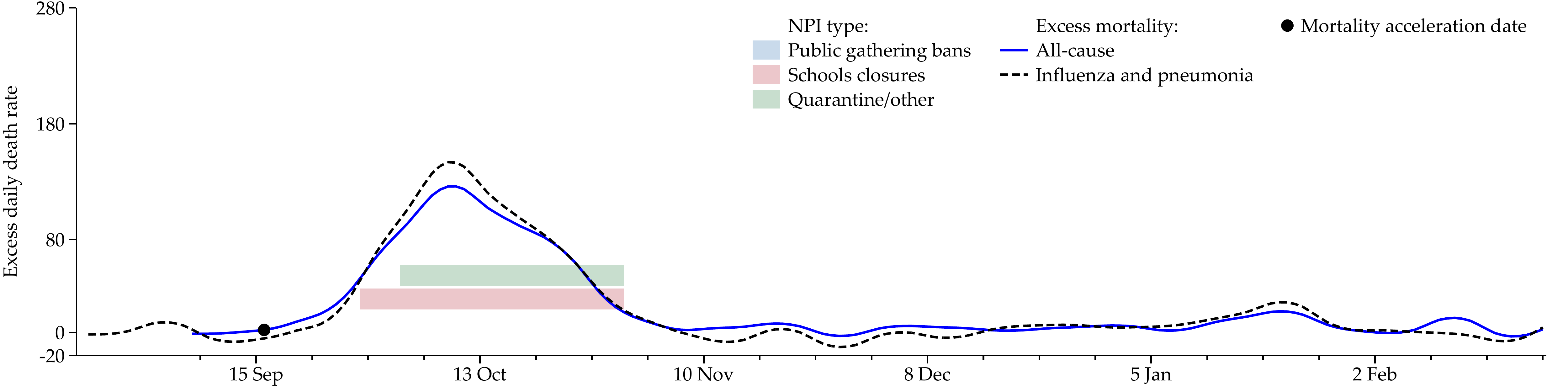}}\hfill
    \subfloat[Memphis]{\includegraphics[width=0.9\textwidth]{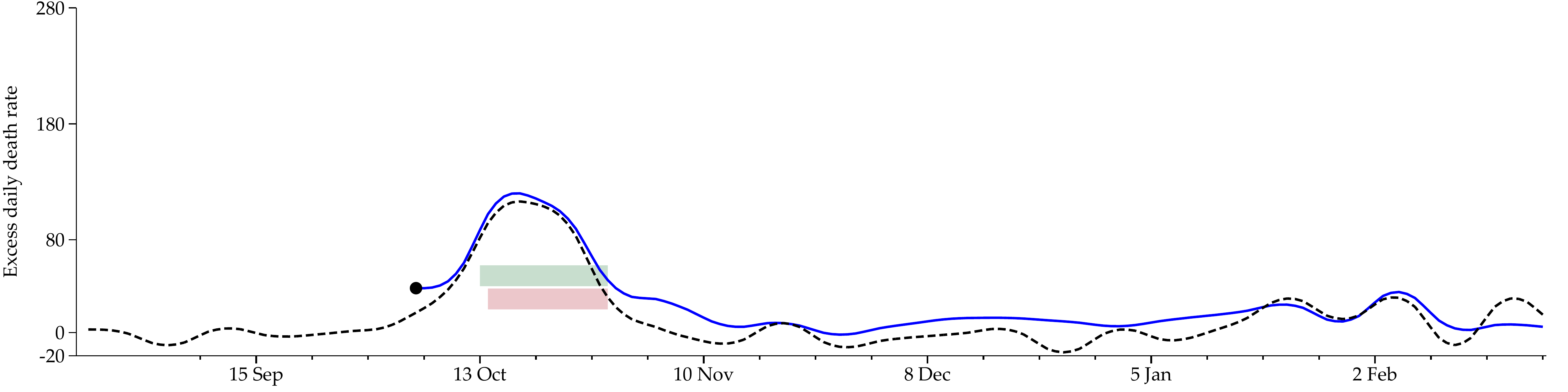}}\hfill
    \subfloat[Milwaukee]{\includegraphics[width=0.9\textwidth]{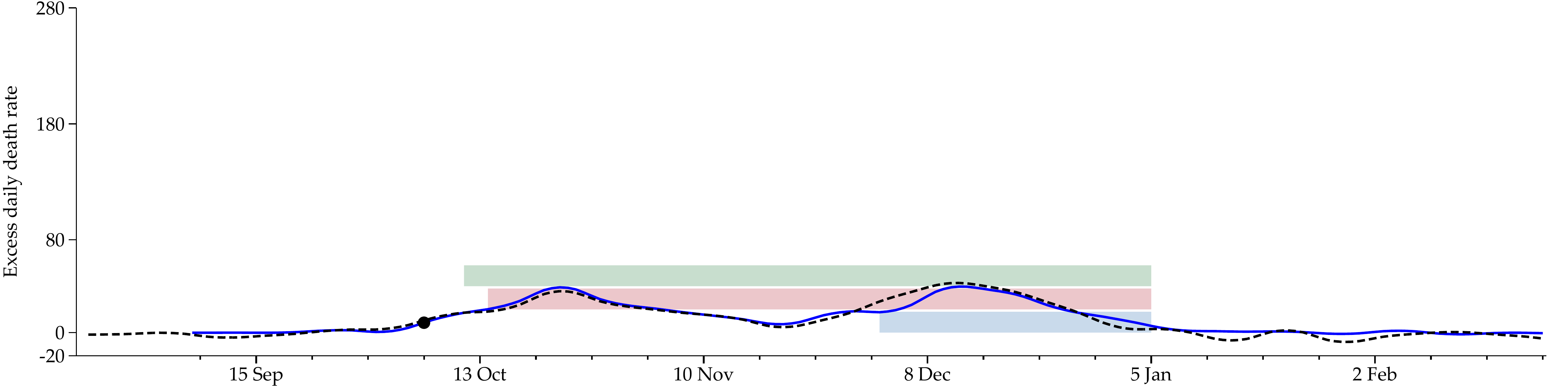}}\hfill
    \subfloat[Minneapolis]{\includegraphics[width=0.9\textwidth]{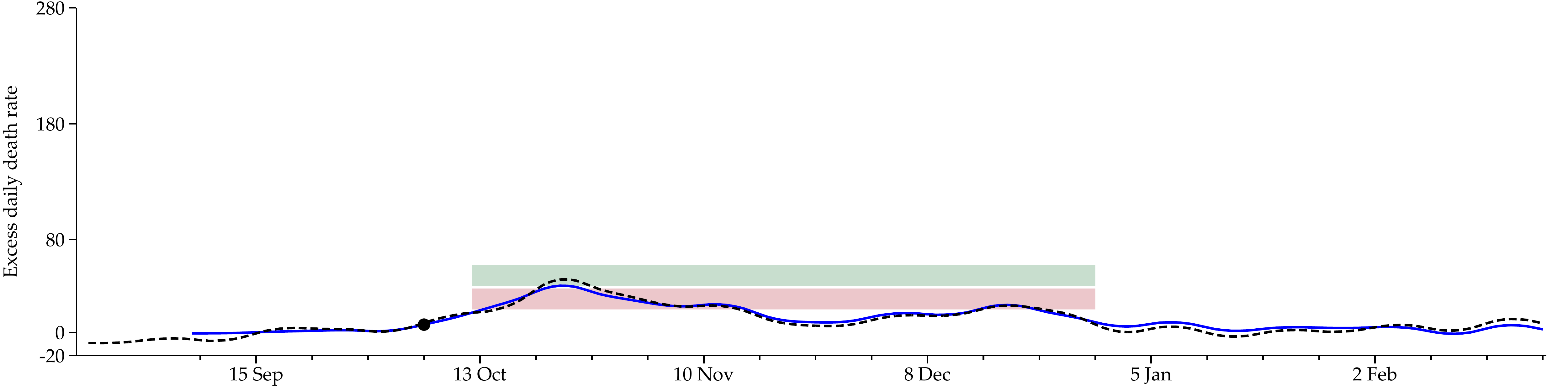}}\hfill
\CitiesCaption
\end{figure}

\begin{figure}[ht]\ContinuedFloat\centering
    \subfloat[Nashville]{\includegraphics[width=0.9\textwidth]{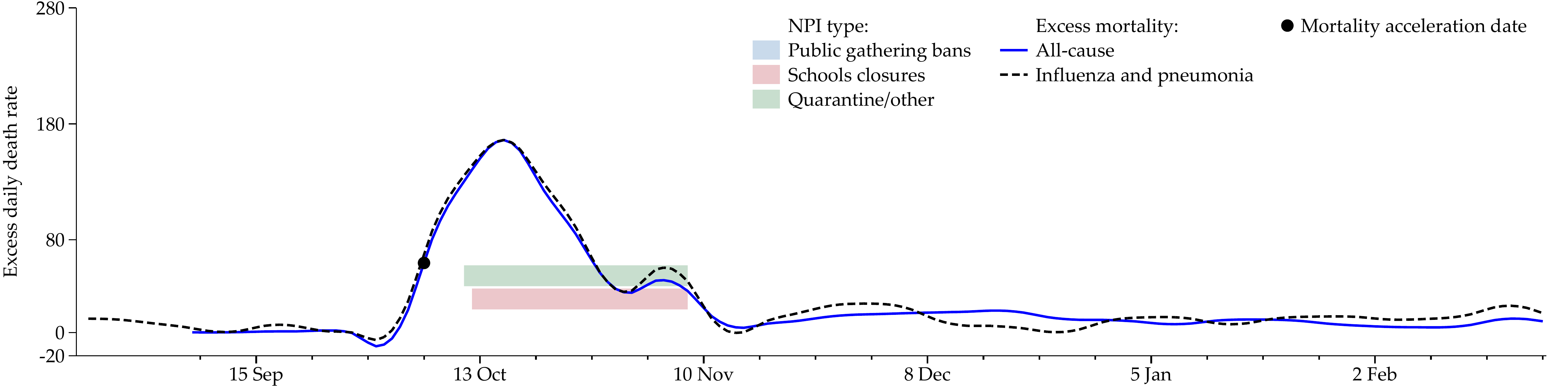}}\hfill
    \subfloat[New Haven]{\includegraphics[width=0.9\textwidth]{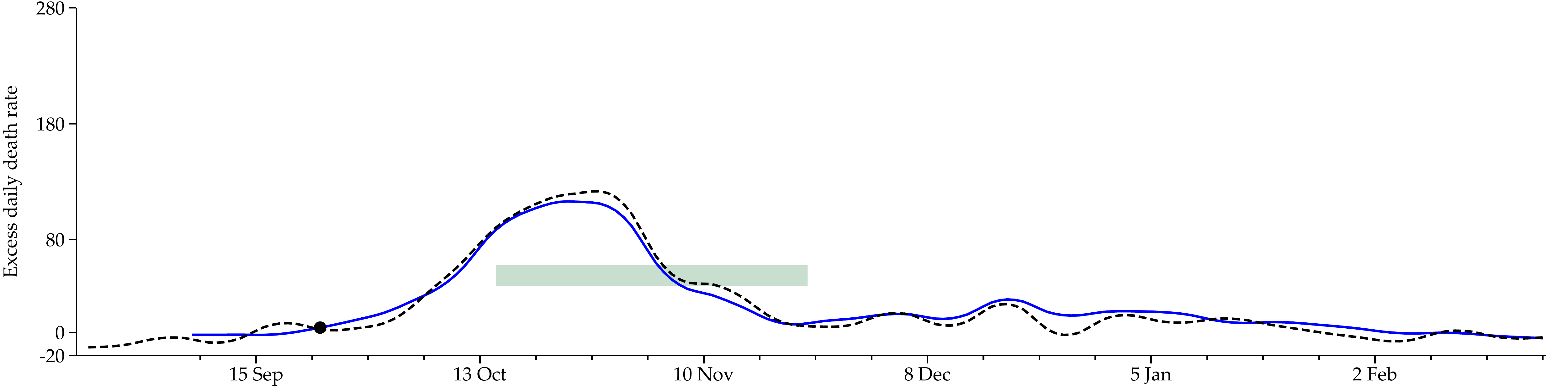}}\hfill
    \subfloat[New Orleans]{\includegraphics[width=0.9\textwidth]{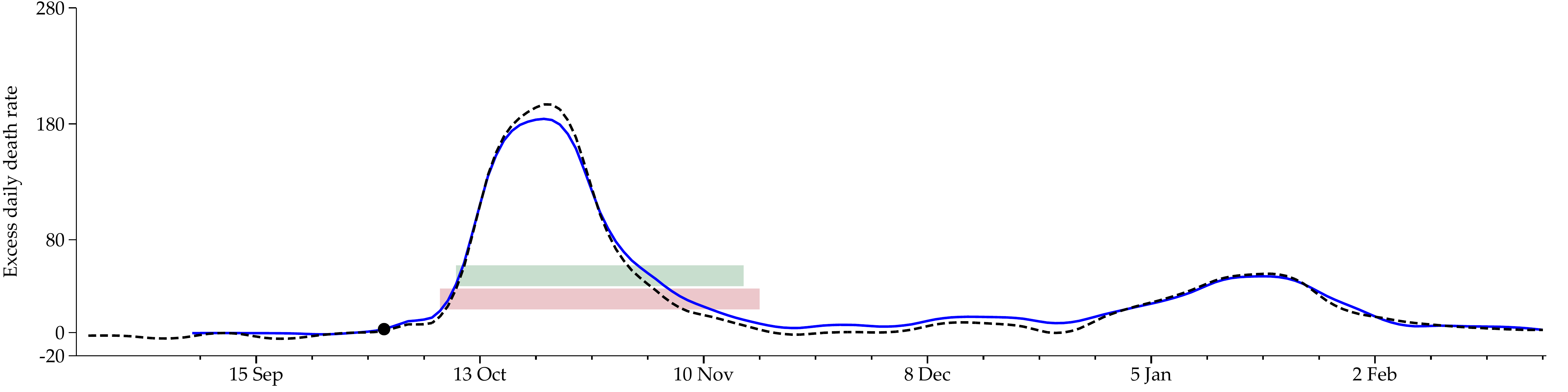}}\hfill
    \subfloat[New York City]{\includegraphics[width=0.9\textwidth]{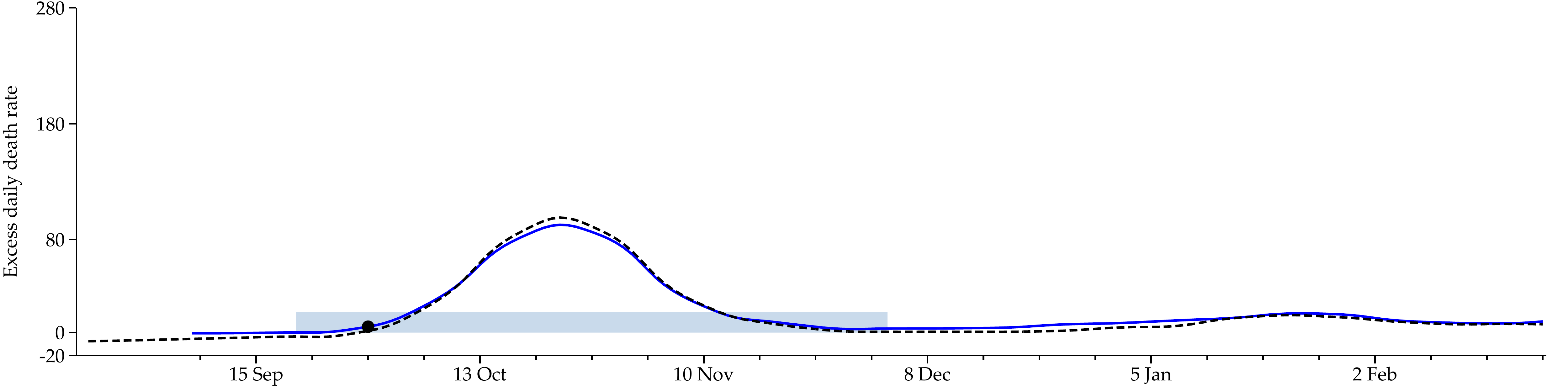}}\hfill
\CitiesCaption
\end{figure}

\begin{figure}[ht]\ContinuedFloat\centering
    \subfloat[Newark]{\includegraphics[width=0.9\textwidth]{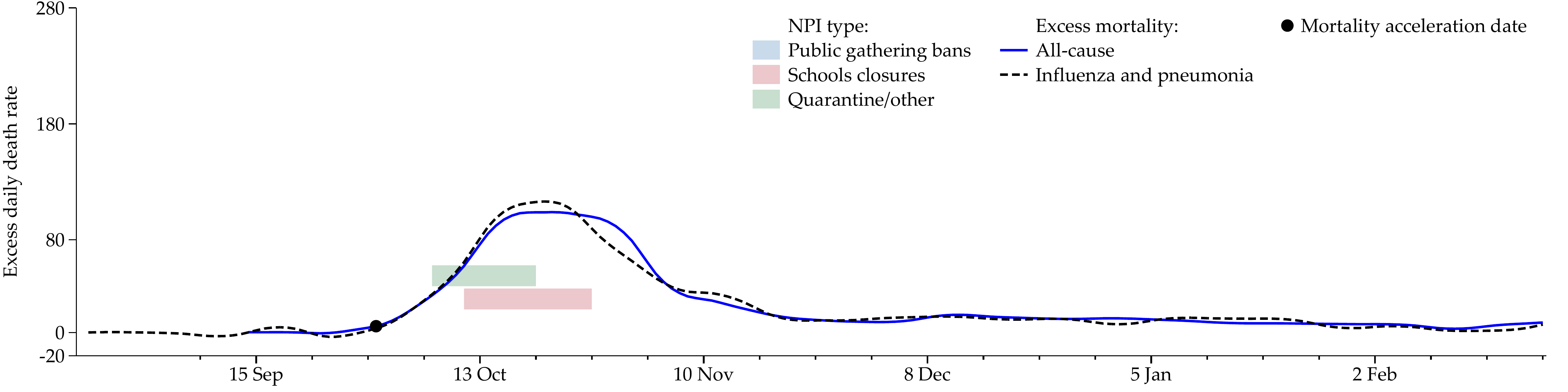}}\hfill
    \subfloat[Oakland]{\includegraphics[width=0.9\textwidth]{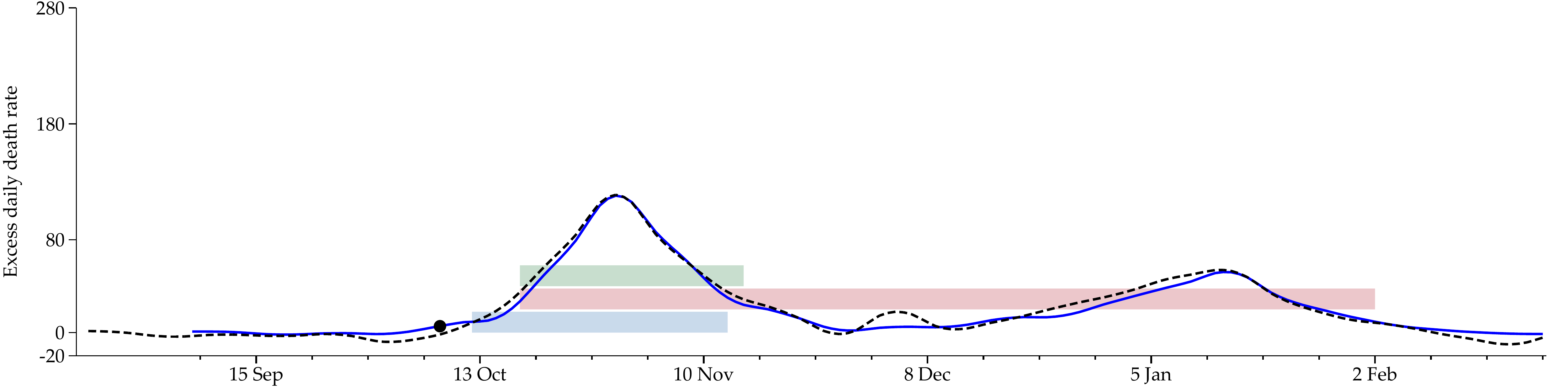}}\hfill
    \subfloat[Omaha]{\includegraphics[width=0.9\textwidth]{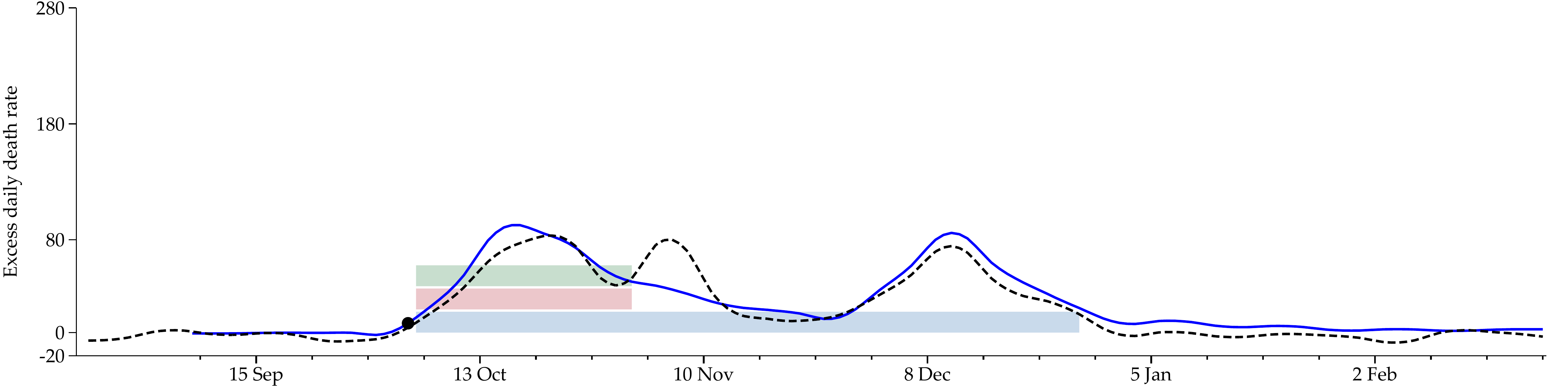}}\hfill
    \subfloat[Philadelphia]{\includegraphics[width=0.9\textwidth]{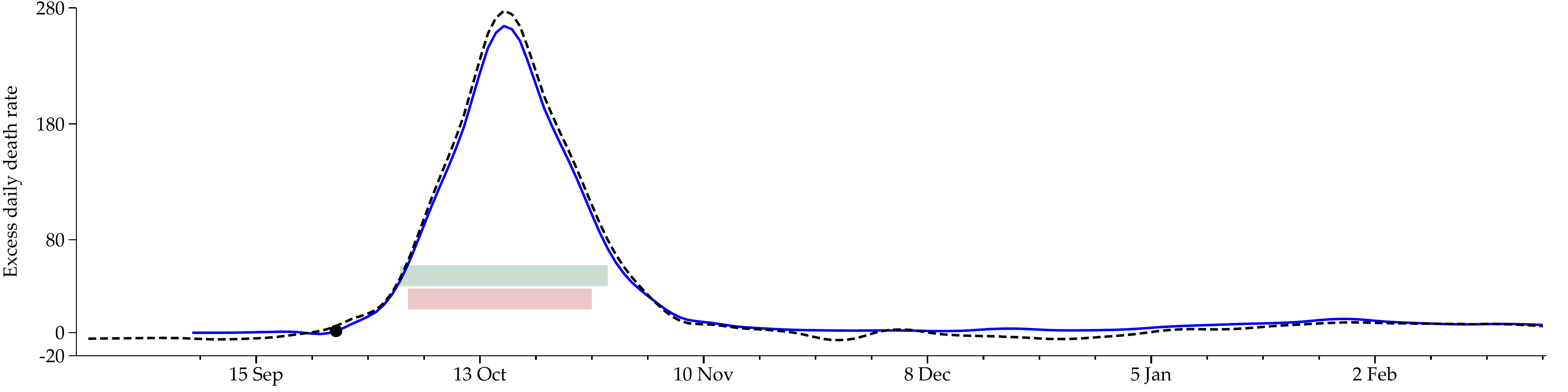}}\hfill
\CitiesCaption
\end{figure}

\begin{figure}[ht]\ContinuedFloat\centering
    \subfloat[Pittsburgh]{\includegraphics[width=0.9\textwidth]{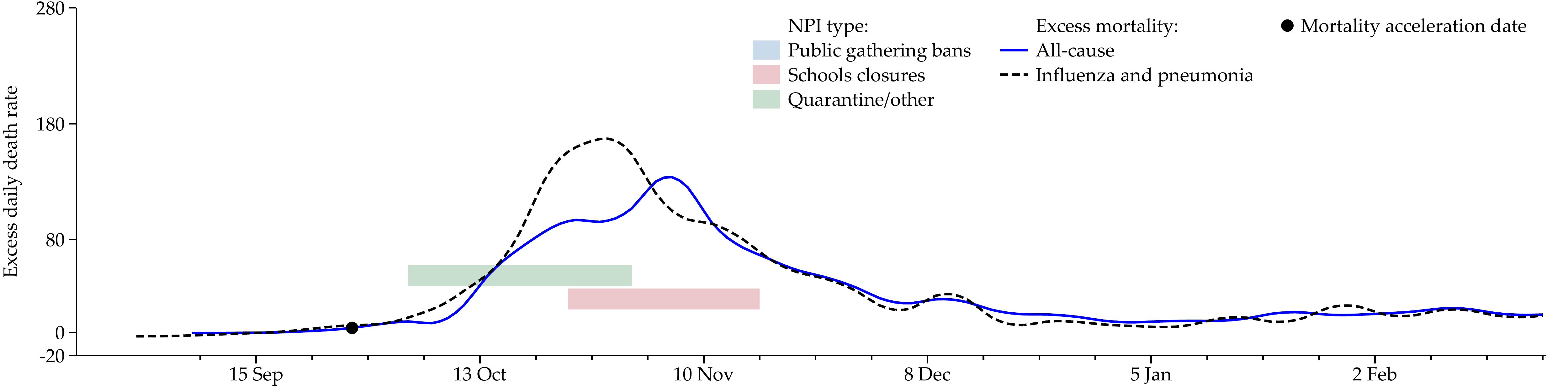}}\hfill
    \subfloat[Portland]{\includegraphics[width=0.9\textwidth]{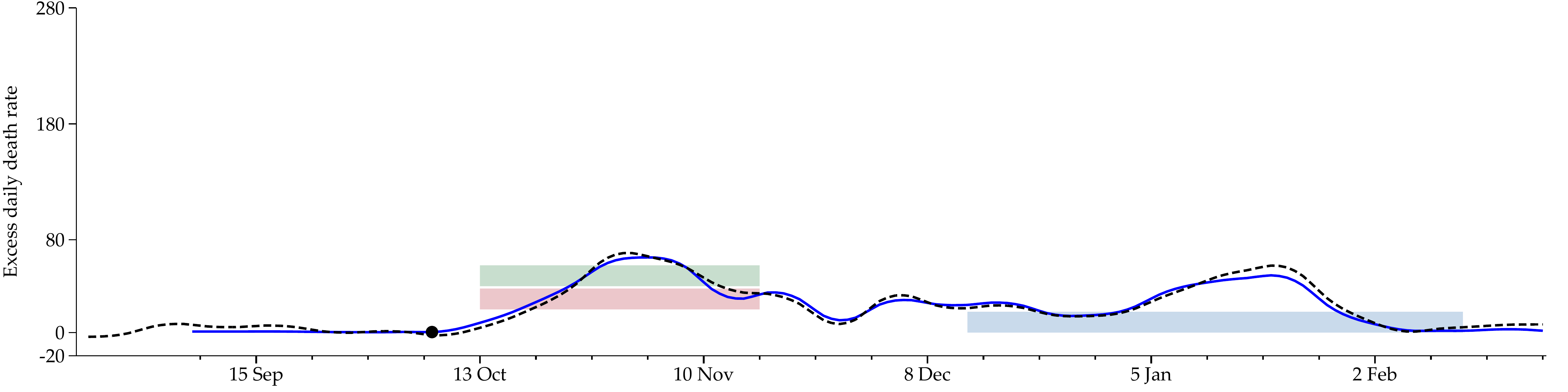}}\hfill
    \subfloat[Providence]{\includegraphics[width=0.9\textwidth]{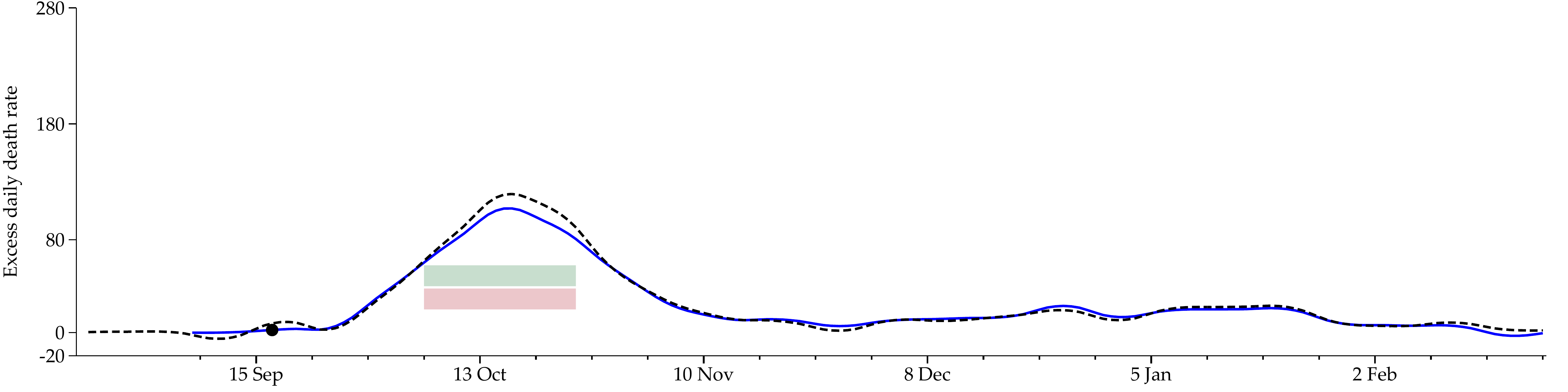}}\hfill
    \subfloat[Richmond]{\includegraphics[width=0.9\textwidth]{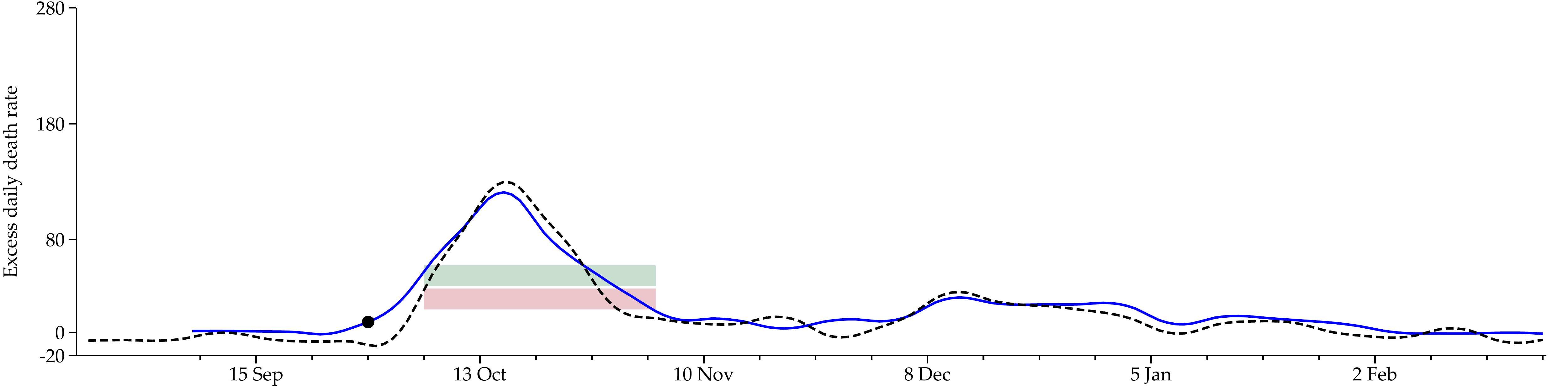}}\hfill
\CitiesCaption
\end{figure}

\begin{figure}[ht]\ContinuedFloat\centering
    \subfloat[Rochester]{\includegraphics[width=0.9\textwidth]{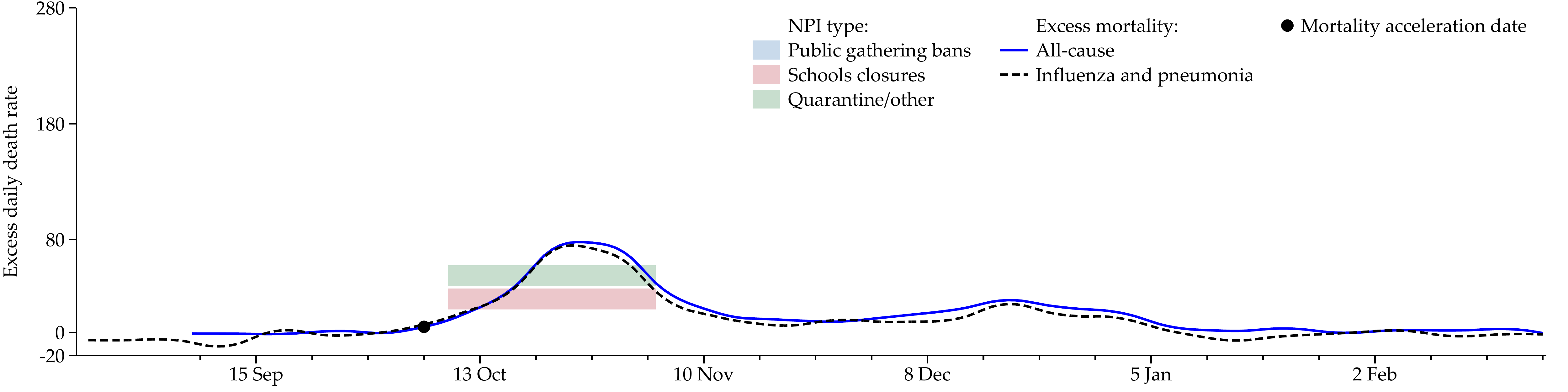}}\hfill
    \subfloat[Saint Louis]{\includegraphics[width=0.9\textwidth]{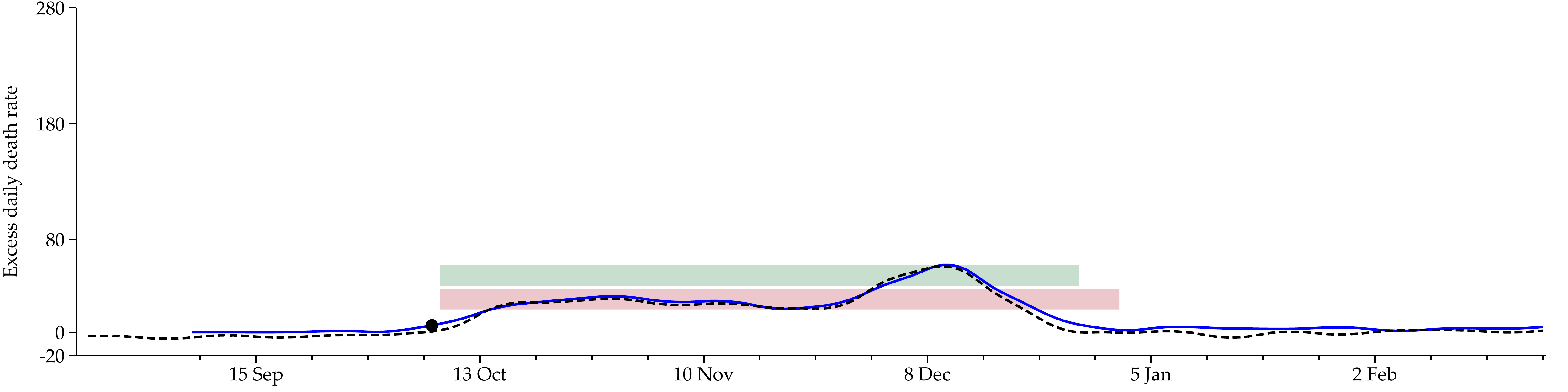}}\hfill
    \subfloat[Saint Paul]{\includegraphics[width=0.9\textwidth]{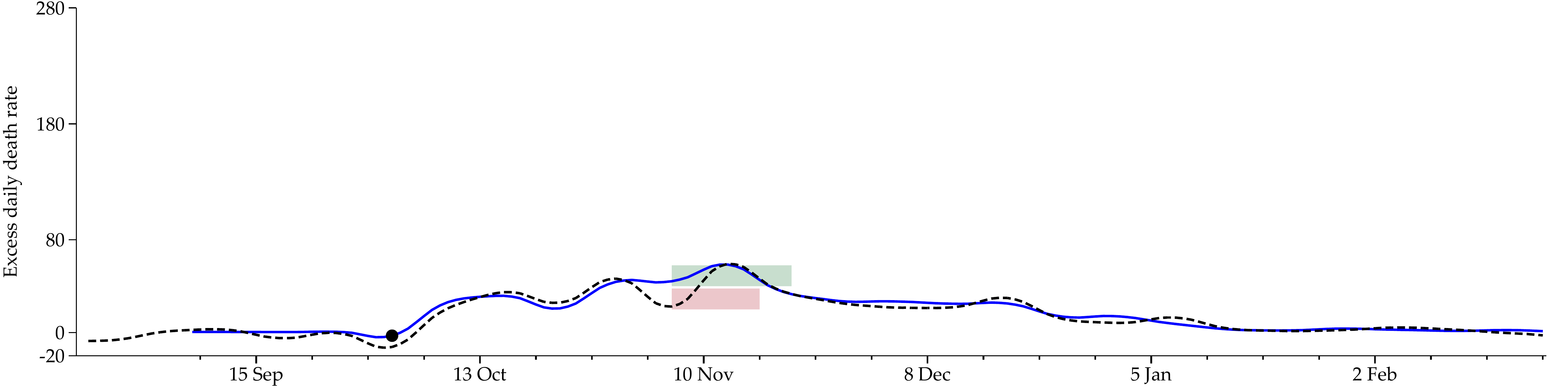}}\hfill
    \subfloat[San Francisco]{\includegraphics[width=0.9\textwidth]{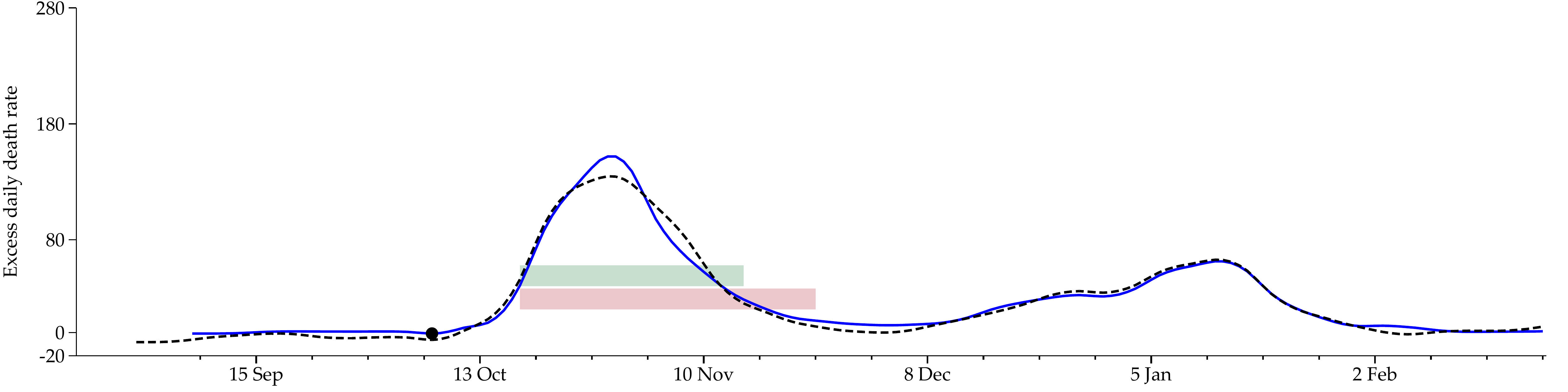}}\hfill
\CitiesCaption
\end{figure}

\begin{figure}[ht]\ContinuedFloat\centering
    \subfloat[Seattle]{\includegraphics[width=0.9\textwidth]{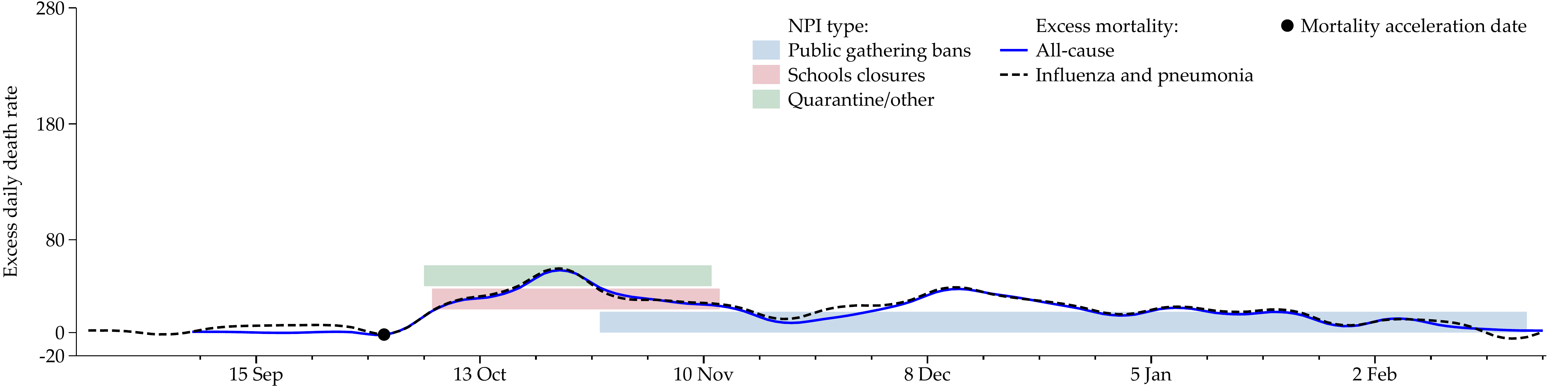}}\hfill
    \subfloat[Spokane]{\includegraphics[width=0.9\textwidth]{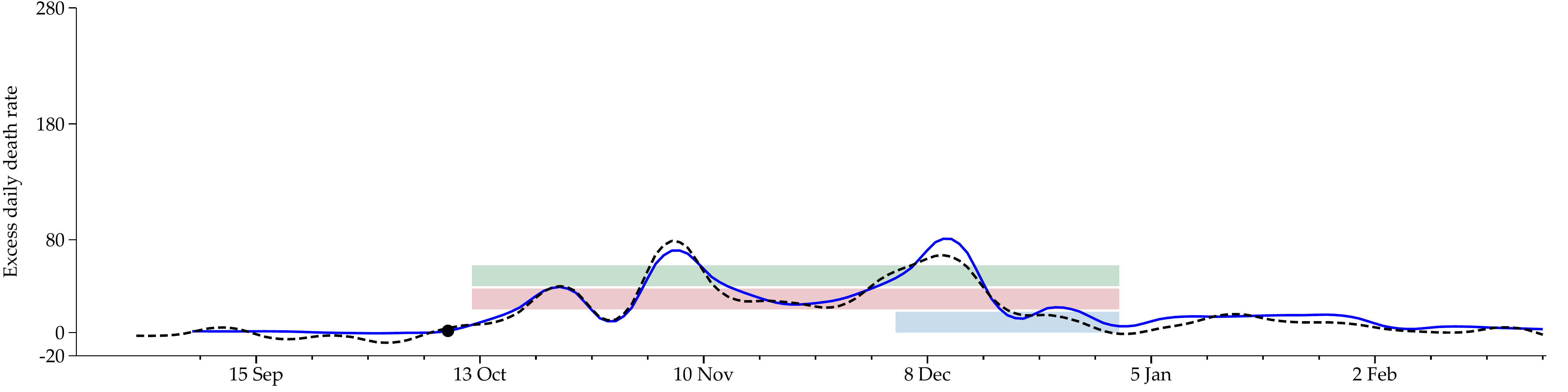}}\hfill
    \subfloat[Syracuse]{\includegraphics[width=0.9\textwidth]{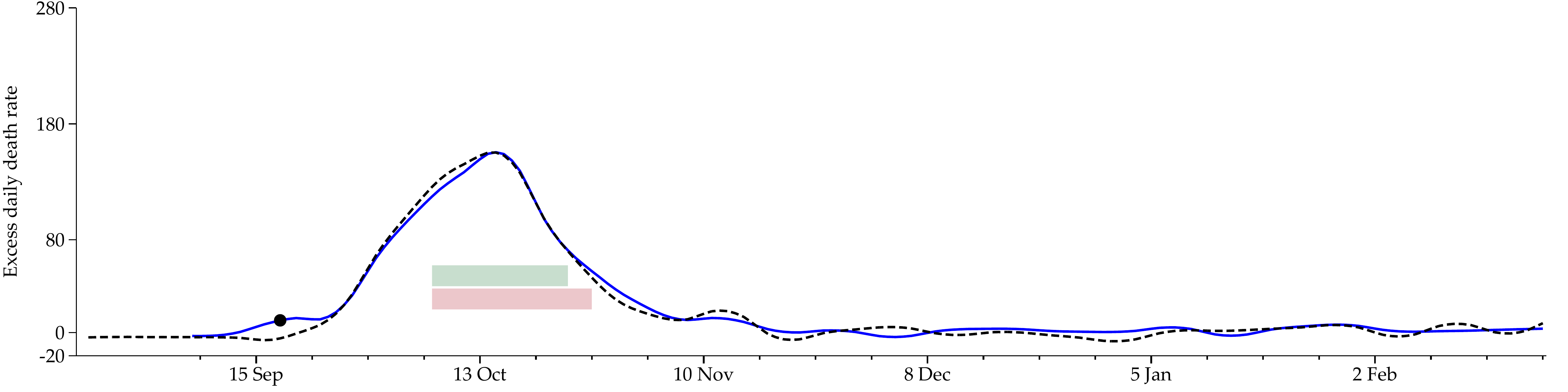}}\hfill
    \subfloat[Toledo]{\includegraphics[width=0.9\textwidth]{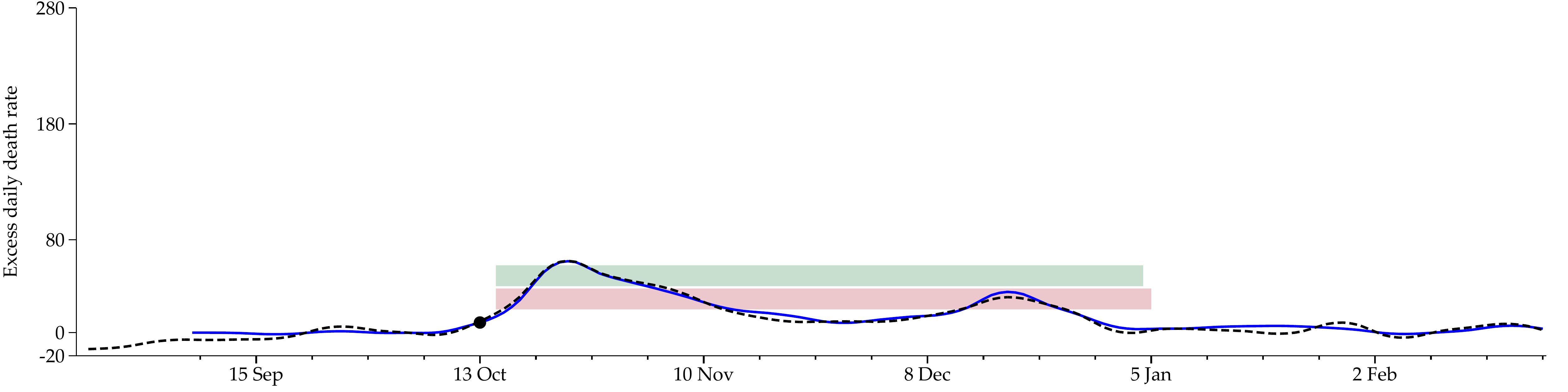}}\hfill
\CitiesCaption
\end{figure}

\begingroup
\makeatletter
\setlength{\@fptop}{0pt}
\setlength{\@fpbot}{0pt plus 1fil}
\makeatother

\begin{figure}[p]\ContinuedFloat\centering
    \subfloat[Washington]{\includegraphics[width=0.9\textwidth]{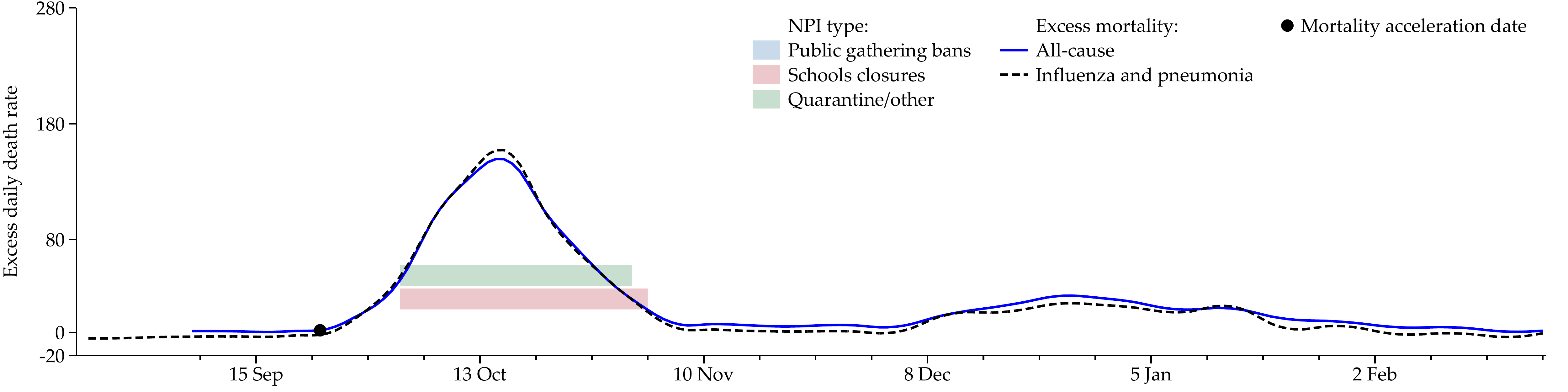}}\hfill
    \subfloat[Worcester]{\includegraphics[width=0.9\textwidth]{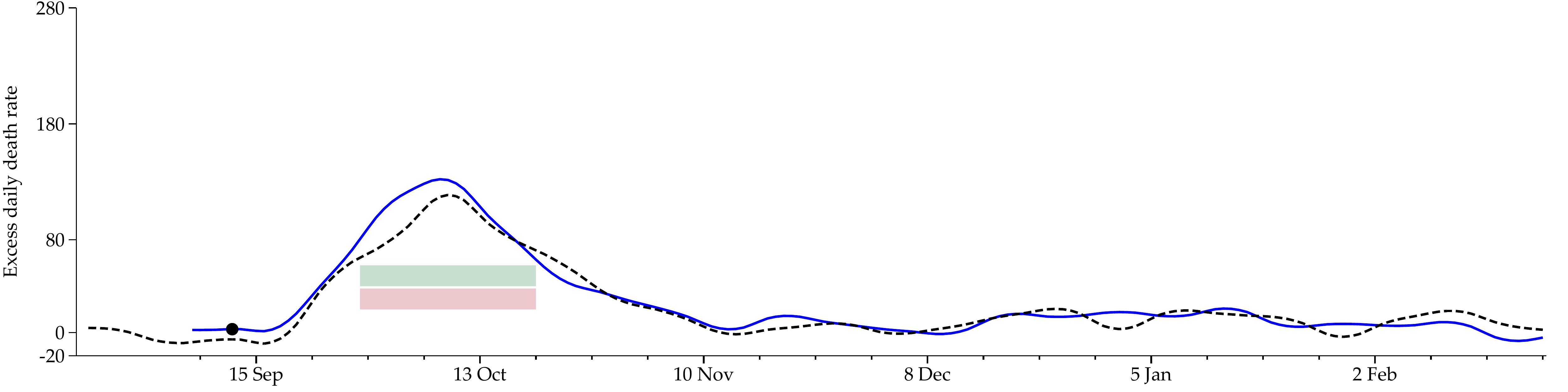}}\hfill
\CitiesCaption
\end{figure}
\clearpage
\endgroup

\end{appendices}

\end{document}